\input amstex
\magnification 1200
\TagsOnRight
\def\qed{\ifhmode\unskip\nobreak\fi\ifmmode\ifinner\else
 \hskip5pt\fi\fi\hbox{\hskip5pt\vrule width4pt
 height6pt depth1.5pt\hskip1pt}}
\NoBlackBoxes \baselineskip 19 pt
\parskip 6 pt
\def\stretch {\noalign{\medskip}}
\define \bC {\bold C}
\define \bCp {\bold C^+}
\define \bCm {\bold C^-}
\define \bCpb {\overline{\bold C^+}}
\define \bCmb {\overline{\bold C^-}}
\define \ds {\displaystyle}
\define \bR {\bold R}
\define \bm {\bmatrix}
\define \endbm {\endbmatrix}


\centerline {\bf THE INVERSE SCATTERING PROBLEM}
\centerline {\bf FOR THE MATRIX SCHR\"ODINGER EQUATION}

\vskip 10 pt
\centerline {Tuncay Aktosun}
\vskip -8 pt
\centerline {Department of Mathematics}
\vskip -8 pt
\centerline {University of Texas at Arlington}
\vskip -8 pt
\centerline {Arlington, TX 76019-0408, USA}
\vskip -8 pt
\centerline {aktosun\@uta.edu}

\centerline {Ricardo Weder\plainfootnote{$^\dagger$}
{Fellow Sistema Nacional de Investigadores}}
\vskip -8 pt
\centerline {Departamento de F\'\i sica Matem\'atica}
\vskip -8 pt
\centerline {Instituto de Investigaciones en
Matem\'aticas Aplicadas y en Sistemas}
\vskip -8 pt
\centerline {Universidad Nacional Aut\'onoma de M\'exico}
\vskip -8 pt
\centerline {Apartado Postal 20-126, IIMAS-UNAM, M\'exico DF 01000, M\'exico}
\vskip -8 pt \centerline {weder\@unam.mx}

\noindent {\bf Abstract}: The matrix Schr\"odinger equation is considered
on the half line with the general selfadjoint boundary condition at
the origin described by two boundary matrices satisfying
certain appropriate conditions. It is assumed that the matrix
potential is integrable, is selfadjoint, and has a finite
first moment. The corresponding scattering
data set is constructed, and such scattering data sets are characterized
by providing a set of necessary and sufficient conditions
assuring the existence and uniqueness of the correspondence
between the scattering data set
and the input data set containing the potential and boundary matrices.
The work presented here provides a generalization of the classical
result by Agranovich and Marchenko from the Dirichlet boundary condition
to the general selfadjoint
boundary condition.
The theory presented is illustrated with various explicit
examples.


\vskip -8 pt
\par\noindent {\bf Short title:} Half-line matrix Schr\"odinger
equation

\newpage

\centerline  {\bf TABLE OF CONTENTS}

\noindent 1. INTRODUCTION \dotfill \quad 4

\noindent 2. THE MATRIX SCHR\"ODINGER EQUATION \dotfill 17

\noindent 3. SOME MATHEMATICAL PRELIMINARIES \dotfill 22

3a. Vectors in $\bC^n$ and matrices acting on $\bC^n$ \dotfill 22

3b. Banach and Hilbert spaces \dotfill 24

3.c Useful inequalities \dotfill 27

3.d Hardy spaces \dotfill 28

3.e Other useful Banach spaces \dotfill 31

3.f Integral operators with kernels depending on a sum \dotfill 33

3.g Further results in Banach and Hilbert spaces \dotfill 37

3.h Other miscellaneous results \dotfill 40

\noindent 4. THE FADDEEV CLASS AND THE MARCHENKO CLASS \dotfill 42

\noindent 5. THE CHARACTERIZATION OF THE SCATTERING DATA \dotfill 51

\noindent 6. EQUIVALENTS FOR SOME CHARACTERIZATION CONDITIONS \dotfill 55

\noindent 7. ALTERNATE CHARACTERIZATIONS OF THE SCATTERING DATA \dotfill 59

\noindent 8. ANOTHER CHARACTERIZATION OF THE SCATTERING DATA \dotfill 66

\noindent 9. THE SOLUTION TO THE DIRECT PROBLEM \dotfill 68

\noindent 10. SOME RELEVANT RESULTS RELATED TO THE DIRECT PROBLEM \dotfill 72

\noindent 11. BOUND STATES \dotfill 80

\noindent 12. FURTHER PROPERTIES OF THE SCATTERING DATA \dotfill 91

\noindent 13. THE MARCHENKO INTEGRAL EQUATION \dotfill 96

\noindent 14. THE BOUNDARY MATRICES \dotfill 100

\noindent 15. THE EXISTENCE AND UNIQUENESS IN THE DIRECT PROBLEM ..........~\quad 103

\noindent 16. THE SOLUTION TO THE INVERSE PROBLEM \dotfill 129

\noindent 17. ADDITIONAL RESULTS RELATED TO THE INVERSE PROBLEM \dotfill 155

\noindent 18. FURTHER RESULTS RELATED TO THE INVERSE PROBLEM \dotfill 168

\noindent 19. INVERSE PROBLEM BY USING ONLY THE SCATTERING MATRIX \dotfill 188

\noindent 20. PARSEVAL'S EQUALITY \dotfill 192

\noindent 21. ALTERNATE CHARACTERIZATION VIA LEVINSON'S THEOREM \dotfill 198

\noindent 22. THE GENERALIZED FOURIER MAP \dotfill 214

\noindent 23. AN ALTERNATE METHOD TO SOLVE THE INVERSE PROBLEM \dotfill 226

\noindent 24. A STAR GRAPH \dotfill 245

\noindent 25. THE SCHR\"ODINGER EQUATION ON THE FULL LINE \dotfill 247

\noindent 26. SOME EXPLICIT EXAMPLES \dotfill  \quad 249

\noindent Acknowledgments \dotfill  \quad 280

\noindent References \dotfill  \quad 281


\newpage

\noindent {\bf 1. INTRODUCTION}
\vskip 3 pt

In the direct scattering problem for the half-line Schr\"odinger equation,
given the potential and the boundary condition we determine the
spectrum of the corresponding Schr\"odinger operator, namely we determine the
scattering matrix and the bound-state information consisting
of the bound-state energies and the bound-state normalization
constants. In the associated inverse problem, given the scattering
matrix and the bound-state information, we determine the corresponding
potential and the boundary condition. We refer to
the data set consisting of the potential and the boundary condition
as the input data set and denote it by
$\bold D.$ We refer to the data set consisting of the scattering
matrix and the bound-state information as the scattering data set
and denote it by $\bold S.$ Thus, we can view the direct
scattering problem as the mapping $\Cal D: \bold D\mapsto \bold S$
and view the inverse scattering problem as the mapping $\Cal D^{-1}: \bold S\mapsto \bold D.$

There are four aspects related to the direct and inverse problems. These are the existence,
uniqueness, construction, and characterization. In the existence
aspect in the direct problem, given $\bold D$ in a specified class
we determine whether a corresponding $\bold S$ exists in some
specific class. Thus, the domain of $\Cal D$ as well as
its range must be determined. The uniqueness
aspect is concerned with whether there exists a unique $\bold S$ corresponding
to a given $\bold D$ in the domain of $\Cal D,$ or two or more distinct sets $\bold S$
may correspond to the same $\bold D.$
The construction deals with
the recovery of $\bold S$ from $\bold D.$
For the inverse problem to be well defined,
one needs to show that the range of the direct scattering map $\Cal D$
coincides with the domain of the inverse scattering map $\Cal D^{-1}.$
In the inverse
problem the existence problem deals with the existence of some
$\bold D$ corresponding to $\bold S$ specified in a particular class,
which must coincide with the range of the direct scattering map.
The uniqueness deals with the question whether $\bold D$ corresponding to
a given $\bold S$ in the range of $\Cal D$ is
unique, and the construction consists of the recovery of
$\bold D$ from $\bold S.$
After the existence and uniqueness aspects in the direct and
inverse problems are settled, one now turns the attention to the
characterization problem, which consists of
the identification of the class to which $\bold D$ belongs
and the identification of the class to which $\bold S$ belongs so that
there is a one-to-one correspondence between $\bold D$ and $\bold S.$

The characterization aspect in the direct and inverse problems
is usually the most difficult to establish. This is not surprising because
the establishment of the characterization includes the establishment
of the existence and the uniqueness in both
the direct and inverse problems. Our goal in this monograph
is to establish a characterization for the matrix
Schr\"odinger operator on the half line with the general selfadjoint
boundary condition. We do so in such a way that our characterization
result
naturally also holds in the scalar case, it holds for any selfadjoint
boundary condition, it yields the construction in the
corresponding direct and inverse problems,
and it reveals how the individual conditions in the
characterization affect the existence, uniqueness, and construction.
We also provide some comments and explicit examples to clarify
various issues so that our approach can be useful in establishing
characterizations for other direct and inverse problems.

The only viable characterization in the literature for the matrix Schr\"odinger operator
on the half line can be found in the seminal work by Agranovich and
Marchenko [2]. However, the
analysis in [2] is restricted to the Dirichlet boundary condition,
and hence our study can be viewed as a generalization of
the characterization in [2]. It is ironic that
a characterization in the scalar case valid for a general selfadjoint
boundary condition does not exist and cannot exist in the way
a scattering matrix is defined in the existing literature. As indicated in Section~4
of [7], as a result of defining [8,14,37,38] the scattering matrix in one way
with the Dirichlet boundary condition and in a different way
with a non-Dirichlet boundary condition, it is impossible to have
the uniqueness aspect unless separate characterizations are developed
in the Dirichlet case and in the non-Dirichlet case, respectively.
In the scalar case, it is known [7] that,
in the absence of bound states, an
input data set consisting of a real-valued potential with the Dirichlet
boundary condition and another input data set consisting of a real-valued potential with the Neumann
boundary condition may correspond to the same scattering matrix.
The reader is referred to Section~4 of and Example~6.3 of [7]
for further details. In our solution to the characterization
problem given in this monograph
we do not encounter such a nonuniqueness issue because
we define the scattering matrix in a unique way,
without defining it in one way in the Dirichlet case and
in another way in the non-Dirichlet case. Actually,
we define the scattering matrix in such a way that
the associated Schr\"odinger operator
for the unperturbed
problem
has the Neumann boundary condition.
This definition is motivated by the theory of quantum graphs [29,30,42].
In fact, in the matrix
case, a boundary condition could be partly Dirichlet and partly
non-Dirichlet, and the Dirichlet boundary condition
in the matrix case is really a very special case and can be
referred to as the purely Dirichlet case [9]. For further
details we refer the reader to Section~4 of [9].

We analyze the existence, uniqueness, reconstruction, and
characterization
issues related to the relevant direct and inverse problems under the assumption
that $\bold D$ belongs to the Faddeev class
and $\bold S$ belongs to the Marchenko class.
The Faddeev class consists of input data sets
$\bold D$ as in (4.1), where the potential $V$ and the
boundary matrices $A$ and $B$ are as specified in Definition~4.1.
The Marchenko class consists of scattering data sets $\bold S$
as in (4.2), where
the scattering matrix $S$ and the bound-state data
$\{\kappa_j,M_j\}_{j=1}^N$ are as specified in Definition~4.5.
In [2] the inverse problem for
(2.1) is studied in the special case
when $A=0$ and $B=I,$ with $0$ being the $n\times n$ zero
matrix and $I$ denoting the $n\times n$ identity matrix,
and when
the potential $V$ is not necessarily integrable, i.e.
when the potential $V$ appearing in (2.1)
satisfies (2.2), is Lebesgue measurable, and satisfies
$\int_0^\infty dx\, x\, |V(x)|<+\infty$
instead of satisfying (2.3).
A characterization of the corresponding scattering data was presented
in [2]. Our work provides a generalization of
the characterization of [2] to the case with the general selfadjoint boundary condition. When a non-Dirichlet boundary condition is
used at $x=0,$ the integrability of the potential
is necessary, and that is why the integrability of
$V(x)$ in (2.3) is crucial. In particular, to be able to
define the regular solution $\varphi(k,x)$ appearing in (9.5),
it is necessary that the potential is integrable
at $x=0.$ For further details on this issue we refer the reader
to Theorem~1.2.1 of [2] and also [45]. In Chapter~26 we illustrate
this issue with an explicit example.

Let us mention also the relevant references [23-25], where
the direct and inverse problems for (2.1)
are formally studied with the general selfadjoint boundary
condition, not as in (2.4)-(2.6) but in a form equivalent to (2.4)-(2.6). However,
the study in [23-25] lacks the large-$k$ analysis beyond the
 leading term and also lacks the small-$k$ analysis of the scattering data,
 which are both essential for the
analysis of the relevant inverse problem. Thus, our study can also
be considered as a complement to the work by Harmer [23-25].
In our monograph we use results from
previous work [2,5,6,9,42-44], in particular [2,5,9,42].

The matrix Schr\"odinger equation on the half line with the general selfadjoint boundary
condition has many applications in quantum mechanics,
especially in the scattering of particles
with internal structures such as spin [14,40]
and in the scattering on quantum graphs [11-13,18,19,21,22,26,28-36].
A particular phenomenon governed by a matrix Schr\"odinger equation
is a star graph, i.e. a one-vertex graph with a finite
number of semiinfinite edges.
In this case, a homogeneous boundary condition linear
in the wavefunction and its derivative is imposed
at the vertex, and the dynamics on each edge
is governed by the Schr\"odinger equation.
Physically, a star graph represents a finite number
of very thin quantum wires connected at the vertex.
The study of quantum wires has physical relevance to the design of elementary gates
in quantum computing and in nanotubes
for microscopic electronic devices, where string of atoms may form a star graph.
The consideration of the general boundary condition at the vertex, rather than just the
Dirichlet boundary condition, is relevant.
For quantum graphs it is crucial that the boundary conditions
at the vertices link the values of the wavefunction and its derivative
arriving from different edges.
An important case is the Kirchhoff boundary condition,
which amounts to the continuity of the wavefunction at the vertex
and also that at the vertex the sum of the derivatives
of the wavefunction from all the edges is zero, which expresses the conservation of current
at the vertex.
Actually, a quantum graph is an idealization
of quantum wires with very small cross sectional areas,
where such wires meet at the vertices.
The quantum graph is obtained in the zero limit for the cross
sectional area.
The boundary conditions at the vertices of
the graph depends on how the limit is taken.
 From this point of view, it is relevant to
 consider all possible selfadjoint boundary
 conditions since such conditions may result in various
 limiting procedures.


Our monograph is organized as follows. In Chapter~2 we introduce
the matrix Schr\"odinger equation on the half line, introduce the
related $n\times n$ matrix potential $V(x),$ and list the properties
of the potential so that we can study the
corresponding direct and inverse problems. The matrix potential
is required to be hermitian as stated in (2.2) and also
required to satisfy the
so-called $L^1_1$-condition given in (2.3).
The existence of the first moment for the potential in (2.3)
ensures that the number of bound states can at most be finite,
which makes the analysis of the corresponding direct and
inverse problems manageable. In this chapter we also introduce
the general selfadjoint boundary condition at $x=0$ in terms
of a pair of constant $n\times n$ matrices denoted by $A$ and
$B.$ The boundary condition is given in (2.4), where the
 boundary matrices satisfy (2.5) and (2.6).
The boundary condition is unchanged if the boundary matrices are post
multiplied by
an invertible $n\times n$ matrix $T.$ In Proposition~2.1 we show
that the boundary matrices $A$ and $B$ are uniquely determined
 modulo post multiplication by $T.$

Chapter~3 contains a summary of
various mathematical definitions, notations, and results used in our monograph.
In our monograph we deal with vector-valued functions as well as matrix-valued
functions. Our vectors may be row vectors or column vectors with $n$ components.
In Section~3.a we review the standard vector norm, the standard scalar product, and
the standard matrix norm in $\bC^n.$ In our monograph we use the standard
matrix norm defined in (3.11) instead of another matrix norm used
in [2]. Since all matrix norms are equivalent in a finite dimensional
vector space, we could use any other matrix norm than that defined
in (3.11). Our choice of the standard matrix norm given in (3.11)
is motivated by the fact that we would like to have the inequalities
(3.12)-(3.14) as well as the equalities given in (3.15). With any other matrix norm
some constants depending on $n$ may need to be used in (3.12)-(3.14)
to retain the inequalities, and similarly some constants depending on
$n$ may be needed to retain the equalities in (3.15).
In Section~3.b we review the basic facts about Banach spaces and Hilbert spaces,
as our vector-valued and matrix-valued functions belong to
various Banach spaces or various Hilbert spaces.
In our monograph we especially use the Banach space
$L^1(\bR^+),$ consisting of complex-valued
integrable functions of a real-valued independent
variable $x\in \bR^+.$ Similarly, we use
$L^2(\bR^+),$ the Hilbert space of square-integrable
functions, and we also use
$L^\infty(\bR^+),$ the Banach space of bounded functions.
Without explicitly mentioning we assume that
all our functions are Lebesgue measurable and
the integrals are assumed to be Lebesgue integrals.
In Section~3.c we list various inequalities
in various Banach spaces and Hilbert spaces, including
some inequalities involving convolutions.
In Section~3.4 we summarize certain basic facts
on Hardy spaces that we use later on.
The pointwise bounds stated in Propositions~3.1 and
3.2 turn out to be very useful later on, especially in solving
some Riemann-Hilbert problems related to the characterization
of scattering data sets.
In Section~3.e we introduce various
Banach spaces that become useful in solving
various Riemann-Hilbert problems related to
the characterization
of scattering data sets.
In Section~3.f we summarize basic facts on
integral operators whose kernels depend
on a sum. Our key integral
equation, namely, the Marchenko
integral equation has such a kernel as well
as various other integral equations used
in our characterization of scattering
data sets. In Section~3.g we provide a summary of
certain
results in Banach and Hilbert spaces. Such results
are used in the characterization presented in
Sections~8 and 23.

In Chapter~4 we introduce the Faddeev class
of input data sets $\bold D$ and the Marchenko class
of scattering data sets, in
Definition~4.1 and Definition~4.5, respectively.
These two classes conveniently allow us to summarize
one of our main characterization results,
Theorem~5.1 in Chapter~5,
by saying that there is a one-to-one correspondence
between the Faddeev class and the Marchenko class.
In order to formulate various characterization results
in an efficient manner,
we list a set of properties for
a scattering data set $\bold S$ in Definition~4.2,
and these properties are identified
by using Arabic numerals, namely
$(\bold 1),$ $(\bold 2),$ $(\bold 3),$ and $(\bold 4).$
Actually, there are two versions of $(\bold 3),$ denoted
by $(\bold 3_a)$ and $(\bold 3_b).$ There are
five versions of $(\bold 4),$ denoted
by $(\bold 4_a),$ $(\bold 4_b),$ $(\bold 4_c),$ $(\bold 4_d),$ $(\bold 4_e).$
We also provide a second set of properties for
a scattering data set $\bold S$ in Definition~4.3,
and those properties are identified
by using Roman numerals, namely
$(\bold I),$ $(\bold {III}),$ $(\bold V),$ and $(\bold{VI}).$
Actually, there are three versions of $(\bold {III}),$ denoted
by $(\bold {III}_a),$ $(\bold {III}_b),$ and $(\bold {III}_c).$ There are
eight versions of $(\bold V),$ denoted
by $(\bold V_a),$ $\bold V_b),$ $(\bold V_c),$ $\bold V_d),$ $\bold V_e),$ $(\bold V_f),$ $\bold V_g),$ $(\bold V_h).$

In Chapter~5 we present one of our main characterization results. A stated
in Theorem~5.1 we show that for each input data set $\bold D$
in the Faddeev class there exists and uniquely exists
a corresponding scattering data set $\bold S$ in the Marchenko class.
We also remark that by replacing the $L^1_1$-condition
by the $L^1_p$-condition, where
$p$ is an integer greater than one,
the characterization result of Theorem~5.1 remains valid in a subclass.
In other words, let us modify the definition of
the Faddeev class so that the potential $V(x)$ satisfies
(2.3) with $(1+x)$ replaced by $(1+x)^p,$ and
let us also modify
the definition
of the Marchenko class so that so that $F_s'(y)$
satisfies (4.8) of
$(\bold 2)$ with $(1+y)$ replaced by $(1+y)^p.$
Then, the characterization result stated in Theorem~5.1
remains valid. Informally speaking, we then
obtain the characterization in the $L^1_p$-class with $p=2,3,4,\dots.$

In Chapter~6 we analyze the interconnections among
the properties listed in Definitions~4.2 and 4.3.
We show that the two versions of $(\bold 3),$ namely $(\bold 3_a)$
and $(\bold 3_b),$ are
equivalent. We also show that the five versions of
of $(\bold 4),$ namely $(\bold 4_a),$
$(\bold 4_b),$ $(\bold 4_c),$ $(\bold 4_d),$ and $(\bold 4_e),$
are all equivalent.
We show that the three versions of $(\bold{III}),$
namely $(\bold{III}_a),$ $(\bold{III}_b),$ $(\bold{III}_c),$
are also equivalent. We further show that all the eight versions of
$(\bold V),$ namely $(\bold V_a),$ $(\bold V_b),$ $(\bold V_c),$ $(\bold V_d),$ $(\bold V_e),$ $(\bold V_f),$ $(\bold V_g),$ $(\bold V_h),$
are equivalent. We then
show that $(\bold 3_a)$ is equivalent to
the combination of $(\bold{III}_a)$ and $(\bold V_a).$
This last equivalence also yields other equivalences between
the two versions of
$(\bold 3)$ and all possible 24 combinations of
$(\bold{III})$ and $(\bold V).$
Informally speaking, we then have the equivalence
between $(\bold 1,\bold 2,\bold 3,\bold 4)$ and
$(\bold 1,\bold 2,\bold {III}+\bold V,\bold 4).$

With the help of
various equivalences established in Chapter 6, in Chapter~7
we are able to present various equivalent versions of
the characterization result of Theorem~5.1.
The result in Theorem~5.1 is stated when the
scattering data set $\bold S$ belongs to the
Marchenko class, i.e. when
$\bold S$ satisfies $(\bold 1,\bold 2,\bold 3_a,\bold 4_a).$
In Chapter~7 we present various characterization
results, where the scattering data $\bold S$ satisfies
some six versions of
$(\bold 1,\bold 2,\bold {III}+\bold V,\bold 4).$
Although the six characterization
results presented in Theorems~7.1-7.6 are
equivalent, some of these characterization versions
may have certain advantages over others.
For example, the characterizations stated in
Theorems~7.1 and 7.2 allows us to verify the characterization conditions
without having to solve the Marchenko integral equation first.
Having so many equivalent versions of the characterization
allows us to have many different options for methods in the analysis
of the corresponding inverse scattering problem, and also it allows us
to understand how various different methods are connected to each other.
For example, in some cases it may be advantageous to solve an
integral equation rather than a corresponding Riemann-Hilbert problem or vice versa.
It may also be more convenient to look for a
solution to an integral equation
in the class of bounded and integrable functions
rather than in the class of bounded and square-integrable
functions. It may be more convenient to look
for a solution to a homogenous Riemann-Hilbert problem
in a more restricted class
rather than in a Hardy class.
In Chapter~7, in Theorems~7.9 and 7.10 we present two
equivalent versions of yet another
characterization of the scattering data, where this new
characterization is based on using Levinson's theorem.
The details of this characterization are provided
in Chapter~21, where it is also shown that
the two versions of this characterization
are equivalent to the previous characterization
of Theorem~5.1 and all its equivalents.

In Chapter~8 we provide another characterization for the
scattering data so that it uniquely corresponds to
an input data set in the Faddeev class. This characterization
is summarized in Theorem~8.1
and is different from the previous characterizations and its details
are developed in Chapter~23. In our monograph we do not show the
equivalence of this new characterization with the previous
characterizations because that equivalence is still an open
problem.
It is our feeling that showing such an equivalence will
reveal interconnections among various different areas of mathematical
analysis, by not only contributing to the analysis of
inverse scattering problems but perhaps by contributing to
the field of analysis at large.

In Chapter~9 we provide an outline of the
solution to the direct problem starting from a potential
and a pair of boundary matrices. Namely, starting with
an input data set $\bold D$ in the Faddeev class
we indicate how all the relevant quantities are constructed,
among which are various solutions to the
Schr\"odinger equation such as the Jost solution
$f(k,x),$ the
regular solution $\varphi(k,x),$ the physical solution
$\Psi(k,x),$ and the normalized
bound-state matrix solutions $\Psi_j(x).$ Other relevant quantities
constructed from the input data set include
the Jost matrix $J(k),$ the scattering matrix $S(k),$
the bound-state energies $-\kappa_j^2,$ the normalization
matrices $M_j.$ The construction is outlined in Chapter~9.

In Chapter~10, we summarize the properties of various
quantities constructed from the input data set
in the Faddeev class. In particular, we present
certain properties of the quantity
$K(x,y)$ constructed from the input data set $\bold D.$
Such properties later helps us to
establish that the constructed scattering data set $\bold S$
belongs to the Marchenko class. The quantity
$K(x,y)$ plays a key role also in the analysis of the
inverse scattering problem because
as we see later it corresponds to the
unique solution to the Marchenko integral equation (13.1).

In Chapter~11 various properties related to
the bound states are established. In other words,
starting with an input data set $\bold D,$
various quantities are constructed related to
the discrete spectrum of the
corresponding Schr\"odinger operator and
their properties are established in Chapter~11.

In Chapter~12 certain properties of the constructed
scattering matrix are established, especially related to
the quantity $F_s(y),$ which is related to
the constructed scattering matrix $S(k)$ through
the relationship (4.7). In other words, starting from
an input data set $\bold D$ in the Faddeev class
we construct the corresponding scattering data set
$\bold S$ and in Chapter~12 we analyze
the properties of $F_s(y),$ which plays a key role
in the solution to the inverse problem.

In Chapter~13 the Marchenko integral equation
is derived and it is shown that
$K(x,y)$ constructed from the input scattering data
set $\Cal D$ actually satisfies the Marchenko equation.
Further, it is shown that $K_x(x,y)$ satisfies
the derivative Marchenko integral
equation given in (13.7). Let us mention that
the derivative Marchenko equation (13.7) has a prominent role
as the Marchenko integral equation (13.1)
in the inverse scattering problem with the general selfadjoint
boundary condition (2.4). This is because
both $\psi(0)$ and $\psi'(0)$ appear in the
boundary condition. The presence of
$\psi'(0)$ in the boundary condition (2.4)
makes the derivative Marchenko equation (13.7) as important
as the Marchenko equation (13.1). In the Dirichlet case
studied in [2], the boundary condition
(2.4) reduces to having $\psi(0)=0,$ and hence
the absence of
$\psi'(0)$ in (2.4) in the Dirichlet case
also
diminishes the prominence of the derivative Marchenko integral equation
(13.7) compared to the Marchenko equation (13.1).

In Chapter~14 we show that the boundary matrices $A$ and $B$
in the input data set $\bold D$ appear in the large-$k$ asymptotics
of the constructed scattering matrix $S(k).$ The fundamental
result given in (14.2) shows how $A$ and $B$ are related to
the large-$k$ limit of $S(k)$ and the matrix quantity
$K(0,0),$ where $K(x,y)$ is the solution to the Marchenko equation
(13.1).

In Chapter~15 we establish various results
related to the properties listed in
Definition~4.2 and Definition~4.3.
One consequence of the
results in Chapter~15 is that if we construct
the scattering data set $\bold S$ from the input data set
$\bold D,$ then the constructed $\bold S$ satisfies
the properties $(\bold 1,\bold 2,\bold 3,\bold 4)$
listed in Definition~4.2 and
the properties $(\bold I,\bold {III},\bold V,\bold{VI})$
listed in Definition~4.3.

In Chapter~16 we analyze the inverse problem
of the construction of $\bold D$ from
a scattering data set $\bold S$
satisfying one or more of the properties
listed in Definitions~4.2 and 4.3.
In particular, we show that
the Marchenko integral equation (13.1)
is uniquely solvable and
its solution $K(x,y)$ has certain
important properties.
In this chapter we also show that
the derivative Marchenko integral
equation (13.7) is uniquely solvable
and its unique solution is
given by $K_x(x,y),$ the $x$-partial derivative
of the solution $K(x,y)$ to the Marchenko equation.
In Chapter~16 we also show how, starting from
a scattering data set $\bold S,$ we can construct
the boundary matrices $A$ and $B$
uniquely modulo a postmultiplication by
an invertible matrix $T.$
This is achieved by solving
the system given in (16.70) and
we show that (16.70) is solvable
when the scattering data set $\bold S$ satisfies
the properties $(\bold 1)$ and $(\bold 4)$
of Definition~4.2.
In Chapter~16 we provide various other results
to be used later on to obtain
the characterization based on the method
given in Chapter~23.

In Chapter~18 we obtain various results
to show how some of the properties listed
in Definitions~4.2 and 4.3 are related to each other.
In other words, the results presented
in Chapter~18 allow us to prove the
 equivalencies stated in Chapter~6.
 In particular, it shows
 how $(\bold 3)$ is equivalent to
 the two combined properties
 $(\bold{III})$ and $(\bold V),$
 and it also shows
 how all the eight versions of
 $(\bold V)$ are all equivalent.

 In Chapter~19 we analyze the
 inverse scattering problem when the
 bound-state data is missing in the
 scattering data set $\bold S.$ It shows that
 under certain circumstances we can
 supplement a scattering matrix with
 some appropriate bound-state
 data set so that the resulting scattering data set
 $\bold S$ corresponds to an input data set
$\bold D$ in the Faddeev class.

Parseval's equality is a
fundamental equation equivalent to the completeness
relation for the physical solution
$\Psi(k,x)$ and the normalized bound-state
matrix solutions $\Psi_j(x)$ of the Schr\"odinger
equation with the boundary condition (2.4).
In Chapter~20 we show that
Parseval's equality holds when the
corresponding scattering data set $\bold S$
satisfies certain minimal conditions.
We also show how Parseval's
equality is related to the Marchenko equation
by showing that the two are equivalent
under some mild conditions on the scattering
data set $\bold S.$

Levinson's theorem is a fundamental result
in the scattering theory and it shows how
the scattering matrix is intrinsically related
to the bound states. For the matrix
Schr\"odinger equation (2.1) with the
boundary condition (2.4), the corresponding
statement of Levinson's theorem
is summarized in (21.5). In Chapter~21 the details
are provided for
the two equivalent characterizations presented
in Theorems~7.9 and 7.10, utilizing Levinson's theorem.
In Chapter~21 also the equivalence is shown
between the characterization using Levinson's
theorem and the previous characterization
given in Theorem~5.1 and all
its equivalents based on the results
of Chapter~6.

In Chapter~22 we introduce the generalized Fourier
map $\bold F$ and establish its various relevant properties.
In particular, we show that
the generalized Fourier map is unitary, and we also
establish various properties of its adjoint map
$\bold F^\dagger.$  Such properties are needed to
establish another characterization result,
developed in Chapter~23, whose two equivalent versions
are summarized in Theorems~8.1 and
8.2. In Section~22 we also
show the orthonormality relations among
the physical solution $\Psi(k,x)$ and
the bound-state matrix solutions
 $\Psi_j(x),$
 where the result is given in Proposition~22.4.

In Chapter~23 we present the alternate method
to solve the inverse problem of constructing
the input data set $\bold D$ from a scattering data set
$\bold S.$ A summary of the method
is provided in the beginning of Chapter~22.
Based on this alternate method, a characterization
of the scattering data set $\bold S$ is obtained so that
there is a unique corresponding input data set $\bold D$
in the Faddeev class. As already indicated,
the resulting characterization is summarized
in Theorem~8.1.
As mentioned earlier, it is an open problem
to show that this new characterization
is equivalent to the characterization provided
in Theorem~5.1.

We briefly mention two
applications of our results in
the field of quantum graphs.
In Chapter~24 we show that a matrix Schr\"odinger equation with
a diagonal potential matrix is equivalent to having a star graph.
In Chapter~25 we show that a $2\times 2$ matrix Schr\"odinger
equation is unitarily equivalent to
a Schr\"odinger equation on the full line with a point interaction at $x=0.$

In Chapter~26 various explicitly solved examples are
provided for the inverse problem
when the scattering matrix $S(k)$ is a rational function of $k.$
Broadly speaking we have developed three versions of
the characterization of the scattering data.
The first version is based on Theorem~5.1 and its various
equivalent versions based on the results of
Section~6 and summarized in Theorems~7.1-7.6.
The second version utilizes Levinson's theorem and
the corresponding characterization is stated in
Theorem~7.9 and its equivalence in Theorem~7.10.
The third version is provided in Theorem~8.1.
In various examples provided in Chapter~26,
we illustrate how a scattering data set may comply with
these three versions of the characterization and
how the failure of any one of our characterization conditions
affects the solution to the inverse problem.

\newpage
\noindent {\bf 2. THE MATRIX SCHR\"ODINGER EQUATION}
\vskip 3 pt

In this chapter we introduce the
matrix Schr\"odinger equation (2.1), the potential $V,$ and the
boundary matrices $A$ and $B$ used to describe the selfadjoint boundary condition.

Consider the matrix Schr\"odinger equation on the half line
$$-\psi''+V(x)\,\psi=k^2\psi,\qquad x\in\bR^+,\tag 2.1$$
where $\bR^+:=(0,+\infty),$ the prime denotes the derivative with respect to
the spatial coordinate $x,$
$k^2$ is the complex-valued spectral parameter,
the potential $V$ is an $n\times n$ selfadjoint matrix-valued function of $x$ and
belongs to class $L_1^1(\bR^+),$ and $n$ is any positive integer.
We assume that $n$ is fixed and is known.
The selfadjointness of $V$ is expressed as
$$V(x)=V(x)^\dagger,\qquad x\in\bR^+,\tag 2.2$$
where the dagger denotes the matrix adjoint (complex
conjugate and matrix transpose).
We equivalently say hermitian to describe a selfadjoint matrix.
We remark that, unless we are in the scalar case, i.e. unless $n=1,$
the potential in not necessarily real valued.
The condition $V\in L_1^1(\bR^+)$
means that
each entry of the matrix $V$ is Lebesgue measurable on
$\bR^+$ and
$$\int_0^\infty
dx\,(1+x)\,|V(x)|<+\infty,\tag 2.3$$
where $|V(x)|$ denotes the operator matrix norm.
Clearly, a matrix-valued function belongs
to $L^1_1(\bR^+)$
if and only if each entry of that matrix belongs
to $L^1_1(\bR^+).$

The wavefunction $\psi(k,x)$ appearing in (2.1)
may be either an $n\times n$ matrix-valued function
or it may be a column vector with $n$
components. We use $\bC$ for the complex plane,
$\bR$ for the real line $(-\infty,+\infty),$
$\bR^-$ for the left-half line $(-\infty,0),$
$\bCp$ for the open upper-half complex plane,
$\bCpb$ for $\bCp\cup\bR,$ $\bCm$ for the open lower-half complex plane, and
$\bCmb$ for $\bCm\cup\bR.$

We are interested in studying (2.1) with an $n\times n$
 selfadjoint matrix potential $V$
in $L^1_1(\bR^+)$ under the general
selfadjoint boundary condition at $x=0.$
There are various equivalent formulations [5,9,23-25,29,30] of
the general
selfadjoint boundary condition at $x=0,$ and we find it
convenient
to state it [5,9] in terms of two
constant $n\times n$ matrices $A$ and $B$ as
$$-B^\dagger\psi(0)+A^\dagger\psi'(0)=0,\tag 2.4$$
where $A$ and $B$ satisfy
$$-B^\dagger A+A^\dagger B=0,\tag 2.5$$
$$A^\dagger A+B^\dagger B>0.\tag 2.6$$
The condition in (2.6) means that
the $n\times n$ matrix $(A^\dagger
A+B^\dagger B)$ is positive definite.
One can easily verify that (2.4) remains invariant if
the boundary matrices $A$ and $B$
are replaced with $AT$ and $BT,$ respectively,
where $T$ is an arbitrary $n\times n$ invertible matrix.
The details of this invariance is provided
in Proposition~2.1.
We express this fact by saying that
the selfadjoint boundary condition in (2.4) is uniquely determined
by the matrix pair $(A,B)$ modulo an invertible
matrix $T,$ and we equivalently state
 that (2.4) is equivalent to
the knowledge of $(A,B)$ modulo $T.$

The matrix $(A^\dagger
A+B^\dagger B)$ appearing in (2.6) is selfadjoint,
and thus from (2.6) it follows that
there exists a unique positive
definite matrix $E$ defined as
$$E:=(A^\dagger A+B^\dagger B)^{1/2},\tag 2.7$$
in such a way that $E$ is selfadjoint and invertible, and hence
$$E=E^\dagger,\qquad (E^\dagger)^{-1}(A^\dagger A
+B^\dagger B)E^{-1}=I.\tag 2.8$$
 From (1.16) of [5] we have
$$AE^{-2}A^\dagger+BE^{-2}B^\dagger=I,
\quad BE^{-2}A^\dagger-AE^{-2}B^\dagger=0.
\tag 2.9$$

The following proposition shows that the boundary matrices $A$ and $B$
appearing in (2.4)-(2.6) are uniquely defined modulo post multiplication by an invertible matrix $T.$

\noindent {\bf Proposition 2.1} {\it
 Let $A_j$ and $ B_j$ for $ j=1,2$  be $ n\times n$ matrices.  Assume that}
$$
 A_1^\dagger\,  A_1 + B_1^\dagger B_1 >0.
\tag 2.10$$
{\it Let}
$$
 L_j:=\left\{ (Z_1, Z_2) \in \bC^{2n}:\ - B_j^\dagger\, Z_1+ A_j^\dagger\, Z_2=0, \ j=1,2\right\}.
\tag 2.11$$
{\it Then, $L_1=L_2$ if and only if there is an invertible matrix $T$ such that}
$$
A_2 = A_1 T,\quad  B_2=  B_1 T.
\tag 2.12$$

\noindent PROOF: It is immediate that if (2.12) holds, then $L_1=L_2.$ On the hand, assume that  $L_1=L_2$. Let $\bold D_j$
 for $j=1,2$ be the  operators from $ \bC^{2n}$ into $\bC^n$ defined as
$$
 \bold D_j(Z_1,Z_2):=   - B_j^\dagger Z_1+ A_j^\dagger\, Z_2,  \qquad
  Z_1,Z_2 \in \bC^n, \quad  j=1,2.
\tag 2.13$$
 We have $L_1=L_2$ if and only if
$$
 \text{\rm Ker}[\bold D_1]=    \text{\rm Ker}[\bold D_2].
\tag 2.14$$
 We will prove that (2.12) is satisfied. For that purpose, we first prove that the range of $\bold D_1$ is equal to $\bC^n$. Suppose that for some $W \in \bC^n$,
 we have
$$
\langle W,- B_1^\dagger\, Z_1+ A_1^\dagger\, Z_2\rangle=0, \qquad Z_1, Z_2 \in \bC^n,
\tag 2.15$$
where $\langle\cdot,\cdot\rangle$ is the standard scalar product
in $\bC^n.$
In (2.15) choosing $Z_1= - B \,Z$ and $Z_2= A \,Z$
with $ Z \in   \bC^n,$ we obtain
$$
\langle( A^\dagger\,  A+ B^\dagger B) W, Z \rangle =0, \qquad Z \in  \bC^n.
\tag 2.16$$
Choosing $Z$ as $Z= ( A^\dagger\,  A+ B^\dagger B) W,$
from (2.16) we get
$$
\langle ( A^\dagger\,  A+ B^\dagger B) W, ( A^\dagger\,  A+ B^\dagger B) W \rangle=0,
\tag 2.17$$
 and then, by (2.10), we obtain $W=0$. Since the only vector in $\bC^n$ that is orthogonal to the range of $\bold D_1$ is the zero vector, the range of $\bold D_1$ is equal to $\bC^n.$
 Since the range of $\bold D_1$ is $\bC^n,$ for any $U \in \bC^n$ there exist $Z_1$
 and $Z_2$ in $\bC^n$ such that
$$
 U= - B_1^\dagger\, Z_1+ A_1^\dagger Z_2.
\tag 2.18$$
 We denote by $\bold E$ the operator from   $\bC^n$  into  $\bC^n$
 defined as
$$
\bold E \,U= - B_2^\dagger\, Z_1+ A_2^\dagger\, Z_2.
\tag 2.19$$
Let us check that $\bold E\,U$ is well defined, i.e. that it is independent of the particular $(Z_1,Z_2)$ that we use in (2.18) to represent $U$. So, suppose that for some other   $ \tilde Z_1$ and $\tilde Z_2$ in $\bC^n$ we have
$$
 U= - B_1^\dagger\, \tilde Z_1 + A_1^\dagger\, \tilde Z_2.
\tag 2.20$$
 It follows from (2.18) and (2.20) that
$$
 (Z_1-\tilde{Z_1}, Z_2-\tilde{Z_2}) \in \text{\rm Ker}\,  \bold D_1.
\tag 2.21$$
 But by (2.14)
$$
 (Z_1-\tilde Z_1, Z_2-\tilde Z_2) \in \text{\rm Ker}[\bold D_2],
\tag 2.22$$
 and then
$$
 - B_2^\dagger\, Z_1+ A_2^\dagger\, Z_2=  - B_2^\dagger\, \tilde Z_1+ A_2^\dagger \,\tilde Z_2,
\tag 2.23$$
 which proves that $\bold E \,U$ is well defined. We will also denote by $\bold  E$ the $n \times  n$ matrix associated to the operator (2.19).
 Let us prove that $\bold E$ is invertible. Hence, suppose that for some $U \in \bC^n$ we have  $\bold E\, U=0.$ Then, choosing $Z_1$ and $Z_2$ as in (2.18),
 we have
 $$
 \bold E\, U=  - B_2^\dagger Z_1+ A_2^\dagger Z_2=0.
\tag 2.24$$
 This means that $(Z_1,Z_2) \in \text{\rm Ker}[\bold D_2]$. But by (2.14),   $(Z_1,Z_2) \in \text{\rm Ker}[\bold D_1],$ and consequently we obtain
$$
 U=  - B_1^\dagger\, Z_1+ A_1^\dagger\, Z_2= \bold D_1 (Z_1,Z_2)=0,
\tag 2.25$$
 and, hence, $\bold E$ is invertible.
  Taking $ Z_2=0$ in (2.18) we get
$$
 \bold E \,U= \bold E \,(- B^\dagger_1\, Z_1) = - B_2^\dagger \,Z_1, \qquad Z_1 \in \bC^n.
 \tag 2.26$$
 Hence,
$$
 \bold E B_1^\dagger= B_2^\dagger.
\tag 2.27$$
 In the same way, taking in (2.18) $Z_1=0$ we prove that
$$
 \bold E A_1= A_2.
\tag 2.28$$
 Denoting $T= \bold E^\dagger$ and taking the adjoint of (2.27) and (2.28) we obtain (2.12).
\qed

\newpage
\noindent {\bf 3. SOME MATHEMATICAL PRELIMINARIES}
\vskip 3 pt

In this chapter we present certain mathematical results
needed in later chapters.

\vskip 10 pt
\noindent {\bf 3.a Vectors in $\bC^n$ and matrices acting on $\bC^n$}
\vskip 3 pt

For a scalar-valued function $f(y),$ we have the absolute value given by
$$|f(y)|:=\sqrt{|f(y)|^2}=\sqrt{f(y)^\ast \,f(y)},\tag 3.1$$
where the asterisk denotes the complex conjugation. For vector-valued functions with $n$ components, we use the absolute-value notation to denote the length of the vector with
$n$ components, whether it is a row vector or a column vector.
For a vector-valued function $f(y),$ which is a column vector with $n$ components given by
$$f(y)=\bm f_1(y)\\
\vdots\\ f_n(y)\endbm,\tag 3.2$$
we have
$$|f(y)|:=\sqrt{|f_1(y)|^2+\cdots |f_n(y)|^2}=\sqrt{f(y)^\dagger \,f(y)},\tag 3.3$$
where the dagger denotes the matrix adjoint.
For a vector-valued function $f(y),$ which is a row vector with $n$ components given by
$$f(y)=\bm f_1(y)&
\cdots& f_n(y)\endbm,\tag 3.4$$
we have
$$|f(y)|:=\sqrt{|f_1(y)|^2+\cdots +|f_n(y)|^2}=\sqrt{f(y)^\ast \,f(y)^T}=
\sqrt{f(y)\,f(y)^\dagger},\tag 3.5$$
where the superscript $T$ denotes the matrix transpose.
Thus, for any vector-valued function with $n$ components, we have
$$|f(y)|=|f(y)^\dagger|=|f(y)^T|=|f(y)^\ast|.\tag 3.6$$

For the standard scalar product in $\bC^n,$ we have
$$\langle f(y),g(y)\rangle :=f_1(y)^\ast\,g_1(y)+\cdots+f_n(y)^\ast\,g_n(y).\tag 3.7$$
Thus, for two column vectors $f(y)$ and $g(y)$ we have
$$\langle f(y),g(y)\rangle=f(y)^\dagger\ g(y),\tag 3.8$$
and for two row vectors $f(y)$ and $g(y)$ we have
$$\langle f(y),g(y)\rangle=f(y)^\ast\ g(y)^T.\tag 3.9$$
Thus, for the vector-valued function $f(y),$ whether it is a column vector or a row vector,
we have
$$|f(y)|=\sqrt{\langle f(y),f(y)\rangle}.\tag 3.10$$

If $A(y)$ is an $n\times n$ matrix-valued function and if $f(y)$ is a column vector with
$n$ components, then $A(y)\,f(y)$ is a column vector with $n$ components. We define the matrix norm of $A(y),$ denoted by $|A(y)|,$ as
$$|A(y)|:=\sup_{|f(y)|=1} |A(y)\,f(y)|.\tag 3.11$$
We then get the standard inequality
$$|A(y)\,f(y)|\le |A(y)|\,|f(y)|.\tag 3.12$$
If $A(y)$ and $B(y)$ are two $n\times n$ matrix-valued functions, we get
$$|A(y)\,B(y)|\le |A(y)|\,|B(y)|.\tag 3.13$$
If $g(y)$ is a row vector with $n$ components, we then get
$$|g(y)\,A(y)|\le |g(y)|\,|A(y)|.\tag 3.14$$
We also get
$$|A(y)^\dagger|=|A(y)^\ast|=|A(y)^T|=|A(y)|.\tag 3.15$$
Since we use $|A(y)|$ to denote the matrix norm of the $n\times n$ matrix $A(y)$
and we also use $|f(y)|$ to denote the standard norm in $\bC^n$ for the
vector $f(y)$ with $n$ components, the notation $|\cdot |$ becomes clearer by
checking whether it applies to a matrix or to a vector.

\vskip 10 pt
\noindent {\bf 3.b Banach and Hilbert
spaces}
\vskip 3 pt

 We introduce some elementary concepts from linear operator theory in Banach and in Hilbert spaces over the field of complex numbers. For a thorough presentation of this subject see [27].

 Recall that a Banach space $\Cal B$ over the complex numbers is a vector space that has a norm $||\cdot ||_{\Cal B}$ and that is complete.
 The norm satisfies the following conditions:
\item{(a)} We have
 $|| Y ||_{\Cal B}\ge 0$ for all
  $Y\in \Cal B,$ and the equality holds if and only if $Y$ is the zero vector in
  $\Cal B.$

 \item{(b)} For any complex number $\alpha$ and for any vector
 $Y\in\Cal B$ we have $||\alpha Y||_{\Cal B}=|\alpha|\, ||Y||_{\Cal B}.$

 \item{(c)} For any pair of vectors $Y$ and $Z$ in
 $\Cal B,$ we have
 $||Y+Z||_{\Cal B}\le ||Y||_{\Cal B}+||Z||_{\Cal B}.$

The completeness of $\Cal B$ means that
every Cauchy sequence in $\Cal B$ is convergent, i.e.
for any sequence $\{Y^{(l)}\}_{l=1}^\infty$ with elements
in $\Cal B$ that has the property $||Y^{(l)}-Y^{(m)}||_{\Cal B}\to 0$
as $l,m\to +\infty$ there exists an element $Y\in\Cal B$
such that $||Y^{(l)}-Y||_{\Cal B}\to 0$ as $n\to+\infty.$

Let us now consider the particular case of operators between Hilbert spaces. Recall that a Hilbert space, $\Cal H,$ is a Banach space that has a scalar product, $(\cdot, \cdot)_{\Cal H}$ such that the norm is derived from the scalar product as, $\|  Y\|_{\Cal H}= ( ( Y,Y)_{\Cal H})^{1/2}$. We take the scalar product antilinear-linear and it satisfies the following properties:

\item{(a)} For any pair of vectors
$Y$ and $Z$ in $\Cal H$ we have
$(Y,Z)_{\Cal H}=(Z,Y)^\ast_{\Cal H},$ where the asterisk denotes complex conjugation.

\item{(b)} For any pair of vectors
$Y$ and $Z$ in $\Cal H$ and any complex number $\alpha$ we have
$(Y,\alpha\,Z)_{\Cal H}=\alpha \,(Y,Z)_{\Cal H}.$ By the antilinearity, we mean
$(\alpha\,Y,Z)_{\Cal H}=\alpha^\ast \,(Y,Z)_{\Cal H}.$

\item{(c)} For any three vectors $X,$ $Y,$ and $Z$ in $\Cal H$ we have
$(X,Y+Z)_{\Cal H}=(X,Y)_{\Cal H}+(X,Z)_{\Cal H}.$

\item{(d)} We have
$(Y,Y)_{\Cal H}\ge 0$ for all
  $Y\in \Cal B,$ and the equality holds if and only if $Y$ is the zero vector in
  $\Cal H.$

For a vector-valued function $f(y),$ viewed as the vector $f$ in the Banach space
$L^1(\bR^+),$ we have the length of $f$ defined as
$$||f||_1:=\int_0^\infty dy\, |f(y)|.\tag 3.16$$
Note that (3.16) holds whether $f(y)$ is a column vector or a row vector.
Assume that the $n\times n$ matrix-valued
operator $\bold O$ acts on $L^1(\bR^+).$ Then, for a
vector-valued function $f(y),$ which is a column-vector with
$n$ components in $L^1(\bR^+),$ we have $(\bold O f)(y)$ also a column
vector with $n$ components. Then,
the operator norm $||\bold O||_1$ is
given by
$$||\bold O||_1:=\sup_{||f||_1=1} ||\bold O f||_1=\sup_{||f||_1=1} \int_0^\infty
dy\,|(\bold O f)(y)| .\tag 3.17$$
We have the standard inequality
$$||\bold O f||_1\le ||\bold O||_1\,||f||_1.\tag 3.18$$
Since we use $||\bold O||_1$ to denote the operator norm of $\bold O$ on $L^1(\bR^+)$
and we also use $|| f||_1$ to denote the standard norm in $L^1(\bR^+)$ for the
vector $f$ with $n$ components, the notation $||\cdot ||_1$ becomes clearer by
checking whether it applies to a matrix or to a vector.

In the Hilbert space $L^2(\bR^+)$ we have the standard scalar product
$$\left( f,g\right)_2:=\int_0^\infty dy\, \left[ f_1(y)^\ast \,g_1(y)+\cdots +
f_n(y)^\ast\,g_n(y)\right]=\int_0^\infty dy\,\langle f(y),g(y)\rangle.\tag 3.19$$
For a vector-valued function $f(y),$ viewed as the vector $f$ in the Hilbert space
$L^2(\bR^+),$ we have the norm of $f$ defined as
$$||f||_2:=\left[\int_0^\infty dy\, |f(y)|^2\right]^{1/2}.\tag 3.20$$
Note that (3.20) holds whether $f(y)$ is a column vector or a row vector.
Comparing (3.19) and (3.20) we see that
the standard norm $||\cdot||_2$ in $L^2(\bR^+)$ given in
(3.20) is induced by the standard
scalar product in (3.19) via
$$||f||_2=\sqrt{\left( f,f\right)_2}.\tag 3.21$$

Assume that the $n\times n$ matrix-valued
operator $\bold O$ acts on $L^2(\bR^+).$ Then, for a
vector-valued function $f(y),$ which is a column-vector with
$n$ components in $L^2(\bR^+),$ we have $(\bold O f)(y)$ also a column
vector with $n$ components. Then,
the operator norm $||\bold O||_2$ is
given by
$$||\bold O||_2:=\sup_{||f||_2=1} ||\bold O f||_2=\sup_{||f||_2=1}\left[ \int_0^\infty
dy\,|(\bold O f)(y)|^2\right]^{1/2} .\tag 3.22$$
We have the standard inequality
$$||\bold O f||_2\le ||\bold O||_2\,||f||_2.\tag 3.23$$
Since we use $||\bold O||_2$ to denote the operator norm of $\bold O$ on $L^2(\bR^+)$
and we also use $|| f||_2$ to denote the standard norm in $L^2(\bR^+)$ for the
vector $f$ with $n$ components, the notation $||\cdot ||_2$ becomes clearer by
checking whether it applies to a matrix or to a vector.
Further results on the operator norm
are given in Section~3.g.

For a vector-valued function $f(y),$ viewed as the vector $f$ in the Banach space
$L^\infty(\bR^+),$ we have the norm of $f$ defined as
$$||f||_\infty:=\matrix \phantom{xx}\\
\operatorname{ess\,sup}\\
{y\in\bR^+} \endmatrix \, |f(y)| .\tag 3.24$$
Note that (3.24) holds whether $f(y)$ is a column vector or a row vector.

The closure of a set is obtained by adding all the limit points to the set.
A subspace of $L^2(\bR^+)$ is dense
if its closure is equal to $L^2(\bR^+).$
We recall that there are various dense subspaces of
$L^2(\bR^+).$ For example, $C_0(\bR^+),$ the subspace of
continuous functions with compact support in $\bR^+$ is dense in
$L^2(\bR^+).$ Another dense subspace of
$L^2(\bR^+)$ is $C^\infty_0(\bR^+),$ the subspace of infinitely
differentiable functions with compact support in $\bR^+.$
Yet another dense subspace of $L^2(\bR^+)$ is
$L^1(\bR^+)\cap L^2(\bR^+).$ A subspace of $L^2(\bR^+)$
is dense if and only if
its orthogonal
complement
is $\{0\}.$

\vskip 10 pt
\noindent {\bf 3.c Useful inequalities}
\vskip 3 pt

The Cauchy-Schwarz inequality in $\bC^n$ is given by
$$|\langle f(y),g(y)\rangle |\le |f(y)|\,|g(y)|.\tag 3.25$$
The Cauchy-Schwarz inequality in $L^2(\bR^+)$ is given by
$$|\left( f,g\right) |\le ||f||_2\,|| g||_2.\tag 3.26$$
We have the H\"older inequality
$$||fg||_1\le ||f||_2\,||g||_2,\tag 3.27$$
which can informally be stated as the product of
two square-integrable quantities is integrable.
We also have
$$||fg||_1\le ||f||_1\,||g||_\infty,\tag 3.28$$
which can informally be stated as the product of
an integrable quantity and a bounded quantity is integrable.
We have Young's inequality for products given by
$$|f(y)\,g(y)|\le \ds\frac{1}{2}\left[|f(y)|^2+|g(y)|^2\right],\tag 3.29$$
$$||f\,g||_1\le \ds\frac{1}{2}\left[||f||_2^2+||g||_2^2\right],\tag 3.30$$
the latter of which can informally be stated as the product of two
square-integrable quantities is integrable.
The properties in (3.25)-(3.30) hold for scalar, vector-valued, matrix-valued,
and matrix-valued operator quantities.

We have the standard triangle inequalities
$$|f(y)+g(y)|\le |f(y)|+|g(y)|,\tag 3.31$$
$$||f+g||_1\le ||f||_1+||g||_1,\tag 3.32$$
$$||f+g||_2\le ||f||_2+||g||_2,\tag 3.33$$
$$||f+g||_\infty\le ||f||_\infty+||g||_\infty,\tag 3.34$$
all holding for scalar, vector-valued, matrix-valued,
and matrix-valued operator quantities.

Recall that the convolution $f\ast g$ is defined as
$$(f\ast g)(y):=\int_{-\infty}^\infty dz\,f(y-z)\,g(z).\tag 3.35$$
We have the symmetry
$$(g\ast f)(y)=(f\ast g)(y),\tag 3.36$$
as well as Young's inequalities for convolutions
$$||f\ast g||_1\le ||f||_1\,||g||_1,\tag 3.37$$
$$||f\ast g||_2\le ||f||_1\,||g||_2,\tag 3.38$$
$$||f\ast g||_\infty\le ||f||_1\,||g||_\infty,\tag 3.39$$
all holding for scalar, vector-valued, matrix-valued,
and matrix-valued operator quantities.

\vskip 10 pt
\noindent {\bf 3.d Hardy spaces}
\vskip 3 pt

We use $\bold H^2(\bCp)$ to denote the Hardy space of all complex-valued
functions $f(k)$ that are analytic
in $k\in\bCp$ with a finite norm defined as
$$||f||_{\bold H^2(\bCp)}:=\sup_{\rho >0}\left[\int_{-\infty}^\infty d\alpha\,|f(\alpha+i\rho)|^2\right]^{1/2}.\tag 3.40$$
Thus, $f(k)$ is square integrable along all lines in $\bCp$ that are
parallel to the real axis, i.e.
$$\int_{-\infty}^\infty d\alpha\,|f(\alpha+i\rho)|^2\le \left( ||f||_{\bold H^2(\bCp)}\right)^2,
\qquad \rho\in\bR^+.\tag 3.41$$
The value of $f(k)$ for $k\in\bR$ is defined [41] to be
the non-tangential limit of $f(k+i\rho)$ as $\rho \to 0^+,$ and in particular
$$f(k):=\lim_{\rho\to 0^+} f(k+i\rho),\qquad k\in\bR.\tag 3.42$$
It is known
[41] that such a non-tangential limit exists a.e. in $k\in\bR$ and belongs to
$L^2(\bR)$ in the sense that
$$\lim_{\rho\to 0^+}\int_{-\infty}^\infty dk\,|f(k+i\rho)-f(k)|^2=0.\tag 3.43$$
It is known that $f(k)$ belongs to $\bold H^2(\bCp)$ if and only if
there exists a corresponding function $g(x)$ belonging to $L^2(\bR^+)$ in such a way that
$$f(k)=\int_0^\infty dx\, g(x)\, e^{ikx},\tag 3.44$$
and that
$$||f||_{\bold H^2(\bCp)}=\sqrt{2\,\pi}\, ||g||_2.\tag 3.45$$

The following pointwise estimate is useful for functions belonging to
a Hardy space $\bold H^2(\bCp).$

\noindent {\bf Proposition 3.1} {\it Assume that
$f(k)$ belongs to the Hardy space $\bold H^2(\bCp).$ Then, we have}
$$|f(k)|\le \ds\frac{||f||_{\bold H^2(\bCp)}}{\sqrt{2\,\pi} \,\sqrt{2\,|k|\,\sin\theta}},\qquad k\in\bCp,\tag 3.46$$
{\it where $||f||_{\bold H^2(\bCp)}$ is the norm
of $f$ defined in (3.40), and
$|k|$ and $\theta$ are the absolute value
 and the argument of the point $k\in\bC^+,$ i.e. $k=|k|\,e^{i\theta}$
 is the polar representation of the complex number $k$
 with $\theta\in(0,\pi).$
 Thus, $f(k)$ satisfies the pointwise estimate}
$$|f(k)|\le \ds\frac{C}{\sqrt{|k|\,\sin\theta}},\qquad k\in\bCp,\tag 3.47$$
{\it where $C$ is a generic constant.}

 \noindent PROOF: In terms of the real and imaginary part of
 $k\in\bC^+,$ denoted by
 $k_R$ and $k_I,$ respectively, we have $k=k_R+i k_I.$ Thus, from (3.44)
 we obtain
$$f(k)=\int_0^\infty dx \, g(x)\,e^{i k_R x -k_I x}.\tag 3.48$$
Applying the Cauchy-Schwarz inequality on the integrand in (3.48) we get
$$|f(k)|\le \int_0^\infty dx\, |g(x)|\, e^{-k_I x}
\le \sqrt{\int_0^\infty dx\, |g(x)|^2}
\sqrt{\int_0^\infty dx\, e^{-2 k_I x}},\tag 3.49$$
 yielding
$$|f(k)|\le ||g||_2 \,\ds\frac{1}{\sqrt{2 k_I}}.\tag 3.50$$
Using $k_I=|k|\,\sin \theta$ and (3.45) in (3.50) we obtain
(3.46). When $f(k)$ belongs to the Hardy space, its norm
defined in (3.40) is finite, and hence (3.46) yields (3.47).
\qed

In a similar way, we use $\bold H^2(\bCm)$ to denote the Hardy space of all complex-valued
functions $f(k)$ that are analytic
in $k\in\bCm$ with a finite norm defined as
$$||f||_{\bold H^2(\bCm)}:=\sup_{\rho >0}\left[\int_{-\infty}^\infty d\alpha\,|f(\alpha-i\rho)|^2\right]^{1/2}.\tag 3.51$$
Thus, $f(k)$ is square integrable along all lines in $\bCm$ that are
parallel to the real axis, i.e.
$$\int_{-\infty}^\infty d\alpha\,|f(\alpha-i\rho)|^2\le \left( ||f||_{\bold H^2(\bCm)}\right)^2,
\qquad \rho\in\bR^+.\tag 3.52$$
The value of $f(k)$ for $k\in\bR$ is defined [40] to be
the non-tangential limit of $f(k-i\rho)$ as $\rho \to 0^+,$ and in particular
$$f(k):=\lim_{\rho\to 0^+} f(k-i\rho),\qquad k\in\bR.\tag 3.53$$
Such a non-tangential limit exists a.e. in $k\in\bR$ and belongs to
$L^2(\bR)$ in the sense that
$$\lim_{\rho\to 0^+}\int_{-\infty}^\infty dk\,|f(k-i\rho)-f(k)|^2=0.\tag 3.54$$
It is known that $f(k)$ belongs to $\bold H^2(\bCm)$ if and only if
there exists a corresponding function $g(x)$ belonging to $L^2(\bR^-)$ in such a way that
$$f(k)=\int_{-\infty}^0 dx\, g(x)\, e^{ikx},\tag 3.55$$
and that
$$||f||_{\bold H^2(\bCm)}=\sqrt{2\,\pi}\, ||g||_2.\tag 3.56$$

The following pointwise estimate is useful for functions belonging to
the Hardy space $\bold H^2(\bCm).$ Its proof is similar to the proof of
Proposition~3.1 and hence is omitted.

\noindent {\bf Proposition 3.2} {\it Assume that
$f(k)$ belongs to the Hardy space $\bold H^2(\bCm).$ Then, we have}
$$|f(k)|\le \ds\frac{||f||_{\bold H^2(\bCm)}}{\sqrt{2\,\pi} \,\sqrt{2\,|k|\,\sin\theta}},\qquad k\in\bCm,\tag 3.57$$
{\it where $||f||_{\bold H^2(\bCm)}$ is the norm
of $f$ defined in (3.51), and
$|k|$ and $\theta$ are the absolute value
 and the argument of the point $k\in\bC^+,$ i.e. $k=|k|\,e^{i\theta}$
 is the polar representation of the complex number $k$
 with $\theta\in(-\pi,0).$
 Thus, $f(k)$ satisfies the pointwise estimate}
$$|f(k)|\le \ds\frac{C}{\sqrt{|k|\,|\sin\theta|}},\qquad k\in\bCm,\tag 3.58$$
{\it where $C$ is a generic constant.}

\vskip 10 pt
\noindent {\bf 3.e Other useful Banach spaces}
\vskip 3 pt

We use $\hat L^1(\bCp)$ to denote the
Banach space of all complex-valued functions $f(k)$ that
are analytic in $k\in\bCp$ with the finite
norm $||f||_{\hat L^1(\bCp)}$
in such a way that there exists
a corresponding function $g(x)$ belonging to $L^1(\bR^+)$
satisfying
$$f(k)=\int_0^\infty dx\, g(x)\, e^{ikx}.\tag 3.59$$
We define the norm
$||f||_{\hat L^1(\bCp)}$ as
$$||f||_{\hat L^1(\bCp)}:=\sqrt{2\,\pi}\,||g||_1.\tag 3.60$$
Let us remark that if $f(k)$ belongs to
$\hat L^1(\bCp)$ then $f(k)$ is continuous in $k\in\bR$ and we have $f(k)=o(1)$ as
$k\to\infty$ in $\bCpb.$

We use $\hat L^1(\bCm)$ to denote the
Banach space of all complex-valued functions $f(k)$ that
are analytic in $k\in\bCm$ with the finite
norm $||f||_{\hat L^1(\bCm)}$
in such a way that there exists
a corresponding function $g(x)$ belonging to $L^1(\bR^-)$
satisfying
$$f(k)=\int_{-\infty}^0 dx\, g(x)\, e^{ikx}.\tag 3.61$$
We define the norm
$||f||_{\hat L^1(\bCm)}$ as
$$||f||_{\hat L^1(\bCm)}:=\sqrt{2\,\pi}\,||g||_1.\tag 3.62$$
Note that if $f(k)$ belongs to
$\hat L^1(\bCm)$ then $f(k)$ is continuous in $k\in\bR$ and we have $f(k)=o(1)$ as
$k\to\infty$ in $\bCmb.$

We use $\hat L^1_{\infty}(\bCp)$ to denote the
Banach space of all complex-valued functions $f(k)$ that
are analytic in $k\in\bCp$ with the finite
norm $||f||_{\hat L^1_{\infty}(\bCp)}$
in such a way that there exists
a corresponding function $g(x)$ belonging to $L^1(\bR^+)\cap L^\infty(\bR^+)$
satisfying
$$f(k)=\int_0^\infty dx\, g(x)\, e^{ikx}.\tag 3.63$$
We define the norm
$||f||_{\hat L^1_\infty(\bCp)}$ as
$$||f||_{\hat L^1_\infty(\bCp)}:=\sqrt{2\,\pi}\left(||g||_1+
||g||_\infty\right).\tag 3.64$$
Note that if $f(k)$ belongs to
$\hat L^1_{\infty}(\bCp)$ then $f(k)$ must also belong to
$\bold H^2(\bCp).$ In other words, we have
$\hat L^1_{\infty}(\bCp)\subset \bold H^2(\bCp).$

In a similar way, we use $\hat L^1_{\infty}(\bCm)$ to denote the
Banach space of all complex-valued functions $f(k)$ that
are analytic in $k\in\bCm$ with the finite
norm $||f||_{\hat L^1_{\infty}(\bCm)}$
in such a way that there exists
a corresponding function $g(x)$ belonging to $L^1(\bR^-)\cap L^\infty(\bR^-)$
satisfying
$$f(k)=\int_{-\infty}^0 dx\, g(x)\, e^{ikx}.\tag 3.65$$
We define the norm
$||f||_{\hat L^1_\infty(\bCm)}$ as
$$||f||_{\hat L^1_\infty(\bCm)}:=\sqrt{2\,\pi}\left(||g||_1+
||g||_\infty\right).\tag 3.66$$
We remark that $\hat L^1_{\infty}(\bCm)\subset \bold H^2(\bCm).$

The definitions for
$\bold H^2(\bCp),$ $\bold H^2(\bCm),$ $\hat L^1(\bCp),$ $\hat L^1(\bCm),$
$\hat L^1_{\infty}(\bCp),$ and
$\hat L^1_{\infty}(\bCm)$ given above
can naturally be extended to vector-valued
or matrix-valued functions
This is achieved
by requiring that each entry of
such functions belongs to the appropriate space and by replacing the
absolute value used in the scalar case by the operator matrix norm.

For example, the quantity $X(y)$ appearing in
$(\bold 4_c)$ of Definition~4.3
is a row vector with $n$ components.
It belongs to $L^1(\bR^+)\cap L^\infty(\bR^+),$ and hence
we can assume that it is bounded and integrable
in $y\in\bR$ with the understanding that $X(y)=0$ for $y\in\bR^-.$
The quantity $\hat X(k)$ appearing in
$(\bold 4_d)$ of Definition~4.3 is also a
row vector with $n$ components,
and it belongs to $\hat L^1_{\infty}(\bCp).$
In fact,
$X(y)$ is related to
$\hat X(k)$
via a Fourier transform as
$$X(y)=\ds\frac{1}{2\pi} \int_{-\infty}^\infty dk\,
 \hat X(k)\,e^{-iky},\qquad y\in\bR.\tag 3.67$$
The quantity $\hat X(k)$
belongs to $\hat L^1_{\infty}(\bCp),$ and hence it is
analytic in $k\in\bCp,$ continuous in $k\in\bCpb,$
uniformly $o(1)$ as $k\to\infty$ in $\bCpb,$ and given by
$$\hat X(k)=\int_0^\infty dy\,X(y)\,e^{iky},\qquad k\in\bR.\tag 3.68$$

We are also interested in
those vector-valued $\hat X(k)$
for which the corresponding $X(y)$ belong to $L^2(\bR^-)$
and vanish for $y\in\bR^+.$ For such functions $\hat X(k)$
belonging to $\bold H^2(\bCm)$ we still have
(3.67) holding but (3.68) must be replaced with
$$\hat X(k)=\int_{-\infty}^0 dy\,X(y)\,e^{iky},\qquad k\in\bR.\tag 3.69$$
Such functions $X(y)$ and $\hat X(k)$ are related to $(\bold{III}_a)$ and $(\bold{III}_b)$ in Definition~4.3.

\vskip 10 pt
\noindent {\bf 3.f Integral operators with kernels depending on a sum}
\vskip 3 pt

Some general results pertaining to integral operators whose kernels
depend on a sum are listed in the following three propositions for easy referencing.
These results will be used to analyze various integral equations arising
in the analysis of the inverse problem, including the Marchenko integral equation
(13.1).

\noindent {\bf Proposition 3.3} {\it Let $\epsilon$ be some fixed
number in the interval $[0,+\infty).$ Assume that $\Cal A(y)$ is an $n\times n$
matrix-valued function integrable
in $y\in(\epsilon,+\infty).$ Then:}

\item{(a)} {\it The operator $
\Cal A:\ X(y)\mapsto \int_\epsilon^\infty dz\,X(z)\,\Cal A(z+y)$
is compact on $L^1(\epsilon,+\infty),$ where
$X(y)$ is a row vector with $n$ components.}

\item{(b)} {\it The linear integral equation given by}
$$X(y)+\Cal A(y)+\int_\epsilon^\infty dz\,X(z)\,\Cal A(z+y)=0,\tag 3.70$$
{\it has a unique solution
in $L^1(\epsilon,+\infty)$ if
and only if the only solution in $L^1(\epsilon,+\infty)$ to
the corresponding homogeneous equation
$$X(y)+\int_\epsilon^\infty dz\,X(z)\,\Cal A(z+y)=0,\tag 3.71$$
is the trivial solution.}

\item{(c)} {\it If further $\Cal A(y)$ is assumed
to be bounded in $y\in(\epsilon,+\infty),$ then
any solution $X(y)$ in $L^1(\epsilon,+\infty)$ to (3.71)
must also belong to $L^\infty(\epsilon,+\infty).$ In particular,
any solution to (3.71) in $L^1(\epsilon,+\infty)$ must also belong to
$L^2(\epsilon,+\infty).$}

\item{(d)} {\it If further $\Cal A(y)$ is assumed
to be bounded in $y\in(\epsilon,+\infty),$ then
any solution $X(y)$ in $L^1(\epsilon,+\infty)$ to (3.70) must also belong to $L^\infty(\epsilon,+\infty).$ In particular,
any solution to (3.70) in $L^1(\epsilon,+\infty)$ must also belong to
$L^2(\epsilon,+\infty).$}

\noindent PROOF: The compactness in (a) is established
in Lemma~3.3.1 of [2]. The result in (b) is a result
of the compactness established in (a). The result in (c)
is obtained with the help of (3.39), as follows.
When $\Cal A(y)$ is bounded and
$X(y)$ is integrable, from
(3.71) we obtain
$$|X(y)|\le \int_\epsilon^\infty dz\, |X(z)|\,|\Cal A(z+y)\le C
\int_\epsilon^\infty dz\, |X(z)|<+\infty
,\tag 3.72$$
for some constant $C.$ Thus, $X(y)$ is bounded. Then, being in $L^1(\epsilon,+\infty)$
and $L^\infty(\epsilon,+\infty),$ the quantity $X(y)$ belongs to
$L^p(\epsilon,+\infty)$ for all $p$ with $1\le p\le +\infty.$
In particular, it belongs to $L^2(\epsilon,+\infty).$ Thus, the
proof of (c) is complete.
 From (3.70) we obtain
$$|X(y)|\le |\Cal A(y)|+\int_\epsilon^\infty dz\, |X(z)|\,|\Cal A(z+y)|\le C+C
\int_\epsilon^\infty dz\, |X(z)|<+\infty
,\tag 3.73$$
and hence any solution $X(y)$ in $L^1(\epsilon,+\infty)$
to (3.70) also belongs to $L^\infty(\epsilon,+\infty),$
and hence in particular to $L^2(\epsilon,+\infty).$
  \qed

The next result is the analog of Proposition~3.3, but for operators
acting on $L^2(\epsilon,+\infty).$

\noindent {\bf Proposition 3.4} {\it Let $\epsilon$ be some fixed
number in the interval $[0,+\infty).$ Assume that $\Cal A(y)$ is an $n\times n$
matrix-valued function square integrable
in $y\in(\epsilon,+\infty).$ Then:}

\item{(a)} {\it The operator $
 \Cal A: \ X(y)\mapsto \int_\epsilon^\infty dz\,X(z)\,\Cal A(z+y)$
is compact on $L^2(\epsilon,+\infty),$ where
$X(y)$ is a row vector with $n$ components.}

\item{(b)} {\it The linear integral equation given by
(3.70) has a unique solution
in $L^2(\epsilon,+\infty)$ if
and only if the only solution in $L^2(\epsilon,+\infty)$ to
the corresponding homogeneous equation (3.71)
is the trivial solution.}

\item{(c)} {\it Any solution $X(y)$ in $L^2(\epsilon,+\infty)$ to (3.71)
must also belong to $L^\infty(\epsilon,+\infty).$}

\item{(d)} {\it If further $\Cal A(y)$ is assumed
to be bounded in $y\in(\epsilon,+\infty),$ then
any solution $X(y)$ in $L^2(\epsilon,+\infty)$ to (3.70) must also belong to $L^\infty(\epsilon,+\infty).$}

\noindent PROOF: For the proof of (a), we proceed as follows.
Let us first prove that the operator
$\Cal A$ is bounded on
$L^2(\epsilon,+\infty).$
Since $X(y)$ and $\Cal A(y)$ belong to
$L^2(\epsilon,+\infty),$ we have their respective $L^2$-Fourier transforms
$$\hat X(k):=\int_\epsilon^\infty dy\,X(y)\,e^{iky},\tag 3.74$$
$$\hat \Cal A(k):=\int_\epsilon^\infty dy\,\Cal A(y)\,e^{-iky},\tag 3.75$$
with the understanding that
$X(y)=0$ and
$\Cal A(y)=0$ for $y\in(-\infty,\epsilon).$
Thus, we can write (3.74) and (3.75) as
$$\hat X(k)=\int_{-\infty}^\infty dy\,X(y)\,e^{iky},\tag 3.76$$
$$\hat \Cal A(k)=\int_{-\infty}^\infty dy\,\Cal A(y)\,e^{-iky},\tag 3.77$$
with
$\hat X$ and $\hat\Cal A$ belonging to
$L^2(\bR).$ Since the Fourier transform is a bijection on
$L^2(\bR),$ from (3.76) and (3.77) yield
$$X(y)=\ds\frac{1}{2\pi}\int_{-\infty}^\infty dk\,\hat X(k)\
e^{-iky},\tag 3.78$$
$$\Cal A(y)=\ds\frac{1}{2\pi}\int_{-\infty}^\infty dk\,\hat \Cal A(k)\
e^{iky}.\tag 3.79$$
%
%
%
%
%
 From (3.78) and (3.79), with the help of
 (11.37), we obtain
$$ \int_\epsilon^\infty dz\,X(z)\,\Cal A(z+y)=
\int_{-\infty}^\infty dz\,X(z)\,\Cal A(z+y)=
\ds\frac{1}{2\pi}\int_{-\infty}^\infty dk\,\hat X(k)\,
\hat \Cal A(k)\,
e^{iky}.\tag 3.80$$
 From the second equality in
(3.80) we obtain
$$||X\ast \Cal A||_2=\ds\frac{1}{2\pi}\,|| \hat X\,\hat\Cal A||_2,\tag 3.81$$
 from which we obtain
$$||X\ast \Cal A||_2\le  \ds\frac{1}{2\pi}\,|| \hat X||_2\, ||\hat\Cal A||_2
=||X||_2\,||\Cal A||_2.\tag 3.82$$
 From (3.82) we conclude that
the operator $\Cal A$ is bounded on $L^2(\epsilon,+\infty).$
Let $\{ \Cal A^{(l)}\}_{l=1}^\infty$
be a sequence of $n \times n$ matrix-valued functions
belonging to $C^\infty_0(\epsilon,+\infty)$
converging to $\Cal A(y)\in L^2(\epsilon,+\infty),$ i.e.
$$\lim_{l \to +\infty} || \Cal A^{(l)}-\Cal A||_2=0.\tag 3.83$$
Let $\Cal A^{(l)}$ be the operator
on $L^2(\epsilon,+\infty)$ for the
mapping
$X(y)\mapsto \int_\epsilon^\infty dz\,X(z)\,\Cal A^{(l)}(z+y).$
 From (3.83) we then conclude that
 the sequence of operators
$\{ \Cal A^{(l)}\}_{l=1}^\infty$  converge in norm to the operator
$\Cal A.$ Consequently, to prove that the operator
$\Cal A$ is compact, it is enough to prove that the operator
$\Cal A^{(l)}$ is compact. It is enough to prove that
$\Cal A^{(l)}$ is a Hilbert-Schmidt operator on
$L^2(\epsilon,+\infty).$ We have
$$\aligned
\int_\epsilon^\infty dy \int_\epsilon^\infty \, dz\, |\Cal A^{(l)}(y+z)|^2 & =   \int_\epsilon^\infty dy \int_{\epsilon+y}^\infty \, dz\, |\Cal A^{(l)}(z)|^2 \\
&
\leq
\int_\varepsilon^\infty\,dy\, \frac{1}{(1+y)^2} \int_{\varepsilon+y}^\infty \, dz\, (1+ z)^2\,|\Cal A^{(l)}(z)|^2 .\endaligned\tag 3.84$$
Because $\Cal A^{(l)}(y)$ is compactly supported in
$L^2(\epsilon,+\infty),$ the last integral in (3.84) is finite
and hence $\Cal A^{(l)}$ is Hilbert-Schmidt and hence also compact
on $L^2(\epsilon,+\infty).$ Thus, the proof of (a) is complete.
Let us now turn to the proof of (b).
The result in (b) is a result
of the compactness established in (a). The result in (c)
is obtained without needing
the boundedness of
$\Cal A(y),$ as follows.
 Using (3.29) in
(3.71) we obtain
$$|X(y)|\le \int_\epsilon^\infty dz\, |X(z)|\,|\Cal A(z+y)|\le \ds\frac{1}{2}
\int_\epsilon^\infty dz\, \left(|X(z)|^2+|\Cal A(z+y)|^2\right)<+\infty
,\tag 3.85$$
for some constant $C.$ Thus, $X(y)$ is bounded, establishing (c). Let us
now turn to the proof of (d).
 From (3.70) we obtain
$$\aligned
|X(y)|&\le |\Cal A(y)|+\int_\epsilon^\infty dz\, |X(z)|\,|\Cal A(z+y)|
\\
&\le |\Cal A(y)|+\ds\frac{1}{2}
\int_\epsilon^\infty dz\, \left(|X(z)|^2+|\Cal A(z+y)|^2\right),\endaligned
\tag 3.86$$
and hence from (3.86) we
conclude (d). \qed

The following result is the analog of Proposition~3.4
for $y\in\bR^-.$

\noindent {\bf Proposition 3.5} {\it Let $\Cal A(y)$ be an $n\times n$
matrix-valued function square integrable
in $y\in\bR^-.$ Then:}

\item{(a)} {\it The operator $X(y)\mapsto \int_{-\infty}^0 dz\,X(z)\,\Cal A(z+y)$
is compact on $L^2(\bR^-),$ where
$X(y)$ is a row vector with $n$ components.}

\item{(b)} {\it Any solution $X(y)$ in $L^2(\bR^-)$ to the linear
homogeneous integral equation}
$$-X(y)+\int_{-\infty}^0 dz\,X(z)\,\Cal A(z+y)=0,\tag 3.87$$
{\it must also belong to $L^\infty(\bR^-).$}

\noindent PROOF: The proof is similar to the proofs of (a) and (c) of
Proposition~3.4. \qed

\vskip 10 pt
\noindent {\bf 3.g Further results in Banach and Hilbert spaces}
\vskip 3 pt

Let $\Cal B_1$ and $\Cal B_2$ be Banach spaces over the complex numbers. We consider linear operators (operators in short) from $\Cal B_1$
into $\Cal B_2$ that are only defined in a linear subspace in $\Cal B_1$.

\noindent {\bf Definition 3.6} {\it
An operator $L$ from  $\Cal B_1$ into $\Cal B_2$ is a linear function from a linear subspace $\text{\rm Dom}[L]$ of $\Cal B_1$ into $\Cal B_2$. We call $\text{\rm Dom}[L]$ the domain of $L$. Then, for all complex numbers $\alpha_1$ and $\alpha_2$ we have}
$$L(\alpha_1\, Y_1+ \alpha_2\, Y_2)= \alpha_1\, L Y_1+ \alpha_2\,  LY_2, \qquad
Y_1,Y_2 \in \text{\rm Dom}[L].
\tag 3.88$$

In our monograph we only consider operators that are densely defined in
$\Cal B_1$. That is to say, we assume that the closure of the domain
of $L$, denoted by $\overline{\text{\rm Dom}[L]},$ is equal to $\Cal B_1.$

An important class of operators is the class of closed operators defined below.

\noindent {\bf Definition 3.7} {\it
An operator $L$ from $\Cal B_1$ into $\Cal B_2$ is said to be a closed
operator
if for every sequence $\{Y^{(l)}\}_{l=1}^\infty$ with
 elements in $\text{\rm{Dom}}[L]$ converging to an element $Y\in \Cal B_1,$
 i.e. $Y:=\lim_{l\to +\infty} Y^{(l)},$
 and such that $\{L Y^{(l)}\}_{l=1}^\infty$ converging to an element
$W\in \Cal B_2,$ i.e. $W:=\lim_{l\to +\infty} LY^{(l)},$ the following is true:
$Y$ belongs to $\text{\rm Dom}[L]$
and $W=LY.$}

Next, we introduce the class of closable operators.

\noindent {\bf Definition 3.8} {\it
 An operator $L$ from $\Cal B_1$ into $\Cal B_2$ is said to be closable, if for every  sequence $\{Y^{(l)}\}_{l=1}^\infty$ with
 elements in $\text{\rm Dom}[L]$ converging to zero, i.e.
 $\lim_{l\to +\infty} Y^{(l)}=0,$ and such that
$\{L Y^{(l)}\}_{l=1}^\infty$ converging to an element
$W\in \Cal B_2,$ i.e. $W:=\lim_{l\to +\infty} LY^{(l)},$
we then have $W=0$.}

Next we define an order relation between two operators.

\noindent {\bf Definition 3.9} {\it
Let $L_1$ and $L_2$ be operators from  $\Cal B_1$ into $\Cal B_2$. We say that $ L_1 \subset L_2$ if $ \text{\rm Dom}[L_1]\subset \text{\rm Dom}[L_2]$ and if $L_1 Y= L_2Y$ for all $ Y \in \text{\rm Dom}[L_1].$}

When $ L_1 \subset L_2$ we say that $L_1$ is a restriction of $L_2$ or that $L_2$ is an extension of $L_1$. Observe that $ L_1 \subset L_2$ means intuitively that $L_1$ acts in the same way as $L_2$ but with a smaller domain.

A closable operator $L$ has always a closed extension that we call the closure of $L$, which we denote by $\overline{L}$. The domain of $\overline{L},$
denoted by
$\text{\rm Dom}[\overline{L}]$ consists of all vectors
$Y\in\Cal B_1$ that are the limits of sequences $\{Y^{(l)}\}_{l=1}^\infty$ with elements
in the domain of $L$ in such a way that each corresponding sequence
 $\{L\,Y^{(l)}\}_{l=1}^\infty$ converges to an element $W\in\Cal B_2$
 and we have
%
%
%
   $$
  \overline{L}\, Y= W, \qquad Y \in \text{\rm Dom}[\overline{L}].\tag 3.89
  $$
 Note that since $L$ is closable, if $Y = 0$, then $W= 0$ and then, we do not reach the contradiction $ \overline{L}\, 0 \neq 0.$ Observe that if $L$ is closable, then, $ \overline{L}$ is the smallest closed extension of $L$, i.e. if $D$ is any closed extension of $L$ then $\overline{L} \subset D$.

 \noindent {\bf Definition 3.10} {\it
  An operator $L$ from  $\Cal B_1$, into $\Cal B_2$ is bounded if for some constant $C$ we have}
  $$
  ||L Y||_{\Cal B_2} \leq C\,   || Y||_{\Cal B_1}, \qquad Y \in \text{\rm Dom}[L],
  \tag 3.90
  $$
 {\it and we define the operator norm of $L$ from $\Cal B_1$ into $\Cal B_2,$ denoted by $||L||_{\Cal B_2\,\Cal B_1}$  as }
  $$
 || L||_{\Cal B_2\,\Cal B_1} = \matrix \\
 \text{\rm sup}\\
 || Y||_{\Cal B_1}=1\endmatrix \| L Y  \|_{\Cal B_2}.\tag 3.91
  $$

   Since we assume that  the domain of  $L$ is dense, $L$ can be uniquely extended (taking the closure)  to a  bounded operator with domain $\Cal B_1$. Hence, we will always assume that the bounded operators have
   as their domain the whole Banach space.

\noindent {\bf Definition 3.11} {\it
Let $\Cal H$ be a Hilbert space.
The adjoint of an operator $L$ from $\Cal H$ into $\Cal H $, which we denote by $L^\dagger$, is defined as the map $L^\dagger: Y\mapsto W$ with the
domain given by}
$$
\text{\rm Dom}[L^\dagger] := \{  Y \in \Cal H: \ (Y, L V)_{\Cal H}=  (W,V)_{\Cal H}
\text{ for all } V \in \text{\rm Dom}[L] \}.\tag 3.92
 $$

Since $\text{\rm Dom}[L]$ is dense in
$\Cal H,$ the operator  $L^\dagger$ is well defined in the sense that there is at most only one $W\in\Cal H$ such that  $(Y, L V)_{\Cal H}=  ( W,V)_{\Cal H}$ for any $V \in \text{\rm Dom}[L]$.

\noindent {\bf Definition 3.12} {\it Let $\Cal H$ be a Hilbert space,
and let $L$ be an operator from $\Cal H$ into $\Cal H$. Then, we say that $L$   is symmetric if
 $ L \subset L^\dagger.$ We say that $L$ is selfadjoint if
 $
 L= L^\dagger.
 $}
%
%

 In applications in differential operators it is relatively simple to verify that an operator $L$ is symmetric. However, it is usually a rather delicate issue
 to verify that it is also selfadjoint, i.e. to verify that
 $\text{\rm Dom}[L^\dagger]= \text{\rm Dom}[L].$

\noindent {\bf Definition 3.13} {\it Let $\Cal H_1$ and
$\Cal H_2$ be two Hilbert spaces.
 An operator $U$ from $\Cal H_1$ into $\Cal H_2$ is said to be isometric if}
 $$
 \| U Y \|_ {\Cal H_2}=  \| Y \|_{\Cal H_1},\qquad Y \in \Cal H_1.\tag 3.93
 $$

It is known [27] that $U$ is isometric if and only if we have
 $$
 (U Y_1,UY_2)_ {\Cal H_2}= ( Y_1,Y_2)_ {\Cal H_1}, \qquad Y_1,Y_2 \in \Cal H_1.
 \tag 3.94
 $$
 Moreover,  $U$ is isometric if and only if $ U^\dagger\, U= I_ {\Cal H_1},$ where by $ I_ {\Cal H_1}$ we denote the identity operator in $\Cal H_1$.

\noindent {\bf Definition 3.14} {\it Let $\Cal H_1$ and
$\Cal H_2$ be two Hilbert spaces.
 An operator $U$ from $\Cal H_1$ into $\Cal H_2$ is unitary if it is isometric and onto, where the onto property
 is expressed as $\text{\rm Ran} [U]= \Cal H_2$.}

It is know [27] that $U$ is unitary if and only if $ U^\dagger\, U= I_ {\Cal H_1}$ and  $ U\, U^\dagger= I_ {\Cal H_2}$, with $I_ {\Cal H_1}$
and $I_ {\Cal H_2}$ denoting the identity operators on
$\Cal H_1$ and $\Cal H_2,$ respectively.

\vskip 10 pt
\noindent {\bf 3.h Other miscellaneous results}
\vskip 3 pt

We recall [1] that the Sobolev space $\bold H^1(\bR^+)$ consists of all vectors in $L^2(\bR^+)$ with also their first derivatives in  $L^2(\bR^+)$. In other words,
it consists of all square-integrable functions $X(y)$
defined for $y\in\bR^+$ where the derivative
$X'(y)$ exists in the distribution sense
and belongs to $L^2(\bR^+)$. We remark that $\bold H^1(\bR^+)$
is a Hilbert space with the scalar product
$$(X,Y)_{\bold H^1(\bR^+)}:=(X,Y)_2+(X',Y')_2.
\tag 3.95$$

We recall that the Riemann-Lebesgue lemma states that the Fourier
transform of an integrable function vanishes at infinity.
For example, if $g(x)$ appearing in (3.59) belongs to
$L^1(\bR^+),$ then $f(k)$ appearing in (3.59) vanishes
as $k\to\pm\infty$ according to the Riemann-Lebesgue lemma.

Let us introduce
$$\sigma(x):=\int_x^\infty dz\,|V(z)|,\quad
\sigma_1(x):=\int_x^\infty dz\,z\,|V(z)|,\qquad x\ge 0.\tag 3.96$$
Note that $\sigma(x)$ and $\sigma_1(x)$
are nonincreasing functions
of $x\in[0,+\infty),$ and it is readily seen that
$\sigma(0)$ and $\sigma_1(0)$ are both
finite when the potential satisfies (2.3).
As also shown on p. 68 of [2], we
have
$$x\,\sigma(x)\le \sigma_1(x),\quad \int_x^\infty dz\,\sigma(z)
\le \sigma_1(x).\tag 3.97$$
With the help of (3.96) we get
$$\int_0^\infty dz\,[\sigma(z)]^2\le \sigma(0)
\int_0^\infty dz\,\sigma(z)=\sigma(0)\,\sigma_1(0)<+\infty,\tag 3.98$$
$$\int_0^\infty dz\,z\,[\sigma(z)]^2\le
\int_0^\infty dz\,\sigma_1(z)\,\sigma(z)
\le \sigma_1(0)\int_0^\infty dz\,\sigma(z)=[\sigma_1(0)]^2<+\infty.\tag 3.99$$
The inequalities in (3.97)-(3.99) are used later on in some proofs.

\newpage
\noindent {\bf 4. THE FADDEEV CLASS AND THE MARCHENKO CLASS}
\vskip 3 pt

In this chapter, in preparation for a characterization of the scattering data,
 we introduce the Marchenko class of scattering data sets
 and the Faddeev class of input data sets. Our characterization
 basically consists of showing that there is a one-to-one correspondence
 between the Marchenko class and the Faddeev class.

In the direct scattering problem related to (2.1) and (2.4),
our input data set $\bold D$ is given by
$$\bold D:=\{V,A,B\},\tag 4.1$$
with the understanding that $V$ is equivalent to
the knowledge of the $n\times n$ matrix $V(x)$ for $x\in\bR^+,$ the boundary
matrices $A$ and $B$ are two constant $n\times n$
matrices and they are defined up to a postmultiplication by an invertible
$n\times n$ matrix $T.$
In the inverse scattering problem our scattering data set $\bold S$ is given by
$$\quad \bold S:=\left\{S,\{\kappa_j,M_j\}_{j=1}^N\right\},\tag 4.2$$
with the understanding that
$S$ is equivalent to the knowledge of the $n\times n$ scattering
matrix $S(k)$ specified
for $k\in\bR,$ the
$\kappa_j$ are $N$ distinct positive numbers related to the bound-state
energies $-\kappa_j^2,$ the $M_j$ are $N$
constant $n\times n$ matrices related to the
normalizations of matrix-valued bound-state wavefunctions, and
each $M_j$ is nonnegative and hermitian and has rank $m_j$
for some integer between $1$ and $n.$
Thus, we use $N$ to denote the number of bound states
without counting the multiplicities. We remark that
the possibility $N=0$ is included in our
consideration, in which case the scattering data set
consists of the scattering matrix alone. The finiteness of
$N$ is guaranteed [9] by (2.3). The integer $m_j$
corresponds to the multiplicity of the bound state
associated with $k=i\kappa_j,$ and hence
the integer $\Cal N$ defined as
the sum of the ranks of $N$
matrices $M_j,$ i.e.
$$\Cal N:=\ds\sum_{j=1}^N m_j,\tag 4.3$$
corresponds the total number of bound states including the multiplicities.

In our analysis of the direct problem
related to (2.1) and (2.4), we
assume that our input data set $\bold D$ belongs to the {\it Faddeev class}
defined below.

\noindent {\bf Definition 4.1} {\it The input data set $\bold D$ given in
(4.1) is said to belong to the Faddeev class $\Cal F$ if
the potential $V$ satisfies (2.2) and (2.3) and the matrices
$A$ and $B$ satisfy (2.5) and (2.6).
In other words,
$\bold D$ belongs to the Faddeev class if the
$n\times n$ matrix-valued
potential $V$ appearing in (2.1)
is hermitian and belongs to class
$L^1_1(\bR^+)$ and the constant $n\times n$
matrices $A$ and $B$ appearing in
(2.4) satisfy (2.5) and (2.6).}

In our monograph, we provide various equivalent formulations of the
characterization of the scattering data set
$\bold S$ given in (4.2) when the corresponding $\bold D$ in (4.1)
belongs to the Faddeev class. In order to state those various conditions
in an efficient manner, we first introduce a set of
properties that are all indicated with an Arabic numeral.

\noindent {\bf Definition 4.2} {\it The properties $(\bold 1)$, $(\bold 2)$, $(\bold 3_a)$,
$(\bold 3_b)$, $(\bold 4_a)$, $(\bold 4_b)$, $(\bold 4_c)$, $(\bold 4_d)$, $(\bold 4_e)$ for the scattering data set $\bold S$ in (4.2) are defined as follows:}

\item{$(\bold 1)$} {\it The scattering matrix $S(k)$ satisfies}
$$S(-k)=S(k)^\dagger=S(k)^{-1},\qquad k\in\bR,\tag 4.4$$
{\it and there exist constant $n\times n$ matrices
$S_\infty$ and $G_1$ in such a way that}
$$S(k)=S_\infty+\ds\frac{G_1}{ik}+o\left(\ds\frac{1}{k}\right),\qquad k\to\pm\infty.
\tag 4.5$$
{\it The quantity $S(k)-S_\infty$ is the Fourier transform
of an $n\times n$ matrix $F_s(y)$ in such a way that
$F_s(y)$ is bounded in $y\in\bR$ and integrable in $y\in\bR^+.$ Thus, the constant matrix
$S_\infty$ is obtained from the
scattering matrix $S(k)$ via}
$$S_\infty:=\lim_{k\to\pm\infty} S(k),\tag 4.6$$
{\it and the quantity $F_s(y)$ is related to
$S(k)$ as}
$$F_s(y):=\ds\frac{1}{2\pi} \int_{-\infty}^\infty dk\,[S(k)-S_\infty]\,e^{iky},
\qquad y\in\bR.
\tag 4.7$$

\item{$(\bold 2)$} {\it For the matrix $F_s(y)$ defined in
(4.7), the derivative $F_s'(y)$ exists a.e. for $y\in\bR^+$ and it satisfies}
$$\int_0^\infty dy\,(1+y)\,|F'_s(y)|<+\infty,\tag 4.8$$
{\it where we recall that the norm in (4.8) is the
operator norm of a matrix.}

\item{$(\bold 3_a)$} {\it The physical solution $\Psi(k,x)$ satisfies
the boundary condition (2.4), i.e.
$$-B^\dagger \Psi(k,0)+A^\dagger \Psi'(k,0)=0,\qquad k\in\bR.\tag 4.9$$
We clarify this
property as follows: The scattering matrix appearing in $\bold S$ yields a
particular $n\times n$
matrix-valued solution $\Psi(k,x)$
to (2.1) known as the physical solution given in (9.4)
and also yields a pair of matrices $A$ and $B$ (modulo an invertible
matrix) satisfying (2.5) and (2.6).
Our statement $(\bold 3_a)$ is equivalent
to saying that (2.4) is satisfied if we use in (2.4) the quantities
$\Psi(k,x),$ $A,$ and $B$ constructed from $S(k)$ appearing in $\bold S.$}

\item{$(\bold 3_b)$} {\it The Jost matrix $J(k)$ satisfies}
$$J(-k)+S(k)\,J(k)=0,\qquad k\in\bR.\tag 4.10$$
{\it We clarify this property as follows:
The scattering matrix $S(k)$ given in $\bold S$
yields a Jost matrix
$J(k)$ constructed as in (9.2), unique up to a post multiplication by an invertible matrix.
Using the scattering matrix $S(k)$ given in
$\bold S$ and the Jost matrix constructed from $S(k),$ we find that (4.10)
is satisfied.}

\item{$(\bold 4_a)$} {\it The Marchenko equation (13.1) at $x=0$ given by
$$K(0,y)+F(y)+\int_0^\infty dz\,K(0,z)\,F(z+y)=0,\qquad y\in\bR^+,\tag 4.11$$
has a unique solution $K(0,y)$ in
$L^1(\bR^+).$
Here, $F(y)$ is the $n\times n$
matrix related
to $F_s(y)$ given in (4.7) as}
$$F(y):=F_s(y)+\ds\sum_{j=1}^N M_j^2 \,e^{-\kappa_j y},
\qquad y\in\bR^+.\tag 4.12$$

\item{$(\bold 4_b)$} {\it The only solution in $L^1(\bR^+)$ to the
homogeneous Marchenko integral equation at $x=0$ given by
$$K(0,y)+\int_0^\infty dz\,K(0,z)\,F(z+y)=0, \qquad y\in\bR^+,
\tag 4.13$$
which is the homogeneous version of
the Marchenko
equation at $x=0$ given by (4.11), is the trivial solution $K(0,y)\equiv 0.$
Here, $F(y)$ is the quantity defined in (4.12).}

\item{$(\bold 4_c)$} {\it The only integrable solution
$X(y),$ which is a row vector with $n$
integrable components in $y\in\bR^+,$ to the linear homogeneous integral equation}
$$X(y)+\int_0^\infty dz\,X(z)\,F(z+y)=0,\qquad y\in\bR^+,\tag 4.14$$
{\it is the trivial solution $X(y)\equiv 0.$
Here, $F(y)$ is the quantity defined in (4.12).}

\item{$(\bold 4_d)$} {\it
The only solution $\hat X(k)$ to the system}
$$\cases \hat X(i\kappa_j)\,M_j=0,\qquad j=1,\dots,N,\\
\stretch
\hat X(-k)+\hat X(k)\,S(k)=0,\qquad k\in\bR,\endcases\tag 4.15$$
{\it where $\hat X(k)$ is a row vector
with $n$ components belonging
to the class $\hat L^1(\bCp),$ is the trivial
solution $\hat X(k)\equiv 0.$
}

\item{$(\bold 4_e)$} {\it
The only solution $h(k)$ to the system}
$$\cases M_j\, h(i\kappa_j)=0,\qquad j=1,\dots,N,\\
\stretch
h(-k)+S(k)\,h(k)=0,\qquad k\in\bR,\endcases\tag 4.16$$
{\it where $h(k)$ is a column vector
with $n$ components belonging
to the class $\hat L^1(\bCp),$ is the trivial
solution $h(k)\equiv 0.$
}

Let us make some comments on the condition $(\bold 1)$ of Definition~4.2 above.
The property (4.5) implies that $S(k)-S_\infty$ is square integrable in $k\in\bR$
and hence also implies $F_s(y)\in L^2(\bR^+),$ and in fact (4.5)
contains even more information. The quantity $G_1$ and hence (4.5) itself
are used to construct the boundary matrices $A$ and $B$ appearing in (2.4)-(2.6).
On the other hand, for the construction of the potential and various other quantities
the square integrability of $S(k)-S_\infty$ is sufficient. So that we can use such
a weaker condition to obtain various results, we specify the property
$(\bold I)$ in the following definition. We also introduce
various other properties, which are denoted by using a Roman numeral.

\noindent {\bf Definition 4.3} {\it The properties $(\bold I)$, $(\bold {III}_a)$,
$(\bold {III}_b)$, $(\bold {III}_c)$, $(\bold {V}_a)$, $(\bold {V}_b)$, $(\bold {V}_c)$, $(\bold {V}_d)$, $(\bold {V}_e)$, $(\bold {V}_f)$, $(\bold {V}_g)$, $(\bold {V}_h)$, and $(\bold {VI})$ for the scattering data set $\bold S$ in (4.2) are defined as follows:}

\item{$(\bold I)$} {\it The scattering matrix $S(k)$ satisfies
(4.4), the quantity $S_\infty$ defined in
(4.6) exists, the quantity $S(k)-S_\infty$ is square integrable
in $k\in\bR,$ and the quantity $F_s(y)$ defined in
(4.7) is bounded in $y\in\bR$ and integrable in $y\in\bR^+.$}

\item{$(\bold {III}_a)$} {\it For the function $F_s(y)$ given in (4.7),
the derivative
$F_s'(y)$ for $y\in\bR^-$
can be written as the sum of two functions, one of which
is integrable and the other is square integrable in $y\in\bR^-.$
Furthermore, the only solution
$X(y),$ which is a row vector with $n$ square-integrable
components in $y\in\bR^-,$ to the linear
homogeneous integral equation}
$$-X(y)+\int_{-\infty}^0 dz\,X(z)\,F_s(z+y)=0,\qquad
y\in\bR^-,\tag 4.17$$
{\it is the trivial solution $X(y)\equiv 0.$
}

\item{$(\bold {III}_b)$} {\it The only solution $\hat X(k)$ to the homogeneous Riemann-Hilbert problem}
$$-\hat X(-k)+\hat X(k)\,S(k)=0,\qquad k\in\bR,\tag 4.18$$
{\it where $\hat X(k)$ is a row vector
with $n$ components belonging
to the class $\bold H^2(\bCm),$ is the trivial
solution $\hat X(k)\equiv 0.$}

\item{$(\bold {III}_c)$} {\it The only solution $h(k)$ to the homogeneous Riemann-Hilbert problem}
$$-h(-k)+S(k)\, h(k)=0,\qquad k\in\bR,\tag 4.19$$
{\it where $h(k)$ is a column vector
with $n$ components belonging
to the class $\bold H^2(\bCm),$ is the trivial
solution $h(k)\equiv 0.$}

\item{$(\bold {V}_a)$} {\it Each of the $N$ normalized bound-state matrix solutions $\Psi_j(x)$
constructed as in (9.8) satisfies the boundary condition (2.4), i.e.
$$-B^\dagger \Psi_j(0)+A^\dagger \Psi'_j(0)=0,
\qquad j=1,\dots,N.\tag 4.20$$
We clarify this statement as follows. The scattering matrix $S(k)$ and the bound-state data $\{\kappa_j,M_j\}_{j=1}^N$
given in $\bold S$ yield $n\times n$ matrices $\Psi_j(x)$ as in
(9.8),
where each $\Psi_j(x)$ is a solution to
(2.1) at $k=i\kappa_j.$
As stated
in $(\bold 3_b)$ of Definition~4.2, the scattering matrix given in $\bold S$ yields a pair of matrices $A$ and $B$ (modulo an invertible
matrix) satisfying (2.5) and (2.6). The statement $(\bold {V}_a)$ is equivalent
to saying that (2.4) is satisfied if we use in (2.4) the quantities
$\Psi_j(x),$ $A,$ and $B$ constructed from the quantities appearing in $\bold S.$}

\item{$(\bold {V}_b)$} {\it The normalization matrices
$M_j$ appearing in $\bold S$ satisfy}
$$J(i\kappa_j)^\dagger M_j=0,
\qquad j=1,\dots,N.\tag 4.21$$
{\it We clarify this condition as follows: As indicated in $(\bold 3_b)$
in Definition~4.2,
the scattering matrix $S(k)$ given in $\bold S$
yields a Jost matrix
$J(k).$ Using in (4.21) the matrix $M_j$ given in
$\bold S$ and the Jost matrix constructed from $S(k),$ at each
$\kappa_j$-value listed in $\bold S$ the matrix
equation (4.21) holds.}

\item{$(\bold {V}_c)$} {\it The linear homogeneous integral equation
$$X(y)+\int_0^\infty dz\,X(z)\,F_s(z+y)=0,\qquad y\in\bR^+,\tag 4.22$$
has precisely $\Cal N$ linearly independent row vector
solutions with $n$ components which are integrable in $y\in\bR^+.$
Here $F_s(y)$ is the matrix defined in (4.7).
}

\item{$(\bold {V}_d)$} {\it The homogeneous Riemann-Hilbert problem given by}
$$\hat X(-k)+\hat X(k)\,S(k)=0,\qquad k\in\bR,\tag 4.23$$
{\it has precisely $\Cal N$ linearly independent
row vector solutions whose $n$ components belong
to the class $\hat L^1(\bCp).$
}

\item{$(\bold {V}_e)$} {\it The homogeneous Riemann-Hilbert problem given by}
$$h(-k)+S(k)\,h(k)=0,\qquad k\in\bR,\tag 4.24$$
{\it has precisely $\Cal N$ linearly independent
column vector solutions whose $n$ components belong
to the class $\hat L^1(\bCp).$
}

\item{$(\bold {V}_f)$} {\it The number of linearly independent square-integrable
solutions
$X(y)$ to (4.22) in $y\in\bR^+$ is equal to the nonnegative integer
$\Cal N$ given in (4.3). The matrix $F_s(y)$
appearing in the kernel of (4.22)
is defined in (4.7).}

\item{$(\bold {V}_g)$} {\it The homogeneous Riemann-Hilbert problem given
in (4.23)
has precisely $\Cal N$ linearly independent
row vector solutions whose $n$ components belong
to the class $\bold H^2(\bCp),$ where
$\Cal N$ is the nonnegative integer specified in (4.3).}

\item{$(\bold {V}_h)$} {\it The homogeneous Riemann-Hilbert problem given
in (4.24)
has precisely $\Cal N$ linearly independent
row vector solutions whose $n$ components belong
to the class $\bold H^2(\bCp),$ where
$\Cal N$ is the nonnegative integer specified in (4.3).}

\item{$(\bold {VI})$} {\it The scattering matrix $S(k)$ is continuous in $k\in\bR.$}


As a result of the square integrability of
$S(k)-S_\infty$ stated in $(\bold I)$ of Definition~4.3
it follows that the Fourier transform $F_s(y)$ given in (4.7)
is square integrable in $y\in\bR.$ For easy
referencing we state the result in the following proposition.

\noindent {\bf Proposition 4.4} {\it Consider a scattering data set $\bold S$
as in (4.2), which consists of
an $n\times n$ scattering matrix $S(k)$ for $k\in\bR,$ a set of $N$ distinct
 positive constants $\kappa_j,$ and a set of
$N$ constant $n\times n$ hermitian and nonnegative matrices
$M_j$ with respective positive ranks $m_j,$ where $N$ is a nonnegative integer.
If $\bold S$ satisfies either $(\bold I)$
of Definition~4.3 or
$(\bold 1)$ of Definition~4.2,
then $F_s(y)$ defined in (4.7)
is bounded in $y\in\bR,$
integrable in $y\in\bR^+,$ and
square integrable in $y\in\bR.$}

Having defined various properties for the scattering data set $\bold S,$
we are able to formulate various characterizations for $\bold S$ so that
it corresponds to a unique input data set $\bold D$ in the Faddeev class
specified in Definition~4.1. Such a typical characterization has
the form $(\bold 1,\bold 2,\bold 3,\bold 4),$ by which we mean
$\bold S$ satisfies the properties $(\bold 1),$ $(\bold 2),$ either of the
two properties $(\bold 3_a)$ or $(\bold 3_b),$ and
any one of the five properties listed as $(\bold 4_a),$ $(\bold 4_b),$ $(\bold 4_c),$ $(\bold 4_d),$ $(\bold 4_e).$ Another typical characterization
has the form $(\bold 1,\bold 2,\bold {III}+\bold V,\bold 4),$ by which we mean
$\bold S$ satisfies $(\bold 1,\bold 2,\bold 3,\bold 4),$ except for the
fact that instead of satisfying either of the two properties
$(\bold 3_a)$ or $(\bold 3_b),$ it instead satisfies two
other properties, the first being one of the three properties
$(\bold {III}_c)$, $(\bold {III}_c)$, $(\bold {III}_c)$
and the second being one of the eight properties
$(\bold {V}_a)$, $(\bold {V}_b)$, $(\bold {V}_c)$, $(\bold {V}_d)$, $(\bold {V}_e),$ $(\bold {V}_f),$ $(\bold {V}_g,)$ $(\bold {V}_h).$
The notation used, although maybe awkward, enables us to streamline all the characterizations of $\bold S.$ One such characterization, namely
that $\bold S$ satisfying $(\bold 1,\bold 2,\bold 3_a,\bold 4_a),$
enables us to define the {\it Marchenko class}
of scattering data sets as follows.

\noindent {\bf Definition 4.5} {\it Consider a scattering data set $\bold S$
as in (4.2), which consists of
an $n\times n$ scattering matrix $S(k)$ for $k\in\bR,$ a set of $N$ distinct
 positive constants $\kappa_j,$ and a set of
$N$ constant $n\times n$ hermitian and nonnegative matrices
$M_j$ with respective positive ranks $m_j,$ where $N$ is a nonnegative integer.
We say that $\bold S$ belongs to the Marchenko class $\Cal M$ if $\bold S$
satisfies
$(\bold 1,\bold 2,\bold 3_a,\bold 4_a),$ i.e. if
$\bold S$ satisfies the four properties
$(\bold 1),$ $(\bold 2),$ $(\bold 3_a),$ $(\bold 4_a)$
specified in Definition~4.2.}

As stated in Theorem~5.1 in the next chapter we prove that if $\bold D$ belongs to
the Faddeev class then there exists and uniquely exists a
corresponding
scattering data set $\bold S$ in the Marchenko class. We provide
the steps to construct $\bold S$ when $\bold D$ is given.
Furthermore, we show that,
for each scattering data set $\bold S$ in the Marchenko class,
there exists and uniquely exists a corresponding input data set $\bold D$
in the Faddeev class. We provide
the steps to construct $\bold D$ when $\bold S$ is given.

Let us comment on the condition $(\bold 3_a)$ of Definition~4.2,
which states that the physical solution $\Psi(k,x)$ constructed from
the scattering data set $\bold S$ must satisfy the boundary
condition (2.4) where $A$ and $B$ are the boundary matrices constructed
 from $\bold S.$ One may then question why we do not include in Definition~4.5
a separate condition that the bound-state solutions $\Psi_j(x)$
constructed as in (9.8) are required to satisfy the boundary
condition (2.4). The answer is given in the following proposition, which
indicates that the fulfilment of $(\bold 3_a)$ actually implies that
the constructed bound-state solutions
indeed satisfy the boundary
condition (2.4).


\noindent {\bf Proposition 4.6} {\it Consider a scattering data set $\bold S$
as in (4.2), which consists of
an $n\times n$ scattering matrix $S(k)$ for $k\in\bR,$ a set of $N$ distinct
 positive constants $\kappa_j,$ and a set of
$N$ constant $n\times n$ hermitian and nonnegative matrices
$M_j$ with respective positive ranks $m_j,$ where $N$ is a nonnegative integer.
Assume that $\bold S$ belongs to the Marchenko class, i.e. it satisfies the
conditions $(\bold 1)$, $(\bold 2)$, $(\bold 3_a)$, $(\bold 4_a)$
stated in Definition~4.5. Then, we also have $(\bold V_a)$ holding.
We clarify this statement as follows. The scattering matrix $S(k)$ and the bound-state data $\{\kappa_j,M_j\}_{j=1}^N$
given in $\bold S$ yield $n\times n$ matrices $\Psi_j(x)$ as in
(9.8),
where each $\Psi_j(x)$ is a solution to
(2.1) at $k=i\kappa_j.$
As stated
in $(\bold 3_a)$ of Definition~4.2, the scattering matrix given in $\bold S$ yields a pair of matrices $A$ and $B$ (modulo an invertible
matrix) satisfying (2.5) and (2.6). Our statement $(\bold V_a)$ is equivalent
to saying that (2.4) is satisfied if we use in (2.4) the quantities
$\Psi_j(x),$ $A,$ and $B$ constructed from the quantities appearing in $\bold S.$}

\noindent PROOF: The result follows from Proposition~18.2(a). \qed

Let us comment on Proposition~4.6. It states that, if the scattering
data set $\bold S$ belongs to the Marchenko class, then $(\bold 3_a)$ in Definition~4.2
implies $(\bold V_a)$ of Definition~4.3. As we show in Example~26.5, if
$\bold S$ does not satisfy $(\bold 3_a),$ and hence
if $\bold S$ does not belong to the Marchenko class, it may still be possible that
$(\bold V_a)$ holds even though $(\bold 3_a)$ does not hold.
We remark that, in the absence of bound states, the condition
$(\bold V_a)$ stated in Proposition~4.6 becomes redundant.

\newpage
\noindent {\bf 5. THE CHARACTERIZATION OF THE SCATTERING DATA}
\vskip 3 pt

Next we present one of our main results by showing that
the four conditions given in Definition~4.5 for the Marchenko class
form a characterization of the scattering data sets $\bold S$
that have one-to-one
correspondence with the input data sets $\bold D$
in the Faddeev class specified in Definition~4.1.

\noindent {\bf Theorem 5.1} {\it Consider a scattering data set $\bold S$
as in (4.2), which consists of
an $n\times n$ scattering matrix $S(k)$ for $k\in\bR,$ a set of $N$ distinct
 positive constants $\kappa_j,$ and a set of
$N$ constant $n\times n$ hermitian and nonnegative matrices
$M_j$ with respective positive ranks $m_j,$ where $N$ is a nonnegative integer.
Consider also an input data set
$\bold D$ as in (4.1) consisting of an $n\times n$ matrix potential
$V$ satisfying (2.2) and (2.3) and a pair of constant
$n\times n$ matrices $A$ and $B$ satisfying (2.5) and (2.6).
Let $\Cal F$ be the Faddeev class of input data sets $\bold D,$ as specified
in Definition~4.1. Let $\Cal M$ be the Marchenko class of
scattering data sets $\bold S,$ as specified in
Definition~4.5.
Then, we have the following:}

\item{(a)} {\it For each $\bold D\in\Cal F,$ there exists and uniquely
exists a scattering data set $\bold S\in\Cal M.$}

\item{(b)} {\it Conversely,
for each $\bold S\in\Cal M,$ there exists
and uniquely
exists an input data set $\bold D\in\Cal F,$ where $A$ and $B$ are uniquely determined up to a postmultiplication
by an invertible $n\times n$ matrix $T.$}

\item{(c)} {\it Let $\tilde {\bold S}$ be the scattering data set corresponding
to $\bold D$ given in (b), where $\bold D$ is constructed from
the scattering data set $\bold S.$ Then, we must have $\tilde \bold S=\bold S,$ i.e.
the scattering data set constructed from $\bold D$ must be equal to
the scattering data set used to construct $\bold D.$}

\item{(d)} {\it
The characterization outlined in (a)-(c) can equivalently be stated as follows.
A set $\bold S$ as in (4.2) is the scattering data set
corresponding to an input data set $\bold D$ in the Faddeev class
if and only if
$\bold S$ satisfies $(\bold 1)$, $(\bold 2)$, $(\bold 3_a)$, and $(\bold 4_a)$
stated in Definition~4.5.}

\noindent PROOF: First, when $\bold D$ belongs to
the Faddeev class, the corresponding $\bold S$ belongs to
the Marchenko class, as proved in Theorem~15.10.
Let us now turn to the inverse problem. The
unique construction of $\bold D$ from
$\bold S$ is outlined in Chapter~16,
and in particular the uniqueness of the construction
is proved in Proposition~16.2. As indicated
in Proposition~16.1(a), when $(\bold I)$ is satisfied, the
Marchenko equation (13.1) is uniquely solvable and
hence its solution $K(x,y)$ is uniquely constructed
for each $x\in\bR^+.$ The condition $(\bold 4_a)$ assures that
the Marchenko equation (13.1) is uniquely solvable also
at $x=0.$ The potential $V(x)$ is then uniquely constructed from
$K(x,y)$ as in (10.4). As indicated in Proposition~16.11(a)
the condition $(\bold 1)$ ensures that the constructed
potential $V(x)$ satisfies (2.2), and as indicated in Proposition~16.11(a)
the condition $(\bold 2)$ ensures that the constructed
potential $V(x)$ satisfies (2.3). The boundary matrices
$A$ and $B$ are uniquely constructed (modulo $T$),
as indicated in Proposition~16.9, and they satisfy (2.5) and (2.6).
The physical solution $\Psi(k,x)$ is constructed
as in (9.4). As indicated in Proposition~16.11(b),
the constructed $\Psi(k,x)$ satisfies (2.1), and
the condition $(\bold 3_a)$ assures that
$\Psi(k,x)$ satisfies the boundary condition (2.4).
We still need to show that, in the direct scattering problem
the constructed $\bold D$ yields the same
$\bold S$ that is used as input to the inverse scattering problem.
Mathematically speaking, we have $\bold D=\Cal D^{-1}(\bold S)$
and we need to show that $\Cal D(\bold D)=\bold S.$
Let us use a tilde to denote any quantity constructed
by using $\Cal D$ as input in the direct scattering map.
Thus, we have $\tilde {\bold S}:=\Cal D(\bold D)$ and we would like to show
that $\tilde {\bold S}=\bold S.$
The proof is obtained as follows.
We claim that the Jost solution $f(k,x)$ constructed from
$\bold S$ as in (10.6) is the same as the Jost solution
$\tilde f(k,x)$ obtained from $V$ as in [2,5,24]
$$\tilde f(k,x)=e^{ikx}I+\ds\frac{1}{k}\int_x^\infty dy\,
\left[\sin k(y-x)\right]\,V(y)\,\tilde f(k,y).\tag 5.1$$
This is true because, as indicated in Proposition~16.11(a),
$f(k,x)$ satisfies (2.1) and (9.1) when $V$ is used
as the potential in (2.1). It is already known [2,5,24]
that $\tilde f(k,x)$ given in (5.1) satisfies (2.1) and (9.1) when $V$ is used
as the potential in (2.1).
Thus, as the solution
$\tilde f(k,x)$ to (5.1) is unique,
we have $\tilde f(k,x)\equiv f(k,x).$
We then have the equivalence of the Jost matrices,
i.e. $\tilde J(k)\equiv J(k)$ because
both $J(k)$ and $\tilde J(k)$ are constructed
as in (9.2) using the same $f(k,x)$ and the same
$A$ and $B$ as input.
 As (9.3) implies, we must
also have the equivalence of the
scattering matrices, i.e. $\tilde S(k)\equiv S(k).$ Then, from (4.6)
it follows that we have the equivalence
of the large $k$-limits of the scattering matrix,
i.e. $\tilde S_\infty=S_\infty.$ Because of (4.7), we get
the equivalence $\tilde F_s(y)\equiv F_s(y).$
We already have $\tilde K(x,y)\equiv K(x,y),$
where $\tilde K(x,y)$ constructed from $\tilde f(k,x),$
or equivalently from $f(k,x),$ as
in (10.1). For each $x\in[0,+\infty),$ as indicated in Proposition~16.1(b),
the quantity
$K(x,y)$ is integrable in $y\in\bold R^+.$ Thus, we can
view the Marchenko integral equation (13.1) with
$K(x,y)$ as input and $F(y)$ as the unknown. As indicated in the proof
of Theorem~13.2, the Marchenko equation then yields $F(y)$ uniquely
with the properties stated in Proposition~16.7.
Consequently, we have $\tilde F(y)\equiv F(y).$ Then, from (4.7) and
(4.12) we obtain
$$\ds\sum_{j=1}^{\tilde N} \tilde M_j^2\, e^{-\tilde \kappa_j y}=\ds\sum_{j=1}^N M_j^2\, e^{-\kappa_j y},\qquad y\in\bold R^+.\tag 5.2$$
Since (5.2) holds for all $y\in\bR^+,$
we can use a recursive argument to prove that
$\tilde N=N$ and also
 $\tilde \kappa_j=\kappa_j$
for $j=1,\dots,N.$ Because $\kappa_j$-values are
distinct, the functions $e^{-\kappa_j y}$ are linearly
independent in $y\in\bR^+,$ and hence
(5.2) also yields $\tilde M_j^2=M_j^2.$ Then, from the
nonnegativity of $M_j$ it follows that we must have
$\tilde M_j=M_j.$ Thus, we have proved that $\tilde{\bold S}=\bold S.$ \qed

Let us now provide some remarks on the four
characterization conditions appearing in Definition~4.5.
First, let us make some remarks concerning
$(\bold 1)$ and $(\bold 4_a)$ of Definition~4.2.
As shown in Example~26.5, even if the unitarity of $S(k)$ in $(\bold 1)$ does not hold, corresponding to
$\bold S$ we may be able to construct
a unique $\bold D$ in the Faddeev class. However, the scattering data set
 $\tilde \bold S$ corresponding to
$\bold D$ may not agree with $\bold S,$ a point emphasized in the proof
of Theorem~5.1. Concerning the symmetry relation of $(\bold 1)$, namely
$S(-k)=S(k)^\dagger$ for $k\in\bR$ stated in the first equality in
(4.4), we have the following remark.
If that symmetry relation does not hold, the hermitian property of
various constructed quantities fails, and the constructed potential may not be hermitian
and the constructed Schr\"odinger operator may no longer be selfadjoint.
The condition that $F_s(y)$ is bounded and integrable in
$y\in\bR$ appearing in $(\bold 1)$ assures that the Marchenko integral operator
related to (13.1) is compact and the Marchenko integral
equation (13.1), for each $x\in\bR^+,$ has a unique solution
$K(x,y)$ for $y\in(x,+\infty)$ and the potential $V(x)$
can be constructed. However, without the additional assumption
$(\bold 4_a)$, it is possible that, for $x=0,$ the Marchenko integral equation may
not have a solution and hence the constructed potential may be singular
at $x=0,$ as illustrated in Examples~26.7 and 26.14. The property
(4.5) of $(\bold 1)$ enables us to construct the boundary matrices
$A$ and $B.$

Let us now comment on $(\bold 2)$ of Definition~4.2.
The condition $(\bold 2)$ assures us that the constructed
potential $V$ belongs to class $L^1_1(\bR^+).$ Let us remark that
the characterization result
stated in Theorem~5.1 involving the Faddeev class and
the Marchenko class still holds if we modify the definitions
of the Faddeev class and the Marchenko class in such a way that
we replace the $L^1_1(\bR^+)$ condition given in (2.3) by
$$\int_0^\infty
dx\,(1+x)^p\,|V(x)|<+\infty,\tag 5.3$$
and replace (4.8) by
$$\int_0^\infty dy\,(1+y)^p\,|F'_s(y)|<+\infty,\tag 5.4$$
where $p$ is any positive integer. We can summarize this by stating that
the $L^1_1(\bR^+)$ characterization provided in our monograph
extends to the $L^1_p(\bR^+)$ characterization for any positive integer
$p.$ In fact, the idea behind the $L^1_2(\bR)$ characterization
given by Deift and Trubowitz in [16]
for the full-line scalar Schr\"odinger equation
was to modify the $L^1_1(\bR)$ characterization
given by Faddeev in [20].

\newpage
\noindent {\bf 6. EQUIVALENTS FOR SOME CHARACTERIZATION CONDITIONS}
\vskip 3 pt

The following result shows that
the properties $(\bold 4_a),$ $(\bold 4_b),$ $(\bold 4_c),$ $(\bold 4_d),$
$(\bold 4_e)$ presented in Definition~4.2 are all equivalent.

\noindent {\bf Proposition 6.1} {\it Consider a scattering data set $\bold S$
as in (4.2), which consists of
an $n\times n$ scattering matrix $S(k)$ for $k\in\bR,$ a set of $N$ distinct
 positive constants $\kappa_j,$ and a set of
$N$ constant $n\times n$ hermitian and nonnegative matrices
$M_j$ with respective positive ranks $m_j,$ where $N$ is a nonnegative integer.
Assume that $\bold S$ satisfies $(\bold I)$ of Definition~4.3. Then, the property $(\bold 4_a)$
is equivalent
to any of the four properties $(\bold 4_c)$, $(\bold 4_d)$, $(\bold 4_e)$, $(\bold 4_b)$.}

\noindent PROOF: The
equivalences among $(\bold 4_c)$, $(\bold 4_d)$, and $(\bold 4_e)$ are established in
(c) and (d) of Proposition~15.3.
Note that $(\bold 4_c)$ and $(\bold 4_b)$ are equivalent
because, comparing (4.14) and (4.13) we see that each row of the matrix solution $K(0,y)$ to (4.13)
is a row vector solution
$X(y)$ to (4.14) and, conversely, each
row vector solution $X(y)$ is a row of the matrix solution $K(0,y)$ to (4.13).
Finally, the equivalence of $(\bold 4_a)$ and $(\bold 4_b)$ is established
in Proposition~16.1(c). \qed

In the next proposition we show the equivalences among
$(\bold{III}_a),$ $(\bold{III}_b),$ and $(\bold{III}_c)$
of Definition~4.3.

\noindent {\bf Proposition 6.2} {\it Consider a scattering data set $\bold S$
as in (4.2), which consists of
an $n\times n$ scattering matrix $S(k)$ for $k\in\bR,$ a set of $N$ distinct
 positive constants $\kappa_j,$ and a set of
$N$ constant $n\times n$ hermitian and nonnegative matrices
$M_j$ with respective positive ranks $m_j,$ where
$N$ is a nonnegative integer.
Assume that $\bold S$ satisfies
$(\bold I)$ of Definition~4.3. Then, the properties $(\bold{III}_a),$ $(\bold{III}_b),$ and $(\bold{III}_c)$ are equivalent.}

\noindent PROOF: The result is a direct consequence of Proposition~15.11. \qed

Next we show that, when the scattering data set $\bold S$ belongs
 to the Marchenko class, it is possible to replace $(\bold 3_a)$ and
 $(\bold 4_a)$ by various equivalent conditions.

\noindent {\bf Proposition 6.3} {\it Consider a scattering data set $\bold S$
as in (4.2), which consists of
an $n\times n$ scattering matrix $S(k)$ for $k\in\bR,$ a set of $N$ distinct
 positive constants $\kappa_j,$ and a set of
$N$ constant $n\times n$ hermitian and nonnegative matrices
$M_j$ with respective positive ranks $m_j,$ where $N$ is a nonnegative integer.
Assume that $\bold S$ satisfies the three conditions
$(\bold 1)$, $(\bold 2)$, $(\bold 4_a)$
stated in Definition~4.2. Then, the condition $(\bold 3_a)$ for the Marchenko class
is equivalent to $(\bold 3_b)$ of Definition~4.2}

\noindent PROOF: Note that $(\bold 3_a)$ in Definition~4.2 is given in (4.9), where
where $A$ and $B$ are the boundary matrices constructed from $\bold S$
as described in Proposition~16.9 and $\Psi(k,x)$ is the physical
solution constructed as in (9.4). Using (9.4) in (4.9) we obtain
$$-\left[ B^\dagger \,f(-k,0)-A^\dagger \, f'(k,0)\right]
-\left[ B^\dagger \,f(-k,0)-A^\dagger \, f'(k,0)\right] S(k)=0.\tag 6.1$$
With the help of (9.2) we can construct the Jost matrix $J(k),$
and then we can write (6.1) in terms of the constructed
$J(k)$ constructed as
$$-J(k)^\dagger-J(-k)^\dagger S(k)=0,\qquad k\in\bR.\tag 6.2$$
As a result of the first equality in (4.4), we can replace
$S(k)$ by $S(-k)^\dagger$ and hence rewrite (4.23) as
$$-\left[J(-k)+S(k)\,J(k)\right]^\dagger=0,\qquad k\in\bR,\tag 6.3$$
which is equivalent to (4.10). \qed

In the next proposition we show that
the property $(\bold 3_a)$ is equivalent to
the combination of the two properties $(\bold{III}_a)$
and $(\bold{V}_a).$

\noindent {\bf Proposition 6.4} {\it Consider a scattering data set $\bold S$
as in (4.2), which consists of
an $n\times n$ scattering matrix $S(k)$ for $k\in\bR,$ a set of $N$ distinct
 positive constants $\kappa_j,$ and a set of
$N$ constant $n\times n$ hermitian and nonnegative matrices
$M_j$ with respective positive ranks $m_j,$ where
$N$ is a nonnegative integer.
Assume that $\bold S$ satisfies
$(\bold 1)$, $(\bold 2)$, and $(\bold 4_a)$
of Definition~4.2. Then, the property $(\bold 3_a)$ is equivalent to
the combination of the two properties $(\bold{III}_a)$
and $(\bold{V}_a).$}

\noindent PROOF: The result is a direct consequence of Proposition~18.2. \qed

Next we show that the properties $(\bold V_a),$ $(\bold V_b),$ $(\bold V_c),$ $(\bold V_d),$ $(\bold V_e),$ $(\bold V_f),$ $(\bold V_g),$ $(\bold V_h)$
listed in Definition~4.3 are all equivalent.

\noindent {\bf Proposition 6.5} {\it Consider a scattering data set $\bold S$
as in (4.2), which consists of
an $n\times n$ scattering matrix $S(k)$ for $k\in\bR,$ a set of $N$ distinct
 positive constants $\kappa_j,$ and a set of
$N$ constant $n\times n$ hermitian and nonnegative matrices
$M_j$ with respective positive ranks $m_j,$ where
$N$ is a nonnegative integer.
Assume that $\bold S$ satisfies
$(\bold 1)$, $(\bold 2)$, and $(\bold 4_a)$
of Definition~4.2 as well as
$(\bold{III}_a)$. Then, the property $(\bold V_a)$
appearing in Proposition~4.6 is equivalent
to any of the seven properties $(\bold V_b)$, $(\bold V_c)$, $(\bold V_d)$, $(\bold V_e)$, $(\bold V_f)$, $(\bold V_g)$, $(\bold V_h)$.}

\noindent PROOF: The equivalences among $(\bold V_c)$,
$(\bold V_d)$, $(\bold V_e)$ are established in
(a) and (b) of Proposition~15.7.
The equivalences among $(\bold V_f)$,
$(\bold V_g)$, $(\bold V_h)$ are given
in (a) and (b) of Proposition~15.6. By (e) and (f) of Proposition~18.3 we know
that $(\bold V_a)$ and $(\bold V_b)$ are equivalent.
Since $(\bold{III}_a)$ is also satisfied, by Proposition~6.4 we know that
the combination of $(\bold{III}_a)$ and $(\bold V_a)$ is equivalent to
$(\bold 3_a)$. However, then $\bold S$ satisfies
all the four conditions $(\bold 1)$, $(\bold 2)$, $(\bold 3_a)$, $(\bold 4_a)$, in which case we know that Proposition~18.8 implies that $(\bold V_c)$ and
$(\bold V_f)$ are equivalent and
 we also know that Proposition~15.8 implies that
$(\bold V_c)$ is satisfied.
We also know from Proposition~18.7 that $(\bold V_f)$ implies that $(\bold V_b)$ is satisfied.
Thus, the proof is complete. \qed

The following result provides various other properties
equivalent to the combination of $(\bold 3_a)$ and $(\bold 4_a)$ on $\bold S$
stated in Definition~4.2.

\noindent {\bf Proposition 6.6} {\it Consider a scattering data set $\bold S$
as in (4.2), which consists of
an $n\times n$ scattering matrix $S(k)$ for $k\in\bR,$ a set of $N$ distinct
 positive constants $\kappa_j,$ and a set of
$N$ constant $n\times n$ hermitian and nonnegative matrices
$M_j$ with respective positive ranks $m_j,$ where $N$ is a nonnegative integer.
Assume that $\bold S$ satisfies the four conditions
$(\bold 1)$, $(\bold 2)$, $(\bold 3_a)$, $(\bold 4_a)$
of Definition~4.5, i.e. that $\bold S$ belongs to the Marchenko
class, and hence $\bold S$ correspond to a unique
input data set $\bold D$ in the Faddeev class. Then, it is possible to make any combination of
the following changes in $(\bold 3_a)$ and $(\bold 4_a)$
in such a way that each modified scattering data set
set still uniquely corresponds
to the original set $\bold D.$}

\item{(a)} {\it One can replace $(\bold 3_a)$ by $(\bold 3_b).$}

\item{(b)} {\it One can replace $(\bold 4_a)$ by any one of
the four conditions $(\bold 4_b)$, $(\bold 4_c)$, $(\bold 4_d)$, $(\bold 4_e).$}

\item{(c)} {\it One can replace $(\bold 3_a)$ by two conditions, the first of which
    is any of the three conditions $(\bold{III}_a),$ $(\bold{III}_b),$ $(\bold{III}_c),$ and the second of which is any of the eight
    conditions $(\bold V_a),$ $(\bold V_b),$  $(\bold V_c),$ $(\bold V_d),$  $(\bold V_e),$ $(\bold V_f),$ $(\bold V_g),$ $(\bold V_h).$}

\noindent PROOF: We note that
(a) follows from Proposition~6.3, (b) follows from Proposition~6.1,
and (c) follows from Propositions~6.2, 6.4, and 6.5. \qed

\newpage
\noindent {\bf 7. ALTERNATE CHARACTERIZATIONS OF THE SCATTERING DATA}
\vskip 3 pt

In this chapter we provide two alternate
sets of characterizations of the scattering data set $\bold S$ in
the Marchenko class in Definition~4.5
so that it corresponds to
the input data set $\bold D$ belonging to
the Faddeev class introduced in Definition~4.1.

First, in Theorems~7.1-7.6 we present six different versions of
a characterization equivalent to the characterization given in Theorem~5.1.

Next, in Theorems~7.9 and 7.10 we present two different
versions of a different characterization based on
the use of Levinson's theorem,
where the details of this characterization are developed in Chapter~21.

A slight drawback of using
$(\bold 3_a)$ of Definition~4.2
as a
characterization condition is that, from the
given scattering data set $\bold S,$
one first needs to construct the boundary matrices $A$ and $B$
as well as
the physical solution
$\Psi(k,x).$
Using the equivalent statements presented in Proposition~6.6 for
the characterization conditions $(\bold 3_a)$ and $(\bold 4_a)$,
it is possible to
arrange various equivalent formulations of the
characterization of $\bold S$ belonging to the Marchenko class.
For example, we can assemble a set of
five conditions
so that they will form the characterization
in the general selfadjoint case, generalizing the
 characterization presented in the
 Dirichlet case in the seminal work [2].
In this formulation, the emphasis on the conditions is on the Fourier transform of
the scattering matrix. These five conditions can directly be checked
without first having to construct the boundary matrices $A$ and $B$ and
the physical solution $\Psi(k,x).$ The proof is omitted because
the result directly follows from Proposition~6.6.

\noindent {\bf Theorem~7.1} {\it  Consider a scattering data set $\bold S$
as in (4.2), which consists of
an $n\times n$ scattering matrix $S(k)$ for $k\in\bR,$ a set of $N$ distinct
 positive constants $\kappa_j,$ and a set of
$N$ constant $n\times n$ hermitian and nonnegative matrices
$M_j$ with respective positive ranks $m_j,$ where $N$ is a nonnegative integer. The set $\bold S$
is the scattering data set
corresponding to a unique input data set $\bold D$ as in (4.1)
in the Faddeev class
specified in Definition~4.1
if and only if $\bold S$ satisfies the five conditions
$(\bold 1)$, $(\bold 2),$ $(\bold{III}_a)$, $(\bold 4_c)$, $(\bold V_c)$.}

It is possible to have another characterization
of the scattering data by modifying the condition
$(\bold V_c)$ of Theorem~7.1 and
by replacing it with
$(\bold V_f)$
and this is done in the next theorem. Again the proof is omitted because
the result directly follows from Proposition~6.6.

\noindent {\bf Theorem~7.2} {\it  Consider a scattering data set $\bold S$
as in (4.2), which consists of
an $n\times n$ scattering matrix $S(k)$ for $k\in\bR,$ a set of $N$ distinct
 positive constants $\kappa_j,$ and a set of
$N$ constant $n\times n$ hermitian and nonnegative matrices
$M_j$ with respective positive ranks $m_j,$ where $N$ is a nonnegative integer. The set $\bold S$
is the scattering data set
corresponding to a unique input data set $\bold D$ as in (4.1)
in the Faddeev class
specified in Definition~4.1
if and only if $\bold S$ satisfies the five conditions
consisting of $(\bold 1)$, $(\bold 2),$ $(\bold{III}_a)$, $(\bold 4_c)$,
 and $(\bold V_f)$.}

In the next theorem, again with the help of
Propositions~6.6,
we present an alternate set of five characterization
conditions on the scattering data set $\bold S,$ where the emphasis
on the conditions is on the scattering matrix itself.
We again omit the proof because the theorem is
a direct consequence of Propositions~6.6.

\noindent {\bf Theorem~7.3} {\it Consider a scattering data set $\bold S$
as in (4.2), which consists of
an $n\times n$ scattering matrix $S(k)$ for $k\in\bR,$ a set of $N$ distinct
 positive constants $\kappa_j,$ and a set of
$N$ constant $n\times n$ hermitian and nonnegative matrices
$M_j$ with respective positive ranks $m_j,$ where $N$ is a nonnegative integer. The set $\bold S$
is the scattering data set
corresponding to a unique input data set $\bold D$ as in (4.1)
in the Faddeev class
specified in Definition~4.1
if and only if $\bold S$ satisfies the five conditions
consisting of $(\bold 1)$, $(\bold 2),$ $(\bold{III}_b)$, $(\bold 4_d)$,
 and $(\bold V_d)$.}

It is possible to have still another characterization
of the scattering data by modifying the condition
$(\bold V_d)$ of Theorem~7.3 and
by replacing it with
$(\bold V_g)$
and this is done in the next theorem.
The proof is again omitted because the result is actually
a corollary of Proposition~6.6.

\noindent {\bf Theorem~7.4} {\it  Consider a scattering data set $\bold S$
as in (4.2), which consists of
an $n\times n$ scattering matrix $S(k)$ for $k\in\bR,$ a set of $N$ distinct
 positive constants $\kappa_j,$ and a set of
$N$ constant $n\times n$ hermitian and nonnegative matrices
$M_j$ with respective positive ranks $m_j,$ where $N$ is a nonnegative integer. The set $\bold S$
is the scattering data set
corresponding to a unique input data set $\bold D$ as in (4.1)
in the Faddeev class
specified in Definition~4.1
if and only if $\bold S$ satisfies the five conditions
consisting of if and only if $\bold S$ satisfies the five conditions
consisting of $(\bold 1)$, $(\bold 2),$ $(\bold{III}_b)$, $(\bold 4_d)$,
 and $(\bold V_g)$.}

We next present another characterizations of the scattering data,
which is a direct consequence of
Proposition~6.6, and hence by omitting the proof.

\noindent {\bf Theorem~7.5} {\it Consider a scattering data set $\bold S$
as in (4.2), which consists of
an $n\times n$ scattering matrix $S(k)$ for $k\in\bR,$ a set of $N$ distinct
 positive constants $\kappa_j,$ and a set of
$N$ constant $n\times n$ hermitian and nonnegative matrices
$M_j$ with respective positive ranks $m_j,$ where $N$ is a nonnegative integer. The set $\bold S$
is the scattering data set
corresponding to a unique input data set $\bold D$ as in (4.1)
in the Faddeev class
specified in Definition~4.1
if and only if $\bold S$ satisfies the following five conditions
consisting of $(\bold 1)$, $(\bold 2),$ $(\bold{III}_c)$, $(\bold 4_e)$,
 and $(\bold V_e)$.}

It is possible to have still another characterization
of the scattering data by modifying the condition
$(\bold V_g)$ of Theorem~7.4 and
by replacing it with
$(\bold V_h)$
and this is done in the next theorem.
Again the proof is omitted because the result directly follows from
Proposition~6.6.

\noindent {\bf Theorem~7.6} {\it  Consider a scattering data set $\bold S$
as in (4.2), which consists of
an $n\times n$ scattering matrix $S(k)$ for $k\in\bR,$ a set of $N$ distinct
 positive constants $\kappa_j,$ and a set of
$N$ constant $n\times n$ hermitian and nonnegative matrices
$M_j$ with respective positive ranks $m_j,$ where $N$ is a nonnegative integer. The set $\bold S$
is the scattering data set
corresponding to a unique input data set $\bold D$ as in (4.1)
in the Faddeev class
specified in Definition~4.1
if and only if $\bold S$ satisfies the four conditions
consisting of $(\bold 1)$, $(\bold 2),$ $(\bold{III}_c)$, $(\bold 4_e)$,
 and $(\bold V_h)$.}

 Next, we present a different characterization based on the use of Levinson's
 theorem. Levinson's theorem is given in Theorem~21.1
and the details of the derivation of this characterization are
 presented in Chapter~21.
 Before we state the characterization and one of its equivalent, we introduce
 the property $(\bold L)$ in the following definition.

 \noindent {\bf Definition 7.7} {\it
 The set of conditions $(\bold L)$ for the scattering data set $\bold S$  in (4.2) is defined as follows:}

 \item {$(\bold L)$}  {\it We say that
 $\bold S$ satisfies the property $(\bold L)$ if
the scattering matrix $S(k)$ in $\bold S$ is continuous for $ k \in \bold R$ and Levinson's theorem (21.5) is satisfied with $\mu$, $n_D$, and $ \Cal N$
coming from $\bold S.$ Here,
$\mu$ is the algebraic (and geometric) multiplicity of the eigenvalue $+1$
  of the zero-energy scattering matrix $S(0)$, $n_D$ is the algebraic (and geometric) multiplicity of the eigenvalue $-1$ of  the hermitian matrix
   $S_\infty$ defined in (4.6), and
   $ \Cal N$ is the nonnegative integer
   equal to the sum of the ranks $m_j$ of the matrices $M_j$
   appearing in $\bold S.$}

    We also introduce  three new  properties,
    namely, $(\bold 4_{c,2}), (\bold 4_{d,2}),$ and $(\bold 4_{e,2})$ that we need
    in the characterizations using
    Levinson's theorem. Actually, they resemble the respective
    properties $(\bold 4_{c}),$
     $(\bold 4_{d}),$ and $(\bold 4_e)$ of Definition 4.2, but
     involving $L^2(\bR^+)$ and $\bold H^2(\bCp)$, respectively,
     instead of $L^1(\bR^+)$ and $\hat L^1(\bCp).$

\noindent {\bf Definition 7.8} {\it
   The properties $(\bold 4_{c,2}),$ $(\bold 4_{d,2}),$ and $(\bold 4_{e,2}) $ for the scattering data set $\bold S$ in (4.2) are defined as follows:}

\item{$(\bold 4_{c,2})$} {\it The only square-integrable solution $X(y),$ which is a row vector with $n$ square-integrable components in $ y \in \bold R^+$, to the linear homogeneous integral  equation}
$$
   X(y)+ \int_0^\infty\, dz \, X(z)\, F(z+y)=0,\qquad y\in\bR^+,
\tag 7.1$$
   {\it is the trivial solution $ X(y)\equiv 0$. Here, $F(y)$ is the quantity defined in (4.12).}

   \item{$(\bold 4_{d,2})$} {\it
The only solution $\hat X(k)$ to the system}
$$\cases
\hat X(i\kappa_j)\,M_j=0,\qquad j=1,\dots,N,\\
\hat X(-k)+\hat X(k)\,S(k)=0,\qquad k\in\bR,
\endcases
\tag 7.2$$
{\it where $\hat X(k)$ is a row vector
with $n$ components belonging
to the Hardy space $\bold H^2(\bCp),$ is the trivial
solution $\hat X(k)\equiv 0.$}

\item{$(\bold 4_{e,2})$} {\it
The only solution $h(k)$ to the system}
$$
\cases
 M_j\, h(i\kappa_j)=0,\qquad j=1,\dots,N,\\
h(-k)+S(k)\,h(k)=0,\qquad k\in \bR,
\endcases
\tag 7.3$$
{\it where $h(k)$ is a column vector
with $n$ components belonging
to the Hardy space $\bold H^2(\bCp),$  is the trivial
solution $h(k)\equiv 0.$}

 Next we present the characterization of the scattering data
 utilizing Levinson's theorem.

\noindent {\bf Theorem 7.9} {\it
 Consider a scattering data set $\bold S$
as in (4.2), which consists of
an $n\times n$ scattering matrix $S(k)$ for $k\in\bold R,$ a set of $N$ distinct
 positive constants $\kappa_j,$ and a set of
$N$ constant $n\times n$ nonnegative, hermitian matrices
$M_j$ with respective positive ranks $m_j,$ where $N$ is a nonnegative integer. The set $\bold S$
is the scattering data set
corresponding to a unique input data set $\bold D$ as in (4.1)
in the Faddeev class
specified in Definition~4.1
if and only if $\bold  S$ satisfies the four properties consisting of $(\bold 1)$ and $ (\bold 2)$ of Definition~4.2, $(\bold 4_{e,2})$ of Definition~7.8, and $(\bold L)$ of Definition~7.7.
}

 \noindent PROOF: The proof is given at the end of Chapter~21. \qed

A characterization equivalent to
the one given in Theorem~7.9 is presented next.
In this equivalent characterization, the condition
$(\bold 4_{e,2})$ of Theorem~7.9 is replaced with
either of
$(\bold 4_{e,2})$ and
$(\bold 4_{e,2})$ of Definition~7.8.

\noindent {\bf Theorem 7.10} {\it
 Consider a scattering data set $\bold S$
as in (4.2), which consists of
an $n\times n$ scattering matrix $S(k)$ for $k\in\bold R,$ a set of $N$ distinct
 positive constants $\kappa_j,$ and a set of
$N$ constant $n\times n$ hermitian and nonnegative matrices
$M_j$ with respective positive ranks $m_j,$ where $N$ is a nonnegative integer. The set $\bold S$
is the scattering data set
corresponding to a unique input data set $\bold D$ as in (4.1)
in the Faddeev class
specified in Definition~4.1
if and only if $\bold  S$ satisfies the following  conditions consisting of $(\bold 1), (\bold 2)$ of Definition 4.2, $(\bold L)$ of Definition~7.7, and either
condition $(\bold 4_{c,2})$  or condition    $(\bold 4_{d,2})$     of
Definition~7.8.}

 \noindent PROOF: The proof
 follows from Proposition~21.2 and Theorem~7.9. \qed

Let us comment on Levinson's theorem and the
property $(\bold L)$ used in Theorem~7.7.
We know from Theorem~21.1, which was proved in Theorem~9.3 of [9],
that if the scattering data set $\bold S$
containing $S(k)$ belongs to the Marchenko class, then
the change in the argument
of the determinant of $S(k)$ as $k$ moves from $k=+\infty$ to $k=0^+$
is related to the nonnegative integer $\Cal N$ given in (4.3).
This result, mathematically related to an argument principle, is known as Levinson's
theorem.
The proof of (21.5) given in [9]
essentially uses the property stated in $(\bold 3_b)$,
and it has an important consequence; namely, in the Marchenko class
of scattering data set,
the scattering matrix $S(k)$ itself reveals
$\Cal N$ even though in general it does not reveal any further information
on the bound-states. In other words, in general, neither $\kappa_j$-values nor
$M_j$ appearing in (4.2), and in fact not even $N$ itself, can be extracted from
$S(k).$ As seen from
(21.5), given $S(k),$ one can extract the nonnegative integer
$\mu$ as the multiplicity of the eigenvalue
$+1$ of $S(0),$ extract the nonnegative integer $n_{\text{\rm D}}$ as
the multiplicity of the eigenvalue $-1$ of $S_\infty,$ extract
the positive integer $n$ from the size of the $n\times n$ matrix
$S(k),$ and also evaluate the left-hand side of
(21.5) directly from $S(k).$ Thus, using (21.5) we obtain the value of
$\Cal N$ associated with the given $S(k).$ In case the value of
$\Cal N$ predicted by (21.5) turns out to be negative, we know that
the given $S(k)$ cannot be a part of the
scattering data set $\bold S$ belonging to the Marchenko class no matter how
we choose $\{\kappa_j,M_j\}_{j=1}^N$ in
$\bold S$ given in (4.2). If the value of
$\Cal N$ predicted by (21.5) turns out to be zero, then we know that
we cannot have any bound states in $\bold S$ if it must belong to
the Marchenko class. If the value of
$\Cal N$ predicted by (21.5) turns out to be one, then
we know that we must have exactly one bound state of multiplicity one
if $S(k)$ is to be a part of the scattering data $\bold S$ in the
Marchenko class.
In other words, we must have $N=1$ and $M_1$ must have rank one.
There may be certain restrictions on the positive constant
$\kappa_1$ and the nonnegative, hermitian, rank-one matrix $M_1$ so
that $\bold S$ belongs to the Marchenko class.
If the value of
$\Cal N$ predicted by (21.5) turns out to be two, then
we may be able to choose $\{\kappa_j,M_j\}_{j=1}^N$ in
$\bold S$ given in (4.2) in two different ways.
In the first possibility we could have $N=1,$
the value of $\kappa_1$ could be chosen as
a positive constant, and $M_1$ could be chosen
as a nonnegative, hermitian matrix of rank two.
In the second possibility we could have $N=2,$ the values of
$\kappa_1$ and $\kappa_2$ could be chosen as two distinct positive constants, and
and $M_1$ and $M_2$ could be chosen as
two nonnegative, hermitian matrices of rank one.
As shown in some of the examples in Section~26, not all these choices may be viable
and the condition of being in the Marchenko class may restrict
some of these choices.
Clearly, the same argument applies in how many different ways we might be able
to choose
$\{\kappa_j,M_j\}_{j=1}^N$ in
$\bold S$ when the value of $\Cal N$ predicted by (21.5) is three or higher,
but again the constraints for belonging to the Marchenko class may restrict
some of the available choices.
In Section~26 various illustrative examples are provided to
indicate how $S(k)$ predicts $\Cal N$ via (21.5) and how such a restriction
and other restrictions play a role on the bound-state information for
 the scattering data set
$\bold S$ to belong to the Marchenko class.

\newpage
\noindent {\bf 8. ANOTHER CHARACTERIZATION OF THE SCATTERING DATA}
\vskip 3 pt

In this chapter we provide yet another
 characterization of the scattering data set $\bold S$ in
the Marchenko class in Definition~4.5
so that it corresponds to
the input data set $\bold D$ belonging to
the Faddeev class introduced in Definition~4.1.

This new characterization has some similarities and differences
compared to the characterization given in Section~5 and the alternate
characterizations given in Section~7. Related to this
new characterization, the construction of the potential
in the solution to the inverse problem is the same as in the
previous characterizations; namely, one constructs the potential by solving the
Marchenko equation. Hence, the conditions $(\bold 1)$, $(\bold 2)$, $(\bold 4_a)$
of Definition~4.5
in the first characterization, the conditions
$(\bold 1)$, $(\bold 2)$, $(\bold 4_c)$ in the alternate characterizations
of Section~7, and
the conditions  $(\bold I),$
$(\bold 2),$
$(\bold 4_c)$ in this new
characterization are essentially used to construct
the potential. This new characterization differs from
the earlier ones in regard to the
satisfaction of the boundary condition by the physical solution
$\Psi(k,x)$ and by the bound-state matrix solutions $\Psi_j(x).$
It is based on the alternate solution to the inverse problem as
summarized in Section~23.
It
uses six conditions, where five of the conditions
are already listed in Definitions~4.2 and 4.3;
namely, $(\bold I),$
$(\bold 2),$ $(\bold 4_c),$ either
of $(\bold V_e)$ or $(\bold V_h),$ and $(\bold {VI}).$
It also uses the condition $(\bold A),$
which is not listed in Definitions~4.2 and 4.3.
The condition $(\bold A),$ stated in the following
theorem,
somehow resembles $(\bold{III}_c)$ of Proposition~4.3, but there are also
some major differences.
In $(\bold{III}_c)$ a solution is sought to
the homogeneous Riemann-Hilbert problem (4.19)
as a column vector with $n$ components where each of those
components belongs to $\bold H^2(\bCm),$
and the only solution is expected to be the trivial solution
$h(k)\equiv 0.$
On the other hand, in
$(\bold A)$ one solves a nonhomogeneous
Riemann-Hilbert problem and the
solution is sought as a column vector where each of the
$n$ components belongs the Hardy space
$\bold H^2(\bCp),$ and certainly the
corresponding solution is in general nontrivial
and such a solution is not required to be unique.
The condition $(\bold {VI}),$
which is the continuity of the scattering matrix $S(k),$
is mainly needed to prove that
the physical solution satisfies the boundary condition.

\noindent {\bf Theorem 8.1} {\it Consider a scattering data set $\bold S$
as in (4.2), which consists of
an $n\times n$ scattering matrix $S(k)$ for $k\in\bR,$ a set of $N$ distinct
 positive constants $\kappa_j,$ and a set of
$N$ constant $n\times n$ hermitian and nonnegative matrices
$M_j$ with respective positive ranks $m_j,$ where $N$ is a nonnegative integer. The set $\bold S$
is the scattering data set
corresponding to a unique input data set $\bold D$ as in (4.1)
in the Faddeev class
specified in Definition~4.1
if and only if $\bold S$ satisfies the following six conditions:
$(\bold 2)$ and $(\bold 4_c)$ of Definition~4.2,
$(\bold I),$ $(\bold {VI}),$ and either one of
$(\bold V_e)$ or $(\bold V_h)$ of Definition~4.3, and
the following condition named $(\bold A):$}

\item{$(\bold A)$} {\it Consider
the nonhomogeneous
Riemann-Hilbert problem given by}
$$h(k)+S(-k)\, h(-k)=g(k),\qquad k\in\bR,\tag 8.1$$
{\it where
the nonhomogeneous
term $g(k)$
belongs to a dense
subset $\overset{\circ}\to \Upsilon$
of the vector space $\Upsilon$ of column vectors with $n$ square-integrable
components and satisfying $g(-k)=S(k)\,g(k)$ for $k\in\bR.$
Then, for each such given
$g(k),$ the equation (8.1) has a solution $h(k)$
as a column vector
with $n$ components belonging
to the Hardy space $\bold H^2(\bCp).$}

\noindent PROOF: The proof is given in Section~23, after the
proof of Proposition~23.6. \qed

\newpage
\noindent {\bf 9. THE SOLUTION TO THE DIRECT PROBLEM}
\vskip 3 pt

In this chapter we summarize the solution to the
direct scattering problem of obtaining
the scattering data set $\bold S$ given in (4.2) from the input
data set $\bold D$ given in (4.1). We assume that
$\bold D$ belongs to the Faddeev class specified in Definition~4.1.
The proofs and details will be provided in later chapters. We postpone the proof that
$\bold S$ belongs to the Marchenko class specified in
Definition~4.5 when $\bold D$ belongs to the Faddeev class, because that proof
will be given in Theorem~15.10.

The relevant existence and uniqueness in the construction of $\bold S$
 from $\bold D$ are implicit in each step described below.

\item{(a)} When our input data set $\bold D$ belongs to the Faddeev class,
regardless of the boundary matrices
$A$ and $B,$ the
matrix Schr\"odinger equation (2.1)
possesses the $n\times n$ matrix-valued Jost solution
$f(k,x)$
satisfying the asymptotic condition
$$f(k,x)=e^{ikx} [I+o(1)],\qquad x\to+\infty.\tag 9.1$$
The existence and uniqueness of $f(k,x)$ as well
as its relevant properties are summarized
in Proposition~10.1.

\item{(b)} In terms of the boundary matrices $A$ and $B$ in
$\bold D$ and the Jost solution $f(k,x)$
obtained in (a), we construct
the Jost matrix $J(k),$
an $n\times n$ matrix-valued function of $k,$ as
$$J(k):=f(-k^\ast,0)^\dagger B-f'(-k^\ast,0)^\dagger A,\qquad k\in\bR,\tag 9.2$$
where the asterisk denotes complex conjugation.
When $\bold D$ belongs to the Faddeev class, the relevant
properties of the Jost matrix are summarized
in Proposition~10.2. The redundant appearance of $k^\ast$
instead of $k$ in (9.2) when $k\in\bR$ is useful
in extending the Jost matrix analytically from
$k\in\bR$ to $k\in\bCp.$

\item{(c)} In terms of the Jost matrix $J(k),$
uniquely obtained from $\bold D$ as indicated in (9.2),
we construct the scattering matrix $S(k),$
an $n\times n$ matrix-valued function of $k,$ as
$$S(k):=-J(-k)\,J(k)^{-1},\qquad k\in\bR.\tag 9.3$$
When $\bold D$ belongs to the Faddeev class, the relevant
properties of $S(k)$ are summarized
in Proposition~10.3.

\item{(d)} In terms of the Jost solution $f(k,x)$
obtained in (a) and the scattering matrix $S(k)$
obtained in (c), we construct the physical solution
$\Psi(k,x)$ as
$$\Psi(k,x):=f(-k,x)+f(k,x)\,S(k),\qquad k\in\bR.\tag 9.4$$
In Proposition~10.5, we show that
the $n\times n$ matrix-valued $\Psi(k,x)$ is a
solution to (2.1) and
satisfies the boundary condition (2.4)
and we also summarize the relevant properties
of $\Psi(k,x).$

\item{(e)} Instead of constructing the physical solution
via (9.4), one can alternatively construct it in an equivalent way as follows:
When our input data set $\bold D$ belongs to the Faddeev class, as indicated
in Proposition~10.4, the
matrix Schr\"odinger equation (2.1)
possesses a unique $n\times n$ matrix-valued solution
$\varphi(k,x)$
satisfying the initial conditions [5]
$$\varphi(k,0)=A,\quad \varphi'(k,0)=B,\tag 9.5$$
where $A$ and $B$ are the matrices appearing in (2.4)-(2.6)
and (4.1). The solution $\varphi(k,x)$ is known as
the regular solution because it is entire in
$k$ for each fixed $x\in\bR^+.$ In terms of the regular solution
$\varphi(k,x)$ and the Jost matrix $J(k)$ appearing in (9.2) we
can introduce the physical solution as
$$\Psi(k,x)=-2ik\,\varphi(k,x)\,J(k)^{-1}.\tag 9.6$$
One can show that the expressions given in (9.4) and (9.6)
are equivalent, and this can be shown by using the relationship given in
(3.5) of [5], i.e.
$$\varphi(k,x)=\ds\frac{1}{2ik}\,f(k,x)\,J(-k)-\ds\frac{1}{2ik}\,f(-k,x)\,J(k),
\tag 9.7$$
where we recall that $f(k,x)$ is the Jost solution appearing in (9.1).

\item{(f)} As indicated in Proposition~10.2,
the determinant of the Jost matrix $J(k)$
has an analytic extension from $k\in\bR$ to
$k\in\bCp$ and that determinant is nonzero
in $\bCp$ except perhaps at a finite number of
$k$-values on the positive imaginary axis.
We assume that there are $N$ such $k$-values occurring
at $k=i\kappa_j$ and use $m_j$ to denote the multiplicity of
the zero of $\det[J(k)]$ at $k=i\kappa_j.$
Thus, the set of $N$ distinct values of
positive $\kappa_j$ appearing in $\bold S$
is uniquely constructed from
$\bold D.$ Furthermore, the positive
integers $m_j$ are also uniquely determined by
$\bold D$ as a result of the
unique construction of
$J(k).$ It is possible that
the determinant of $J(k)$ never vanishes
in $k\in\bCp,$ in which case we have $N=0.$
When $N=0,$ the scattering data set $\bold S$ given in (4.2) consists
of $S(k)$ alone, and the summation term appearing in (4.12)
is absent.

\item{(g)} At $k=i\kappa_j,$ as shown
in Proposition~11.4, (2.1) has $m_j$ linearly
independent column vector-valued solutions, where we recall that
the values of $m_j$ are already uniquely determined from $\bold D,$
as indicated in the previous step.
It is possible to rearrange those linearly
independent column vector solutions into
an $n\times n$ matrix
$\Psi_j(x),$ in such a way that
$\Psi_j(x)$ can be uniquely constructed as
$$\Psi_j(x):=f(i\kappa_j,x)\,M_j,\qquad j=1,\dots,N,\tag 9.8$$
where $M_j$ is an $n\times n$ nonnegative hermitian matrix of rank $m_j.$
The unique construction of $M_j$ is given in (11.22) in terms of
the projection matrix $P_j$ appearing in (11.1) and the matrix
$B_j^{-1/2},$ where $B_j$ is defined in (11.3). The relevant
properties of $P_j$ are given in
(11.1) and those of $B_j^{-1/2}$ are summarized in Proposition~11.2.

\item{(h)} As a part of the direct problem,
when $\bold D$ belongs to the Faddeev class, we show that
the collection of the set of
$N$ matrices $\Psi_j(x)$ given in (9.8) and the physical
solution $\Psi(k,x)$ given in
(9.4) satisfy the orthonormalization
condition and the completeness condition (Parseval's equality), which are
summarized in Proposition~22.2 and Proposition~20.2, respectively.

\item{(i)} As a part of the direct problem,
when $\bold D$ belongs to the Faddeev class, we verify that
$\bold S$ belongs to the Marchenko class, by showing that the four conditions listed in
Definition~4.5 are satisfied by $\bold S.$
This is done in Theorem~15.10.

\item{(j)} We show that the equivalence statements given in Proposition~6.6,
hold when the input data set $\bold D$ belongs to the Faddeev class.

\newpage
\noindent {\bf 10. SOME RELEVANT RESULTS RELATED TO THE DIRECT PROBLEM}
\vskip 3 pt

In this chapter we elaborate and justify some of the steps outlined in Section~3
for the unique construction of
the scattering data set $\bold S$ when our input data set
$\bold D$ belongs to the Faddeev class. The remaining steps outlined in Section~3 will
be proved in later chapters.

\noindent {\bf Proposition 10.1} {\it Assume that the input data set $\bold D$ appearing in (4.1) belongs to
the Faddeev class specified in Definition~4.1. Then, regardless of the boundary
matrices $A$ and $B,$ (2.1) has a unique
$n\times n$ matrix-valued solution
$f(k,x),$ known as the Jost solution, satisfying the asymptotics (9.1). Furthermore,
we have:}

\item{(a)} {\it For each fixed $x\in[0,+\infty),$
the quantity $f(k,x)$ has an analytic
extension from $k\in\bR$ to $k\in\bCp$ and that extension
is continuous in $k\in\bCpb.$}

\item{(b)} {\it The quantity $K(x,y)$ defined as}
$$K(x,y):=\ds\frac{1}{2\pi}\int_{-\infty}^\infty dk\,[f(k,x)-e^{ikx}I]\,e^{-iky},
\tag 10.1$$
{\it vanishes when $y<x,$ i.e.
$$K(x,y)=0,\qquad x>y,\tag 10.2$$
and it is related to the potential via}
$$K(x,x)=\ds\frac{1}{2}\int_x^\infty dz\,V(z),\qquad x\ge 0,\tag 10.3$$
$$V(x)=-2\ds\frac{d K(x,x)}{dx},\qquad x\in\bR^+.\tag 10.4$$
{\it The constant $n\times n$ matrix $K(0,0)$ is well defined, hermitian, and related to the potential
as}
$$K(0,0)=\ds\frac{1}{2}\int_0^\infty dz\,V(z).\tag 10.5$$
{\it We remark that we use $K(x,x)$ to denote $K(x,x^+)$ and
use $K(0,0)$ to denote $K(0,0^+).$}

\item{(c)} {\it The Jost solution $f(k,x)$ has the representation}
$$f(k,x)=e^{ikx}I+ \int_x^\infty dy\,K(x,y)\,e^{iky},
\tag 10.6$$
{\it where $K(x,y)$ is the quantity given in (10.1).}

\item{(d)} {\it The quantity $K(x,y)$ appearing in (10.1) satisfies}
$$|K(x,y)|\le \ds\frac {1}{2}\, e^{\sigma_1(x)}\,\sigma\left(\ds\frac{x+y}{2}\right),\qquad
y\ge x\ge 0,\tag 10.7$$
{\it where $\sigma(x)$ and $\sigma_1(x)$ are the scalar quantities defined in (3.96).
Hence, the quantity $K(x,y)$ is integrable in $y\in[x,+\infty)$
for each fixed $x\ge 0.$}

\item{(e)} {\it The quantity $K(x,y)$ appearing in (10.1) is continuous
in $(x,y)$
in the region $0\le x\le y$.}

\item{(f)} {\it The quantity $K_x(x,y),$ i.e. the $x$-derivative
of $K(x,y),$ exists a.e. and satisfies}
$$K_x(x,y)=0,\qquad x>y,\tag 10.8$$
$$|K_x(x,y)|\le \ds\frac{1}{4}\,\bigg| V\left(\ds\frac{x+y}{2}\right)\bigg|+
\ds\frac{1}{2}\,e^{\sigma_1(x)}\,\sigma\left(\ds\frac{x+y}{2}\right)\,
\sigma(x),\qquad 0\le x\le y.
\tag 10.9$$
{\it Hence, the quantity $K_x(x,y)$ is integrable in $y\in[x,+\infty)$
for each fixed $x\ge 0.$}

\noindent PROOF: We refer the reader to Theorem~1.3.1 of [2]
for (a)-(d), in particular to (1.3.11) of [2] for (10.3).
and to Lemma~1.3.1 of [2] for (f).
We have the following remarks to complete the proof.
The hermitian property of
$K(0,0)$ in (10.5) is directly obtained by using in (10.5) the hermitian
property of the potential $V(x).$
The properties
that, for each fixed $x\ge 0,$ the quantities
 $K(x,y)$ and $K_x(x,y)$ are each integrable
 in $y\in[x,+\infty)$ is proved as follows.
 From
(2.3) and (3.96) we know that $|V(y)|$
and $\sigma(y)$ are both in $L^1(\bR^+).$
Furthermore, as a result of (2.3) we know that
$\sigma(x)$ and $\sigma_1(x)$ are both finite for each $x\ge 0.$
Then, (10.7) and (10.9) imply that $K(x,y)$and $K_x(x,y),$
respectively, are integrable in $y\in[x,+\infty)$
for each fixed $x\ge 0.$
The continuity stated in (e) can be proved as follows.
The matrix $K(x,y)$ satisfies the integral equation
(1.3.6) of [2]. That
integral equation can be solved by the method of
successive approximation, by representing
$K(x,y)$ as a uniformly convergent infinite series, where
each term is continuous in the region $0\le x\le y.$
As a result of the uniform convergence, $K(x,y)$
is continuous in the same region, proving (e).
The property (10.8) follows from (10.2).
The estimate (10.9) can be found in (1.3.9) of [2]
and can be obtained by solving
the integral equation (1.3.8) of [2]
iteratively. In general, in the Faddeev class,
the potential $V(x)$ is not continuous in $x\in\bR^+$
and hence $V(x)$ exists a.e. in $x\in\bR^+.$ Then, applying
the method of successive approximation to
(1.3.8) of [2], it follows that
the quantity
$K_x(x,y)$ in general exists only a.e.
\qed

We remark that the Jost solution
$f(k,x)$ appearing in (9.1) is only affected by the
potential $V$ in the input data set $\bold D$ and not
by the boundary matrices $A$ and $B$
appearing in (2.4)-(2.6).

Next we present the relevant properties of the Jost matrix
$J(k)$ introduced in (9.2).

\noindent {\bf Proposition 10.2} {\it Assume that the input data set $\bold D$ appearing in (4.1) belongs to
the Faddeev class specified in Definition~4.1. Let
$J(k)$ be the corresponding Jost matrix constructed as in
(9.2). Then:}

\item{(a)} {\it The Jost matrix $J(k)$ is invertible for
$k\in\bR\setminus\{0\}.$ It is either invertible at
$k=0$ or it has a simple zero at $k=0.$
The matrix $J(k)^{-1}$ has at most a simple pole at $k=0,$
and hence $k\,J(k)^{-1}$ is continuous in $k\in\bR.$}

\item{(b)} {\it The Jost matrix $J(k)$ has an analytic extension from
$k\in\bR$ to $k\in\bCp$ in such a way that $J(k)$ is continuous in $k\in\bCpb.$
Furthermore, we have}
$$J(k)=-ikA+B+K(0,0) \,A+o(1),\qquad k\to\infty \text{ in }\bCpb,\tag 10.10$$
{\it where $A$ and $B$ are the boundary matrices in the input data set
$\bold D$ and $K(0,0)$ is the constant $n\times n$ matrix given in (10.5).}

\item{(c)} {\it The determinant of
$J(k)$ in $\bCp$ is nonzero, except at a finite number of distinct
$k$-values on the positive imaginary axis, say at
$k=i\kappa_j$ for $j=1,\dots,N,$ where the multiplicity of
the zero at $k=i\kappa_j$ is denoted by
$m_j.$ The matrix $J(k)^{-1}$ is meromorphic
in $k\in\bCp$ with simple poles at $k=i\kappa_j$ for $j=1,\dots,N.$
If $N=0,$ then the determinant of $J(k)$ does not vanish in $k\in\bCpb\setminus \{0\}.$}

\item{(d)} {\it The multiplicity $m_j$ of the zero $k=i\kappa_j$ for the
determinant of $J(k)$ is equal to the nullity of the matrix
$J(i\kappa_j)^\dagger.$}

\item{(e)} {\it We have}
$$J(k)=J(i\kappa_j)+(k-i\kappa_j)\,\dot J(i\kappa_j)+O\left((k-i\kappa_j)^2\right),
\qquad k\to i\kappa_j \text{ in } \bCp,\tag 10.11$$
$$J(k)^{-1}=\frac{N_j}{k-i\kappa_j}+L_j+O\left(k-i\kappa_j\right),
\qquad k\to i\kappa_j \text{ in } \bCp,\tag 10.12$$
{\it for some constant $n\times n$ matrices $N_j$ and $L_j,$ where
an overdot indicates the $k$-derivative.}

\item{(f)} {\it The constant $n\times n$
matrices $N_j$ and $L_j,$ appearing in (10.12) satisfy}
$$ N_j\,J(i\kappa_j)=0, \quad L_j\, J(i\kappa_j)+N_j\,\dot J(i\kappa_j)=I.
\tag 10.13$$

\item{(g)} {\it We have $J(k)^{-1}=O(k)$ as $k\to\infty$ in $k\in\bCpb.$}

\noindent PROOF: A proof of (a) can be found in Theorems~5.1 and 6.3 of [5].
For the first statement in
(b) we refer the reader to Theorem~3.1(a) of [9].
The large-$k$ asymptotics of the Jost matrix $J(k)$ in (10.10)
is known from
(7.11) of [9].
Let us now prove (c). For this, let us first argue that
the number of zeros of $\det[J(k)]$ in $\bCpb$ must be finite,
and the argument is as follows.
By (3.1) of [9] we know that any zero of $\det[J(k)]$ in $\bCp$ yields a bound state
for (2.1) and (2.4), and by the first paragraph of
Section~8 of [9] we know that such a zero can only occur on the positive
imaginary axis in $\bCp.$
 From (a) we know that
$\det[J(k)]$ is nonzero for $k\in\bR\setminus\{0\}.$ By (6.5) of [9]
we know that $\det[J(k)]$ either does not
vanish at $k=0$ or its zero at $k=0$ has a finite multiplicity.
Furthermore, by (7.17) of [9] we know that
$\det[J(k)]$ does not vanish as $k\to\infty$ in $\bCpb.$
 From (b)
we know that $\det[J(k)]$ is analytic in $k\in\bCp$ and continuous
in $k\in\bCpb.$ Hence,
there cannot be any accumulation points for the zeros of
$\det[J(k)]$ in $\bCpb$ and the number of zeros of
$\det[J(k)]$ in $\bCpb$ must be finite.
Having proved that the number of zeros of
$\det[J(k)]$ in $\bCpb$ is finite, the rest of (c)
follows from Theorems 8.4 and 8.4 of [9].
Similarly, (d) follows from
Theorems 8.4 and 8.4 of [9].
As for the proof of (e), we remark that
the results in (10.11) and (10.12) follow from
(8.26) and (8.32), respectively, of [9]. Concerning (f), we
obtain (10.13) by using
(10.10) and (10.12) in the identity $J(k)^{-1} J(k)=I.$
Finally, we note that
(g) follows from (5.8) and (7.13) of [9]. Thus, the proof is complete.
\qed

Next we present some relevant properties of the scattering matrix $S(k)$ introduced
in (9.3).

\noindent {\bf Proposition 10.3} {\it Assume that the input data set $\bold D$ appearing in (4.1) belongs to
the Faddeev class specified in Definition~4.1. Then:}

\item{(a)} {\it The scattering matrix $S(k)$ defined in (9.3) is continuous
for $k\in\bR,$ and it satisfies (4.4).}

\item{(b)} {\it The large-$k$ asymptotics of $S(k)$ is given by}
$$S(k)=S_\infty+\ds\frac{G(k)}{ik}+O\left(\ds\frac{1}{k^2}\right),\qquad k\to\pm\infty,
\tag 10.14$$
{\it where $S_\infty$ is the constant $n\times n$ matrix defined in (4.6) and uniquely
determined by the boundary matrices $A$ and $B$ as}
$$S_\infty=\ds\lim_{k\to \pm\infty} \left[-(B+ikA)(B-ikA)^{-1}\right],\tag 10.15$$
{\it and the matrix $G(k)$ is continuous and uniformly bounded for $k\in\bR.$ In fact, we have}
$$G(k)=G_1+G_2(k),\tag 10.16$$
{\it where $G_1$ is the constant $n\times n$
matrix appearing in (4.5) and defined as}
$$G_1:=W_1+\ds\frac{1}{2}\int_0^\infty dz\,V(z)\,S_\infty+\ds\frac{1}{2}\int_0^\infty dz\,S_\infty\,V(z),\tag 10.17$$
{\it with $W_1$ being some constant $n\times n$ hermitian matrix,
$S_\infty$ the constant matrix appearing in (4.6) and (10.14), and $G_2(k)$ the $n\times n$ matrix defined as}
$$G_2(k):=\ds\frac{1}{2}\int_0^\infty dz\,V(z)\,e^{-2ikz}+
\ds\frac{1}{2}\int_0^\infty dz\,S_\infty\,V(z)\,S_\infty\,e^{2ikz},\tag 10.18$$
{\it with the property that $G_2(k)=o(1)$ as $k\to\pm\infty.$}

\item{(c)} {\it The constant matrices $S_\infty$ and $G_1,$
defined in (4.6) and (10.17), respectively,
are both hermitian.}

\item{(d)} {\it Each eigenvalue of the constant matrix $S(0)$ is either
$+1$ or $-1.$}

\noindent PROOF: The continuity of $S(k)$ is stated in Proposition~3.3 of [9].
We quote the property in (4.4) directly from (3.11) of [9]. The large-$k$ asymptotics
in (10.14) is given in Theorem~7.6 of [9], where the properties of
$G(k)$ is obtained from (10.17) and (10.18) by using the fact that
$V(x)$ is integrable in $x\in\bR^+$ and hence the Riemann-Lebesgue
lemma on the right-hand side of (10.18) yields
$G_2(k)=o(1)$ as $k\to\pm\infty.$ The hermitian property of $S_\infty$
directly follows from (4.4) and (10.14). Since $W_1,$ $S_\infty,$ and
$V(x)$ are all hermitian, from (10.17) we see that
$G_1$ is also hermitian. The fact that the eigenvalues
of $S(0)$ can only be $+1$ or $-1$ is proved in Proposition~6.3 of [9].
\qed

The relevant properties of the
regular solution $\varphi(k,x)$ appearing in (9.5)
are given in the following proposition.

\noindent {\bf Proposition 10.4} {\it Assume that the input data set $\bold D$ appearing in (4.1) belongs to
the Faddeev class specified in Definition~4.1. Then, (2.1) has a unique solution
$\varphi(k,x),$ known as the regular solution,
satisfying the initial conditions given in (9.5).
For each fixed
$x\in[0,+\infty),$ the regular solution
$\varphi(k,x)$ is entire in $k$ and it satisfies
as $k\to\infty$ in
$\bCpb$}
$$\aligned
-2ik\, \varphi(k,x) \,e^{ikx}=&-ik\left(1+e^{2ikx}
\right)A+\left(1-e^{2ikx}
\right)B\\
\stretch
& +\ds\frac{1}{2}
\int_0^x dz\,\left(1+e^{2ikz}
\right)\left(1-e^{2ik(x-z)}
\right)V(z)\,A+O\left(\ds\frac{1}{k}\right).\endaligned\tag 10.19$$

\noindent PROOF: The regular solution $\varphi(k,x)$
satisfies the integral relation given
in (3.7) of [5], which is
$$\varphi(k,x)=A\,\cos kx+B\,\ds\frac{\sin kx}{k}+\ds\frac{1}{k}\int_0^x dz\,[\sin k(x-z)]
V(z)\,\varphi(k,z).\tag 10.20$$
By iterating (10.20) one can establish [5] the existence and uniqueness of
the regular solution and also prove that $\varphi(k,x)$ is entire in $k$
for each fixed $x\in[0,+\infty).$
Multiplying both sides of (10.20) with $e^{ikx},$ after some simplification,
we obtain
$$\aligned
e^{ikx}\varphi(k,x)=&\ds\frac{1}{2}\left( 1+e^{2ikx}\right)
A-\ds\frac{1}{2ik}\left( 1-e^{2ikx}\right)B\\
\stretch
&+\ds\frac{1}{2ik}
\int_0^x dz\,[e^{2ik(x-z)}-1]
V(z)\,[e^{ikz}\varphi(k,z)].\endaligned\tag 10.21$$
We obtain (10.19) from (10.21) via iteration. \qed

Next, we establish
some relevant properties of the physical solution
$\Psi(k,x)$ appearing in (9.4) and in (9.6).

\noindent {\bf Proposition 10.5} {\it Assume that the input data set $\bold D$ appearing in (4.1) belongs to
the Faddeev class specified in Definition~4.1. Let $\Psi(k,x)$ be the physical solution defined in (9.4).
Then:}

\item{(a)} {\it The representation (9.6) for the physical solution
$\Psi(k,x)$ is equivalent to the representation given in (9.4).}

\item{(b)} {\it For each fixed $x\in[0,+\infty),$ the quantity
$\Psi(k,x)$ is continuous in $k\in\bR$ and
meromorphic in $k\in\bCp$ with simple poles coinciding
with the poles of $J(k)^{-1}$ as indicated in (10.12), i.e.
simple poles at $k=i\kappa_j$ for $j=1,\dots,N.$}

\item{(c)} {\it The physical solution $\Psi(k,x)$ satisfies (2.1) and
the boundary condition (2.4).}

\item{(d)} {\it For each fixed $x\in[0,+\infty),$ we have}
$$e^{ikx}\,\Psi(k,x)=
W_2+W_3\,e^{2ikx}
+O\left(\ds\frac{1}{k}\right),\qquad k\to\infty \text{ in } \bCpb,
\tag 10.22$$
{\it for some constant $n\times n$ matrices $W_2$ and $W_3.$}

\noindent PROOF: We establish (a) in a straightforward manner
with the help of (9.3) and (9.7). Let us now turn to
the proof of (b). As indicated in Theorem~3.1 of [5],
$J(k)^{-1}$ is continuous in $k\in\bR\setminus\{0\}$ with a possible simple pole
at $k=0.$ We know from Proposition~10.4 that $\varphi(k,x)$ is entire in $k$
for each fixed $x\in[0,+\infty).$ Furthermore, as indicated in
Proposition~10.2(c) the matrix $J(k)^{-1}$ is meromorphic
in $\bCp$ with simple poles at $k=i\kappa_j$ for $j=1,\dots,N.$
With the help of (9.6),
(10.12), and (10.19) we conclude (b).
The proof of (c) is obtained as follows.
 From Proposition~10.1 we know that each column of the Jost solution
 $f(k,x)$ satisfies (2.1). Since $k$ appears as $k^2$ in (2.1),
 each column of $f(-k,x)$ is also a solution
to (2.1). As seen from (9.4), each column of $\Psi(k,x)$ is a linear
combination of columns of $f(k,x)$ and $f(-k,x).$ Hence, the $n\times n$
matrix $\Psi(k,x)$ is a solution to (2.1).
 From (2.5) and (9.5) we see that the regular solution
 $\varphi(k,x)$ satisfies the selfadjoint boundary condition given in (2.4), and
 from (9.6) we see that each column of the physical solution
 $\Psi(k,x)$
 is a linear combination of columns of $\varphi(k,x).$ Thus, the physical solution
 also satisfies the boundary condition (2.4).
Hence, the proof of (c) is complete.
Let us now turn to the proof of (d).
 From (10.8), (10.9), and
(7.12) of [9], we respectively have
$$J_0(k)^{-1}=W_4+O\left(\ds\frac{1}{k}\right),
\qquad k\to\infty \text{ in } \bCpb,\tag 10.23$$
$$A\,J_0(k)^{-1}=-\ds\frac{1}{ik}\,W_5+O\left(\ds\frac{1}{k^2}\right),
\qquad k\to\infty \text{ in } \bCpb,\tag 10.24$$
$$J(k)^{-1}=J_0(k)^{-1}\left[I+O\left(\ds\frac{1}{k}\right)\right],
\qquad k\to\infty \text{ in } \bCpb,\tag 10.25$$
where $W_4$ and $W_5$ are some constant $n\times n$ matrices and
$J_0(k)$ is the Jost matrix corresponding to the input
data set $\bold D$ when $V$ is the zero potential.
Using (10.19) and (10.25) on the right-hand side of
(9.6), with the help of (10.23) and (10.24) and
after some simplification, we obtain (10.22). \qed

\newpage
\noindent {\bf 11. BOUND STATES}
\vskip 3 pt

In this chapter, we continue our analysis of the direct
scattering problem and continue
the justification of the steps outlined in Section~3 when our
input data set $\bold D$ given in (4.1) belongs to the Faddeev class
specified in Definition~4.1. We prove
various relevant
properties of the corresponding scattering data set $\bold S$ given
in (4.2), especially those properties related to
the bound-state information in $\bold S.$

Concerning the bound states of
the Schr\"odinger operator associated with (2.1) and (2.4), we have
the following basic facts. By
definition, a bound state is a column vector solution to (2.1)
which also satisfies the boundary condition (2.4). Because of the
selfadjointness of the Schr\"odinger operator, a bound state, if
it exists,
must occur when the spectral parameter $k^2$ is real.
When $k^2>0$ or $k^2=0,$ none of $2n$ linearly independent
column vector solutions to (2.1) are square integrable,
as argued in the first paragraph of
Section~8 of [9]. When $k^2<0,$
i.e. when $k$ is on the positive imaginary axis in $\bCp,$
we have the following argument. Among the $(2n)$ linearly
independent column-vector solutions to (2.1),
only $n$ of them are square integrable
in $x\in\bR^+,$ and the columns of
$f(k,x)$ are such solutions.
Among the $(2n)$ linearly-independent column-vector solutions
to (2.1), only $n$ of them satisfy (2.4), and
the columns of the regular solution $\varphi(k,x)$
are such solutions. Thus, a particular $k$-value on the
positive imaginary axis in $\bC$ corresponds to a bound state
provided a column vector at that $k$-value can be expressed as
a linear combination of the columns of $f(k,x)$ and
also of the columns of $\varphi(k,x).$
It turns out that such $k$-values
correspond to the zeros of the determinant of
the Jost matrix $J(k)$ given in (9.2),
which occur at $k=i\kappa_j$ for $j=1,\dots,N,$
as indicated in Proposition~10.2(c). Furthermore,
at $k=i\kappa_j$ there are exactly $m_j$ linearly-independent
column vectors satisfying both (2.1) and (2.4). This is elaborated
in Propositions~11.2 and 11.3. Below we provide a summary
of the basic facts on the bound states for the relevant Sch\"odinger operator,
and for the proof and further details on the bound states
we refer the reader to [9].

\noindent {\bf Proposition 11.1} {\it Assume that the input data set $\bold D$ appearing in (4.1) belongs to
the Faddeev class. Then:}

\item{(a)} {\it The bound states corresponding to the Schr\"odinger
operator related to (2.1) and (2.4)-(2.6) occur only at
the $k$-values on the positive imaginary axis in $\bC$ where the determinant of Jost matrix $J(k)$ given in (9.2) vanishes. Such $k$-values
are denoted by $k=i\kappa_j$ for $j=1,\dots,N,$ where the
$\kappa_j$ are distinct positive numbers and $N$ is a nonnegative integer.
If $N$ is zero, then there are no bound states.}

\item{(b)} {\it For each $k=i\kappa_j$ there are exactly $m_j$ linearly
independent square-integrable column-vector solutions to (2.1) that also satisfy (2.4).
Here, $m_j$ is the positive integer equal to the dimension of the kernel of
$J(i\kappa_j)^\dagger.$ The positive integer $m_j.$}

Let $\text{\rm Ker}[J(i\kappa_j)^\dagger]$ denote the kernel of
the matrix $J(i\kappa_j)^\dagger.$
We use $P_j$ to denote the orthogonal projection matrix onto
$\text{\rm Ker}[J(i\kappa_j)^\dagger].$ Then, $P_j$
is an $n\times n$ hermitian, idempotent matrix, i.e. we have
$$P_j^\dagger=P_j,\quad P_j^2=P_j,\qquad j=1,\dots,N.\tag 11.1$$

Having defined the orthogonal projections $P_j,$ let us now define the normalization
matrices $M_j$ at each bound state with $k=i\kappa_j.$
Proceeding as on pp. 60--61 of [2], we
define the constant $n\times n$ matrices $A_j$ and
$B_j$ as
$$A_j:=\int_0^\infty dx\,f(i\kappa_j,x)^\dagger f(i\kappa_j,x),\qquad j=1,\dots,N,\tag 11.2$$
$$B_j:=(I-P_j)+P_jA_j\,P_j,\qquad j=1,\dots,N,\tag 11.3$$
where $f(k,x)$ is the Jost solution appearing in (9.1)
and $P_j$ is the hermitian projection matrix appearing
in (11.1). We remark that the definitions of $A_j$ and $B_j$
in (11.2) and (11.3), respectively, are the same as
the corresponding definitions appearing on pp. 61--62 of [2].
Because [2] uses the Dirichlet boundary condition instead of
(2.4), the matrix $B_j$ appearing in (11.3) is a generalization of
the corresponding matrix in [2].
The properties of $B_j$ are similar to those
in [2] and are listed in the following proposition.

\noindent {\bf Proposition 11.2} {\it Assume that the input data set $\bold D$ appearing in (4.1) belongs to
the Faddeev class. Then:}

\item{(a)} {\it The matrix $B_j$ defined in (11.3) is hermitian and positive definite.}

\item{(b)} {\it There exists a unique $n\times n$ matrix $B_j^{1/2}$ such that
$B_j^{1/2}B_j^{1/2}=B_j.$ In fact,
$B_j^{1/2}$ is also hermitian and positive definite.}

\item{(c)} {\it The matrices $B_j$ and $B_j^{1/2}$ are both invertible. The inverse of $B_j^{1/2},$
denoted by $B_j^{-1/2},$ is also hermitian and positive definite.}

\item{(d)} {\it Each of the matrices $B_j,$ $B_j^{-1},$ $B_j^{1/2},$ and $B_j^{-1/2}$
commutes with the projection matrix
$P_j$ given in (11.1).}

\noindent PROOF: A condensed proof in the Dirichlet case
can be found on pp. 60--61 of [2].
For the sake of establishing the notation and clarity, we provide
a short proof.
Using (11.1) in (11.3), we see that $B_j=B_j^\dagger,$
and hence $B_j$ is hermitian. For any column vector $v\in\bC^n,$ let us use
$||v||_1$ to denote the standard length of $v,$ i.e. let $||v||_1:=\sqrt{v^\dagger v}.$
Note that $B_j$ is positive definite if $v^\dagger B_j v>0$ for any nonzero
vector $v\in\bC^n.$ From (11.1) and (11.3) it follows that
$$v^\dagger B_j v=|(I-P_j)v|^2+\int_0^\infty dx\,
\left|f(i\kappa_j,x)\,P_j\,v\right|^2.\tag 11.4$$
 From (11.4) we see that $v^\dagger B_j v\ge 0$ and that we have $v^\dagger B_j v=0$
 if and only if $v=P_j v$ and $f(i\kappa_j,x)\,v\equiv 0.$ On the other hand, as seen from (9.1),
we have $f(i\kappa_j,x)\,v=e^{-\kappa_j x}v+o(1)$ as $x\to+\infty,$ and hence $f(i\kappa_j,x)\,v\equiv 0$
if and only if $v=0.$ Thus, we have completed the proof of (a). The existence
of $B_j^{1/2}$ stated in (b) directly follows from (a).
Since $B_j$ is hermitian and positive definite,
all the eigenvalues
of $B_j$ are real and in fact positive. Thus, $B_j$ can be diagonalized,
with the help of a unitary matrix $U,$ into a diagonal
matrix $D_j,$ where $D_j=U^\dagger B_j U.$ It is
clear that there exists a unique matrix $D_j^{1/2}$ such
$D_j=D_j^{1/2} D_j^{1/2}$ and that $D_j^{1/2}$ is invertible
with the inverse denoted by $D_j^{-1/2}.$ Consequently, we have
$B_j^{1/2}=U D_j^{1/2} U^\dagger$ and that $B_j^{1/2}$ is hermitian and
positive definite. Thus, we have proved (b). The invertibility of $B_j$ and
$B_j^{1/2}$ directly follows from the invertibility of
$D_j$ and $D_j^{1/2},$ respectively, and in fact we have
$B_j^{-1}=U D_j^{-1} U^\dagger$ and $B_j^{-1/2}=U D_j^{-1/2} U^\dagger.$
Since $D_j^{-1/2}$ is a diagonal matrix with positive entries, it follows that
$B_j^{-1/2}$ is hermitian and positive definite. Thus, the proof of (c)
is complete. From (11.1) and
(11.3) it directly follows that $P_jB_j=B_j P_j.$ Multiplying the
latter equation by $B_j^{-1}$ on the left and on the right, we establish $P_jB_j^{-1}=B_j^{-1} P_j.$ Since
$B_j=UD_j U^\dagger,$ in $P_jB_j=B_j P_j$
let us replace $B_j$ with $UD_j U^\dagger.$ With some minor algebra,
this yields $(U^\dagger P_j U)D_j=D_j(U^\dagger P_j U).$ Thus,
$U^\dagger P_j U$ commutes with the diagonal matrix $D_j.$ One can directly verify
that $U^\dagger P_j U$ must also commute with each of the diagonal matrices
$D_j^{1/2}$ and $D_j^{-1/2}.$ From $(U^\dagger P_j U)D_j^{1/2}=D_j^{1/2}(U^\dagger P_j U),$
using $B_j^{1/2}=U D_j^{1/2} U^\dagger$ and some minor algebra we get $P_j B_j^{1/2}=B_j^{1/2}P_j.$ Multiplying $P_j B_j^{1/2}=B_j^{1/2}P_j$ by $B_j^{-1/2}$
on the left and on the right,
we also establish $P_j B_j^{-1/2}=B_j^{-1/2}P_j.$ \qed

Let us now clarify the relationship between the Jost solution $f(k,x)$ and the
regular solution $\varphi(k,x)$ at a bound-state value $k=i\kappa_j.$

\noindent {\bf Proposition 11.3} {\it Assume that the input data set $\bold D$ appearing in (4.1) belongs to
the Faddeev class. Let $P_j$ be the projection matrix appearing in (11.1),
where the columns of $P_j$ belong to $\text{\rm{Ker}}[J(i\kappa_j)^\dagger].$
Let $f(k,x)$ and $\varphi(k,x)$ be the Jost
solution and the regular solution appearing in (9.1) and
(9.5), respectively. Then:}

\item{(a)} {\it Corresponding to $P_j,$
there exists a unique $n\times n$ matrix $Q_j$ whose columns belonging to
$\text{\rm{Ker}}[J(i\kappa_j)]$ in such a way that}
$$f(i\kappa_j,x)\,P_j=\varphi(i\kappa_j,x)\,Q_j.\tag 11.5$$

\item{(b)} {\it The matrix $Q_j$ can be expressed explicitly
in various equivalent forms such as}
$$Q_j=E^{-2}(A^\dagger+iB^\dagger)\left[f(i\kappa_j,0)-if'(i\kappa_j,0)\right]P_j,\tag 11.6$$
$$Q_j=E^{-2}\left[A^\dagger f(i\kappa_j,0)+B^\dagger\,f'(i\kappa_j,0)\right]P_j,\tag 11.7$$
{\it where $A,$ $B,$ and $E$ are the constant $n\times n$ matrices appearing in
(2.4)-(2.9).}

\item{(c)} {\it The matrix $Q_j$ can also be expressed explicitly
as}
$$Q_j=-2i \kappa_j N_j A_j P_j,\tag 11.8$$
{\it where $N_j$ and $A_j$ are the matrices appearing in (10.12) and (11.2),
respectively.}

\noindent PROOF: The existence of $Q_j$ and the fact that
the columns of $Q_j$ belong to the kernel of
$J(i\kappa_j)$ directly follow from Theorem~10.1(d) of
[9]. To prove that $Q_j$ appearing in
(11.5) has the form given in (11.6), it is enough to prove that
each side of (11.5) satisfies (2.1) at $k=i\kappa_j$ and that
both sides agree at $x=0$ and that the $x$-derivatives of both sides agree
at $x=0,$ due to the fact that the relevant initial-value problem
has a unique solution. Since every column of $f(i\kappa_j,x)$ and of $\varphi(i\kappa_j,x)$
satisfies (2.1) at $k=i\kappa_j,$ it is clear that
each side of (11.5) satisfies (2.1) at $k=i\kappa_j.$
Recall that $P_j$ is the orthogonal projection onto
$\text{\rm Ker}[J(i\kappa_j)^\dagger],$
and hence we have
$J(i\kappa_j)^\dagger P_j=0.$
With the help of (2.9), (9.2), (9.5), and the fact that
$J(i\kappa_j)^\dagger P_j=0,$ one can directly verify that
the right-hand side of (11.5) at $x=0$ with $Q_j$ as in (11.6)
has the value equal to $f(i\kappa_j,0)\,P_j.$
Similarly, one can directly verify that
the $x$-derivative of the
right-hand side of (11.5) at $x=0$ with $Q_j$ as in (11.6)
has the value equal to $f'(i\kappa_j,0)\,P_j.$ Thus, (11.6) is established.
With the help of (9.2) and
$J(i\kappa_j)^\dagger P_j=0,$ one can simplify (11.6) to
(11.7). Let us now turn to the proof of (c). From the first equality in (10.13)
it follows that $J(i\kappa_j)^\dagger N_j^\dagger=0,$ and
hence the columns of
$N_j^\dagger$ belong to $\text{\rm Ker}[J(i\kappa_j)^\dagger].$
Thus, we have $P_j N_j^\dagger=N_j^\dagger,$ yielding
$$N_j=N_j\, P_j,\tag 11.9$$
where we have used the first equality in (11.1).
With the help of (11.9), we see that the right-hand side of (11.8) satisfies
$$-2i \kappa_j\, N_j\, A_j\, P_j=-2i \kappa_j\, N_j P_j\, A_j\, P_j.\tag 11.10$$
 From (11.2) and (11.5) we get
$$P_j A_j P_j=\int_0^\infty dx\,[\varphi(i\kappa_j,x)\,Q_j]^\dagger \, [\varphi(i\kappa_j,x)\,Q_j].\tag 11.11$$
Using (8.22) of [9] at $\kappa=\kappa_j,$ multiplying the resulting equation
on the left and on the right by $P_j,$ and integrating the
resulting equation over $x\in\bR^+,$ we obtain
$$-2i\kappa_j\,P_j A_j P_j=P_j\left[-f(i\kappa_j,0)^\dagger \,\dot f'(i\kappa_j,0)
+f'(i\kappa_j,0)^\dagger \,\dot f(i\kappa_j,0)\right]P_j,\tag 11.12$$
where we recall that $A_j$ is the matrix in (11.2).
The adjoint of (11.12) yields
$$2i\kappa_j\,P_j A_j P_j=P_j\left[-\dot f'(i\kappa_j,0)^\dagger\,f(i\kappa_j,0)
+\dot f(i\kappa_j,0)^\dagger\,f'(i\kappa_j,0)\right]P_j,\tag 11.13$$
where we have used $P_j^\dagger=P_j$ and
$A_j^\dagger=A_j,$ as seen from (11.1) and (11.2), respectively.
 From (9.5) and (11.5) we see that
$$f(i\kappa_j,0)\,P_j=A\,Q_j,\quad f'(i\kappa_j,0)\,P_j=B\,Q_j,\tag 11.14$$
where $A$ and $B$ are the boundary matrices appearing in (2.4)-(2.6).
Using (11.14) in (11.13) we get
$$2i\kappa_j\,P_j A_j P_j=P_j\left[-\dot f'(i\kappa_j,0)^\dagger\,A
+\dot f(i\kappa_j,0)^\dagger\,B\right]Q_j.\tag 11.15$$
 From the second equality of (8.11) of [9] and from the first
 equality of (8.12) of [9], we see that
$$\dot f(i\kappa_j,0)^\dagger=-\ds\frac{df(-k^\ast,0)^\dagger}{dk}\bigg|_{k=i\kappa_j},
\quad \dot f'(i\kappa_j,0)^\dagger=-\ds\frac{df'(-k^\ast,0)^\dagger}{dk}\bigg|_{k=i\kappa_j}.
\tag 11.16$$
 From (9.2) and (11.16) it follows that
$$\dot f'(i\kappa_j,0)^\dagger\,A
-\dot f(i\kappa_j,0)^\dagger\,B=\dot J(i\kappa_j),\tag 11.17$$
where we recall that an overdot indicates the $k$-derivative. Thus, using
(11.17) in (11.15) we obtain
$$-2i\kappa_j\, P_j\, A_j \,P_j =P_j\, \dot J(i\kappa_j)\,Q_j.\tag 11.18$$
Using (11.18) on the right-hand side of (11.10) we obtain
$$-2i\kappa_j\, N_j\, A_j\, P_j =N_j \,P_j \,\dot J(i\kappa_j)\,Q_j.\tag 11.19$$
With the help of (11.9), we write (11.19) as
$$-2i\kappa_j\, N_j\, A_j\, P_j =N_j\, \dot J(i\kappa_j)\,Q_j.\tag 11.20$$
Using the second equality of (10.13) on the right-hand side of (11.20)
we obtain
$$-2i\kappa_j\, N_j\, A_j\, P_j =\left[I-L_j\, J(i\kappa_j)\right]Q_j.\tag 11.21$$
By (a) we know that the columns of
$Q_j$ belong to $\text{\rm Ker}[J(i\kappa_j)],$ and hence
 we have $J(i\kappa_j)\,Q_j=0.$ Thus,
(11.21) simplifies to (11.8). \qed

As on p. 61 of [9], we define the normalization matrices $M_j$ as
$$M_j:=B_j^{-1/2} P_j,\qquad j=1,\dots,N,\tag 11.22$$
where $P_j$ and $B_j$ are the matrices appearing in (11.1) and
(11.3), respectively. Recall that the existence of
$B_j^{-1/2}$ and its basic properties are stated in Proposition~11.2.
Let us now consider the $n\times n$ matrix-valued
function $\Psi_j(x)$ defined in (9.8).
Some relevant properties of
$\Psi_j(x)$ are indicated in the following proposition.

\noindent {\bf Proposition 11.4} {\it Assume that the input data set $\bold D$ in (4.1) belongs to
the Faddeev class specified in Definition~4.1. Then:}

\item{(a)} {\it The normalization matrix $M_j$ defined in (11.22) is hermitian and nonnegative
    and has rank
equal to $m_j,$ which is the dimension of $\text{\rm{Ker}}[J(i\kappa_j)^\dagger].$}

\item{(b)} {\it At each bound state $k=i\kappa_j,$ the matrix $\Psi_j(x)$ defined
in (9.8) satisfies (2.1) and the boundary condition (2.4).}

\item{(c)} {\it The matrix $\Psi_j(x)$ is normalized in the sense that}
$$\int_0^\infty dx\, \Psi_j(x)^\dagger\, \Psi_j(x)=P_j,\qquad j=1,\dots,N,\tag 11.23$$
{\it where $P_j$ is the projection matrix appearing in (11.1). Furthermore,
the matrices $\Psi_j(x)$ for $j=1,\dots,N$ are orthogonal in the sense that}
$$\int_0^\infty dx\, \Psi_l(x)^\dagger\, \Psi_j(x)=0,\qquad l\ne j.\tag 11.24$$

\item{(d)} {\it For each $j=1,\dots,N,$ the normalized bound-state matrix solution
$\Psi_j(x)$ defined in (9.8)
yields $m_j$ linearly-independent column-vector
solutions to (2.1) at $k=i\kappa_j,$
where $m_j$ is equal to the rank of
the normalization matrix $M_j$ given in (11.22).}

\item{(e)} {\it Any column-vector solution to (2.1) at $k=i\kappa_j$
satisfying (2.4) can be written
as $\Psi_j(x)\,v$ for some column vector $v$ in $\bC^n.$
Equivalently stated, any eigenfunction of the Schr\"odinger operator
associated with (2.1) and (2.4) can be written as
$\Psi_j(x)\,v$ for some $v\in\bC^n$ and some positive
integer $j$ with $1\le j\le N.$}

\noindent PROOF: The projection matrix
$P_j$ appearing in (11.1) has $m_j$ linearly independent columns, where $m_j$ is the dimension of $\text{\rm{Ker}}[J(i\kappa_j)^\dagger].$ Thus, the rank of $P_j$ is $m_j.$
 From (11.1) we know that $P_j$ is hermitian.
By Proposition~11.2(c) we know that $B_j^{-1/2}$ is hermitian and invertible.
Thus, from (11.22) we conclude that $M_j$ is hermitian and has rank equal to $m_j.$
The nonnegativity of $M_j$ follows from (11.1), (11.22), the positive definiteness of $B^{-1/2}_j,$ and $B_j^{-1/2}P_j=P_j B_j^{-1/2},$ where the last two
properties are assured by Proposition~11.2.
Let us now prove (b).
With the help of (11.5), (11.22), and Proposition~11.2(d), we can write
(9.8) as
$$\Psi_j(x)=\varphi(i\kappa_j,x)\,Q_j P_j B^{-1/2}_j,\qquad j=1,\dots,N.\tag 11.25$$
Since $\varphi(k,x)$ is
an $n\times n$ matrix-valued solution
to (2.1),
each column of $\varphi(i\kappa_j,x)$ satisfies (2.1) at $k=i\kappa_j.$
Hence, the right-hand side of (11.25) satisfies the corresponding matrix
Schr\"odinger equation at $k=i\kappa_j.$
Using (9.5) in (2.4), with the help of (2.5) we conclude that (2.4) is
satisfied by the right-hand side of (11.25) and hence also by $\Psi_j(x).$
Let us now turn to the proof of (c).
Using (11.25) we write the left-hand side of (11.23) as
$$\int_0^\infty dx\, \Psi_j(x)^\dagger\, \Psi_j(x)=B_j^{-1/2}\, P_j
\int_0^\infty dx\,[\varphi(i\kappa_j,x)\,Q_j]^\dagger \, [\varphi(i\kappa_j,x)\,Q_j]P_j B_j^{-1/2},\tag 11.26$$
where we have used the fact that
$P_j$ and $B_j^{-1/2}$ are hermitian.
Using (11.11) on the right-hand side of (11.26) we obtain
$$\int_0^\infty dx\, \Psi_j(x)^\dagger\, \Psi_j(x)=B_j^{-1/2} P_j\, P_j\, A_j \,P_j
P_j\, B_j^{-1/2}.\tag 11.27$$
 With the help of (11.3) we replace $P_jA_jP_j$ on the right-hand side
 of (11.27) by $B_j-(I-P_j)$ and hence obtain
 $$\int_0^\infty dx\, \Psi_j(x)^\dagger\, \Psi_j(x)=B_j^{-1/2}\, P_j \,[B_j-I+P_j]\,
P_j \,B_j^{-1/2}.\tag 11.28$$
Next, with the help of Proposition~11.2(d) and the second equality in (11.1),
we simplify the right-hand side of (11.28) and obtain (11.23).
The proof of (11.24) will be given in the proof of Proposition~22.4(a).
Let us now prove (d).
Each of the $n$ columns of $f(i\kappa_j,x)$ satisfies
(2.1) at $k=i\kappa_j.$ From (9.1) we get
$$f(i\kappa_j,x)=e^{-\kappa_j x}\left[I+o(1)\right],\qquad x\to +\infty,\tag 11.29$$
and hence from (11.29)
we conclude that
$f(i\kappa_j,x)$ yields $n$ linearly-independent
column-vector solutions to (2.1) at $k=i\kappa_j.$
Thus, from (9.8) we conclude that
$\Psi_j(x)$ yields as many linearly-independent
column-vector solutions as the rank of the matrix $M_j.$ By (a) we know that
the rank of $M_j$ is $m_j,$ and hence we conclude that
$\Psi_j(x)$ yields $m_j$ linearly-independent column-vector
solutions to (2.1) at $k=i\kappa_j.$ Hence, the proof of
(d) is complete. Let us now prove (e).
 From Theorem~8.1(c) of [9] it follows that any bound-state column-vector
solution to (2.1) at $k=i\kappa_j$ must be a linear combination of
the columns of $\Psi_j(x).$ Thus, any column-vector solution to
(2.1) at $k=i\kappa_j$ satisfying (2.4) must be a linear
combination of the columns of $\Psi_j(x),$
indicating that such a column-vector solution must be
of the form $\Psi_j(x)\,v$ for some constant column vector in $\bC^n.$
By definition, an eigenvector of the Schr\"odinger operator
is a square-integrable column vector solution to (2.1) satisfying
the boundary condition (2.4). Thus, an eigenvector of
 the Schr\"odinger operator must have the form
 $\Psi_j(x)\,v$ for some $v\in\bC^n$ and some positive
integer $j$ with $1\le j\le N.$ This completes the proof of (e).
 \qed

The following result is needed later on in the derivation of the Marchenko integral equation (13.1).

\noindent {\bf Proposition 11.5} {\it Assume that the input data set $\bold D$ appearing in (4.1) belongs to
the Faddeev class specified
in Definition~4.1. Then, at each bound state with $k=i\kappa_j,$ the Jost solution
$f(k,x)$ appearing in (9.1) and the regular solution $\varphi(k,x)$
appearing in (9.5) are related to each other as}
$$f(i\kappa_j,x)\,M_j^2=-2i\kappa_j\,\varphi(i\kappa_j,x)\,N_j,
\qquad j=1,\dots,N,\tag 11.30$$
{\it where $M_j$ is the normalization matrix defined in
(11.22) and $N_j$ is the residue matrix appearing in (10.12).}

\noindent PROOF: Using (9.8) and (11.25), after multiplication with $M_j$ on the
right, we obtain
$$f(i\kappa_j,x)\,M_j^2=\varphi(i\kappa_j,x)\,Q_j\, P_j\, B_j^{-1/2}\,M_j.\tag 11.31$$
Using (11.3), (11.8)-(11.10), (11.22), and Proposition~11.2(d),
we simplify the right-hand side
of (11.31) and obtain (11.30). \qed

The following result is also later needed in the derivation of the Marchenko integral equation (13.1).

\noindent {\bf Proposition 11.6} {\it Assume that the input data set $\bold D$ appearing in (4.1) belongs to
the Faddeev class specified in Definition~4.1. Let $\Psi(k,x)$ be the physical solution appearing
 in (9.4) and (9.6),
$f(k,x)$ be the Jost
solution appearing in (9.1), and $M_j$ be the normalization matrix given in (11.22).
Then, for $y>x\ge 0$ we have}
$$\ds\frac{1}{2\pi}\int_{-\infty}^\infty dk\,\Psi(k,x)\,e^{iky}=-\sum_{j=1}^N
f(i\kappa_j,x)\,M_j^2\, e^{-\kappa_j y}.\tag 11.32$$

\noindent PROOF:
For each fixed
$x\in[0,+\infty),$ let us consider the quantity $e^{ikx}\Psi(k,x)-(W_2+W_3\,e^{2ikx})$
appearing in (10.22). By Proposition~10.5, that quantity is continuous in $k\in\bR,$
behaves like $O(1/k)$ as $k\to\infty$ in $\bCpb,$ and is meromorphic
in $\bCp$ with simple poles at $k=i\kappa_j$ for $j=1,\dots,N.$ Thus,
for $y>x\ge 0,$ with the help of residues we have
$$\ds\frac{1}{2\pi}\int_{-\infty}^\infty dk\, \left[
e^{ikx}\Psi(k,x)-(W_2+W_3\,e^{2ikx})\right]e^{ik(y-x)}=
i\ds\sum_{j=1}^N \text{Res}\left[e^{ikx}\Psi(k,x)\,e^{ik(y-x)},i\kappa_j\right],
\tag 11.33$$
where the notation $\text{Res}[g(k),i\kappa_j]$ is used to denote
the residue of a function $g(k)$ at $k=i\kappa_j.$
With the help of (9.6), (10.12), and Proposition~10.5(b), we evaluate
each residue on the right-hand side of (11.33) as
$$\text{Res}\left[e^{ikx}\Psi(k,x)\,e^{ik(y-x)},i\kappa_j\right]=2i\kappa_j\,
e^{-\kappa_j y}\,\varphi(i\kappa_j,x)\,N_j.\tag 11.34$$
Using (11.30) we can write
(11.34) as
$$\text{Res}\left[e^{ikx}\Psi(k,x)\,e^{ik(y-x)},i\kappa_j\right]=-
e^{-\kappa_j y}f(i\kappa_j,x)\,M_j^2.\tag 11.35$$
We have
$$\ds\frac{1}{2\pi}\int_{-\infty}^\infty dk\,(W_2+W_3\,e^{2ikx})\,e^{ik(y-x)}=
W_2\, \delta(y-x)+W_3\,\delta(y+x),\tag 11.36$$
where $\delta(x)$ is the Dirac delta distribution given by
$$\delta(x):=\ds\frac{1}{2\pi}\int_{-\infty}^\infty dk\,e^{ikx},\qquad x\in\bR.\tag 11.37$$
When $y>x\ge 0,$ the right-hand side of (11.36) vanishes. Thus,
for $y>x\ge 0,$ using (11.35) and (11.36) in (11.33) we obtain (11.32). \qed

\newpage
\noindent {\bf 12. FURTHER PROPERTIES OF THE SCATTERING DATA}
\vskip 3 pt

In this chapter we continue to elaborate
on the steps outlined in Section~3 to solve the
direct scattering problem, by establishing various
 properties of the scattering data set $\bold S$ corresponding
 to an input data set $\bold D$ in the Faddeev class.
  Toward that goal, we obtain some relevant
properties of $F_s(y)$ and $F(y)$ defined in (4.7) and (4.12),
respectively.

\noindent {\bf Theorem 12.1} {\it Let the scattering data set $\bold S$ in (4.2)
 correspond to
the input data set $\bold D$ in (4.1) that belongs to
the Faddeev class specified in Definition~4.1, where
$S(k)$ is the corresponding scattering matrix
defined in (9.3), the $\kappa_j$ are the distinct positive constants appearing in (9.8)
related to the bound states,
and $M_j$ are the $n\times n$ normalization matrices appearing in (4.12)
and (11.22).
Let $F_s(y)$ and $F(y)$ be the $n\times n$ matrix-valued
functions defined in (4.7) and (4.12), respectively.
We then have the following:}

\item{(a)} {\it The matrices $F_s(y)$ and $F(y)$ are both hermitian, i.e.
$F_s(y)^\dagger=F_s(y)$ for each $y\in\bR^+\cup \bR^-$ and $F(y)^\dagger=F(y)$ for each $y\in\bR^+.$}

\item{(b)} {\it The derivative matrix $F'_s(y)$ is hermitian, i.e.
$F'_s(y)^\dagger=F'_s(y)$ for each $y\in\bR^+\cup \bR^-.$}

\item{(c)} {\it The matrices
$F_s(y)$ and $F(y)$ are continuous and bounded in $y\in\bR^+,$
and they both vanish as $y\to+\infty.$}

\item{(d)} {\it The matrix $F_s(y)$ is continuous and bounded in $y\in\bR^-,$
and it vanishes as $y\to-\infty.$}

\item{(e)} {\it The matrices $F_s(y)$ and
$F(y)$ are each bounded and
integrable in $y\in\bR^+.$}

\item{(f)} {\it For $y\in\bR^-$ the matrix $F_s(y)$
can be written as a sum as}
$$F_s(y)=F_s^{(1)}(y)+F_s^{(2)}(y),\qquad y\in\bR^-,\tag 12.1$$
{\it where $F_s^{(1)}(y)$ is bounded and integrable for $y\in\bR^-$ and
$F_s^{(2)}(y)$ is bounded and square integrable for $y\in\bR^-.$
Consequently, $F_s(y)$ itself is bounded and square integrable in $y\in\bR^-.$}

\item{(g)} {\it The matrix $F_s(y)$ has a jump discontinuity at $y=0,$
which is given by}
$$F_s(0^+)-F_s(0^-)=G_1\tag 12.2$$
{\it where $G_1$ is the constant matrix appearing in (10.17).
Hence, $F_s'(y)$ contains a delta-function term at $y=0,$
which is given by $G_1\,\delta(y).$}

\item{(h)} {\it For $y\in\bR^-,$ the matrix $F'_s(y)$
can be written as a sum of
two matrix-valued functions, one of which is integrable and the other
is square integrable.}

\noindent PROOF: Note that $S_\infty$ appearing in (4.6) and
(4.7) corresponds to the scattering matrix
when the potential is zero. Hence, from Proposition~10.3(a) it follows
that
both $S(k)$ and $S_\infty$ satisfy (4.4). With the help of
the first equality in (4.4) for $S(k)$ and $S_\infty,$ from (4.7) we conclude that
$F_s(y)$ is hermitian. The matrix $M_j$ given in (11.22)
is hermitian because the matrix $B_j^{-1/2}$ is hermitian as stated
Proposition~11.2(c) and the matrix $P_j$ is hermitian as indicated by the first
equality in (11.1). Since $\kappa_j$ appearing in (4.12) is positive
for each $j=1,\dots,N,$ we also conclude that $F(y)$ is hermitian.
Thus, (a) holds.
 From (4.7) we
have
$$F_s'(y)=\ds\frac{1}{2\pi}\int_{-\infty}^\infty dk\,
ik\left[S(k)-S_\infty\right] e^{iky}.\tag 12.3$$
Using the first equality in (4.4) for $S(k)$ and $S_\infty,$ from (12.3) we conclude that
$F_s(y)$ is hermitian, and hence (b) is proved.
For $y\in\bR^+,$ it is readily seen that the summation part in (4.12)
is continuous, bounded, and integrable
on $\bR^+$ and it vanishes as $y\to+\infty.$ Thus, it is sufficient to prove (c)-(f) only for $F_s(y).$
Let us decompose $S(k)-S_\infty$ as
$$S(k)-S_\infty=\left[S(k)-S_\infty+\ds\frac{i\,G(k)}{k+i}\right]-
\ds\frac{i\,G(k)}{k+i},\tag 12.4$$
where $G(k)$ is the matrix appearing in (10.14) and (10.16).
Using (12.4) in (4.7) we see that
$$F_s(y)=\ds\frac{1}{2\pi}\int_{-\infty}^\infty dk\,
\left[S(k)-S_\infty+\ds\frac{i\,G(k)}{k+i}\right]e^{iky}-
\ds\frac{1}{2\pi}\int_{-\infty}^\infty dk\,
\ds\frac{i\,G(k)}{k+i}\,e^{iky}.\tag 12.5$$
With the help of Proposition~10.3, we conclude that the term
in the brackets on the right-hand side
in (12.4)
is continuous in $k\in\bR$ and is $O(1/k^2)$ as $k\to\pm\infty.$
Hence, that term is both integrable and square integrable
in $k\in\bR.$ Thus, the first integral in (12.5) is continuous
in $y\in\bR,$ vanishes as $y\to\pm\infty,$ and is square
integrable in $y\in\bR.$ It is also bounded in $y\in\bR$ as
a result of the continuity in $y\in\bR$
and the zero asymptotics
as $y\to\pm\infty.$ Thus, we only need to show that (c)-(f) hold
for the second integral in (12.5).
 From Proposition~10.3(b) we know that $G(k)$ is the sum
 of $G_1$ and $G_2(k)$ given in
 (10.17) and (10.18), respectively.
Hence, with the help of (10.17) and (10.18), we are able to evaluate the second integral
in (12.5) explicitly by using the residues, and we get
$$-\ds\frac{1}{2\pi}\int_{-\infty}^\infty dk\,
\ds\frac{i\,G(k)}{k+i}\,e^{iky}=\cases g_+(y),\qquad y>0,\\
g_-(y),\qquad y<0,\endcases\tag 12.6$$
where we have defined
$$g_+(y):=
-\ds\frac{1}{2}\int_{y/2}^\infty dz\, V(z)\, e^{-(2z-y)},\tag 12.7$$
$$g_-(y):=-G_1\, e^y-\ds\frac{1}{2}\int_0^\infty dz\, V(z)\, e^{-(2z-y)}
-\ds\frac{1}{2}\int_0^{-y/2} dz\, S_\infty V(z)\,S_\infty\, e^{2z+y}.\tag 12.8$$
Our proof for (c)-(f) will be complete if we can show that $g_+(y)$ given in (12.7) is
continuous and bounded in $y\in\bR^+,$ vanishes as $y\to+\infty,$ and
is integrable on $y\in\bR^+$ and also that $g_-(y)$ given in (12.8)
is
continuous and bounded in $y\in\bR^-,$ vanishes as $y\to-\infty,$ and
is integrable on $y\in\bR^-.$
We will show that (2.3) guarantees the aforementioned properties.
In terms of $\sigma(x)$ and $\sigma_1(x)$ defined in (3.96), from (12.7)
we obtain
$$||g_+(y)||\le \ds\frac{1}{2}\int_{y/2}^\infty dz\,|V(z)|\le \ds\frac{1}{2}\,\sigma\left(\ds\frac{y}{2}\right).\tag 12.9$$
 From (2.3), (3.96), and (12.9)
 it follows that $g_+(y)$ is bounded and integrable
 in $y\in\bR^+$ and vanishes as $y\to+\infty.$
Let us now prove the continuity of $g_+(y)$ in $\bR^+.$ From (12.7) we see
that
$$e^{-y} g_+(y)=
-\ds\frac{1}{2}\int_{y/2}^\infty dz\, V(z)\, e^{-2z},\tag 12.10$$
where the integrand is integrable as a result of (2.3). Thus,
by the Lebesgue differentiation theorem, the right-hand side is
a continuous function of $y.$ Since $e^y$ is continuous,
the function $g_+(y),$ being a product of two continuous functions,
is also continuous in $y\in\bR^+.$
In a similar manner, one can prove that
$g_-(y)$
is
continuous and bounded in $y\in\bR^-,$ vanishes as $y\to-\infty,$ and
is integrable on $y\in\bR^-.$
Note that the first term, $-G_1\, e^y,$ on the right-hand side of (12.8) is readily
seen to satisfy these four
properties. For the second term on the right-hand side
of (12.8), by taking $e^y$
outside the integral, we see that
the second integral is the product of $e^y$ with a constant
$n\times n$ matrix, and hence it readily satisfies all the
four properties.
As for the third term on the right-hand side of (12.8), let us break it
into two terms as
$$\aligned
-\ds\frac{1}{2}\int_0^{-y/2} dz\, S_\infty V(z)\,S_\infty\, e^{2z+y}=&
-\ds\frac{1}{2}\,e^{y/2}\int_0^{-y/4} dz\, S_\infty V(z)\,S_\infty\, e^{2z+y/2}\\
\stretch
&-\ds\frac{1}{2}\int_{-y/4}^{-y/2} dz\, S_\infty V(z)\,S_\infty\, e^{2z+y}.
\endaligned\tag 12.11$$
For the first term on the right-hand side of (12.11) we have
$$\aligned
\bigg|-\ds\frac{1}{2}\,e^{y/2}\int_0^{-y/4} dz\, S_\infty V(z)\,S_\infty\, e^{2z+y/2}
\bigg|\le &\ds\frac{1}{2}\,e^{y/2} |S_\infty|^2\int_0^{-y/4} dz\,|V(z)|\\
\stretch\le &\ds\frac{1}{2}\,e^{y/2} |S_\infty|^2\sigma(0).
\endaligned\tag 12.12$$
For the second term on the right-hand side of (12.11) we get
$$\aligned
\bigg|
-\ds\frac{1}{2}\int_{-y/4}^{-y/2} dz\, S_\infty V(z)\,S_\infty\, e^{2z+y}
\bigg|\le &\ds\frac{1}{2}\,|S_\infty|^2\int^{-y/2}_{-y/4} dz\,|V(z)|
\\
\stretch= &\ds\frac{1}{2}\, |S_\infty|^2\left[\sigma\left(-\ds\frac{y}{4}\right)
-\sigma\left(-\ds\frac{y}{2}\right)\right].
\endaligned\tag 12.13$$
The left-hand side of (12.12) is bounded and integrable in
$y\in\bR^-$ and it vanishes as
$y\to-\infty$ because that left-hand side is bounded by a constant multiple of
$e^{y/2}$ on $\bR^-.$
 From (12.13) we see that its left-hand side is bounded by a constant multiple of
$\sigma(-y/4),$ and we already know from (2.3) and the first equality in (3.96)
that $\sigma(-y/4)$ is bounded on $\bR^-$ and vanishes
as $y\to-\infty$ and  we also know from (2.3) and
the second equality in (3.96)
that $\sigma(-y/4)$ is integrable in $\bR^-.$
Hence, the left-hand side of (12.13) is bounded and integrable on $\bR^-$
and it vanishes
as $y\to-\infty.$
In order to complete the proof of our theorem,
we only need to prove that the left-hand side of
(12.11) is continuous in $y\in\bR^-.$ For a given
$y\in\bR^-,$ we can find a constant $a<0$ so that
$a<y.$ We can write the left-hand side of
(12.11) as
$$-\ds\frac{1}{2}\int_0^{-y/2} dz\, S_\infty V(z)\,S_\infty\, e^{2z+y}=
-\ds\frac{1}{2}\,e^{y-a} \int_0^{-y/2} dz\, S_\infty V(z)\,S_\infty\, e^{2z+a}.
\tag 12.14$$
By the Lebesgue differentiation theorem,
the integral on the right-hand side of (12.14) is
a continuous function of $y$ in
$y\in[a,0)$ because the integrand is
integrable as a result of (2.3).
Since $e^{y-a}$ is also continuous, we conclude that
the right-hand side of (12.14) is continuous
in $y\in[a,0)$ for every $a<0.$ Thus, we have completed the proof that
the left-hand side of
(12.11) is continuous in $y\in\bR^-.$ Having completed the proof of
(c)-(f), let us now prove (g). Earlier in the proof, we have already
indicated that the first integral in (12.5) is continuous in $y\in\bR.$
Thus, from (12.5) and (12.6), it follows that the left-hand side of
(12.2) is given by
$$F_s(0^+)-F_s(0^-)=g_+(0)-g_-(0).\tag 12.15$$
Using (12.7) and (12.8) on the right-hand side of (12.15), we get
(12.2). Thus, the proof of (g) is complete. Finally, let us prove (h).
 From (10.14)-(10.18) it follows that
$$ik\left[S(k)-S_\infty\right]-G_1-G_2(k)=O\left(\ds\frac{1}{k}\right),
\qquad k\to\pm\infty.\tag 12.16$$
By Proposition~10.3(a) we know that the scattering matrix
$S(k)$ is continuous in $k\in\bR.$ Hence, from (12.15) we conclude
that the left-hand side of
(12.16) is square integrable in $k\in\bR.$
Using
(10.17), (10.18), (11.37), and (12.16) we conclude that
$$\aligned
\ds\frac{1}{2\pi}\int_{-\infty}^\infty dk\, \left\{ \right.
ik&  \left. \left[S(k)-S_\infty\right]-G_1-G_2(k)\right\} e^{iky}\\
\stretch
 &=F_s'(y)-G_1\,\delta(y)
-\ds\frac{1}{2}\,V\left(\ds\frac{y}{2}\right)-\ds\frac{1}{2}\,
S_\infty\,V\left(-\ds\frac{y}{2}\right)\,S_\infty,\endaligned\tag 12.17$$
with the understanding that $V(x)=0$ for $x\in\bR^-.$
When $y\in\bR^-,$ from (2.3) we know that
the last term on the right-hand side of (12.17) is integrable.
The left-hand side of (12.17) is square integrable
in $y\in\bR$ because it is the Fourier transform
of a square-integrable function of $k\in\bR.$ Thus, from (12.17)
we conclude (h). \qed

\newpage
\noindent {\bf 13. THE MARCHENKO INTEGRAL EQUATION}
\vskip 3 pt

In this chapter, when the
input data set $\bold D$ given in (4.1)
belongs to the Faddeev class, we derive the matrix Marchenko integral equation
and provide the basic properties of its kernel.
We also introduce the derivative Marchenko integral equation
(13.7), whose kernel coincides with the
kernel of (1.1) but whose nonhomogeneous term
differs from the nonhomogeneous term of (1.1).
We remark that the
kernel of (13.1) coincides with the kernel of (13.7)
and the two equations only differ by their nonhomogeneous terms.
In the Dirichlet case, the boundary matrix appearing $A$
in (2.4) is the zero matrix and the boundary matrix
$B$ then can be chosen as the identity matrix $I.$
Thus, in the Dirichlet case, as it is studied
in [2], only the Marchenko equation (13.1)
plays a relevant role in the inverse problem, and the derivative Marchenko equation
(13.7) is hardly relevant to the
inverse problem.
On the other hand, in the case
of the general selfadjoint boundary condition, which we study in this monograph,
the roles of (13.1) and (13.7) are equally important
in the analysis of the inverse problem.

\noindent {\bf Theorem 13.1} {\it Let the scattering data set $\bold S$ in (4.2)
 correspond to
the input data set $\bold D$ in (4.1) that belongs to
the Faddeev class specified in Definition~4.1, and let
$K(x,y)$ be the quantity appearing in (10.1), and
$F_s(y)$ and $F(y)$ be the quantities defined in (4.7) and (4.12),
respectively.
Then, $K(x,y)$ satisfies the Marchenko integral equation given by}
$$K(x,y)+F(x+y)+\int_x^\infty dz\,K(x,z)\,F(z+y)
=0,\qquad 0\le x<y.\tag 13.1$$

\noindent PROOF: When $\bold D$ belongs to
the Faddeev class, the existence of $K(x,y)$ and its properties
are assured by Proposition~10.1. The existence of $F_s(y)$ and
$F(y)$ and their properties are assured by Theorem~12.1.
Let us write (9.4) as
$$\aligned
[f(-k,x)-e^{-ikx}I]+&
[S(k)-S_\infty]e^{ikx}+
[f(k,x)-e^{ikx}I]
[S(k)-S_\infty]\\
\stretch
&=\Psi(k,x)-e^{-ikx}I-S_\infty e^{ikx}-
[f(k,x)-e^{ikx}I]S_\infty.\endaligned\tag 13.2$$
Taking the Fourier transform of both sides of (13.2) and using
(10.1) and (11.37), we obtain
$$\aligned
K(x,y)+&F_s(x+y)+\int_x^\infty dz\,K(x,z)\,F_s(z+y)\\
\stretch
&=
\ds\frac{1}{2\pi}\int_{-\infty}^\infty dk\,\Psi(k,x)\,
e^{iky}-I\delta(y-x)-S_\infty \delta(y+x)-
K(x,-y)\,S_\infty.\endaligned\tag 13.3$$
When $y>x\ge 0,$ with the help of (10.2) we see that
only the first term on the right-hand side of (13.3) is nonzero and
in fact that term is explicitly evaluated in (11.32).
Thus, using (11.32) in (13.3) we get
$$K(x,y)+F_s(x+y)+\int_x^\infty dz\,K(x,z)\,F_s(z+y)
=-\sum_{j=1}^N
f(i\kappa_j,x)\,M_j^2\, e^{-\kappa_j y}.\qquad 0\le x<y.\tag 13.4$$
 From (10.6) we obtain
$$f(i\kappa_j,x)\,M_j^2\,e^{-\kappa_j y}=e^{-\kappa_j(x+y)}
M_j^2+\int_x^\infty dz\,K(x,z)\,e^{-\kappa_j(z+y)}M_j^2.\tag 13.5$$
Using (13.5) on the right-hand side of (13.4), with the help of
(4.12), we write (13.4) as (13.1). \qed

By taking the $x$-derivative
of the Marchenko integral equation (13.1) we obtain the integral equation
$$K_x(x,y)+F'(x+y)-K(x,x)\, F(x+y)+\int_x^\infty dz\,
K_x(x,z)\,F(z+y)=0,
\qquad 0\le x<y,\tag 13.6$$
where we recall that
the subscript $x$ is used to denote the $x$-derivative.
We call the integral equation associated with
(13.6), i.e.
$$L(x,y)+F'(x+y)-K(x,x)\, F(x+y)+\int_x^\infty dz\,
L(x,z)\,F(z+y)=0,
\qquad 0\le x<y,\tag 13.7$$
the derivative Marchenko integral equation.
We remark that (13.7)
along with the Marchenko equation
(13.1) plays a key role in the analysis of the inverse scattering
problem related to
(2.1) with the general selfadjoint boundary
condition (2.4). Its solvability
in the context of the inverse problem
is analyzed in Proposition~16.5.

We have the following further comment
on comparing the Marchenko equation
(13.1) and the derivative Marchenko equation (13.7) when
the input data set $\bold D$ belongs to the Faddeev class.
Concerning the Marchenko equation (13.1),
as pointed out in Theorem~12.1(c) the nonhomogeneous term
$F(x+y)$ in (13.1) is continuous when $y>x\ge 0,$ and as
pointed out in Theorem~12.1(c) Proposition~10.1(e) the solution
$K(x,y)$ to (13.1) is continuous when $y>x\ge 0.$
On the other hand, concerning the derivative
Marchenko equation (13.7),
the nonhomogeneous term contains
$F'(x+y)$ and that term is in general not continuous
when $y>x\ge 0,$ and as indicated in
Proposition~10.1(f) the quantity $K_x(x,y)$
is in general not continuous and
exists a.e. but for each $x\ge 0$ it is
integrable in $y\in[x,+\infty).$

In the next theorem, certain relevant properties of the
kernel of the Marchenko equation (13.1) is presented when
the input data set $\bold D$ belongs to the Faddeev class.

\noindent {\bf Theorem 13.2} {\it Let the scattering data set $\bold S$ in (4.2)
 correspond to
the input data set $\bold D$ in (4.1) that belongs to
the Faddeev class specified in Definition~4.1, and let
$F_s(y)$ and $F(y)$ be the quantities defined in (4.7) and (4.12), respectively.
We then have the following:}

\item{(a)} {\it The matrix $F(y)$ satisfies}
$$|F(y)|\le C\,\sigma\left(\ds\frac{y}{2}\right),\qquad y\in\bR^+,\tag 13.8$$
{where $C$ is a generic constant and $\sigma(x)$ is the quantity
defined in (3.96). Furthermore, we have}
$$\int_0^\infty dy\,|F(y)|<+\infty,\tag 13.9$$
$$\int_0^\infty dy\, (1+y)\,|F(y)|^2<+\infty.\tag 13.10$$

\item{(b)} {\it The derivative $F'(y)$ exists a.e. for
$y\in\bR^+$ and satisfies}
$$\bigg| F'(y)-\ds\frac{1}{4}\,V\left(\ds\frac{y}{2}\right)\bigg|\le C\, \left[\sigma\left(\ds\frac{y}{2}\right)\right]^2,
\qquad y\in\bR^+,\tag 13.11$$
{\it where $C$ is a generic constant and $V(x)$ is the potential
appearing in the data set $\bold D.$}

\item{(c)} {\it The derivative $F'(y)$ satisfies}
$$\int_0^\infty dy\,(1+y)\,|F'(y)|<+\infty.\tag 13.12$$

\noindent PROOF: The inequality in (13.8) can be found in
(3.2.4) of [2], where the main idea behind
the proof in [2] is to view the Marchenko equation (13.1)
with $K(x,y)$ as being the input and $F(x+y)$ as being the unknown
quantity, to use (10.7), and to get the corresponding property
of $F(x+y).$
The inequality in (13.11) is given in
(3.2.7) of [2], where the basic idea behind the proof is to
get the appropriate property of $F'(y)$ from (13.6)
with the help of (13.8) and (10.9).
Thus, it is enough to establish (13.9), (13.10), and (13.12).
To prove (13.9), we integrate both sides of (13.8) over
$y\in\bR^+$ and use the second definition
in (3.97) and obtain
%
%
%
$$\int_0^\infty dy \,|F(y)|\le C\,\int_0^\infty dy\,
\sigma\left(\ds\frac{y}{2}\right)=2\,C\,\sigma_1(0)<+\infty,\tag 13.13$$
which establishes (13.9). Let us now prove (13.10).
Squaring both sides of (13.8) and then multiplying by $(1+y)$
and integrating over $y\in\bR^+,$ we obtain
$$\int_0^\infty dy\, (1+y)\,|F(y)|^2
\le C^2 \int_0^\infty dy\, (1+y)\,\sigma\left(\ds\frac{y}{2}\right)
\le C^2 \left[2 \sigma(0)\,\sigma_1(0)+4 [\sigma_1(0)]^2\right]
<+\infty,\tag 13.14$$
where we have used (3.98) and (3.99). Thus, (13.10) is established.
Let us finally prove (13.12). Since we have
$$|F'(y)|\le \ds\frac{1}{4}\,\bigg|V\left(\ds\frac{y}{2}\right)\bigg|+
\bigg| F'(y)-\ds\frac{1}{4}\,V\left(\ds\frac{y}{2}\right)\bigg|,\tag 13.15$$
after using (13.11) in (13.15),
we can multiply both sides of the resulting
inequality by $(1+y)$
and integrate over $y\in\bR^+$ in order to obtain
$$\int_0^\infty dy \,(1+y)\,|F'(y)|\le
\ds\frac{1}{4}\int_0^\infty dy \,(1+y)\,\bigg|V\left(\ds\frac{y}{2}\right)\bigg|+
C\int_0^\infty dy \,(1+y)\,\left[\sigma\left(\ds\frac{y}{2}\right)\right]^2.\tag 13.16$$
By (2.3) the first integral on the right-hand side in (13.16) is finite.
The finiteness of the second integral
follows from (3.98) and (3.99). Thus, (13.11) is established. \qed

\newpage
\noindent {\bf 14. THE BOUNDARY MATRICES}
\vskip 3 pt

In this chapter, when the input data set $\bold D$ belongs to
the Faddeev class, we show that
the boundary matrices $A$ and $B$ appearing in
(2.4)-(2.6) are related to the large-$k$ limit of the scattering matrix
$S(k).$

\noindent {\bf Proposition 14.1} {\it Let the input data set $\bold D$ in (4.1)
belong to the Faddeev class specified in
Definition~4.1. Let $\bold S$ in (4.2) be the
scattering data set corresponding to $\bold D,$ i.e. $S(k)$ be the scattering
matrix defined as in (9.3), where the Jost matrix $J(k)$ is defined
as in (9.2). Let $S_\infty$ be the constant $n\times n$
matrix appearing in
(4.6), $G_1$ be the constant $n\times n$ matrix appearing in (10.14) and (10.17)
and equivalently
given as}
$$G_1=\lim_{k\to \pm\infty} ik[S(k)-S_\infty],\tag 14.1$$
{\it and $K(0,0)$ be the constant $n\times n$ matrix obtained as
in (10.5) from
$K(x,y)$ appearing in (10.1) and (13.1).
Then, the boundary matrices $A$ and $B$ appearing
in (2.4)-(2.6) and (4.1) satisfy the linear homogeneous matrix system}
$$\cases (I-S_\infty)\,A=0,\\
\stretch
(I+S_\infty)\,B=\left[G_1-S_\infty\,K(0,0)-K(0,0)\,S_\infty\right]A.
\endcases\tag 14.2$$

\noindent PROOF: When
$\bold D$ belongs to the Faddeev class, the scattering matrix
is constructed as in the steps of (a)-(c) of Chapter~9 leading to (9.3). From
(9.3) we see that
$$-J(-k)=S(k)\,J(k),\qquad k\in\bR.\tag 14.3$$
The large-$k$ asymptotics of the Jost matrix $J(k)$ is
given in (10.8), from which we get
$$-J(-k)=-ikA-B-K(0,0) \,A+o(1),\qquad k\to\pm\infty.\tag 14.4$$
 From Proposition~10.3(b) we have
$$S(k)=S_\infty+\ds\frac{G_1}{ik}+o\left(\ds\frac{1}{k}\right),
\qquad k\to\pm\infty.\tag 14.5$$
Using (14.4) and (14.5) in (14.3), we obtain
the expansion
$$-ikA-B-K(0,0)\,A+o(1)=-ik\,S_\infty\,A+S_\infty\,
B+S_\infty\,K(0,0)\,A-G_1 \,A+o(1),
\qquad k\to\pm\infty.\tag 14.6$$
By equating the coefficients of $ik$ in (14.6) we obtain
the first line of (14.2) and by equating the next order terms, i.e.
the constant terms, in (14.6), we obtain
$$(I+S_\infty)\,B=\left[G_1-S_\infty\,K(0,0)-K(0,0)\right]A.
\tag 14.7$$
Since we already have $S_\infty A=A$ from the first line in (14.2),
we can use that identity in (14.7) to obtain the second line of (14.2).
 \qed

The following result is useful
in the analysis of the selfadjoint boundary condition given in (2.4).

\noindent {\bf Proposition 14.2} {\it Let
$\psi(x)$ and $\phi(x)$ be two $n\times n$ matrices satisfying
the boundary condition
(2.4), where the boundary matrices $A$ and $B$ satisfy (2.5) and (2.6).
Then, we have
$$\psi'(0)^\dagger \,\phi(0)-\psi(0)^\dagger \,\phi'(0)=0.\tag 14.8$$
The result remains valid when $\psi(x)$ and $\phi(x)$
are column vectors with $n$ entries.}

\noindent PROOF: Since $\psi(x)$ and $\phi(x)$ satisfy (2.4) we have
$$-B^\dagger\,\psi(0)+A^\dagger \,\psi'(0)=0,\quad
-B^\dagger\,\phi(0)+A^\dagger \,\phi'(0)=0,\tag 14.9$$
 The boundary matrices $A$ and $B$ appearing in
(2.4) and satisfying (2.5) and (2.6) also satisfy
(2.9), where $E$ is the invertible matrix defined in (2.7).
The left-hand side in (14.8) can be evaluated
with the help of the first equality in (2.9) as
$$\aligned
\psi'(0)^\dagger& \,\phi(0)-\psi(0)^\dagger \,\phi'(0)\\
&=
\psi'(0)^\dagger \left[A E^{-2} A^\dagger+B E^{-2} B^\dagger\right]\phi(0)-\psi(0)^\dagger \left[A E^{-2} A^\dagger+B E^{-2} B^\dagger\right]\phi'(0),\endaligned\tag 14.10$$
which can be written as
$$\aligned
\psi'(0)^\dagger \,I\,\phi(0)&-\psi(0)^\dagger \,I\,\phi'(0)\\
=&
\left[A^\dagger \,\psi'(0)\right]^\dagger E^{-2} \left[A^\dagger \phi(0)\right]+
\left[B^\dagger \,\psi'(0)\right]^\dagger\ E^{-2} \left[B^\dagger \phi(0)\right]\\
&-
\left[A^\dagger \,\psi(0)\right]^\dagger E^{-2} \left[A^\dagger \phi'(0)\right]-
\left[B^\dagger \,\psi(0)\right]^\dagger\ E^{-2} \left[B^\dagger \phi'(0)\right].\endaligned\tag 14.11$$
Using (14.9) on the right-hand side of (14.11), we obtain
$$\aligned
\psi'(0)^\dagger \,\phi(0)&-\psi(0)^\dagger \,\phi'(0)\\
=&
\left[B^\dagger \,\psi(0)\right]^\dagger E^{-2} \left[A^\dagger \phi(0)\right]+
\left[B^\dagger \,\psi'(0)\right]^\dagger\ E^{-2} \left[A^\dagger \phi'(0)\right]\\
&-
\left[A^\dagger \,\psi(0)\right]^\dagger E^{-2} \left[B^\dagger \phi(0)\right]-
\left[A^\dagger \,\psi'(0)\right]^\dagger\ E^{-2} \left[B^\dagger \phi'(0)\right].\endaligned\tag 14.12$$
We can rewrite (14.12) by rearranging its right-hand side and we get
 $$\aligned
\psi'(0)^\dagger \,\phi(0)&-\psi(0)^\dagger \,\phi'(0)\\
=&
\psi(0)^\dagger\,B\, E^{-2}\,A^\dagger\, \phi(0)+
\psi'(0)^\dagger\, B\, E^{-2}\,A^\dagger\, \phi'(0)\\
&-
\psi(0)^\dagger\,A\, E^{-2} \,B^\dagger\, \phi(0)-
\psi'(0)^\dagger\,A\, E^{-2} \,B^\dagger\, \phi'(0),\endaligned\tag 14.13$$
or equivalently we get
$$\aligned
\psi'(0)^\dagger& \,\phi(0)-\psi(0)^\dagger \,\phi'(0)\\
&=
\psi'(0)^\dagger \left[B E^{-2} A^\dagger-A\, E^{-2} B^\dagger\right]\phi(0)+\psi(0)^\dagger \left[B E^{-2} A^\dagger-A E^{-2} B^\dagger\right]\phi'(0),\endaligned\tag 14.14$$
Using the second equality in (2.9) on the right-hand side of (14.14),
 we see that the right-hand side vanishes and hence (14.14) yields (14.8).
 We remark that the result in (14.8) also remains valid if
 $\psi(x)$ and $\phi(x)$ are column vectors with $n$ components because
 the left-hand side in (14.8) is well defined in that case and is a scalar.
 \qed

\newpage
\noindent {\bf 15. THE EXISTENCE AND UNIQUENESS IN THE DIRECT PROBLEM}
\vskip 3 pt

In this chapter we provide various results related to
the solution of the direct problem when the input data set $\bold D$ belongs to
the Faddeev class specified in Definition~4.1.
In particular, we provide various results related to the
solvability of the three key integral equations
given in (4.22), (4.14), and (4.17), respectively, as
well as various functional equations in $k\in\bR$
appearing in Definitions~4.2 and 4.3. Such results are
used to prove that if the input data $\bold D$ belongs to
the Faddeev class specified in Definition~4.1, then a
corresponding scattering data set $\bold S$ exists, is unique, and
belongs to the Marchenko class specified in Definition~4.5.
We also provide a proof of the alternate formulations of
the characterization condition $(\bold 4_a)$ stated
in Proposition~4.3, which is established in Proposition~15.4.

In the following proposition we apply Propositions~3.3 to the specific
operators related to (4.22) and (4.14).

\noindent {\bf Proposition 15.1} {\it Consider a scattering data set $\bold S$
as in (4.2), which consists of
an $n\times n$ scattering matrix $S(k)$ for $k\in\bR,$ a set of $N$ distinct
 positive constants $\kappa_j,$ and a set of
$N$ constant $n\times n$ hermitian and nonnegative matrices
$M_j$ with respective positive ranks $m_j,$ where $N$ is a nonnegative integer. Assume that $\bold S$ satisfies $(\bold I)$
of Definition~4.3. Let $F_s(y)$ and $F(y)$ be the matrices defined
in (4.7) and (4.12), respectively. Then we have the following:}

\item{(a)} {\it The integral operator
associated with (4.22) is compact on $L^1(\bR^+).$}

\item{(b)} {\it The integral operator
associated with (4.14) is compact on $L^1(\bR^+).$}

\item{(c)} {\it Any solution $X(y)$ in $L^1(\bR^+)$ to
(4.22) must actually belong to $L^1(\bR^+)\cap L^\infty(\bR^+),$
and in particular to $L^1(\bR^+)\cap L^2(\bR^+).$}

\item{(d)} {\it Any solution $X(y)$ in $L^1(\bR^+)$ to
(4.14) must actually belong to $L^1(\bR^+)\cap L^\infty(\bR^+),$
and in particular to $L^1(\bR^+)\cap L^2(\bR^+).$}

\noindent PROOF: Since the $\kappa_j$-values are all
positive, we see that $F(y)$ defined in (4.12)
is also bounded and integrable
in $y\in\bR^+$ when $F_s(y)$ is bounded and integrable
in $y\in\bR^+,$ which is assured by $(\bold I).$
Thus, from Proposition~3.3 we conclude
that (a)-(d) hold. \qed

Next, we apply Propositions~3.2 to the specific
operator related to (4.17).

\noindent {\bf Proposition 15.2} {\it
Consider a scattering data set $\bold S$
as in (4.2), which consists of
an $n\times n$ scattering matrix $S(k)$ for $k\in\bR,$ a set of $N$ distinct
 positive constants $\kappa_j,$ and a set of
$N$ constant $n\times n$ hermitian and nonnegative matrices
$M_j$ with respective positive ranks $m_j,$ where $N$ is a nonnegative integer. Assume that $\bold S$ satisfies
$(\bold I)$ of Definition~4.3.
Then:}

\item{(a)} {\it The integral operator
associated with (4.17) is compact on $L^2(\bR^-).$}

\item{(b)} {\it Any solution $X(y)$ in $L^2(\bR^-)$ to
(4.17) must actually belong to $L^2(\bR^-)\cap L^\infty(\bR^-).$}

\noindent PROOF: By $(\bold I)$ we already
know that $F_s(y)$ is bounded and square integrable in $y\in\bR.$
 From Proposition~3.5 we then conclude
that (a) and (b) hold. \qed

One consequence of the following result
is the equivalence among
$(\bold 4_c)$, $(\bold 4_d)$, and $(\bold 4_e)$ in Proposition~4.3.

\noindent {\bf Proposition 15.3} {\it Consider a scattering data set $\bold S$
as in (4.2), which consists of
an $n\times n$ scattering matrix $S(k)$ for $k\in\bR,$ a set of $N$ distinct
 positive constants $\kappa_j,$ and a set of
$N$ constant $n\times n$ hermitian and nonnegative matrices
$M_j$ with respective positive ranks $m_j,$ where $N$ is a nonnegative integer. Assume that $\bold S$ satisfies
$(\bold I)$ of Definition~4.3.
 Let $F_s(y)$ and $F(y)$ be the quantities defined
in (4.7) and (4.12), respectively,
where $S_\infty$ is the constant $n\times n$ matrix
defined as in (4.6). Then, we have the following:}

\item{(a)} {\it Any solution $\hat X(k)$ in
$\hat L^1(\bCp)$ to (4.15)
must actually belong to $\hat L^1_\infty(\bCp).$}

\item{(b)} {\it Any solution $h(k)$ in
$\hat L^1(\bCp)$ to (4.16)
must actually belong to $\hat L^1_\infty(\bCp).$}

\item{(c)} {\it The row vector
$\hat X(k)$ with $n$ components
belonging to $\hat L^1(\bCp)$ satisfies (4.15)
if and only if the column vector $h(k)$
with $n$ components belonging to $\hat L^1(\bCp)$ satisfies (4.16),
where $\hat X(k)$ and $h(k)$ are related to each other as
$h(k)=\hat X(-k^\ast)^\dagger.$}

\item{(d)} {\it The row vector
$X(y)$ whose $n$ components
belonging to $L^1(\bR^+)$ is a solution to (4.14)
if and only if
$\hat X(k)\in \hat L^1(\bCp)$ satisfies (4.15),
where $X(y)$ and $\hat X(k)$ are related to each other as
in (3.67) and (3.68).}

\noindent PROOF: From Proposition~15.1(d) we know that
any solution $X(y)$ in $L^1(\bR^+)$
to (4.14) must actually belong to $\hat L^1(\bR^+)\cap L^\infty(\bR^+).$
Since $\hat X(k)$ is related to $X(y)$ as in (3.67) and (3.68),
it follows that we must have $\hat X(k)\in\hat L^1(\bCp)$ and in fact
$\hat X(k)\in\hat L^1_\infty(\bCp).$ Thus (a) is proved.
Actually, (b) is a direct consequence of (c) because
$h(k)=\hat X(-k^\ast)^\dagger.$ On the other hand,
the proof of (c) is obtained by taking the matrix adjoint of
both sides (4.15) and by using $S(k)^\dagger=S(-k),$ which follows from
(4.4). Thus, it only remains to prove (d). As already mentioned,
$X(y)$ belongs to both $\hat L^1(\bR^+)$ and $L^\infty(\bR^+),$
and hence $X(y)$ must in particular belong to $L^2(\bR^+).$
Therefore, it is sufficient to give the proof of (d) by only assuming that
$X(y)\in L^2(\bR^+)$ and hence by only assuming that
$\hat X(k)\in \bold H^2(\bCp).$
Let us first show that if $\hat X(k)\in H^2(\bCp)$ satisfies (4.15), then $X(y)\in L^2(\bR^+)$ given in (3.67) satisfies (4.14).
 From the second line of (4.15) we obtain
$$\ds\frac{1}{2\pi}\int_{-\infty}^\infty  dk\, \hat X(-k)\, e^{iky}
+\ds\frac{1}{2\pi}\int_{-\infty}^\infty  dk\, \hat X(k)\,S(k)\, e^{iky}=0,\tag 15.1$$
Using (3.67) and (4.7) in (15.1), with the help of
(11.36) we obtain
$$X(y)+X(-y)\,S_\infty+\int_{-\infty}^\infty dz\,X(z)\,F_s(z+y),\qquad y\in\bR.\tag 15.2$$
Using $X(y)=0$ for $y\in\bR^-,$ from (15.2) we get
$$X(y)+\int_0^\infty dz\,X(z)\,F_s(z+y)=0,\qquad y\in\bR^+.
\tag 15.3$$
Using (4.12) in (15.3) we get
 $$X(y)+\int_0^\infty dz\,X(z)\,\left[F(z+y)
 -\ds\sum_{j=1}^N M_j^2\,e^{-\kappa_j(z+y)}\right]=0,\qquad y\in\bR^+.
\tag 15.4$$
 With the help of (3.68) we can write (15.4) as
$$X(y)+\int_0^\infty dz\,X(z)\,F(z+y)
 -\ds\sum_{j=1}^N \hat X(i\kappa_j) \,M_j\, M_j\,e^{-\kappa_j y}=0,\qquad y\in\bR^+.
\tag 15.5$$
Each term in the summation is zero because of the first line of (4.15), yielding
(4.14). Thus, we have proved that (4.15) implies (4.14).
Let us now prove the converse, namely, show that
if $X(y)\in L^2(\bR^+)$ satisfies (4.14), then $\hat X(k)$ satisfies
(4.15). It is clear that if $X(y)\equiv 0,$ then the assertion clearly holds.
We can then proceed by assuming that $X(y)$ is a nontrivial solution
to (4.14).
Let us multiply both sides of (4.14)
with $X(y)^\dagger$ and integrate over $y\in\bR$ with the understanding that
$X(y)=0$ for $y\in\bR^-.$ Thus, if $X(y)$ satisfies (4.14) then we have
$$\int_{-\infty}^\infty dy\,\left[X(y)+
\int_{-\infty}^\infty dz\,X(z)\,F(z+y)\right]\,X(y)^\dagger=0.\tag 15.6$$
Using (3.67), (4.7), and (4.12) in (15.6), with the help of (11.37)
we simplify the resulting equation and obtain
$$\ds\frac{1}{2\pi} \int_{-\infty}^\infty
dk\, \left(\hat X(-k)+\hat X(k)\,[S(k)-S_\infty]\right)\hat X(-k)^\dagger+
\sum_{j=1}^N \hat X(i\kappa_j)\,M_j^2 \hat X(i\kappa_j)^\dagger=0.
\tag 15.7$$
Since the matrices $M_j$ are hermitian, we can write
 the summation term in (15.7) as
$$\sum_{j=1}^N \hat X(i\kappa_j)\,M_j^2 \hat X(i\kappa_j)^\dagger
=\sum_{j=1}^N [\hat X(i\kappa_j)\,M_j]\,[\hat X(i\kappa_j)\,M_j]^\dagger
=\sum_{j=1}^N |\hat X(i\kappa_j)\,M_j|^2,
\tag 15.8$$
which indicates that the right-hand side in (15.8) is nonnegative and that
it is zero if and only if we have each vector $\hat X(i\kappa_j)\,M_j$ is equal to
zero for $j=1,\dots,N.$
We can simplify the integral part of (15.7) further by using
$$\ds\frac{1}{2\pi} \int_{-\infty}^\infty dk\,
\hat X(k)\,S_\infty \,\hat X(-k)^\dagger=\int_{-\infty}^\infty
dy \, X(y)\,S_\infty\,X(-y)^\dagger=0,\tag 15.9$$
which follows from (3.68) and the fact that $X(y)=0$ for
$y\in\bR^-.$ Thus, (15.7) is equivalent to
$$\ds\frac{1}{2\pi}\int_{-\infty}^\infty
dk\,  \left(\hat X(-k)+\hat X(k)\,S(k)\right)\hat X(-k)^\dagger+
\sum_{j=1}^N |\hat X(i\kappa_j)\,M_j|^2=0.
\tag 15.10$$
Note that (15.10) is equivalent to
$$\ds\frac{1}{2\pi}\int_{-\infty}^\infty
dk\,  \left(\hat X(k)+\hat X(-k)\,S(-k)\right)\hat X(k)^\dagger+
\sum_{j=1}^N |\hat X(i\kappa_j)\,M_j|^2=0.
\tag 15.11$$
By letting $\hat X_1(k):=\hat X(-k)\,S(-k),$ we notice
that
$$||\hat X_1||_2=||\hat X||_2.\tag 15.12$$
We remark that (15.12) is a consequence of the unitarity
of $S(k)$ and is seen from
$$\aligned \left(\hat X_1,\hat X_1\right)&=
\int_0^\infty dk\,
[\hat X(-k)\,S(-k)][\hat X(-k)\,S(-k)]^\dagger\\
&=
\int_0^\infty dk\,
\hat X(-k)\,S(-k)\,S(-k)^\dagger\,\hat X(-k)^\dagger
\\
&=
\int_0^\infty dk\,
\hat X(-k)\,\hat X(-k)^\dagger=
=
\left(\hat X,\hat X\right)=||\hat X||^2,\endaligned\tag 15.13$$
which is equivalent to
$$||\hat X_1||_2^2=||\hat X||_2^2,\tag 15.14$$
which in turn is equivalent to (15.12).
Writing
the integral term in (15.11) in terms of the scalar product
on $L^2(\bR),$ from (15.11) we obtain
$$\left(\hat X,\hat X\right)+\left(\hat X_1,\hat X\right)
+2\pi
\sum_{j=1}^N |\hat X(i\kappa_j)\,M_j|^2=0.
\tag 15.15$$
The first and third terms in (15.15) are real and in fact nonnegative.
Thus, the second term in (15.15) must be real.
Applying the Schwarz
inequality on that second term we get
$$|\left(\hat X_1,\hat X\right)|\le ||\hat X_1||_2\,||\hat X||_2.\tag 15.16$$
Using (15.12) in (15.16) we get
$$|\left(\hat X_1,\hat X\right)|\le ||\hat X||^2_2.\tag 15.17$$
As indicated earlier, $\left(\hat X_1,\hat X\right)$ is real valued, and
hence (15.17) yields
$$-||\hat X||^2_2\le  \left(\hat X_1,\hat X\right)\le ||\hat X||^2_2.\tag 15.18$$
The first inequality in (15.18) yields
$$(\hat X,\hat X)+\left(\hat X_1,\hat X\right)\ge 0.\tag 15.19$$
Then, from (15.15) and (15.19) we get
$$\cases \sum_{j=1}^N |\hat X(i\kappa_j)\,M_j|^2=0,\\
(\hat X,\hat X)+\left(\hat X_1,\hat X\right)=0.\endcases\tag 15.20$$
The first line in (15.20) holds if and only if
the first line of (5.15) holds. Let us now show that
the second line of (15.20) implies
the second line of (4.15). The second line of
(15.20) implies that
$$-(\hat X_1,\hat X)=(\hat X,\hat X).\tag 15.21$$
Comparing (15.21) with
(15.17) we see that the Schwarz inequality yields an equality,
which happens if $\hat X_1(k)=c\,\hat X(k).$ Then, (15.21) yields
$$-c\,(\hat X,\hat X)=(\hat X,\hat X),\tag 15.22$$
which is possible only if $c=-1,$ as we assume that $\hat X(k)$ is
nonzero. Thus, we must have
$\hat X_1=-\hat X(k),$ or equivalently
$\hat X(-k)\,S(k)=-\hat X(k),$ which yields the second line of (4.15).
Thus, the proof is complete.
\qed

The following proposition shows that
if the scattering data set $\bold S$ given in
(4.2) belongs to the
Marchenko class, then the $\bold S$ satisfies
the properties $(\bold 4_c),$ $(\bold 4_d),$ $(\bold 4_e)$
of Definition~4.2.
Among other implications, it also indicates that if the input data set
$\bold D$ belongs to the Faddeev class, then the only solution
to each of (4.14), (4.15), and (4.16) is the trivial solution.

\noindent {\bf Proposition 15.4} {\it
Consider a scattering data set $\bold S$
as in (4.2), which consists of
an $n\times n$ scattering matrix $S(k)$ for $k\in\bR,$ a set of $N$ distinct
 positive constants $\kappa_j,$ and a set of
$N$ constant $n\times n$ hermitian and nonnegative matrices
$M_j$ with respective positive ranks $m_j,$ where $N$ is a nonnegative integer. Assume that $\bold S$ satisfies
$(\bold 1)$, $(\bold 2)$, $(\bold 3_a)$, and $(\bold 4_a)$ of
Definition~4.5. Let $F(y)$ be the quantity
constructed from $\bold S$ as in (4.12).
Then:}

\item{(a)} {\it The only solution $\hat X(k)$ to (4.15),
as a row vector with $n$ components belonging to
the class $\hat L^1(\bCp),$
is the trivial solution $\hat X(k)\equiv 0.$}

\item{(b)} {\it The only solution $X(y),$ which is a row vector with $n$
integrable components in $y\in\bR^+$ to the integral
equation given in (4.14) is the trivial solution $X(y)\equiv 0.$}

\item{(c)} {\it The only solution $h(k)$ to (4.16),
as a column vector with $n$ components belonging to
the class $\hat L^1(\bCp),$
is the trivial solution $h(k)\equiv 0.$}

\noindent PROOF: It is enough to prove (a) because (b) and (c)
directly follows from (a), as indicated
in Proposition~15.3(c) and Proposition~15.3(d).
 From (4.4) we see that we have
$$S(k)=[S(k)^\dagger]^{-1},\qquad k\in\bR.\tag 15.23$$
As a result of Proposition~5.1(c), the Jost matrix $J(k)$
constructed from $\bold S$ satisfies the properties listed
in Proposition~10.2, and the results in Chapter~11 remain valid,
and in particular Proposition~11.2 holds and $M_j$
appearing in $\bold S$ satisfies (11.22). Since the constructed
$J(k)$ satisfies (4.10), using (4.10) and (15.23) we obtain
$$S(k)=-[J(-k^\ast)^\dagger]^{-1} J(k^\ast)^\dagger,\qquad k\in\bR,\tag 15.24$$
where for convenience we have written $k$ as $k^\ast$ for $k\in\bR$ in (15.24).
The reason for this is that $[J(-k^\ast)^\dagger]^{-1}$ can be extended
 from $k\in\bR$ to $k\in\bCp$ meromorphically with simple poles at $k=i\kappa_j$
 for $j=1,\dots,N,$ as a result of the fact that
 $J(k)$ has a similar extension from $k\in\bR$ to $k\in\bCp,$
 as stated in Proposition~10.2(c).
 Consequently, $[J(k^\ast)^\dagger]^{-1}$ can be extended
 from $k\in\bR$ to $k\in\bCm$ meromorphically with simple poles at $k=-i\kappa_j$
 for $j=1,\dots,N.$ Using (15.24) in the second line of (4.15) we obtain
$$\hat X(k)\, [J(-k^\ast)^\dagger]^{-1}=\hat X(-k)\, [J(k^\ast)^\dagger]^{-1},
\qquad k\in\bR.\tag 15.25$$
Since $\hat X(k)\in \hat L^1(\bCp),$
 it follows that $\hat X(k)$ has an analytic extension from
 $k\in\bR$ to $k\in\bCp,$ it is continuous in $k\in\bCpb,$ and vanishes
 uniformly as $k\to\infty$ in $\bCpb.$
On the other hand, it follows from Proposition~10.2 that
 $[J(-k^\ast)^\dagger]^{-1}$ is meromorphic
 in $k\in\bCp$ with simple poles
 at $k=i\kappa_j$ for $j=1,\dots,N$ and that
 it is continuous for $k\in\bR$ except for a possible simple pole at $k=0.$
  From (10.12) it follows that
$$ [J(-k^\ast)^\dagger]^{-1}=-\ds\frac{N_j^\dagger}{k-i\kappa_j}+O(1),
\qquad k\to i\kappa_j \text{ in } \bCp.
\tag 15.26$$
Let us define
$$\Xi(k):=\cases
k\,\hat X(k)\, [J(-k^\ast)^\dagger]^{-1}+ k\,\ds\sum_{j=1}^N \ds\frac{
2i\kappa_j\,\hat X(i\kappa_j)\,N_j^\dagger}{k^2+\kappa_j^2},\qquad k\in\bCpb,\\
\stretch
k\,\hat X(-k)\, [J(k^\ast)^\dagger]^{-1}+k\,\ds\sum_{j=1}^N \ds\frac{
2i\kappa_j\,\hat X(i\kappa_j)\,N_j^\dagger}{k^2+\kappa_j^2}
,\qquad k\in\bCmb.\endcases\tag 15.27$$
With the help of (15.25) we observe that $\Xi(k)$ is analytic in
$k\in\bCp$ and continuous in $k\in\bCpb.$ Furthermore,
$\Xi(k)=o(k)$ as $k\to\infty$ in $k\in\bCpb$ because
we know that $\hat X(k)=o(1)$ because $\hat X(k)\in \hat L^1(\bCp)$
we know that $[J(-k^\ast)^\dagger]^{-1}=O(1)$ as a result of
Proposition~10.2(g). Similarly, from the second line of
(15.27), with the help of
(15.26) we observe that
$\Xi(k)$ is analytic in
$k\in\bCm,$ continuous in $k\in\bCmb,$
and $o(k)$ as $k\to\infty$ in $\bCmb.$
Thus, $\Xi(k)$ must be entire and in fact a constant row vector.
Then, from the first line of (15.27) we obtain
$$k\,\hat X(k)\, [J(-k^\ast)^\dagger]^{-1}+ k\,\ds\sum_{j=1}^N \ds\frac{
2i\kappa_j\,\hat X(i\kappa_j)\,N_j^\dagger}{k^2+\kappa_j^2}=c,\tag 15.28$$
where $c$ is a constant row vector with $n$ components.
Note that (15.28) yields
$$\hat X(k)=\ds\frac{c\,J(-k^\ast)^\dagger}{k}-\ds\sum_{j=1}^N\left(\ds\frac{2i\kappa_j\,\hat X(i\kappa_j)\,N_j^\dagger\,
J(-k^\ast)^\dagger}{k^2+\kappa_j^2}\right),\qquad k\in\bCpb.\tag 15.29$$
Similarly, from the second line of (15.27) we obtain
$$k\,\hat X(-k)\, [J(k^\ast)^\dagger]^{-1}+ k\,\ds\sum_{j=1}^N \ds\frac{
2i\kappa_j\,\hat X(i\kappa_j)\,N_j^\dagger}{k^2+\kappa_j^2}=c,\tag 15.30$$
which yields
$$\hat X(-k)=\ds\frac{c\,J(k^\ast)^\dagger}{k}-\ds\sum_{j=1}^N\left(\ds\frac{2i\kappa_j\,\hat X(i\kappa_j)\,N_j^\dagger\,
J(k^\ast)^\dagger}{k^2+\kappa_j^2}\right),\qquad k\in\bCmb,\tag 15.31$$
or equivalently
$$\hat X(k)=-\ds\frac{c\,J(-k^\ast)^\dagger}{k}-\ds\sum_{j=1}^N\left(\ds\frac{2i\kappa_j\,\hat X(i\kappa_j)\,N_j^\dagger\,
J(-k^\ast)^\dagger}{k^2+\kappa_j^2}\right),\qquad k\in\bCpb.\tag 15.32$$
Comparing (15.29) and (15.32) we see that $c=0$ and
$$\hat X(k)=-\ds\sum_{j=1}^N\left(\ds\frac{2i\kappa_j\,\hat X(i\kappa_j)\,N_j^\dagger\,
J(-k^\ast)^\dagger}{k^2+\kappa_j^2}\right),\qquad k\in\bCpb.\tag 15.33$$
We will now show that $\hat X(i\kappa_j)\,N_j^\dagger=0.$
 From the first line of (4.15) we know that
 $\hat X(i\kappa_j) \,M_j=0.$ From (11.22) and the fact that
 $P_j$ and $B_j^{-1/2}$ commute, as assured
 by Proposition~11.2(d), it follows that
the first line of (4.15) can be written as
$\hat X(i\kappa_j) \,P_j\,B_j^{-1/2}=0.$
Since $B_j^{-1/2}$ is invertible, as assured by
Proposition~11.2(c), we have
$\hat X(i\kappa_j) \,P_j=0.$
 From (11.9) we have $N_j^\dagger=P_j N_j^\dagger,$
 where we have used the first equality in (11.1).
 Thus, we obtain
 $$\hat X(i\kappa_j)\,N_j^\dagger=\hat X(i\kappa_j)\, P_j\,N_j^\dagger=0,
 \qquad j=1,\dots,N,\tag 15.34$$
 and hence, using (15.34) in (15.33) we conclude that $\hat X(k)\equiv 0.$
 Thus, the proof is complete. \qed

The following proposition shows that
if the scattering data set $\bold S$ given in
(4.2) belongs to the
Marchenko class, then the Jost matrix
$J(k)$ constructed from $\bold S$ as in (9.2) satisfies
a certain useful property.
Among other implications, it also indicates that if the input data set
$\bold D$ belongs to the Faddeev class, then the corresponding
Jost matrix given in (9.2) constructed from $\bold D$
satisfies that useful property.

\noindent {\bf Proposition 15.5} {\it
Consider a scattering data set $\bold S$
as in (4.2), which consists of
an $n\times n$ scattering matrix $S(k)$ for $k\in\bR,$ a set of $N$ distinct
 positive constants $\kappa_j,$ and a set of
$N$ constant $n\times n$ hermitian and nonnegative matrices
$M_j$ with respective positive ranks $m_j,$ where $N$ is a nonnegative integer. Assume that $\bold S$ satisfies
$(\bold 1)$, $(\bold 2)$, $(\bold 3_a)$, and $(\bold 4_a)$ of
Definition~4.5. Let $J(k)$ be the Jost matrix
constructed from $\bold S$ via (9.2).
Let $\kappa_j$ for $j=1,\dots,N$ be the set of
distinct positive numbers related to the zeros of
$\det[J(k)],$ as indicated
in Proposition~10.2(c). Let
$N_j$ for $j=1,\dots,N$ be the set of
$n\times n$ matrices as in the first equality in (10.13),
i.e each $N_j^\dagger$ belongs to the
kernel of
$J(i\kappa_j)^\dagger$ and has rank $m_j,$ as
indicated in
in Proposition~10.2.
Then, we have the following:}

\item{(a)} {\it For each $j=1,\dots,N$ the matrix
$J(k)\, N_j/(k^2+\kappa_j^2)$ is analytic
in $k\in\bCp,$ continuous in $k\in\bCpb,$ and
$O(1/k)$ as $k\to\infty$ as $k\to\infty$ in $\bCpb.$}

 \item{(b)} {\it For each $j=1,\dots,N$ the matrix
$J(k)\,N_j/(k^2+\kappa_j^2)$ belongs to
$\hat L^1(\bCp)$ and in fact to $\hat L^1_\infty(\bCp).$
Hence, $J(k)\,N_j/(k^2+\kappa_j^2)$ also belongs to the Hardy space
$\bold H^2(\bCp).$}

\item{(c)} {\it For each $j=1,\dots,N$ we have}
$$\ds\frac{J(k)\,N_j}{k^2+\kappa_j^2}=
\ds\int_0^\infty dy\, e^{iky}\, \Cal J(y),\tag 15.35$$
{\it where $\Cal J(y)$ belongs to
$L^1(\bR^+)\cap L^\infty(\bR^+)$ and is given by}
$$\aligned \Cal J(y):=&e^{-\kappa_j y}\left[\ds\frac{B+K(0,0)\,A}{2\kappa_j}-
\ds\frac{A}{2}\right]N_j\\
&+\ds\frac{1}{2\kappa_j}
\int_0^\infty
dz\, e^{-\kappa_j \,|y-z|}
\left[K(0,z)^\dagger \,B-K_x(0,z)^\dagger\,B\right]N_j,\qquad y\in\bR^+.\endaligned
\tag 15.36$$
{\it Here, $A$ and $B$ are the boundary matrices constructed
 from $\bold S$ as in Proposition~14.1,
 and $K(x,y)$ is the unique solution to
 the Marchenko equation (13.1).}

 \noindent PROOF: From Definition~4.5 we know that
 $\bold S$ belongs to the Marchenko class and hence from
 Theorem~5.1 we know that $J(k)$ constructed
 from $\bold S$ satisfies Proposition~10.2.
 Thus, $J(k)$ is analytic in $k\in\bCp$ and continuous
in $k\in\bCpb.$ Thus, the matrix $J(k)/(k^2+\kappa_j^2)$
is meromorphic in $k\in\bCp$ with a simple pole at $k=i\kappa_j.$
On the other hand, with the help of (10.12) we conclude
that the matrix $J(k)\, N_j/(k^2+\kappa_j^2)$
is analytic in $k\in\bCp$ and continuous in $k\in\bCpb.$
Furthermore, from (10.10) it follows that
$J(k)\, N_j/(k^2+\kappa_j^2)$ is $O(1/k)$ as $k\to\infty$ in $\bCpb.$
Hence, the proof of (a) is complete.
 From (a) it follows that
$J(k)\, N_j/(k^2+\kappa_j^2)$ belongs to
the Hardy space $\bold H^2(\bCp),$ and hence we have established
the existence
of $\Cal J(y)$ as in (15.35) belonging to $L^2(\bR^+)$ with
$\Cal J(y)=0$ for $y\in\bR^-.$
 From the $x$-derivative of (10.6) we obtain
 $$f'(k,x)=ik\, e^{ikx} I-K(x,x)\,e^{ikx}+\int_x^\infty dy\,K_x(x,y)\,e^{iky},
\qquad k\in\bR,\qquad x\ge 0,
\tag 15.37$$
We remark that the integral term in (15.37) is well defined
because $K_x(x,y)$ for each $x\ge 0$ is integrable
in $y\in\bR^+,$ as assured by Proposition~10.1(f). From (10.6) and
(15.37) we respectively obtain
$$f(k,0)=I+\int_{-\infty}^\infty dy\,K(0,y)\,e^{iky},\tag 15.38$$
$$f'(k,0)=ik\,I-K(0,0)+\int_{-\infty}^\infty dy\,K_x(0,y)\,e^{iky}.\tag 15.39$$
where we have used (10.2) and (10.8).
Then, using (15.38) and (15.39) in (9.2) we write the Jost matrix $J(k)$
as
$$J(k)=B-ik\,A+K(0,0)\,A
+\int_0^\infty dy\,\left[K(0,y)^\dagger\,B-K_x(0,y)^\dagger\,A\right]
e^{iky},\tag 15.40$$
where we have used the fact that
$K(0,0)^\dagger=K(0,0),$ which follows from (2.2) and (10.5).
 From (d) and (f) of Proposition~10.1 we know that
$\left( K(0,y)^\dagger\,B-K_x(0,y)^\dagger\,A\right)$ belongs to
$L^1(\bR^+).$
Let us replace $ikA$ in (15.40) with $[-\kappa_j A+i(k-i\kappa_j)A]$ so that
we have
$$B-ik\,A+K(0,0)\,A=[B+\kappa_j A+K(0,0)\,A]-i(k-i\kappa_j)A,\tag 15.41$$
and hence
$$\ds\frac{B-ik\,A+K(0,0)\,A}
{k^2+\kappa_j^2}=\ds\frac{B+\kappa_j A+K(0,0)\,A}
{k^2+\kappa_j^2}-\ds\frac{iA}{(k+i\kappa_j)}.\tag 15.42$$
Note that
$$\ds\frac{1}{k^2+\kappa_j^2}=
\ds\frac{1}{2\kappa_j}
\left[\ds\frac{i}{k+i\kappa_j}-\ds\frac{i}{k-i\kappa_j}
\right].\tag 15.43$$
On the other hand, we have the explicit expressions
$$\ds\frac{i}{k+i\kappa_j}=\int_0^\infty dy\, e^{-\kappa_j y+iky},
\quad \ds\frac{i}{k-i\kappa_j}=-\int^0_{-\infty} dy\, e^{\kappa_j y+iky}.\tag 15.44
$$
 From (15.43) and (15.44) we get
$$ \ds\frac{1}{k^2+\kappa_j^2}=\ds\frac{1}{2\kappa_j}
\int_{-\infty}^\infty dy\, e^{-\kappa_j |y|}\,e^{iky}.\tag 15.45$$
Using (15.40)-(15.45) we obtain
$$\aligned
\ds\frac{J(k)}{k^2+\kappa_j^2}=&
\ds\frac{1}{2\kappa_j}
\left[B+\kappa_j\,A+K(0,0)\,A\right]
\int_{-\infty}^\infty dy\, e^{-\kappa_j |y|}\,e^{iky}-A
\int_0^\infty dy\, e^{-\kappa_j y}\,e^{iky}\\
&+\ds\frac{1}{2\kappa_j}\int_{-\infty}^\infty dy\,e^{iky}
\int_0^\infty dz\, e^{-\kappa_j |y-z|}\left[K(0,z)^\dagger\,B-K_x(0,z)^\dagger\,A
\right].
\endaligned\tag 15.46$$
Postmultiplying both sides of
(15.46) with $N_j,$ the resulting left-hand side satisfies
the properties listed in (a) and (b), and hence
the integral in the resulting right-hand side vanishes
over $y\in\bR^-.$ This yields
$$\aligned
\ds\frac{J(k)\,N_j}{k^2+\kappa_j^2}=&
\ds\frac{1}{2\kappa_j}
\left[B+\kappa_j\,A+K(0,0)\,A\right]N_j
\int_0^\infty dy\, e^{-\kappa_j y}\,e^{iky}-A\,N_j
\int_0^\infty dy\, e^{-\kappa_j y}\,e^{iky}\\
&+\ds\frac{1}{2\kappa_j}\int_0^\infty dy\,e^{iky}
\int_0^\infty dz\, e^{-\kappa_j |y-z|}\left[K(0,z)^\dagger\,B-K_x(0,z)^\dagger\,A
\right].
\endaligned\tag 15.47$$
Combining the first two integrals in (15.47) into one, we
obtain (15.35) and (15.36).
In order to complete the proof, we need to
show that $\Cal J(y)$ is integrable and bounded in $y\in\bR^+.$
Since $e^{-\kappa_j y}$ is bounded and integrable
in $y\in\bR^+,$ we conclude that
the first terms on the right-hand side is belongs to
$L^1(\bR^+)\cap L^\infty(\bR^+).$
Thus, we only need to prove that the integral term
in (15.36) belongs to
$L^1(\bR^+)\cap L^\infty(\bR^+).$ Note that,
that integral term is essentially the convolution of
$e^{-\kappa_j |y|},$ which is bounded and integrable
in $y\in\bR,$ with the
matrix-valued function $\left[K(0,y)^\dagger\,B-K_x(0,y)^\dagger\,A
\right].$ With the help of
(3.37) and (3.39) we
conclude that that integral term
belongs to $L^1(\bR^+)\cap L^\infty(\bR^+).$
Thus, the proof is complete. \qed

One consequence of the following proposition is the equivalence
among $(\bold V_f)$, $(\bold V_g)$, and $(\bold V_h)$ in Proposition~6.5.

\noindent {\bf Proposition 15.6} {\it Consider a scattering data set $\bold S$
as in (4.2), which consists of
an $n\times n$ scattering matrix $S(k)$ for $k\in\bR,$ a set of $N$ distinct
 positive constants $\kappa_j,$ and a set of
$N$ constant $n\times n$ hermitian and nonnegative matrices
$M_j$ with respective positive ranks $m_j,$ where $N$ is a nonnegative integer.
Let $F_s(y)$ be the quantity defined
in
(4.7). Assume that $\bold S$ satisfies $(\bold I)$
of Definition~4.3.
Then, we have the following:}

\item{(a)} {\it The row vector $X(y)$ whose $n$ components
belonging to $L^2(\bR^+)$ is a solution to (4.22)
if and only if the row vector $\hat X(k)$
with $n$ components in $\bold H^2(\bCp)$ is a solution to (4.23),
where
$\hat X(k)$ and $X(y)$ are related to each other as in (3.67) and (3.68).
We remark that any solution $X(y)$ in $L^2(\bR^+)$ to (4.22)
actually belongs to $L^2(\bR^+)\cap L^\infty(\bR^+).$}

\item{(b)} {\it The row vector $\hat X(k)$ whose $n$ components
belonging to $\bold H^2(\bCp)$ is a solution to (4.23)
if and only if the column vector $h(k)$
with $n$ components in $\bold H^2(\bCp)$ is a solution to (4.24),
where
$\hat X(k)$ and $h(k)$ are related to each other as
$h(k)=\hat X(-k^\ast)^\dagger.$}

\noindent PROOF: Let us first remark that
the fact that any solution $X(y)$ in $L^2(R^+)$ to
(4.22) must belong to $L^2(R^+)\cap L^\infty(\bR^+)$
directly follows from the analog of Proposition~3.2.
We remark that the rest of the proof essentially follows
by repeating the proof of Proposition~15.3, as the following
argument indicates.
If we replace $F(y)$ appearing in (4.14) by
$F_s(y)$ then we get (4.22). Thus, the proof
of Proposition~15.3 can be repeated by ignoring
the portions in that proof related to the bound states.
In fact, the proof of
Proposition~15.3 is given when
$X(y)$ only belongs to $L^2(\bR^+),$ and hence
the results stated in (a) and (b) hold. \qed

In the next proposition
we present the
result of Proposition~15.6 in a more restricted class.
One of its consequences is the equivalences
among $(\bold V_c)$, $(\bold V_d)$, and $(\bold V_e)$ in Proposition~6.5.

\noindent {\bf Proposition 15.7} {\it Consider a scattering data set $\bold S$
as in (4.2), which consists of
an $n\times n$ scattering matrix $S(k)$ for $k\in\bR,$ a set of $N$ distinct
 positive constants $\kappa_j,$ and a set of
$N$ constant $n\times n$ hermitian and nonnegative matrices
$M_j$ with respective positive ranks $m_j,$ where $N$ is a nonnegative integer. Assume that $\bold S$ satisfies $(\bold I)$
of Definition~4.3.
Let $F_s(y)$ be the quantity defined
in
(4.7).
Then, we have the following:}

\item{(a)} {\it The row vector $X(y)$ whose $n$ components
belonging to $L^1(\bR^+)$ is a solution to (4.22)
if and only if the row vector $\hat X(k)$
with $n$ components in $\hat L^1(\bCp)$ is a solution to (4.23),
where
$\hat X(k)$ and $X(y)$ are related to each other as in (3.67) and (3.68).}

\item{(b)} {\it The row vector $\hat X(k)$ whose $n$ components
belonging to $\hat L^1(\bCp)$ is a solution to (4.23)
if and only if the column vector $h(k)$
with $n$ components in $\hat L^1(\bCp)$ is a solution to (4.24),
where
$\hat X(k)$ and $h(k)$ are related to each other as
$h(k)=\hat X(-k^\ast)^\dagger.$}

\item{(c)} {\it Any solution $X(y)$ in
$L^1(\bR^+)$ to (4.22)
must actually belong to $L^1(\bR^+)\cap L^\infty(\bR^+).$}

\item{(d)} {\it Any solution $\hat X(k)$ in
$\hat L^1(\bCp)$ to (4.23)
must actually belong to $\hat L^1_\infty(\bCp).$}

\item{(e)} {\it Any solution $h(k)$ in
$\hat L^1(\bCp)$ to (4.24)
must actually belong to $\hat L^1_\infty(\bCp).$}

\noindent PROOF: We remark that
if $X(y)$ belongs to $L^1(\bR^+),$ from (3.67) and (3.68)
it follows that $\hat X(k)$ belongs to
$\hat L^1(\bCp).$ Furthermore, from
$h(k)=\hat X(-k^\ast)^\dagger$ it follows that
$h(k)$ also belongs to $\hat L^1(\bCp).$
Let us also remark that (c)
directly follows from Proposition~3.1(c), and
(c) implies (d) and (e).
Thus, we only need to prove (a) and (b).
We note that (a) and (b) directly follows from
(a) and (b) of Proposition~15.6 because
$X(y)\in L^1(\bR^+)\cap L^\infty(\bR^+)$ implies
that $X(y)\in L^2(\bR^+)$ and as a result
$\hat X(k)\in \hat L^1_\infty(\bCp)$ implies that
$\hat X(k)\in \bold H^2(\bCp)$ and that
$h(k)\in \hat L^1_\infty(\bCp)$ implies that
$h(k)\in \bold H^2(\bCp).$ Thus, the proof is complete.
\qed

One implication of
the following result
is that if
the scattering data set $\bold S$ belongs to
the Marchenko class then $(\bold V_f),$ $(\bold V_g),$ and
$(\bold V_h)$ of Definition~4.3 are satisfied.
Among its other implications, it also indicates that
if the input data set $\Cal D$ belongs to the
Faddeev class then the number of linearly
independent solutions to each of (4.22), (4.23),
(4.24) is equal to the nonnegative integer
$\Cal N$ appearing in (4.3).


\noindent {\bf Proposition 15.8} {\it
Consider a scattering data set $\bold S$
as in (4.2), which consists of
an $n\times n$ scattering matrix $S(k)$ for $k\in\bR,$ a set of $N$ distinct
 positive constants $\kappa_j,$ and a set of
$N$ constant $n\times n$ hermitian and nonnegative matrices
$M_j$ with respective positive ranks $m_j,$ where $N$ is a nonnegative integer. Assume that $\bold S$ satisfies
$(\bold 1)$, $(\bold 2)$, $(\bold 3_a)$, and $(\bold 4_a)$ of
Definition~4.5. Let $F_s(y)$ be the quantity
constructed from $\bold S$ as in (4.7).
Then:}

\item{(a)} {\it The number of linearly independent
solutions $\hat X(k)$ to (4.23), as row vectors whose $n$ components
belonging to $\bold H^2(\bCp),$ is equal to $\Cal N,$ where
$\Cal N$ is the nonnegative integer defined in (4.3).}

\item{(b)} {\it The number of linearly independent
solutions $X(y)$ to (4.22), as a row vector whose $n$ components
belonging to $L^2(\bR^+),$ is equal to $\Cal N.$}

\item{(c)} {\it The number of linearly independent
solutions $h(k)$ to (4.24), as column vectors whose $n$ components
belonging to $\bold H^2(\bCp),$ is equal to $\Cal N.$}

\noindent PROOF: By Proposition~15.7, we know that the solution $X(y)$ to
(4.22) and the solution $\hat X(k)$ to (4.23) are related
to each other as in (3.67) and (3.68). Furthermore, the same proposition
indicated that the solution $h(k)$ to
(4.24) and the solution $\hat X(k)$ to (4.23) are related
to each other as $h(k)=\hat X(-k^\ast)^\dagger.$
Thus, it is enough to prove (a), and the results in (b) and (c) follow
 from (a). For the proof of (a) we proceed as follows.
We remark that (4.23) is identical to the second line of (4.15).
Hence, we equivalently need to solve the second line of
(4.15) and look for the general solution $\hat X(k)$
belonging to $\bold H^2(\bCp).$
Thus, we can repeat the beginning of the proof of
Proposition~15.4 and show that
if $\hat X(k)\in \bold H^2(\bCp)$ is a solution to
(4.23), then the quantity $\Xi(k)$ defined in
(15.27) is sectionally analytic in $k\in\bC,$ i.e.
it is analytic in $k\in\bCp\cup \bCm.$
However, for the rest of the proof we cannot use the argument
given in the proof of
Proposition~15.4 because $\hat X(k)$ cannot be assumed
continuous in $k\in\bCpb$ and cannot be assumed
to have the behavior $o(1)$ as $k\to\infty$ in $\bCpb.$
So, we proceed as follows. First, we prove that
the domain of analyticity of $\Xi(k)$ given in
(15.27) but with $\hat X(k)\in\bold H^2(\bCp)$ extends
to the entire complex plane $\bC.$
This is done as follows. Since
$\hat X(k)\in\bold H^2(\bCp)$ we have
$\hat X(k+i\epsilon)\to \hat X(k)$ as $\epsilon\to 0^+$ a.e. in
$k\in\bR$ and strongly in
$L^2(\bR).$ For $0<\epsilon<1,$ we use $C_\epsilon$ to denote the
positive boundary of the rectangle in $k\in\bCp$ with
respective corners located at
$a+i\epsilon,$ $b+i\epsilon,$ $b+i,$ $a+i,$
where $a$ and $b$ are some positive parameters with $a<b.$
Similarly, we
use $C_{-\epsilon}$ to denote the positive boundary
of the rectangle in $\bCm$ with respective corners
located at $b-i\epsilon,$ $a-i\epsilon,$ $a-i,$ $b-i.$
Since $\Xi(k)$ defined in (15.27) is analytic
in $k\in\bCp\cup\bCm,$ it follows from the Cauchy integral formula that
for any $k$ inside $C_\epsilon$ we
have
$$\Xi(k)=\ds\frac{1}{2\pi i}\int_{C_\epsilon} dt\,\ds\frac{\Xi(t)}{t-k}+
\ds\frac{1}{2\pi i}\int_{C_{-\epsilon}} dt\,\ds\frac{\Xi(t)}{t-k},
\tag 15.48$$
where the contribution by the second integral is zero.
Let us choose $a$ and $b$ so that in the limit $\epsilon\to 0^+$ we have
$\Xi(a\pm i \epsilon)\to \Xi(a)$ and $\Xi(b\pm i \epsilon)\to \Xi(b).$
Then, letting $\epsilon\to 0^+$ in (15.22) we get
$$\Xi(k)=\ds\frac{1}{2\pi i}\int_{C_0} dt\,\ds\frac{\Xi(t)}{t-k},\tag 15.49$$
where
$C_0$ is the positively oriented
boundary of the rectangle with corners
at $-a-ib,$ $a-ib,$ $a+ib,$ and $-a+ib.$
 From the representation in (15.49), we conclude that
$\Xi(k)$ is analytic in the interior of the
rectangle bounded by $C_0,$ including the segment of
the real axis contained in that rectangle.
Since we can let $a\to -\infty$ and $b\to +\infty,$
we conclude that $\Xi(k)$ is in fact entire in $k.$ With the help of
(15.27) we conclude that $\Xi(k)$ is an odd function of $k$
in $\bold C$ and we have
$$\Xi(-k)=-\Xi(k),\qquad k\in\bC.\tag 15.50$$
Since $\Xi(k)$ defined in (15.27)
has an analytic extension to the entire
complex plane, in its Maclaurin expansion of
$\Xi(k)$ given by
$$\Xi(k)=\ds\sum_{p=1}^\infty \ds\frac{1}{p!}\,\left[\Xi^{(p)}(0)\right]\, k^p,\
\qquad k\in\bC,\tag 15.51$$
where the coefficient
$\Xi^{(p)}(0)$ can be evaluated, with the help of the generalized Cauchy
integral formula as
$$\Xi^{(p)}(0):=\ds\frac{d^p\, \Xi(k)}{dk^p}  \bigg|_{k=0}=
\ds\frac{p!}{2\pi i}\, \int_{\bold T_r}dt\,\ds\frac{\Xi(t)}{t^{p+1}},\qquad p=0,1,2,\dots,
\tag 15.52$$
where $\bold T_r$ is the circle of radius $r$ centered at $k=0$
traversed in the positive direction, with $r:=|k|.$
Because of (15.50), from (15.52) we conclude that
$\Xi^{(p)}(0)=0$ for even values of $p$ in (15.51). We will
now estimate the integral in (15.52). Using (15.50), we have
$$\bigg| \int_{\bold T_r}dt\,\ds\frac{\Xi(t)}{t^{p+1}}\bigg|
\le 2 \int_{\bold T^+_r}|dt|\,\ds\frac{|\Xi(t)|}{r^{p+1}},
\tag 15.53$$
where we use $\bold T_r^+$ to denote the upper semicircle
of $\bold T_r.$ From Proposition~10.2(b) it follows that there exists
some positive number
$r_0$ such that
$$|J(k)|\le C\,|k|,\qquad |k|\ge r_0,\quad k\in\bCpb,
\tag 15.54$$
for some generic constant $C.$
On the other hand, since $\hat X(k)\in \bold H^2(\bCp),$
by (3.46) we have
$$|\hat X(k)|\le \ds\frac{C}{\sqrt{|k|\,\sin \theta}},
\qquad k\in\bCp.
\tag 15.55$$
Using (15.54) and (15.55) in the first line of
(15.27), we have with $k=r\,e^{i\theta}$
$$|\Xi(r\,e^{i \theta})|\le \ds\frac{C \,\sqrt{r}}
{\sqrt{\sin\theta}},\qquad r\ge r_0,\quad k\in \bold T_r^+,\tag 15.56$$
for some generic constant $C.$
Using (15.56) in (15.53) we get the estimate
$$\bigg| \int_{\bold T_r}dt\,\ds\frac{\Xi(t)}{t^{p+1}}\bigg|
\le \ds\frac{C}{r^{p+1/2}} \int_0^{\pi /2}\ds\frac{d\theta}{\sqrt{\sin\theta}},
\tag 15.57$$
where we have used $|dt|=r\,d\theta$ and that
$\sin\theta=\sin(\pi-\theta)$ for $\theta\in(0,\pi/2).$
The integral on the right-hand side of (15.57) is convergent
because the singularity of the integrand at $\theta=0$
is an integrable singularity, as we have
$\sin\theta =\theta+ O(\theta^3)$ as $\theta\to 0.$
In fact, that integral is related to the complete elliptic integral of the first kind and we
have
$$ \int_0^{\pi /2}\ds\frac{d\theta}{\sqrt{\sin\theta}}=
\sqrt{2} \int_0^{\pi/2} \ds\frac{d\theta}{\sqrt{1-\ds\frac{1}{2}\,\sin^2\theta}}
=2.6220\overline{6},\tag 15.58$$
where we us an overline on a digit to
indicate a round off.
Thus, using (15.52), (15.57), and (15.58), and letting
$r\to +\infty,$ we conclude that
$\Xi^{(p)}(0)=0$ for $p=0,1,\dots,$ and hence
 from (15.51) we conclude that
 $\Xi(k)\equiv 0.$
 Then, from (15.27) we obtain
$\hat X(k)$ as in the adjoint of (15.45), i.e.
$$\hat X(k)=-\ds\sum_{j=1}^N\left(\ds\frac{2i\kappa_j\,\hat X(i\kappa_j)\,N_j^\dagger\,
J(-k^\ast)^\dagger}{k^2+\kappa_j^2}\right),\qquad k\in\bCpb.\tag 15.59$$
Let us now investigate the general solution
$\hat X(k)$ to (4.23) we have constructed in (15.59),
and let us show that it contains precisely $\Cal N$ linearly
independent row vector solutions.
For the proof we proceed as follows. From the first equality in (10.13) we know that
$J(i\kappa_j)^\dagger N_j^\dagger=0$ and hence
each column of $N_j^\dagger$ belongs to the
kernel of $J(i\kappa_j)^\dagger.$ From Proposition~10.2(d) we then
conclude that the rank of $N_j^\dagger$ is equal to $m_j,$
which is the nullity of the matrix $J(i\kappa_j)^\dagger.$
Since (4.23) is a linear homogeneous system, we then
conclude that the number of linearly independent solutions
to (4.23) is equal to the sum of the ranks of
$N_j^\dagger$ for all $j=1,\dots,N,$ which is $\Cal N$ given in (4.3).
Let us remark that we can directly infer from the explicit solution
given in
(15.59) that $\hat X(k)$ indeed belongs to
the Hardy space $\bold H^2(\bCp).$ This can be argued
as follows. From Proposition~10(b), (10.11), and the first equality in (10.13)
it follows that $\hat X(k)$ is analytic in $k\in\bCp.$
Then, with the help of Proposition~10(b) we conclude that
$\hat X(k)$ is continuous in $k\in\bCpb.$ Then, from (10.10)
we conclude that $\hat X(k)=O(1/k)$ as $k\to\infty$ in $\bCpb.$
Thus, we can conclude that $\hat X(k)\in\bold H^2(\bCp).$ \qed

The following result is similar
to the one given in Proposition~15.8.
One of its implications
is that if
the scattering data set $\bold S$ belongs to
the Marchenko class then $(\bold V_c),$ $(\bold V_d),$ and
$(\bold V_e)$ of Definition~4.3 are satisfied.
Among its other implications, it also indicates that
if the input data set $\Cal D$ belongs to the
Faddeev class then the number of linearly
independent solutions to each of (4.22), (4.23),
(4.24) is equal to the nonnegative integer
$\Cal N$ appearing in (4.3),
where the solutions are sought in a more restricted
class than that used in Proposition~15.8.

\noindent {\bf Proposition 15.9} {\it
Consider a scattering data set $\bold S$
as in (4.2), which consists of
an $n\times n$ scattering matrix $S(k)$ for $k\in\bR,$ a set of $N$ distinct
 positive constants $\kappa_j,$ and a set of
$N$ constant $n\times n$ hermitian and nonnegative matrices
$M_j$ with respective positive ranks $m_j,$ where $N$ is a nonnegative integer. Assume that $\bold S$ satisfies
$(\bold 1)$, $(\bold 2)$, $(\bold 3_a)$, and $(\bold 4_a)$ of
Definition~4.5. Let $F_s(y)$ be the quantity
constructed from $\bold S$ as in (4.7).
Then:}

\item{(a)} {\it The number of linearly independent
solutions $\hat X(k)$ to (4.23), as row vectors whose $n$ components
belonging to $\hat L^1(\bCp),$ is equal to $\Cal N,$ where
$\Cal N$ is the nonnegative integer defined in (4.3).}

\item{(b)} {\it The number of linearly independent
solutions $X(y)$ to (4.22), as a row vector whose $n$ components
belonging to $L^1(\bR^+),$ is equal to $\Cal N.$}

\item{(c)} {\it The number of linearly independent
solutions $h(k)$ to (4.24), as column vectors whose $n$ components
belonging to $\hat L^1(\bCp),$ is equal to $\Cal N.$}

\noindent PROOF: By Proposition~15.7, we know that the solution $X(y)$ to
(4.22) and the solution $\hat X(k)$ to (4.23) are related
to each other as in (3.67) and (3.68). Furthermore, the same proposition
indicates that the solution $h(k)$ to
(4.24) and the solution $\hat X(k)$ to (4.23) are related
to each other as $h(k)=\hat X(-k^\ast)^\dagger.$
Thus, it is enough to prove (a), and the results in (b) and (c) follow
 from (a). For the proof of (a) we proceed as follows.
We remark that (4.23) is identical to the second line of (4.15).
Hence, we equivalently need to solve the second line of
(4.15) and look for the general solution $\hat X(k)$
belonging to $\hat L^1(\bCp).$
Thus, we can repeat the proof of
Proposition~15.4 from the beginning and obtain the solution
$\hat X(k)$ given as in the adjoint of (15.45), i.e.
$$\hat X(k)=-\ds\sum_{j=1}^N\left(\ds\frac{2i\kappa_j\,\hat X(i\kappa_j)\,N_j^\dagger\,
J(-k^\ast)^\dagger}{k^2+\kappa_j^2}\right),\qquad k\in\bCpb.\tag 15.60$$
Since (4.23) and (4.15) differ from each other in the sense that
the solution to (4.23) does not need to satisfy the first line of
(4.15), the solution $\hat X(k)$ given in (15.60) is the general solution
to (4.23). Let us show that the general
solution $\hat X(k)$ given in (15.60) contains precisely $\Cal N$ linearly
independent row vector solutions. For this we can use the same argument
given in the proof of Proposition~15.6, namely, that
the rank of $N_j^\dagger$ is equal to $m_j,$
which is the nullity of the matrix $J(i\kappa_j)^\dagger.$
Since (4.23) is a linear homogeneous system, then from
(15.60) we see
that the number of linearly independent solutions
to (4.23) is equal to the sum of the ranks of
$N_j^\dagger$ for all $j=1,\dots,N,$ which is $\Cal N$ given in (4.3).
Let us remark that the solution $\hat X(k)$ in (15.59)
and the solution $\hat X(k)$ in (15.60) coincide.
We know from Proposition~15.8 that
$\hat X(k)$ of (15.59) belongs to
the Hardy space $\bold H^2(\bCp).$
Let us now prove that $\hat X(k)$ of (15.60)
indeed belongs to $\hat L^1(\bCp).$
In other words, we would like to prove that
$\hat X(k)$ of (15.60) is analytic in $k\in\bCp$
and that it is the Fourier transform
of some function
$X(y),$ as in (3.67) and (3.68), where $X(y)\in L^1(\bR^+)$ and $X(y)=0$ for $y\in\bR^-.$
In fact, this directly follows using Proposition~15.5 in (15.60). \qed

In the next theorem, we show that if the input data set $\bold D$
given in (4.1) belongs to the Faddeev class specified
in Definition~4.1, then a corresponding
scattering data set $\bold S$ as in (4.2) exists, is unique, and
belongs to the Marchenko class specified in Definition~4.5.

\noindent {\bf Theorem 15.10} {\it For any input data set
$\bold D$ in the Faddeev class specified in Definition~4.1,
there exists and uniquely exists a corresponding scattering data set $\bold S$
as in (4.2) belonging to
the Marchenko class specified in Definition~4.5.}

\noindent PROOF: The existence
and uniqueness of $S(k)$ are implicitly given
in the construction steps (a)-(c) given in
Chapter~9. The existence and uniqueness of
the construction of the Jost function $J(k)$
are implicitly given
in the construction steps (a) and (b) given in
Chapter~9. Then, the existence and uniqueness of
the constants $\kappa_j,$ their multiplicities $m_j,$
and their number $N$
are assured because such quantities are related to
$J(k)$ as described in the step (f) in Chapter~9.
The existence and uniqueness of
the normalization matrices $M_j$ are implicit in the
construction summary given in the step (g)
in Chapter~9. Having proved the existence and uniqueness
of the corresponding $\bold S,$ let us now prove that
$\bold S$ belongs to the Marchenko class, i.e. that
all the four conditions stated in Definition~4.5 are
satisfied.
Let us first prove that
$(\bold 1)$ in Definition~4.5 holds when $\bold D$ belongs to the
Faddeev class. By Proposition~10.3(a) we know that $S(k)$ satisfies
(4.4). The property (4.5) follows from Proposition~10.3(b).
The boundedness of
$F_s(y)$ defined in $y\in\bR$ and its integrability
in $y\in\bR^+$ are given in Theorem~12.1
Thus, $(\bold 1)$ is satisfied. Note that $(\bold 2)$ follows from
Theorem~13.2(c) and the fact that the difference
$F(y)-F_s(y),$ as seen from (4.12), is a linear combination of
exponential functions with negative exponents
in $y\in\bR^+$ given by the right-hand side of (5.2).
The property $(\bold 3_a)$ in Definition~4.5
follows from Proposition~10.5(c). Let us now prove
$(\bold 4_a)$. By Proposition~15.1(b), the Marchenko integral
operator for $x=0$ is compact on $L^1(\bR^+).$
Thus, (4.11) has a unique solution in $L^1(\bR^+)$ if the only solution
in $L^1(\bR^+)$ to
(4.13) is the trivial solution.
By Proposition~15.4(d) we know that the only solution in $L^1(\bR^+)$
to (4.14) is the trivial solution, and hence the only solution
in $L^1(\bR^+)$ to
(4.13) is also the trivial solution.
Hence, $(\bold 4_a)$ holds. \qed

The following proposition is related to the
equivalences among
$(\bold{III}_a)$, $(\bold{III}_b)$, and $(\bold{III}_c)$ of Definition~4.3.

\noindent {\bf Proposition 15.11} {\it Consider a scattering data set $\bold S$
as in (4.2), which consists of
an $n\times n$ scattering matrix $S(k)$ for $k\in\bR,$ a set of $N$ distinct
 positive constants $\kappa_j,$ and a set of
$N$ constant $n\times n$ hermitian and nonnegative matrices
$M_j$ with respective positive ranks $m_j,$ where $N$ is a nonnegative
integer. Assume that $\bold S$ satisfies
$(\bold I)$ of Definition~4.3.
Let $F_s(y)$ be the quantity defined
in
(4.7),
where $S_\infty$ is the constant $n\times n$ matrix
defined as in (4.6). Then, we have the following:}

\item{(a)} {\it The row vector $X(y)$ whose $n$ components
belonging to $L^2(\bR^-)$ is a solution to (4.17)
if and only if $\hat X(k)$ related to $X(y)$ as in (3.67) and (3.69)
is a solution in the Hardy space
$\bold H^2(\bCm)$ to (4.18).}

\item{(b)} {\it The row vector $\hat X(k)$ whose $n$ components
belonging to the Hardy space $\bold H^2(\bCm)$ is a solution to (4.18)
if and only if the column vector $h(k)$ with
$n$ components belonging to $\bold H^2(\bCm)$
satisfies (4.19), where $\hat X(k)$ and $h(k)$ are related to
each other as $h(k)=\hat X(-k^\ast)^\dagger.$}

\noindent PROOF: We remark that the proof of (b) is obtained
by taking the matrix adjoint of both sides of (4.18) and
by using $S(k)^\dagger=S(-k)$ for $k\in\bR,$ which follows
 from (4.4). So, we only need to prove (a). From (3.67) and (3.69) we see that
$X(y)\in L^2(\bR^-)$ if and only if $\hat X(k)\in\bold H^2(\bCm).$
Thus, we first need to show that if $\hat X(k)\in \bold H^2(\bCm)$
satisfies (4.18) then $X(y)\in L^2(\bR^-)$ satisfies
(3.70). After that proof, we need to prove the converse.
First, let us show that (4.18) implies (4.17). From (4.18) we get
$$-\ds\frac{1}{2\pi}\int_{-\infty}^\infty  dk\, \hat X(-k)\, e^{iky}
+\ds\frac{1}{2\pi}\int_{-\infty}^\infty  dk\, \hat X(k)\,S(k)\, e^{iky}=0,\tag 15.61$$
Using (3.67) and (4.7) in (15.61), with the help of
(11.36) we obtain
$$-X(y)+X(-y)\,S_\infty+\int_{-\infty}^\infty dz\,X(z)\,F_s(z+y),\qquad y\in\bR.\tag 15.62$$
Using $X(y)=0$ for $y\in\bR^+,$ from (15.62) we get
$$-X(y)+\int^0_{-\infty} dz\,X(z)\,F_s(z+y)=0,\qquad y\in\bR^-.
\tag 15.63$$
and hence we have proved that (4.18) implies (4.17).
Let us now prove the converse. By assuming that $X(y)$ satisfies
(4.17) we would like to show that
$\hat X(k)$ satisfies (4.18).
We can extend $X(y)$ to $y\in\bR$ by letting
$X(y)=0$ for $y\in\bR^+.$
Let us multiply both sides of (4.17)
with $X(y)^\dagger$ and integrate over $y\in\bR$ with the understanding that
$X(y)=0$ for $y\in\bR^+.$ We get
$$\int_{-\infty}^\infty dy\,\left[-X(y)+
\int_{-\infty}^\infty dz\,X(z)\,F_s(z+y)\right]\,X(y)^\dagger=0.\tag 15.64$$
Using (3.67) and (4.7) in (15.64), with the help of (11.37)
we simplify the resulting equation and obtain
$$\ds\frac{1}{2\pi} \int_{-\infty}^\infty
dk\, \left(-\hat X(-k)+\hat X(k)\,[S(k)-S_\infty]\right)\hat X(-k)^\dagger=0.
\tag 15.65$$
We can simplify the integral part of (15.65) further by using
$$\ds\frac{1}{2\pi} \int_{-\infty}^\infty dk\,
\hat X(k)\,S_\infty \,\hat X(-k)^\dagger=\int_{-\infty}^\infty
dy \, X(y)\,S_\infty\,X(-y)^\dagger=0,\tag 15.66$$
which follows from (3.69) and the fact that $X(y)=0$ for
$y\in\bR^+.$ Thus, (15.65) is equivalent to
$$\int_{-\infty}^\infty
dk\,  \left(-\hat X(-k)+\hat X(k)\,S(k)\right)\hat X(-k)^\dagger=0.
\tag 15.67$$
Note that (15.67) can be written as
$$\int_{-\infty}^\infty
dk\,  \left(-\hat X(k)+\hat X(-k)\,S(-k)\right)\hat X(k)^\dagger=0.
\tag 15.68$$
By letting $\hat X_1(k):=\hat X(-k)\,S(-k),$ as indicated in
(15.13) we know that
(15.12) holds and we have
$$||\hat X_1||_2=||\hat X||_2.\tag 15.69$$
We can write (15.68)
in terms of the scalar product
on $L^2(\bR)$ as
$$-\left(\hat X,\hat X\right)+\left(\hat X_1,\hat X\right)
=0.
\tag 15.70$$
or equivalently as
$$\left(\hat X_1,\hat X\right)=||\hat X||_2^2.\tag 15.71$$
Applying the Schwarz
inequality on the left-hand side of  (15.71) we get
$$|\left(\hat X_1,\hat X\right)|\le ||\hat X_1||_2\,||\hat X||_2,\tag 15.72$$
where the equality holds if and only if $\hat X_1(k)=c\,\hat X(k)$
for some constant $c.$
Using (15.69) in (15.72) we get
$$|\left(\hat X_1,\hat X\right)|\le ||\hat X||^2_2.\tag 15.73$$
Comparing (15.71) and (15.73) we see that we must have
the equality holding in the Schwarz inequality and hence
$$\hat X_1(k)=c\,\hat X(k).\tag 15.74$$
Using (15.74) in (15.71) we determine that either
$\hat X(k)\equiv 0$ or $c=1.$ In the former case
$\hat X(k)$ clearly satisfies (4.18), because (4.18) is a homogeneous
Riemann-Hilbert problem. In the latter case (15.74) with $c=1$
yields
$$\hat X(-k)\,S(-k)=\hat X(k),\qquad k\in\bR,\tag 15.75$$
which is equivalent to (4.18). Hence, the proof is complete. \qed

The following result shows that the only solution to
the linear homogeneous integral equation appearing
in $(\bold{III}_a)$ of Definition~4.3 is the trivial solution.
The result is analogous to a part of Theorem~3.5.1 of
[2]. However, we could not rely on the proof
given in [2] because the proof in [2] assumes that
the quantity $\hat X(k)$ given in (3.68) is analytic
in $k\in\bCp,$ continuous in $k\in\bCpb,$ and uniformly
$o(1)$ as $k\to\infty$ in $\bCpb$ when the quantity
$X(y)$ in (3.68) is only square integrable in $y\in\bR^+.$
In Example~26.2 we illustrate that $\hat X(k)$
may not have such a nice behavior unless
$X(y)$ is also integrable.

\noindent {\bf Proposition 15.12} {\it
Consider a scattering data set $\bold S$
as in (4.2), which consists of
an $n\times n$ scattering matrix $S(k)$ for $k\in\bR,$ a set of $N$ distinct
 positive constants $\kappa_j,$ and a set of
$N$ constant $n\times n$ hermitian and nonnegative matrices
$M_j$ with respective positive ranks $m_j,$ where $N$ is a nonnegative integer. Assume that $\bold S$ satisfies
$(\bold 1)$, $(\bold 2)$, $(\bold 3_a)$, and $(\bold 4_a)$ of
Definition~4.5. Let $F_s(y)$ be the quantity
constructed from $\bold S$ as in (4.7).
Then:}

\item{(a)} {\it The only solution $X(y)$ in $L^2(\bR^-)$ to (4.17)
is the trivial solution $X(y)\equiv 0.$}

\item{(b)} {\it The only solution $\hat X(k)$ in $\bold H^2(\bCm)$ to (4.18)
is the trivial solution $\hat X(k)\equiv 0.$}

\item{(c)} {\it The only solution $h(k))$ in $\bold H^2(\bCm)$ to (4.19)
is the trivial solution $h(k)\equiv 0.$}

\noindent PROOF: By Proposition~15.11 we know that $X(y)\in L^2(\bR^-)$ satisfies
(4.17) if and only if $\hat X(k)\in \bold H^2(\bCm)$ satisfies (4.18), where
$X(y)$ and $\hat X(k)$ are related to each other as in
(3.67) and (3.69). Furthermore, by the same proposition we know that
$h(k)\in\bold H^2(\bCm)$ is a solution to (4.19) if and only if
$\hat X(k)\in\bold H^2(\bCm)$ is a solution to (4.18),
where $h(k)=\hat X(-k^\ast)^\dagger.$
Hence, to prove our proposition, it is
 enough to prove that $\hat X(k)\equiv 0.$
As also indicated in the proof of Proposition~15.4,
the Jost matrix $J(k)$ constructed from the scattering
data set $\bold S$ satisfies (4.10) and possesses all the properties
listed in Proposition~10.2.
Thus, in (4.18) we can replace
$S(k)$ by $-J(-k)\,J(k)^{-1}$ and we get
$$\hat X(-k)+\hat X(k)\,J(-k)\,J(k)^{-1}=0,\qquad k\in\bR,\tag 15.76$$
or equivalently
$$\hat X(-k)\,J(k)+\hat X(k)\,J(-k)=0,\qquad k\in\bR,\tag 15.77$$
where $\hat X(k)$ is analytic in $k\in\bCm$ and
$J(k)$ is analytic in $k\in\bCp$ and continuous in $k\in\bCpb.$
We then consider the sectionally analytic function $\Xi(k)$ defined
on the complex plane $\bC$ as
$$\Xi(k):=\cases \hat X(-k)\, J(k),\qquad k\in\bCp,\\
\stretch
-\hat X(k)\, J(-k),\qquad k\in\bCm.\endcases\tag 15.78$$
Since $\hat X(k)$ belongs to the Hardy space $\bold H^2(\bCm),$
it follows that for any $k\in\bR$ we have
$\hat X(k-i\epsilon)\to \hat X(k)$ as $\epsilon \to 0^+$
a.e. in $k\in\bR$ and strongly in $L^2(\bR).$
As in the proof of Proposition~15.8, for $0<\epsilon<1,$ we
use $C_\epsilon$ to denote the positive boundary
of the rectangle in $\bCp$ with respective corners
located at $a+i\epsilon,$ $b+i\epsilon,$ $b+i,$ $a+i,$
where $a$ and $b$ are some positive parameters with $a<b.$
Similarly, we
use $C_{-\epsilon}$ to denote the positive boundary
of the rectangle in $\bCm$ with respective corners
located at $b-i\epsilon,$ $a-i\epsilon,$ $a-i,$ $b-i.$
Since $\Xi(k)$ defined in (15.78) is analytic
in $k\in\bCp\cup\bCm,$ it follows from the Cauchy integral formula that
for any $k$ inside $C_\epsilon$ we
have
$$\Xi(k)=\ds\frac{1}{2\pi i}\int_{C_\epsilon} dt\,\ds\frac{\Xi(t)}{t-k}+
\ds\frac{1}{2\pi i}\int_{C_{-\epsilon}} dt\,\ds\frac{\Xi(t)}{t-k},
\tag 15.79$$
where the contribution by the second integral is zero.
Let us choose $a$ and $b$ so that in the limit $\epsilon\to 0^+$ we have
$\Xi(a\pm i \epsilon)\to \Xi(a)$ and $\Xi(b\pm i \epsilon)\to \Xi(b).$
Then, letting $\epsilon\to 0^+$ in (15.79) we get
$$\Xi(k)=\ds\frac{1}{2\pi i}\int_{C_0} dt\,\ds\frac{\Xi(t)}{t-k},\tag 15.80$$
where
$C_0$ is the positively oriented
boundary of the rectangle with corners
at $-a-ib,$ $a-ib,$ $a+ib,$ and $-a+ib.$
 From the representation in (15.80), we conclude that
$\Xi(k)$ is analytic in the interior of the
rectangle bounded by $C_0,$ including the segment of
the real axis contained in that rectangle.
Since we can let $a\to -\infty$ and $b\to +\infty,$
we conclude that $\Xi(k)$ is in fact entire in $k.$ With the help of
(15.78) we conclude that $\Xi(k)$ is an odd function of $k$
in $\bold C$ and we have
$$\Xi(-k)=-\Xi(k),\qquad k\in\bC.\tag 15.81$$
Since $\Xi(k)$ defined in (15.78)
has an analytic extension to the entire
complex plane, in its Maclaurin expansion of
$\Xi(k)$ given by
$$\Xi(k)=\ds\sum_{p=1}^\infty \ds\frac{1}{p!}\,\left[\Xi^{(p)}(0)\right]\, k^p,\
\qquad k\in\bC,\tag 15.82$$
where the coefficient
$\Xi^{(p)}(0)$ can be evaluated, with the help of the generalized Cauchy
integral formula as
$$\Xi^{(p)}(0):=\ds\frac{d^p\, \Xi(k)}{dk^p}  \bigg|_{k=0}=
\ds\frac{p!}{2\pi i}\, \int_{\bold T_r}dt\,\ds\frac{\Xi(t)}{t^{p+1}},\qquad p=0,1,2,\dots,
\tag 15.83$$
where $\bold T_r$ is the circle of radius $r$ centered at $k=0$
traversed in the positive direction, with $r:=|k|.$
Because of (15.81), from (15.83) we conclude that
$\Xi^{(p)}(0)=0$ for even values of $p$ in (15.82). We will
now estimate the integral in (15.83). Using (15.81), we have
$$\bigg| \int_{\bold T_r}dt\,\ds\frac{\Xi(t)}{t^{p+1}}\bigg|
\le 2 \int_{\bold T^+_r}|dt|\,\ds\frac{|\Xi(t)|}{r^{p+1}},
\tag 15.84$$
where we use $\bold T_r^+$ to denote the upper semicircle
of $\bold T_r.$ From Proposition~10.2(b) it follows that there exists
some positive number
$r_0$ such that
$$|J(k)|\le C\,|k|,\qquad |k|\ge r_0,\quad k\in\bCpb,
\tag 15.85$$
for some generic constant $C.$
On the other hand, since $\hat X(k)\in \bold H^2(\bCm),$ we have
$\hat X(-k)\in\bold H^2(\bCp),$ and hence
by (3.46) we have
$$|\hat X(-k)|\le \ds\frac{C}{\sqrt{|k|\,\sin \theta}},
\qquad k\in\bCp.
\tag 15.86$$
Using (15.85) and (15.86) in the first line of
(15.78), we have with $k=r\,e^{i\theta}$
$$|\Xi(r\,e^{i \theta})|\le \ds\frac{C \,\sqrt{r}}
{\sqrt{\sin\theta}},\qquad r\ge r_0,\quad k\in \bold T_r^+,\tag 15.87$$
for some generic constant $C.$
Using (15.87) in (15.84) we get the estimate
$$\bigg| \int_{\bold T_r}dt\,\ds\frac{\Xi(t)}{t^{p+1}}\bigg|
\le \ds\frac{C}{r^{p+1/2}} \int_0^{\pi /2}\ds\frac{d\theta}{\sqrt{\sin\theta}},
\tag 15.88$$
where we have used $|dt|=r\,d\theta$ and that
$\sin\theta=\sin(\pi-\theta)$ for $\theta\in(0,\pi/2).$
 From (15.58) we know that
the integral on the right-hand side of (15.88).
 is convergent
Thus, using (15.58), (15.83), (15.88), and letting
$r\to +\infty,$ we conclude that
$\Xi^{(p)}(0)=0$ for $p=0,1,\dots,$ and hence
 from (15.82) we conclude that
 $\Xi(k)\equiv 0,$
 yielding
 $\hat X(k)\equiv 0.$
 Thus, our proof is complete.
 \qed

Among various implications of Proposition~15.12, one
consequence is that
if $\bold D$ belongs to the
Faddeev class then each of $(\bold{III}_a)$, $(\bold{III}_b)$, and
$(\bold{III}_c)$ of Definition~4.3 is satisfied.

\newpage
\noindent {\bf 16. THE SOLUTION TO THE INVERSE PROBLEM}
\vskip 3 pt

In this chapter, given the scattering
data set $\bold S$ in (4.2) belonging to the Marchenko class
specified in Definition~4.5,
 our primary goal is to
show that there exists a unique data set $\bold D$ belonging to
the Faddeev class, with the understanding that the boundary matrices $A$ and $B$
are unique up to a multiplication from the right by an invertible matrix.

We summarize the construction of $\bold D$ from $\bold S$ as follows,
where the existence and uniqueness are implicit at each step:

\item{(a)} From the large-$k$ asymptotics of the
scattering matrix $S(k),$ with the help of
(4.6), we determine the $n\times n$ constant
matrix $S_\infty.$ We then determine the constant
$n\times n$ matrix $G_1$ specified in
(14.1).
As we show in Proposition~16.4, the matrices
$S_\infty$ and $G_1$ are hermitian when $\bold S$ satisfies
the condition $(\bold 1)$ of
Definition~4.5.

\item{(b)} In terms of the quantities in $\bold S,$ we uniquely construct
the $n\times n$ matrix $F_s(y)$ by using (4.7) and
the $n\times n$ matrix $F(y)$ by using (4.12).

\item{(c)} One uses the matrix $F(y)$ as input to the
Marchenko integral equation (13.1).
When $F(y)$ is integrable in $y\in(x,+\infty)$ for each $x\in\bR^+,$
we show in Proposition~16.1 that, for each fixed $x\in\bR^+,$ there exists a solution $K(x,y)$ integrable
in $y\in(x,+\infty)$ to (13.1)
and such a solution is unique.
The solution $K(x,y)$ can be constructed by iterating (13.1).
Even though $K(x,y)$ is constructed for $y>x> 0,$ one can extend
$K(x,y)$ to $y\in\bR^+$ by letting $K(x,y)=0$ for $0\le y<x.$

\item{(d)} Having obtained $K(x,y)$ uniquely from $\bold S,$ one
constructs the potential $V(x)$ via (10.4)
and also constructs the Jost solution
$f(k,x)$ via (10.6). Then, as indicated
in Proposition~16.11,
by using $(\bold I$) of Definition~4.3
and $(\bold 2)$ and $(\bold 4_a)$  of
Definition~4.2, one proves that the constructed $V$ satisfies
(2.2) and (2.3) and that
the constructed $f(k,x)$ satisfies (2.1)
with the constructed potential $V(x).$

\item{(e)} Having obtained $K(x,y)$ uniquely, one also
obtains the $n\times n$ constant matrix $K(0,0)$ via (10.5) and
proves that $K(0,0)$ is hermitian.

\item{(f)} With the help
of the uniquely
constructed $n\times n$ constant hermitian matrices $S_\infty,$
$G_1,$ and $K(0,0),$
one constructs the boundary
matrices $A$ and $B$ by solving the linear homogeneous system (14.2)
in such a way that the rank of the $2n\times n$ matrix $\bm A \\ B\endbm$ is
equal to $n.$ The details of this step are provided in Proposition~16.9.
One proves that the solution pair
of matrices $A$ and $B$ to (14.2) is unique up to a multiplication
on the right by an invertible matrix. One also
proves that any solution to (14.2) satisfies (2.5) and
the full rank of the matrix $\bm A \\ B\endbm$ guarantees that
(2.6) is satisfied. Note that, if $K(0,0)$ were not well defined, then,
as seen from (10.5), the potential $V$ constructed via (10.4) would not
be integrable, which cannot happen if the scattering data set
belongs to the Marchenko class.

\item{(g)} Having constructed
the Jost solution $f(k,x),$ one then
constructs the physical solution $\Psi(k,x)$ via
(9.4) and the normalized bound-state matrices
$\Psi_j(x)$ via (9.8). One then proves that the
constructed matrix $\Psi(k,x)$ satisfies (2.1) and
(2.4) and that the constructed $\Psi_j(x)$
satisfies (2.1) at $k=i\kappa_j$ and also (2.4),
where $A$ and $B$ are the matrices constructed as
explained in the previous step. In the proof that
$\Psi(k,x)$ and $\Psi_j(x)$ each satisfy (2.4), one uses
$(\bold 3_a)$ of Definition~4.5.

The following proposition discusses the unique solvability of the
Marchenko equation (13.1), and it also indicates the equivalence of
$(\bold 4_a)$ and $(\bold 4_b)$ of Definition~4.2.

\noindent {\bf Proposition 16.1} {\it Consider a scattering data set $\bold S$
as in (4.2), which consists of
an $n\times n$ scattering matrix $S(k)$ for $k\in\bR,$ a set of $N$ distinct
 positive constants $\kappa_j,$ and a set of
$N$ constant $n\times n$ hermitian and nonnegative matrices
$M_j$ with respective positive ranks $m_j,$ where $N$ is a nonnegative integer.
Assume that $\bold S$ satisfies $(\bold I)$
of Definition~4.3. Then:}

\item{(a)} {\it For each fixed $x>0,$ the Marchenko integral
operator associated with (13.1) is compact on $L^1(x<y<+\infty),$ the
corresponding homogeneous Marchenko equation has only the trivial
solution in $L^1(x<y<+\infty),$ and the Marchenko equation (13.1)
has a unique solution $K(x,y)$ in $y\in L^1(x<y<+\infty).$
Moreover, for each fixed $x>0,$ the solution $K(x,y)$ to the Marchenko equation
actually belongs to $L^1(x<y<+\infty)\cap L^\infty(x<y<+\infty).$
Hence, $K(x,y)$ in particular belongs to $L^2(x<y<+\infty).$}

\item{(b)} {\it Without any further assumption, the result
in (a) may not hold at $x=0.$ If the scattering data further satisfies
$(\bold 4_a)$ of Definition~4.5, then the result in (a) also holds at $x=0.$
In other words, if the scattering data set
$\bold S$ satisfies $(\bold I)$ and $(\bold 4_a)$, then for each fixed $x\ge 0$
the Marchenko integral equation
(13.1) is uniquely solvable in $L^1(x<y<+\infty),$ and the unique solution
$K(x,y)$ actually belongs to $L^1(x<y<+\infty)\cap L^\infty(x<y<+\infty).$
Hence, $K(x,y)$ in particular belongs to $L^2(x<y<+\infty).$}



\item{(c)} {\it The condition $(\bold 4_a)$ and
 $(\bold 4_b)$ of Proposition~4.2
are equivalent.}

\noindent PROOF: Because the scattering data satisfies
$(\bold I)$, it follows that $F_s(y)$ given in (4.7) is
bounded and integrable in $y\in\bR^+.$ Then, from
(4.12) we see that $F(y)$ given in (4.12) is also
bounded and integrable in $y\in\bR^+.$
Hence, the result of Proposition~3.3 applies. Thus, for
each fixed $x>0,$ the
Marchenko integral operator is compact on $L^1(x<y<+\infty).$
It is proved in
Theorem~3.4.1 on p. 82 of
[2] that the Marchenko equation (13.1) has a unique solution
in $L^1(x<y<+\infty).$ Then, from Proposition~3.3(c) it
follows that the solution $K(x,y)$ to
the Marchenko equation also belongs to $L^\infty(x<y<+\infty).$
The integrability and the boundedness implies
the square integrability, and hence $K(x,y)$
in particular belongs to $L^2(x<y<+\infty).$
Thus, the proof of (a) is complete.
%
%
%
As later shown in Example~26.14,
where $(\bold I)$
is satisfied but $(\bold 4_a)$ is not satisfied,
the result in (a) does not necessarily hold at $x=0.$
The further assumption $(\bold 4_a)$ assures that the result of (a) also holds at
$x=0.$ Thus, the proof of (b) is completed.
As stated in Proposition~3.3(b),
the equivalence of $(\bold 4_a)$ and $(\bold 4_b)$ directly
follows from the fact that the Marchenko
integral operator is compact on $L^1(\bR^+),$
which is already established in (a).
\qed

As Proposition~16.1(a) indicates,
the unique solvability of the Marchenko equation (13.1) on $L^1(x,+\infty)$ for each
$x\in\bR^+$ is solely determined by $S(k)$ satisfying $(\bold I)$ and that the
unique solvability
is unaffected by the bound-state data. In other words, for the unique
solvability of the Marchenko equation,
it does not matter what $N$ is, what the $\kappa_j$-values are,
and what $M_j$ are as long as $N$ is a finite nonnegative
integer, the $\kappa_j$ are distinct and positive, and
the $M_j$ are nonnegative hermitian $n\times n$ matrices
of positive rank.
On the other hand, if we want to relate our scattering data set
$\bold S$ in (4.2) to some data set $\bold D$ in (4.1)
belonging to the Faddeev class,
where $V$ is the potential and $\{\kappa_j,M_j\}_{j=1}^N$
is the bound-state data set, then $\bold S$ must satisfy further
restrictions. In other words,
the potential $V(x)$ constructed from the unique solution
$K(x,y)$ to the Marchenko equation
(13.1) via (10.4) must satisfy (2.2) and (2.3)
and that the physical solution $\Psi(k,x)$
and the bound-state solutions $\Psi_j(x)$
constructed via (9.4) and (9.8), respectively,
must satisfy the boundary condition
(2.4).
We further impose $(\bold 2)$ of Definition~4.2
on the scattering data set $\bold S$
so that (2.3) is satisfied.
We also impose $(\bold 3_a)$  of Definition~4.2
on the scattering data set $\bold S$
so that $\Psi(k,x)$
and $\Psi_j(x)$ each satisfy
(2.4).

The following result shows that when $\bold S$ is a scattering data set
that belongs to the Marchenko class, then the solution to the inverse problem $\bold S\mapsto \bold D$
must be unique, with the understanding that the boundary matrices
$A$ and $B$ are uniquely determined only up to
a postmultiplication by an invertible matrix.

\noindent {\bf Proposition 16.2} {\it Consider a scattering data set $\bold S$
as in (4.2), which consists of
an $n\times n$ scattering matrix $S(k)$ for $k\in\bR,$ a set of $N$ distinct
 positive constants $\kappa_j,$ and a set of
$N$ constant $n\times n$ hermitian and nonnegative matrices
$M_j$ with respective positive ranks $m_j,$ where $N$ is a nonnegative integer. Assume that $\bold S$ belongs to the
Marchenko class specified
in Definition~4.5.
Then, two distinct input data sets $\bold D_1:=\{V_1,A_1,B_1\}$ and
$\bold D_2:=\{V_2,A_2,B_2\}$ in the Faddeev class corresponding to the same $\bold S$
must be related to each other as}
$$V_1(x)\equiv V_2(x),\quad A_1=A_2\,T,\quad B_1=B_2\,T,\tag 16.1$$
{\it where $T$ is an $n\times n$ invertible matrix.}

\noindent PROOF: As shown in Proposition~6.2, the scattering data set
$\bold S$ yields a unique solution $K(x,y)$ to the
Marchenko integral equation, and the potential
$V(x)$ is uniquely constructed from $K(x,y)$ via (10.4).
Hence, we must have $V_1(x)\equiv V_2(x).$
The Jost solution $f(k,x)$ is constructed from
$K(x,y)$ as in (10.6), and hence
the constructed $f(k,x)$ is unique.
The physical solution $\Psi(k,x)$ is constructed from
$f(k,x)$ and $S(k)$ as in (9.4) and hence
we must have
$$\Psi_1(k,x)\equiv \Psi_2(k,x),\quad \Psi'_1(k,x)\equiv \Psi'_2(k,x),\tag 16.2$$
where $\Psi_1(k,x)$ and $\Psi_2(k,x)$ are the constructed physical solutions
associated with $\bold D_1$ and $\bold D_2,$ respectively.
Let $\varphi_1(k,x)$ and $\varphi_2(k,x)$ be the constructed regular solutions
associated with $\bold D_1$ and $\bold D_2,$ respectively.
We then see from (9.5) that
$$\varphi_1(k,0)=A_1,\quad \varphi'_1(k,0)=B_1,\quad
\varphi_2(k,0)=A_2,\quad \varphi_2(k,0)=B_2.\tag 16.3$$
Let $J_1(k)$ and $J_2(k)$
be the constructed Jost matrices
associated with $\bold D_1$ and $\bold D_2,$ respectively.
 From (9.6) we see that
$$\Psi_1(k,x)=-2ik\, \varphi_1(k,x)\,J_1(k),\quad
\Psi_2(k,x)=-2ik\, \varphi_2(k,x)\,J_2(k).\tag 16.4$$
Using (16.3) and (16.4) in (16.2) we obtain
$$A_1\,J_1(k)=A_2\,J_2(k),\quad
B_1\,J_1(k)=B_2\,J_2(k),\qquad k\in\bR\setminus\{0\}.\tag 16.5$$
By Proposition~10.2(a) we know that the constructed Jost matrix is invertible for $k\in\bR\setminus\{0\},$
 from (16.5) we obtain
$$A_1=A_2\,J_2(k)\,J_1(k)^{-1},\quad
B_1=B_2\,J_2(k)\,J_1(k)^{-1},\qquad k\in\bR\setminus\{0\},\tag 16.6$$
confirming the second and third equalities in (16.1)
with $T$ being equal to $J_2(k)\,J_1(k)^{-1}$ for
any real nonzero $k$-value. \qed

Analogous to (3.96), let us define
$$\tau(x):=\int_x^\infty dz\,|F'(z)|,\quad
\tau_1(x):=\int_x^\infty dz\,z\,|F'(z)|,\qquad x\ge 0,\tag 16.7$$
where $F'(y)$ is the derivative of the quantity $F(y)$
appearing in (4.12). Comparing (16.7) with (3.97)-(3.99), we see that
$$x\,\tau(x)\le \tau_1(x),\quad \int_x^\infty dz\,\tau(z)\le \tau_1(x),\tag 16.8$$
$$\int_0^\infty dz\,(1+z)\,[\tau(z)]^2\le \left[\tau(0)+\tau_1(0)\right]
\tau_1(0).\tag 16.9$$
Comparing (4.8) and (16.7) we see that $\tau(0)$ and $\tau_1(0)$ are both finite
when the condition $(\bold 2)$ of Definition~4.2 holds.

In the next proposition, we continue to present certain properties of
the solution $K(x,y)$ to the Marchenko equation (13.1).

\noindent {\bf Proposition 16.3} {\it Consider a scattering data set $\bold S$
as in (4.2), which consists of
an $n\times n$ scattering matrix $S(k)$ for $k\in\bR,$ a set of $N$ distinct
 positive constants $\kappa_j,$ and a set of
$N$ constant $n\times n$ hermitian and nonnegative matrices
$M_j$ with respective positive ranks $m_j,$ where $N$ is a nonnegative integer. Let $F_s(y)$ and $F(y)$ be the quantities
defined in (4.7) and (4.12), respectively.
Assume that $\bold S$ satisfies $(\bold I)$
of Definition~4.3 and that
$F_s'(y)$ is integrable in $y\in\bR^+.$
Then:}

\item{(a)} {\it The quantities $F_s(y)$ and $F(y)$ are continuous
in $y\in\bR^+$ and vanish as $y\to+\infty.$
Furthermore, they have finite limits, $F_s(0^+)$ and $F(0^+),$
respectively, as $y\to 0^+.$
Consequently, they are uniformly bounded in $y\in\bR^+.$}


\item{(b)} {\it The solution $K(x,y)$ to the Marchenko equation satisfies}
$$|K(x,y)|\le C\,\tau(x+y),\qquad 0<x\le y,\tag 16.10$$
{\it where $C$ is a generic constant and $\tau(x)$ is the scalar quantity defined
in (16.7). If $\bold S$ further satisfies $(\bold 4_a)$, then
we have}
$$|K(x,y)|\le C\,\tau(x+y),\qquad 0\le x\le y.\tag 16.11$$


\noindent PROOF: Since
$F_s'(y)$ is assumed to be integrable in
$y\in\bR^+,$ we have
$$F_s(y)=-\int_y^\infty dz\,F_s'(z),\qquad y\in\bR^+.\tag 16.12$$
Using Lebesgue's dominated convergence theorem on (16.12)
we conclude that $F_s(y)$ is continuous in $y\in\bR^+,$
$F_s(y)$ vanishes as $y\to+\infty,$ and $F_s(0^+)$ is finite.
These three properties also hold for $F(y)$ because
$F(y)$ is related to $F_s(y)$ as in (4.12),
where all $\kappa_j$ are positive. Hence, the second sentence of (a)
is valid. Let us now prove (b). From (16.12) we obtain
$$|F(y)|\le \int_y^\infty dz\,|F'(z)|=\tau(y),\qquad y\ge 0.
\tag 16.13$$
 From the Marchenko equation (13.1) we obtain
$$|K(x,y)|\le |F(x+y)|+\int_x^\infty dz\,|K(x,z)|\,|F(z+y)|,
\qquad 0<x\le y.\tag 16.14$$
Using (16.13) in
(16.14) we obtain
$$|K(x,y)|\le \tau(x+y)+\int_x^\infty dz\,|K(x,z)|\,\tau(z+y),
\qquad 0<x\le y.\tag 16.15$$
Since $\tau(x)$ defined in (16.7) is a nonincreasing function
in $x\in\bR^+,$ from (16.15) we get
$$|K(x,y)|\le \tau(x+y)+\tau(x+y)\int_x^\infty dz\,|K(x,z)|,
\qquad 0<x\le y.\tag 16.16$$
As asserted in Proposition~16.1(a), $K(x,y)$ is integrable in $y\in (x,+\infty),$
and hence the integral in (16.16) is bounded, yielding
$$|K(x,y)|\le \tau(x+y)+C\, \tau(x+y),
\qquad 0<x\le y,\tag 16.17$$
for some constant $C.$
 From (16.17) we obtain (16.10) for some generic constant $C.$
 Under the further assumption that
$\bold S$ satisfies $(\bold 4_a),$
the results in (16.14)-(16.17) hold also at $x=0,$ and hence
we conclude (16.11).
Thus, the proof is complete. \qed

In the next proposition, we study the relevant properties of some quantities
related to the scattering matrix $S(k)$ and
the solution $K(x,y)$ to the Marchenko integral equation (13.1).

\noindent {\bf Proposition 16.4} {\it Consider a scattering data set $\bold S$
as in (4.2), which consists of
an $n\times n$ scattering matrix $S(k)$ for $k\in\bR,$ a set of $N$ distinct
 positive constants $\kappa_j,$ and a set of
$N$ constant $n\times n$ hermitian and nonnegative matrices
$M_j$ with respective positive ranks $m_j,$ where $N$ is a nonnegative
integer. Assume that $\bold S$ satisfies
$(\bold I)$ of Definition~4.3. Let $F_s(y)$ and $F(y)$ be the quantities defined
in (4.7) and (4.12), respectively,
where $S_\infty$ is the constant $n\times n$ matrix
defined as in (4.6). Let $K(x,y)$ be the
unique solution to
(13.1), whose existence and uniqueness for each fixed $x>0$
are assured by Proposition~16.1(a). Then:}
\item{(a)} {\it The matrix $S_\infty$ satisfies}
$$S_\infty=S_\infty^\dagger=S^{-1}_\infty,\tag 16.18$$
{\it and hence $S_\infty$ is hermitian and each eigenvalue of
$S_\infty$ is either $+1$ or $-1.$}

\item{(b)} {\it The matrices $F_s(y)$ and
$F(y)$ are hermitian for every $y\in\bR^+.$}

\item{(c)} {\it The $n\times n$ matrix
$K(x,x)$ is hermitian for each fixed $x\in\bR^+.$
Under the further assumption of $(\bold 4_a)$ of Definition~4.2,
the matrix $K(x,x)$ is hermitian for $x\ge 0.$
In particular, the hermitian property of
$K(0,0)$ is assured under the additional assumption of $(\bold 4_a)$.}

\item{(d)} {\it If the scattering data set $\bold S$
satisfies $(\bold 1)$ of Definition~4.2, then the $n\times n$ constant matrix
$G_1$ given in (14.1) is hermitian.}

\noindent PROOF: Note that (4.4) yields (16.18)
and hence $S_\infty$ is hermitian and unitary. Thus, each eigenvalue
of $S_\infty$ must be real and equal to either $+1$ or $-1,$
establishing (a).
Let us now turn to the proof of (b). The hermitian property of
$F_s(y)$ is obtained by using the first equality of (4.4)
as well as (a) in (4.7). Because the matrices $M_j$ are hermitian,
the $\kappa_j$-values are real,
and we have already proved that $F_s(y)$ is hermitian
for every $y\in\bR^+,$ it follows from (4.12) that
$F(y)$ is hermitian for every $y\in\bR^+.$
Thus, the proof of (b) is complete.
The proof of (c) can be found on p. 122 of [2] and based on the
fact that $F(y)$ is hermitian and that the Marchenko equation
(13.1) is uniquely solvable for each $x\in\bR^+,$
as indicated in Proposition~16.1(b). Under the additional assumption of
$(\bold 4_a)$, the Marchenko equation is uniquely solvable for each $x\ge 0$ and
the proof of the hermitian property of
$K(x,x)$ for each $x\ge 0$ again follows from p. 122 of [2].
Let us now turn to the proof of (d). The assumption $(\bold 1)$ assures
the existence of $G_1,$ and the hermitian property of $G_1$ follows from the fact that
$S(k)$ is hermitian as stated in the first equality in
(4.4) and that $S_\infty$ is hermitian as stated in (a)
and we have $ik$ appearing as a factor on the right-hand side of (14.1).
\qed

 From Theorem~15.10 and Proposition~16.4(c) it follows that
the matrix $K(x,x)$ is hermitian for
each $x\ge 0$ when the input data set $\bold D$ in (4.1) belongs to
the Faddeev class.
On the other hand, we remark that, in the Faddeev class,
in general
the solution $K(x,y)$ to the Marchenko integral equation
(13.1) is not hermitian when $0\le x<y.$

As mentioned in Chapter~13, in the analysis of the inverse problem for
(2.1) with the general selfadjoint boundary condition (2.4), the derivative Marchenko equation
(13.7) plays an equally important role as the Marchenko equation (13.1).
Let us further elaborate on this issue. In order to prove that
the physical solution $\Psi(k,x)$ constructed as in (9.4) satisfies
the boundary condition, one needs to construct both
$f(k,x)$ and $f'(k,x),$ where $f(k,x)$ is the Jost solution
to (2.1) constructed as in (10.6) and $K(x,y)$ is obtained
by solving the Marchenko equation (13.1). Thus, unless the boundary condition
(2.4)
is the Dirichlet boundary, the construction
of the physical solution also requires the construction
of $f'(k,x),$ and this requires the analysis of the
derivative Marchenko equation (13.7).
In the special case with the Dirichlet boundary condition
studied in [2], the analysis
of the derivative Marchenko equation (13.7) is not needed
for the satisfaction of the boundary condition.
It turns out that, under appropriate restrictions on our
scattering data, which are all satisfied when the scattering data
belongs to the Marchenko class,
the derivative Marchenko equation (13.7) is uniquely solvable and its unique
solution is given by $K_x(x,y),$ where $K(x,y)$ is
the unique solution to the Marchenko equation (13.1).
A proof of this is given in Lemma~5.3.3 of [2] under the
restriction that the derivative $F'(y)$
is continuous in $y\in\bR^+,$ where $F(y)$ is the quantity defined in
(4.12). In general, when the input data set
$\bold D$ in (4.1) is in the Faddeev class, the constructed
$F'(y)$ in the corresponding
scattering data set $\bold S$ is not necessarily
continuous in
$y\in\bR^+.$ We extend the proof in [2] without requiring the
continuity of $F'(y)$ in $y\in\bR^+,$ by only using
the integrability of $F'(y)$ in $y\in\bR^+.$
 From Definition~4.2 we see that
 the integrability of $F'(y)$ in $y\in\bR^+$
is assured if $\bold S$ satisfies the property $(\bold 2).$

\noindent {\bf Proposition 16.5}
{\it Consider a scattering data set $\bold S$
as in (4.2), which consists of
an $n\times n$ scattering matrix $S(k)$ for $k\in\bR,$ a set of $N$ distinct
 positive constants $\kappa_j,$ and a set of
$N$ constant $n\times n$ hermitian and nonnegative matrices
$M_j$ with respective positive ranks $m_j,$ where $N$ is a nonnegative integer.
Assume that $\bold S$ satisfies $(\bold I)$
 of Definition~4.3. Assume also
that $F_s'(y)$ is integrable
in $y\in\bR^+,$ where $F_s(y)$ is the quantity
defined in (4.7). Then:}

\item{(a)} {\it For each fixed $x>0,$ the derivative Marchenko integral
operator associated with (13.7) is compact on $L^1(x<y<+\infty),$ the
corresponding homogeneous derivative Marchenko equation has only the trivial
solution in $L^1(x<y<+\infty),$ and the derivative Marchenko equation (13.7)
has a unique solution $L(x,y)$ in $L^1(x<y<+\infty).$}

\item{(b)} {\it For each fixed $x>0,$ the unique solution $L(x,y)$
to (13.7) is equal to $K_x(x,y),$ where $K(x,y)$ is
the unique solution to the Marchenko equation (13.1), whose existence and uniqueness
are established in Proposition~16.1(a).}

\item{(c)} {\it Without any further assumption, the result
in (a) may not hold at $x=0.$ If the scattering data further satisfies
$(\bold 4_a)$ in Definition~4.5, then the result in (a) also holds at $x=0.$
In other words, if the scattering data set
$\bold S$ satisfies $(\bold 1)$ and $(\bold 4_a)$ and we also have the integrability of
$F_s'(y)$ in $y\in\bR^+,$ then
(13.7) is uniquely solvable with the solution
$L(x,y)$ which is integrable in $L^1(x<y<+\infty)$ for each
$x\in[ 0,+\infty).$}

\item{(d)} {\it When $(\bold 4_a)$ is also satisfied, the solution
$L(x,y)$ to (13.7) satisfies
$L(x,y)=K_x(x,y)$ for $0\le x\le y,$ where $K(x,y)$ is the solution to (13.1),
whose existence and uniqueness is established in Proposition~16.1(b).}

\noindent PROOF: The proofs of (c) and (d)
under the additional assumption $(\bold 4_a)$
are similar to the proofs of (a) and (b) under the assumption
$(\bold I).$ Thus, we only present the proofs for (a) and (b).
Let us compare the Marchenko equation (13.1) and
the derivative Marchenko equation (13.7). They have the same kernel and they only differ
in their nonhomogeneous terms. The nonhomogeneous term
in (13.1) is $F(x+y)$ and the nonhomogeneous term in (13.7) is
$F'(x+y)-K(x,x)\,F(x+y).$ We will show that both nonhomogeneous terms
belong to $L^1(x<y<+\infty)$ for each fixed $x> 0,$
and as a result, the unique solvability stated in (a)
directly follows from Proposition~16.1(a).
After that, it only remains to prove (b) by showing that the unique solution to
(13.7) is given by $K_x(x,y).$ Let us first prove that
the aforementioned nonhomogeneous terms are both integrable in $y\in(x,+\infty)$
for each fixed $x> 0.$ The integrability of
$F(x+y)$ for $y\in(x,+\infty)$ is assured by the integrability of
$F_s(y)$
in $y\in\bR^+,$ which is ensured by $(\bold I)$, and the fact that
$F(y)$ and $F_s(y)$ are related to each other as in (4.12) where
the $\kappa_j$ are all positive. The integrability
of $F'(x+y)$ is assured by the assumed integrability of
$F'_s(y)$
in $y\in\bR^+$ and again by the fact that
$F(y)$ and $F_s(y)$ are related to each other as in (4.7) where
the $\kappa_j$ are all positive.
The integrability of $K(x,x)\,F(x+y)$ is assured by (3.28) and
by the fact that
$K(x,x)$ is bounded, as assured by Proposition~16.1(a), and
that $F(x+y)$ is integrable when
$y\in(x,+\infty).$ Thus, the nonhomogeneous term
in (13.7) belongs to $L^1(x<y<+\infty)$ for each fixed $x> 0,$
and the proof of (a) is complete.
Hence, it only remains to prove (b) by showing that the unique solution
$L(x,y)$ to (13.7) is given by
$K_x(x,y).$ A proof of this is given in Lemma~5.3.3 of [2]
by assuming that $F_s'(y)$ is continuous in $y\in\bR^+.$ We will extend that
proof by assuming that $F'_s(y)$ is integrable in $y\in\bR^+$
instead of assuming that $F_s'(y)$ is continuous in $y\in\bR^+.$
We proceed as follows. By Proposition~16.1(a) we know that
(13.1) has a unique solution $K(x,y),$ which can be written as
$$K(x,y)=-F(x+y)\left(I+\Cal O_x\right)^{-1} ,\tag 16.19$$
where $\Cal O_x$ is the integral operator on
$L^1(x<y<+\infty)$ defined as
$$\left(  X\,\Cal O_x\right)(y):=\ds\int_x^\infty dz\,X(z)\,F(z+y),\qquad 0< x\le y.
\tag 16.20$$
 From Proposition~16.1(b)
 we know that for each $x\ge 0$
 the Marchenko integral operator is compact and the
 homogeneous Marchenko integral
 equation has only the trivial solution.
 Consequently, the operator
 $(I+\Cal O_x)$ is invertible
 and the inverse $(I+\Cal O_x)^{-1}$ is bounded
 as an operator on $L^1(x<y<+\infty).$
By (a) we are assured that (13.7) has a unique solution
and we use $L(x,y)$ to denote that solution. With the help of
(16.20) we get
$$L(x,y)=\left[F'(x+y)-K(x,x)\,F(x+y)\right]\left(I+\Cal O_x\right)^{-1} ,\tag 16.21$$
where $K(x,x)$ is obtained from (16.19) as $K(x,x^+).$
Let us now approximate $F(y)$ by an appropriate sequence $\{F^{(m)}(y)\}_{m=1}^\infty$ converging to
$F(y).$ We will soon see how to choose the sequence.
Let us use $K^{(m)}(x,y)$ to denote the solution to (13.1) when
$F^{(m)}(y)$ is used there instead of $F(y),$ i.e. we would like
$K^{(m)}(x,y)$ to satisfy
$$K^{(m)}(x,y)+ F^{(m)}(x+y)+\ds\int_x^\infty dz\,
 K^{(m)}(x,z)F^{(m)}(z+y)=0,\qquad 0<x\le y.\tag 16.22$$
 From (13.1) and (16.22) we obtain
 $$\aligned | K^{(m)}(x,y)-K(x,y)|\le & | F^{(m)}(x+y)-F(x+y)|\\
 \stretch
 &+\ds\int_x^\infty dz\, | K^{(m)}(x,z)-K(x,z)|\,| F^{(m)}(x+y)|\\
 \stretch
 &+\ds\int_x^\infty dz\, |K(x,z)|\,| F^{(m)}(x+y)-F(x+y)|.
 \endaligned\tag 16.23$$
 We also would like $K_x^{(m)}(x,y),$ the $x$-derivative
 of $K^{(m)}(x,y),$ to be the solution to (13.7), i.e.
$$\aligned
K_x^{(m)}(x,y)+ F^{(m)\prime}&(x+y)-  K^{(m)}(x,x)\,F^{(m)}(x+y)\\
\stretch
& +\ds\int_x^\infty dz\,
 K_x^{(m)}(x,z)\,F^{(m)}(z+y)=0,\qquad 0< x\le y,\endaligned\tag 16.24$$
where $ F^{(m)\prime}(y)$ denotes the $y$-derivative of $F^{(m)}(y).$ From
(13.7) and (16.24) we obtain the analog of (16.23) given by
$$\aligned | K_x^{(m)}(x,y)-L(x,y)|\le & | F^{(m)\prime }(x+y)-F'(x+y)|\\
& +| K^{(m)}(x,x)-K(x,x)|\,| F^{(m)}(x+y)|\\
\stretch
& +| K(x,x)|\,| F^{(m)}(x+y)-F(x+y)|\\
 \stretch
 &+\ds\int_x^\infty dz\, | K^{(m)}(x,z)-K(x,z)|\,| F^{(m)}(x+y)|\\
 \stretch
 &+\ds\int_x^\infty dz\, |K(x,z)|\,| F^{(m)}(x+y)-F(x+y)|.
 \endaligned\tag 16.25$$
 From (16.23) and (16.25) we see that it is appropriate to choose
the sequence $\{F^{(m)}(y)\}_{m=1}^\infty$ in such a way that
$F^{(m)}(y)$ belongs to $ C^\infty_0(\bR^+),$ is uniformly bounded, i.e. satisfying
$|F^{(m)}(y)|\le c$ for some constant $c$ for all $m\ge 1$ and $y\in\bR^+,$
and converging uniformly to $F(y)$ in every compact subset of $\bR^+,$
and also satisfying
$$\ds\lim_{m\to+\infty}\int_0^\infty dy\,| F^{(m)}(y)-F(y)|=0,\tag 16.26$$
$$\ds\lim_{m\to+\infty}\int_0^\infty dy\,| F^{(m)\prime}(y)-F'(y)|=0.\tag 16.27$$
Our goal is now to prove that
$$K_x(x,y)=L(x,y),\qquad 0< x\le y,
\tag 16.28$$
where we recall that $K_x(x,y)$ is the $x$-derivative of the unique
solution $K(x,y)$ to (13.1).
In analogy with (16.20) let us introduce the sequence
of operators
 on
$L^1(x<y<+\infty)$ given by
$\{\Cal O^{(m)}_x\}_{m=1}^\infty$ and converging to
$\Cal O_x$ as $m\to +\infty.$ We define
$$\left( X\,\Cal O^{(m)}_x \right)(y):=\ds\int_x^\infty dz\,X(z)\,F^{(m)}(z+y),\qquad 0< x\le y.
\tag 16.29$$
Since the constructed sequence $\{F^{(m)}(y)\}_{m=1}^\infty$ converges to
$F(y),$ we have the convergence
$||\Cal O^{(m)}_x-\Cal O_x||_{L^1(x<y<+\infty)}\to 0$ as $m\to+\infty$
in the uniform norm of the bounded operators on $L^1(x<y<+\infty).$
Then, for $m$ large enough, the operator
$\left(I+\Cal O^{(m)}_x\right)$ is invertible and we have
$$\ds\lim_{m\to+\infty} \left\|\left(I+\Cal O^{(m)}_x\right)^{-1}-\left(I+\Cal O_x\right)^{-1}\right\|_{L^1(x<y<+\infty)}=0,\tag 16.30$$
in the operator norm on $L^1(x<y<+\infty).$
To see these, we proceed as follows. The invertibility
of the operator $\left(I+\Cal O_x\right)$
has already been established, as argued below (16.20).
Furthermore, the result in (16.30) follows from the bounded invertibility theorem,
i.e. Theorem~1.16 on p. 196 of [27].
 From (16.30) we get
$$ \ds\lim_{m\to+\infty} F^{(m)}(x+y)\left(I+\Cal O^{(m)}_x\right)^{-1}=F(x+y)
\left(I+\Cal O_x\right)^{-1}.\tag 16.31$$
Having constructed the sequence
$\{F^{(m)}(y)\}_{m=1}^\infty$ converging to
$F(y),$ let us now consider the Marchenko equation
(13.1) but $F(y)$ replaced with $F^{(m)}(y)$ there.
Since $\left(I+\Cal O^{(m)}_x\right)$
is invertible for large $m,$
the unique solution to
the corresponding Marchenko
equation with input
$F^{(m)}(y)$ for large $m$ is, analogous to (16.19), given by
$$K^{(m)}(x,y)=-F^{(m)}(x+y)\left(I+\Cal O_x^{(m)}\right)^{-1} .\tag 16.32$$
Proceeding in a similar manner, for the derivative Marchenko integral equation
(13.7) but $F(y)$ and $F'(y)$ replaced with $F^{(m)}(y)$
and $F^{(m)\prime}(y),$ respectively, there,
for large enough $m$ we obtain
$$K_x^{(m)}(x,y)=\left[ F^{(m)\prime}(x+y)- K^{(m)}(x,x)\, F^{(m)}(x+y)  \right]\left(I+\Cal O^{(m)}_x\right)^{-1}.\tag 16.33$$
In order to obtain (16.33) we have used the
invertibility of the operator
$\left(I+\Cal O^{(m)}_x\right)$ for large $m$
as well as the fact that
the unique solution to the derivative Marchenko equation
(13.7) with input $F^{(m)}(y)$
and $F^{(m)\prime}(y)$ for large $m$
is given by $K_x^{(m)}(x,y),$
which is established in
Lemma~5.3.3 of [2].
 From (16.30) we obtain
$$\aligned
 \ds\lim_{m\to+\infty}
\left[ F^{(m)\prime}(x+y)\right. & \left.- K^{(m)}(x,x)\, F^{(m)}(x+y)  \right]\left(I+\Cal O^{(m)}_x\right)^{-1}\\
&=\left[F'(x+y)-K(x,x)\,F(x+y)\right]
\left(I+\Cal O_x\right)^{-1}.\endaligned\tag 16.34$$
Then, we see that (16.31) and (16.32) yields
$$ \ds\lim_{m\to+\infty}  K^{(m)}(x,y)=K(x,y),\tag 16.35$$
and we also see that (16.21) and (16.34) yields
$$ \ds\lim_{m\to+\infty}  K_x^{(m)}(x,y)=L(x,y).\tag 16.36$$
For large $m,$ we know that  $K_x^{(m)}(x,y)$ for
large $m$ is integrable in $y\in\bR^+$ for each fixed $x\ge 0,$
and hence we have
$$K^{(m)}(x,y)=K^{(m)}(0,y)+\int_0^x dz\,K^{(m)}_x(z,y).\tag 16.37$$
Postmultiplying (16.37) by a test function $\varphi(y)$ belonging to
$C^\infty_0(\bR^+),$ and integrating over $y\in\bR^+$ we obtain
$$\int_0^\infty dy\,K^{(m)}(x,y)\,\varphi(y)=\int_0^\infty dy\,K^{(m)}(0,y)\,\varphi(y)
+\int_0^\infty dy\,\int_0^x dz\,K^{(m)}_x(z,y)\,\varphi(y).\tag 16.38$$
Since $K^{(m)}_x(z,y)$ is integrable in $(z,y)$ in the domain of its integration
in (16.38), we can change the order of integration there and obtain
$$\int_0^\infty dy\,K^{(m)}(x,y)\,\varphi(y)=\int_0^\infty dy\,K^{(m)}(0,y)\,\varphi(y)
+\int_0^x dz\,\int_0^\infty dy\,K^{(m)}_x(z,y)\,\varphi(y).\tag 16.39$$
By letting $m\to+\infty$ in (16.39), with the help of
(16.35) and (16.36), we obtain
$$\int_0^\infty dy\,K(x,y)\,\varphi(y)=\int_0^\infty dy\,K(0,y)\,\varphi(y)
+\int_0^x dz\,\int_0^\infty dy\,L(z,y)\,\varphi(y).\tag 16.40$$
Because $L(x,y)$ is integrable in $(z,y)$ in its domain of integration
in (16.40), we can change the order of integration there and obtain
$$\int_0^\infty dy\,K(x,y)\,\varphi(y)=\int_0^\infty dy\,K(0,y)\,\varphi(y)
+\int_0^\infty dy\,\int_0^x dz\,L(z,y)\,\varphi(y).\tag 16.41$$
because of the arbitrariness of the test function $\varphi(y)$ in (16.41), we
obtain
$$K(x,y)=K(0,y)+\int_0^x dz\,L(z,y),\tag 16.42$$
and by taking the $x$-derivative of both sides of (16.42) we obtain
(16.28). Thus, the proof is complete.
\qed

In the next proposition we obtain a useful bound on the solution
$K_x(x,y)$ to the derivative Marchenko equation (13.7).

\noindent {\bf Proposition 16.6}
{\it Consider a scattering data set $\bold S$
as in (4.2), which consists of
an $n\times n$ scattering matrix $S(k)$ for $k\in\bR,$ a set of $N$ distinct
 positive constants $\kappa_j,$ and a set of
$N$ constant $n\times n$ hermitian and nonnegative matrices
$M_j$ with respective positive ranks $m_j,$ where $N$ is a nonnegative integer.
Assume that $\bold S$ satisfies $(\bold I)$
 of Definition~4.3. Assume also
that $F_s'(y)$ is integrable
in $y\in\bR^+,$ where $F_s(y)$ is the quantity
defined in (4.7). Then:}

\item{(a)} {\it For each fixed $x>0,$ the unique solution
$K_x(x,y)$ to the derivative Marchenko integral
equation (13.7) satisfies}
$$|K_x(x,y)|\le |F'(x+y)|+C,\qquad 0<x\le y,\tag 16.43$$
{\it where $F(y)$ is the quantity defined in (4.12) and
$C$ is a generic constant.}

\item{(b)} {\it Without any further assumption, the result
in (a) may not hold at $x=0.$ If the scattering data further satisfies
$(\bold 4_a)$ in Definition~4.2, then the result in (a) also holds at $x=0,$
i.e. we have}
$$|K_x(x,y)|\le |F'(x+y)|+C,\qquad 0\le x\le y.\tag 16.44$$

\item{(c)} {\it If the scattering data further satisfies
$(\bold 4_a)$ in Definition~4.2, then we have}
$$||K_x(x,\cdot)||_{L^1(x<y<+\infty)}\le C\,
||F'(x+\cdot)-K(x,x)\,F(x+\cdot)||_{L^1(x<y<+\infty)},\qquad x\ge 0,\tag 16.45$$
{\it for some generic constant $C.$}

\noindent PROOF: By Proposition~16.5(b) we know that
$K_x(x,y)$ exists and that $K_x(x,y)$ is integrable
in $y\in(x,+\infty)$ for each fixed $x>0.$
 From (13.6) we obtain
$$|K_x(x,y)|\le |F'(x+y)|
+|K(x,x)|\,|F(x+y)|+\int_x^\infty
dz\,|K_x(x,z)|\,|F(z+y)|,\qquad 0<x\le y.\tag 16.46$$
 From Proposition~16.3(a) we know that $F(y)$ is
uniformly bounded in $y\in\bR^+.$ Hence, (16.46)
yields
$$|K_x(x,y)|\le |F'(x+y)|
+C\, |K(x,x)|+C\int_x^\infty
dz\,|K_x(x,z)|,\qquad 0<x\le y.\tag 16.47$$
 From the aforementioned integrability of $K_x(x,y)$
 we conclude that the integral in (16.47) is finite.
 Furthermore, using (16.11) in the middle term in (16.47)
 we obtain
$$|K_x(x,y)|\le |F'(x+y)|
+C\, \tau(x+y)+C,\qquad 0<x\le y.\tag 16.48$$
where $\tau(x)$ is the scalar quantity defined in (16.7).
Since $\tau(x)$ is a nonincreasing function in $x\in\bR^+,$
we have $\tau(x+y)\le \tau(0)<+\infty,$ where
the finiteness of $\tau(0)$ follows from
the first definition in (16.7) with
the assumption that $F'(y)\in L^1(\bR^+).$
Thus, (16.48) yields (16.44). Hence, the proof of
(a) is complete. If we further have $(\bold 4_a),$ then
 from Propositions~16.3(b), 16.5(c), and 16.5(d),
 it follows that
 (16.43) holds also at $x=0,$ and hence
 (16.44) is justified, and the proof of
 (b) is completed.
 Let us now turn to the proof of (c).
 As in (16.20) let us use
 $\Cal O_x$ to denote the Marchenko
 integral operator on $L^1(x<y<+\infty).$
  From Proposition~16.5 we know that
  the derivative Marchenko integral
  equation is uniquely solvable and the solution
  is given by $K_x(x,y).$ Thus, from (16.21) we obtain
$$K_x(x,y)=\left[F'(x+y)-K(x,x)\,F(x+y)\right]\left(I+\Cal O_x\right)^{-1}.\tag 16.49$$
In the proof of Proposition~16.5 we have seen that the term
$F'(x+y)-K(x,x)\,F(x+y)$ belongs to $L^1(x<y<+\infty),$
holding for each $x\ge 0$ under the additional assumption
of $(\bold 4_a).$ Since $\left(I+\Cal O_x\right)^{-1}$ is bounded
on $L^1(x<y<+\infty),$ from (16.49) we obtain
$$\aligned ||K_x(x,\cdot)||&_{L^1(x<y<+\infty)}
\\ &\le
||F'(x+\cdot)-K(x,x)\,F(x+\cdot)||_{L^1(x<y<+\infty)}\,
||\left(I+\Cal O_x\right)^{-1}||_{L^1(x<y<+\infty)}\\
&\le C\,||F'(x+\cdot)-K(x,x)\,F(x+\cdot)||_{L^1(x<y<+\infty)},
\endaligned\tag 16.50$$
yielding (16.45). \qed

The next proposition provides certain properties of
$F_s(y)$ defined in (4.7) and of its derivative
$F_s'(y)$ given in (12.3).

\noindent {\bf Proposition 16.7} {\it Consider a scattering data set $\bold S$
as in (4.2), which consists of
an $n\times n$ scattering matrix $S(k)$ for $k\in\bR,$ a set of $N$ distinct
 positive constants $\kappa_j,$ and a set of
$N$ constant $n\times n$ hermitian and nonnegative matrices
$M_j$ with respective positive ranks $m_j,$ where $N$ is a nonnegative integer. Assume that
$\bold S$ satisfies $(\bold I)$
of Definition~4.3. Further assume that
$F'_s(y)$ given in (12.3) is integrable in
$y\in\bR^+$ and is the sum of an integrable function
and a square integrable function in $y\in\bR^-.$ Then, we have the following:}

\item{(a)} {\it The matrix $F_s(y)$ given in
(4.7) is continuous in
$y\in\bR\setminus\{0\}.$}

\item{(b)} {\it The limits $F_s(0^+)$ and $F_s(0^-)$ exist and we have
(12.2) satisfied, where $G_1$ is the constant matrix appearing in (4.5).}

\item{(c)} {\it The matrix $F'_s(y)$ can be written as}
$$F'_s(y)=G_1\,\delta(y)+ \overset{\circ}\to F'_s(y),\qquad y\in\bR,\tag 16.51$$
{\it where the regular part $\overset{\circ}\to F'_s(y)$ of $F'_s(y)$ is defined as}
$$\overset{\circ}\to F'_s(y):=\cases F'_s(y),\qquad y\in\bR^+,\\
\stretch
F'_s(y),\qquad y\in\bR^-.\endcases \tag 16.52$$

\noindent PROOF: Since $F'_s(y)$ is integrable in $y\in\bR^+,$ we can conclude
that $F_s(y)$ is continuous in $y\in\bR^+$ and that $F_s(0^+)$ exists.
Since $F'_s(y)$ is the sum of an integrable function and a square-integrable
function in $y\in\bR^-,$ it is locally integrable in $y\in\bR^-$ and hence
$F_s(y)$ is continuous in $y\in\bR^-$ and that $F_s(0^-)$ exists.
Thus, we have proved (a) and the first part of (b). Let us now prove the rest of the proposition.
Since $F_s'(y)$ exists in $y\in\bR \setminus\{0\}$ and the limits
$F_s(0^\pm)$ exist, using integration by parts and interpreting the derivative in the
distribution sense, we conclude that
$$F_s'(y)=\left[F_s(0^+)-F_s(0^-)\right] \delta(y)+\overset{\circ}\to F'_s(y),
\tag 16.53$$
where $\overset{\circ}\to F'_s(y)$ is the quantity defined in (16.53).
It only remains to prove that (12.2) holds, i.e. the coefficient of the delta-distribution
in (16.53) is equal to $G_1.$ Let us define
$$H(k):=\int_{-\infty}^\infty dy\, \overset{\circ}\to F'_s(y)\, e^{-iky}.\tag 16.54$$
The quantity $H(k)$ exists because it corresponds to the Fourier transform
of the sum of an integrable function and a square-integrable function
in $y\in\bR.$ Let us use $H_\infty(k)$ to denote the Fourier transform of the integrable part
of $\overset{\circ}\to F'_s(y)$ and use $H_2(k)$ to denote
the Fourier transform of the square-integrable part
of $\overset{\circ}\to F'_s(y).$ Thus, we have
$$H(k)=H_\infty(k)+H_2(k),\qquad k\in\bR.\tag 16.55$$
We remark that $H_\infty(k)$ is bounded in $k\in\bR$ because
it is the Fourier transform of an integrable
function, and furthermore, with the help of the Riemann-Lebesgue lemma,
it follows that
$$H_\infty(k)=o(1),\qquad k\to\pm\infty.\tag 16.56$$
On the other hand, $H_2(k)$ itself is square integrable
in $k\in\bR$ because it is the Fourier transform of a square-integrable
function. Let us use $\hat F'_s(k)$ to denote the Fourier transform
of $F_s'(y),$ where we have defined
$$\hat F'_s(k):=\int_{-\infty}^\infty dy\, F'_s(y)\, e^{-iky}.\tag 16.57$$
 From (4.5) and (12.3) we see that $\hat F'_s(k)$ is given by
$$\hat F'_s(k)=G_1+H_3(k),\qquad k\in\bR,\tag 16.58$$
where, as seen from (4.5), we have $H_3(k)$
has the behavior of the product of $ik$ and $o(1/k)$ as
$k\to\pm\infty.$ Thus, we obtain
$$H_3(k)=o(1),\qquad k\to\pm\infty.\tag 16.59$$
 From (16.53)-(16.59) we obtain
$$F_s(0^+)-F_s(0^-)+H_\infty(k)+H_2(k)=G_1+H_3(k),\qquad k\in\bR.
\tag 16.60$$
Let us write (16.60) as
$$ F_s(0^+)-F_s(0^-)-G_1+H_s(k)=H_3(k)-H_\infty(k),\qquad k\in\bR.
\tag 16.61$$
With the help of (16.56) and (16.59) we see that the right-hand side of
(16.61) vanishes as $k\to\pm\infty.$
In general, the Fourier transform
of a square-integrable function does not vanish as $k\to\pm\infty$ and hence
we cannot conclude that $H_2(k)=o(1)$ as $k\to\pm\infty.$
On the other hand, since the right-hand side of
(16.61) has the limit zero as $k\to\pm\infty,$ we conclude that
$H_2(k)$ must have a limit as $k\to\pm\infty$ and, since
$H_2(k)$ itself is square integrable in $k\in\bR,$ that limit must be zero.
Thus, the left-hand side of (16.61) in the limit as $k\to\pm\infty$
yields
$$F_s(0^+)-F_s(0^-)-G_1=0,\tag 16.62$$
proving that (12.2) holds, and hence (16.62) asserts that (16.51) holds. \qed

The following result is needed in the proof of Proposition~18.1.

\noindent {\bf Proposition 16.8} {\it Consider a scattering data set $\bold S$
as in (4.2), which consists of
an $n\times n$ scattering matrix $S(k)$ for $k\in\bR,$ a set of $N$ distinct
 positive constants $\kappa_j,$ and a set of
$N$ constant $n\times n$ hermitian and nonnegative matrices
$M_j$ with respective positive ranks $m_j,$ where $N$ is a nonnegative integer.
Let $F_s(y)$ be the quantity defined in (4.7).
Assume that $\bold S$ satisfies
$(\bold I)$
of Definition~4.3. Furthermore, assume that
$F'_s(y)$ given in (12.3) is integrable in
$y\in\bR^+$ and is the sum of an integrable function
and a square integrable function in $y\in\bR^-.$ Then, we have the following:}
$$\int_{-\infty}^\infty dy\,F_s(y)\,F_s(y+x)=-F_s(-x)\,S_\infty-S_\infty\,F_s(x),
\qquad x\in\bR,\tag 16.63$$
$$\int_{-\infty}^\infty dy\,F_s(z+y)\,F_s(y+x)=-F_s(z-x)\,S_\infty-S_\infty\,F_s(x-z),
\qquad x,z\in\bR,\tag 16.64$$
$$\int_{-\infty}^\infty dy\,F'_s(y)\,F_s(y+x)=-F'_s(-x)\,S_\infty+S_\infty\,F'_s(x),
\qquad x\in\bR\setminus\{0\},\tag 16.65$$
$$\int_{-\infty}^\infty dy\,\overset{\circ}\to F'_s(y)\,F_s(y+x)=-\overset{\circ}\to F'_s(-x)\,S_\infty+S_\infty\,\overset{\circ}\to F'_s(x)-G_1\,F_s(x),
\qquad x\in\bR,\tag 16.66$$
{\it where $\overset{\circ}\to F'_s(y)$ is the quantity related
to $F_s'(y)$ as in (16.52) and $G_1$ is the constant matrix
appearing in (4.5).}

\noindent PROOF: Note that (16.63) is obtained from
(16.64) by setting $z=0$ there and hence
we can skip the proof of (16.63). Let us first prove (16.64).
If $(\bold 1)$ holds, then from Proposition~4.4 we can conclude that
$F_s(y)$ is square integrable in $y\in\bR.$ Hence, the integrand
in (16.64), being the product of two square-integrable functions,
is integrable, and as a result the integral
on the left-hand side of (16.64) exists.
Using (4.7) we evaluate the left-hand side of (16.64), and
 with the help of (11.37) we get
$$\int_{-\infty}^\infty dy\,F_s(z+y)\,F_s(y+x)=
\ds\frac{1}{2\pi}\int_{-\infty}^\infty dk\,\left[S(k)-S_\infty\right]
\left[S(-k)-S_\infty\right]e^{ik(z-x)}.\tag 16.67$$
 From (4.4) and (16.60) we have
$$\left[S(k)-S_\infty\right]
\left[S(-k)-S_\infty\right]=-\left[S(k)-S_\infty\right]S_\infty-
S_\infty \left[S(-k)-S_\infty\right],\qquad k\in\bR.\tag 16.68$$
Using (16.68) on the right-hand side of
(16.67), with the help of (4.7) we obtain (16.64). Let us now turn to
the proof of (16.65). With the help of (4.7), (11.37), and (12.3), we
evaluate the left-hand side of (16.65) as
$$\int_{-\infty}^\infty dy\,F'_s(y)\,F_s(y+x)=
\ds\frac{1}{2\pi}\int_{-\infty}^\infty dk\,ik \left[S(k)-S_\infty\right]
\left[S(-k)-S_\infty\right]e^{-ikx}.\tag 16.69$$
Using (16.68) on the right-hand side of (16.69),
with the help of (12.3), we simplify the right-hand side of (16.69) and establish
(16.65). As seen from (16.51), $F'_s(x)$ contains a Dirac-delta
distribution at $x=0,$ and hence
we need to exclude
$x=0$ from (16.65). Finally, (16.66) is obtained from
(16.66) with the help of (16.51) and the fact that
the integrand in (16.66) is integrable. The integrability
of the integrand follows from the fact that
$\overset{\circ}\to F'_s(y)$ is integrable in $y\in\bR^+$ and
is the sum of an integrable function and a square-integrable function
 in $y\in\bR^-$ and that $F_s(y)$ is bounded in $y\in\bR$ and
integrable in $y\in\bR^+$ and is square integrable
 in $y\in\bR^-,$ as asserted by Proposition~4.4. \qed

The next result shows that, if the scattering data set $\bold S$
given in (3.97) satisfies $(\bold 1)$ and $(\bold 4_a)$ of Definition~4.2, then
the boundary matrices appearing in (2.4)-(2.6) can be constructed
by solving (14.2).

\noindent {\bf Proposition 16.9} {\it Consider a scattering data set $\bold S$
as in (4.2), which consists of
an $n\times n$ scattering matrix $S(k)$ for $k\in\bR,$ a set of $N$ distinct
 positive constants $\kappa_j,$ and a set of
$N$ constant $n\times n$ hermitian and nonnegative matrices
$M_j$ with respective positive ranks $m_j,$ where $N$ is a nonnegative integer. Assume
$\bold S$ satisfies
$(\bold 1)$ and $(\bold 4_a)$ specified in Definition~4.2.
Let $A$ and $B$ be any solution
to (14.2) in such a way that
the rank of the $2n\times n$ matrix $\bm A \\ B\endbm$ is
equal to $n.$ Then:}

\item{(a)} {\it Such a solution exists.}

\item{(b)} {\it If $A$ and $B$ make up such a solution, then
$AT$ and $BT$ also constitute such a solution for any
$n\times n$ invertible matrix $T.$}

\item{(c)} {\it Such a solution consisting of $A$ and $B$
satisfies
(2.5) and (2.6).}

\item{(d)} {\it If $(A,B)$ and $(\tilde A,\tilde B)$ are any two solutions
to (14.2), then we must have $\tilde A=AT$ and $\tilde B=BT$
for some
$n\times n$ invertible matrix $T.$}

\noindent PROOF: Let us write (14.2) in the block matrix form as
$$\bm I-S_\infty&0\\
\stretch
S_\infty\,K(0,0)+K(0,0)\,S_\infty-G_1&I+S_\infty\endbm \bm A\\
\stretch B\endbm=\bm 0\\
\stretch 0\endbm.\tag 16.70$$
Note that the coefficient matrix in (16.70) is a block lower-triangular matrix.
By Proposition~16.4(a) we know that $S_\infty$ is hermitian
and hence
it can be diagonalized with the help of a unitary matrix.
The same unitary matrix diagonalizes the block
diagonal matrices $(I-S_\infty)$ and
$(I+S_\infty)$ in the coefficient matrix in (16.70).
Furthermore, by Proposition~16.4(a) we know that
each
eigenvalue of $S_\infty$ is
equal to $+1$ or $-1.$
Thus, with the help of the unitary matrix that diagonalizes
$S_\infty,$ we can transform
the coefficient matrix in (16.70) into a lower-triangular matrix
where exactly $n$ of the diagonal entries are zero and
the remaining $n$ diagonal entries are nonzero.
Thus, the nullity of the coefficient matrix in
(16.70) is exactly $n$ and hence
the general solution of (16.70) contains
$n$ linearly independent columns. Thus, (a) is proved.
 From (14.2) we see that (b) is valid for any invertible
 $n\times n$ matrix $T.$
Let us now turn to the proof of (c).
By Proposition~16.4 we know that
the three constant matrices $K(0,0),$ $S_\infty,$ and $G_1$
appearing in (14.2)
are all hermitian. From the first line of (14.2) we have
$S_\infty A=A,$ and hence
$$A^\dagger S_\infty=A^\dagger,\tag 16.71$$
where we have used $S_\infty^\dagger=S_\infty.$
Since $S_\infty,$ $G_1,$ and $K(0,0)$ are all hermitian,
 from the second line of (14.2) we obtain
$$B^\dagger(I+S_\infty)=A^\dagger [G_1-K(0,0)\,S_\infty-S_\infty\,K(0,0)].
\tag 16.72$$
By multiplying (16.72) on the right by $A$ we obtain
$$B^\dagger(I+S_\infty)A=A^\dagger [G_1-K(0,0)\,S_\infty-S_\infty\,K(0,0)]A.
\tag 16.73$$
Using (16.72) on the right-hand side of (16.73) we get
$$B^\dagger(I+S_\infty)A=A^\dagger (I+S_\infty)\,B.
\tag 16.74$$
Using $S_\infty A=A$ and $A^\dagger S_\infty=A^\dagger,$
respectively, in (16.74) we simplify (16.74) and obtain
$$2B^\dagger A=2 A^\dagger B,\tag 16.75$$
which is equivalent to (2.5).
Next, let us show that (2.6) is satisfied.
Note that we can write the left-hand side of (2.6) as
$$A^\dagger A+B^\dagger B=\bm A^\dagger &B^\dagger \endbm \bm A\\
B\endbm.\tag 16.76$$
We need to show that the matrix product appearing
on the right-hand side of (16.76) is positive. From
the left-hand side of (16.76) we see that
that product is itself hermitian and hence
it has $n$ linearly independent
eigenvectors with all real eigenvalues.
To prove the positivity it is enough to prove that
none of the eigenvalues can be negative or zero.
Let $v$ be an eigenvector of $(A^\dagger \,A+B^\dagger\,B)$
with the corresponding eigenvalue $\lambda.$ Then, from (16.76)
we get
$$\lambda\langle v,v\rangle
=\langle v,(A^\dagger \,A+B^\dagger\,B)\,v\rangle=
\langle \bm A\\
B\endbm v,\bm A\\
B\endbm v\rangle\ge 0,\tag 16.77$$
and hence $\lambda\ge 0.$ On the other hand, we
remark that $\lambda$ cannot be zero
because that would imply that
$\bm A\\
B\endbm v=0$ and hence the kernel of the matrix
$\bm A\\
B\endbm$ would contain a nonzero vector.
That would mean that the nullity of
the matrix would be at least one.
Since the nullity and the rank must add up to
$n,$ the rank of $\bm A\\
B\endbm$ would have to be strictly less than $n,$
contradicting the fact that
the rank of $\bm A\\
B\endbm$ is equal to $n.$
Thus, the proof of (c) is complete.
%
 Let us now prove (d). Let us use
 $\bm A^{(j)}\\
 B^{(j)}\endbm$ and $\bm \tilde A^{(j)}\\
 \tilde B^{(j)}\endbm$ to denote the $j$th column of
 the $2n\times n$ matrices $\bm A\\
 B\endbm$ and $\bm \tilde A\\
 \tilde B\endbm,$ respectively.
 From the proof of (a) we know that the general solution
 to (16.70) contains exactly $n$ linearly independent
 column vector solutions. The $n$ columns of
$\bm A\\
 B\endbm$ must be linearly independent because
 the rank of $\bm A\\
 B\endbm$ is $n.$ Similarly, the $n$ columns of
$\bm \tilde A\\
\tilde B\endbm$ must be linearly independent because
 the rank of $\bm \tilde A\\
\tilde B\endbm$ is $n.$ Thus, we can express
$\bm \tilde A^{(j)}\\
 \tilde B^{(j)}\endbm$ as a linear combination of
 columns of $\bm A\\
 B\endbm$ as
 $$\bm \tilde A^{(j)}\\
 \tilde B^{(j)}\endbm =\sum_{l=1}^n \bm A^{(l)}\\
 B^{(l)}\endbm T_{lj},\qquad j=1,\dots,n,\tag 16.78$$
for some coefficients $ T_{lj}.$ Thus, (16.78) implies that
$$\bm \tilde A\\
 \tilde B\endbm =\bm A\\
 B\endbm T,\tag 16.79$$
where $T$ is the $n\times n$ matrix whose $(l,j)$-entry is equal to
$T_{lj}.$ A similar argument implies that
$$\bm A\\
B\endbm =\bm \tilde A\\
\tilde  B\endbm\tilde T,\qquad j=1,\dots,n,\tag 16.80$$
 for some $n\times n$ matrix $\tilde T.$ From (16.79) and (16.80) we obtain
$$\bm \tilde A\\
 \tilde B\endbm = \bm \tilde A\\
 \tilde B\endbm \tilde T\, T.\tag 16.81$$
Let us multiply (16.81) by
$\bm \tilde A^\dagger &\tilde B^\dagger\endbm,$ which yields
$$\bm \tilde A^\dagger &\tilde B^\dagger\endbm
\bm \tilde A\\
 \tilde B\endbm (\tilde T\, T-I)=0,\tag 16.82$$
 or equivalently
$$( \tilde A^\dagger \tilde A+
 \tilde B^\dagger \tilde B) (\tilde T\, T-I)=0,\tag 16.83$$
  From the proof of (c),
   we know that the matrix $( \tilde A^\dagger \tilde A+
 \tilde B^\dagger \tilde B)$ has rank $n$ and hence is
 invertible. Thus, (16.83) implies that $\tilde T\, T=I,$ and hence
 $T$ and $\tilde T$ are both invertible and are inverses of each other.
 Then, from (16.79) we conclude that
 $\tilde A=A T$ and $\tilde B=BT,$ completing the proof of (d). \qed

As indicated in Proposition~10.1(b), $K(x,x)$ denotes $K(x,x^+),$
where $K(x,y)$ is the solution to the Marchenko equation (13.1). In the next proposition
we analyze the $x$-derivative of
$K(x,x).$

\noindent {\bf Proposition 16.10} {\it Consider a scattering data set $\bold S$
as in (4.2), which consists of
an $n\times n$ scattering matrix $S(k)$ for $k\in\bR,$ a set of $N$ distinct
 positive constants $\kappa_j,$ and a set of
$N$ constant $n\times n$ hermitian and nonnegative matrices
$M_j$ with respective positive ranks $m_j,$ where $N$ is a nonnegative integer. Let $F_s(y)$ and $F(y)$ be the quantities defined
in (4.7) and (4.12), respectively,
where $S_\infty$ is the constant $n\times n$ matrix
defined as in (4.4). Let $K(x,y)$ be the
quantity defined in (10.1). Assume that $\bold S$ satisfies
$(\bold I)$ of Definition~4.3
as well as $(\bold 2)$ and $(\bold 4_a)$ of Definition~4.2. Then, we have}
$$\bigg|\ds\frac{d K(x,x)}{dx}+2\,F'(2x)\bigg|\le c\,
[\tau(x)]^2,\qquad x>0,
\tag 16.84$$
{\it where $c$ is some constant and
$\tau(x)$ is the quantity defined in (16.7). Furthermore, we have}
$$\ds\int_0^\infty dx\,(1+x)\,\bigg|\ds\frac{d K(x,x)}{dx}\bigg|<+\infty.
\tag 16.85$$

\noindent PROOF: The proof of (16.84) can be found in the proof of Lemma~5.3.4
on p. 115
of [2] and in the remark on p. 116 in [2]. Let us now prove (16.85).
Using the triangle inequality, we have
$$\bigg|\ds\frac{d K(x,x)}{dx}\bigg|\le
\bigg|\ds\frac{d K(x,x)}{dx}+2\,F'(2x)\bigg|+2|F'(2x)|.\tag 16.86$$
Multiplying both sides of (16.86) with
$(1+x)$ and integrating over $x\in\bR^+,$ with the help of
(16.84) we obtain
$$\ds\int_0^\infty dx\,(1+x)\,\bigg|\ds\frac{d K(x,x)}{dx}\bigg|\le
c\int_0^\infty dx\,(1+x)\,[\tau(x)]^2+2
\int_0^\infty dx\,(1+x)\,|F'(2x)|
\tag 16.87$$
Because we assume $(\bold 2)$, from (4.9) we see that $\tau(0)$ and
$\tau_1(0)$ are both finite, where $\tau(x)$ and $\tau_1(x)$ are the
quantities defined in (16.7).
Then, we see that (16.9) implies that the first integral
on the right-hand side of (16.87) is finite. Furthermore,
(4.8) implies that the second integral
on the right-hand side of
(16.87) is also finite. Thus, (16.85) holds. \qed

In the next proposition, we study the relevant properties of the
potential and the
Jost solution constructed from the scattering data set $\bold S$
given in (3.97).

\noindent {\bf Proposition 16.11} {\it Consider a scattering data set $\bold S$
as in (4.2), which consists of
an $n\times n$ scattering matrix $S(k)$ for $k\in\bR,$ a set of $N$ distinct
 positive constants $\kappa_j,$ and a set of
$N$ constant $n\times n$ hermitian and nonnegative matrices
$M_j$ with respective positive ranks $m_j,$ where $N$ is a nonnegative integer.
Let $F_s(y)$ and $F(y)$ be the quantities defined
in (4.7) and (4.12), respectively,
where $S_\infty$ is the constant $n\times n$ matrix
defined as in (4.6). Assume that $\bold S$ satisfies
$(\bold I)$ of Definition~4.3
as well as $(\bold 2)$ and $(\bold 4_a)$ of Definition~4.2. Let $K(x,y)$ be the
unique solution to
(13.1), whose existence and uniqueness
are assured by Proposition~16.1(b). Then:}

\item{(a)} {\it The Jost solution $f(k,x)$ constructed as in
(10.6) satisfies (2.1) and (9.1) when the potential $V(x)$
appearing in (2.1) is constructed as in (10.4). Furthermore, the potential $V(x)$
constructed via (10.4)
satisfies (2.2) and (2.3).}

\item{(b)} {\it The physical solution $\Psi(k,x)$ constructed
as in (9.4) satisfies (2.1) when the potential $V(x)$
appearing in (2.1) is constructed as in (10.4).}

\item{(c)} {\it The normalized bound-state matrix solution $\Psi_j(x)$
constructed as in (9.8) satisfies (2.1) with
$k=i\kappa_j$ when the potential $V(x)$
appearing in (2.1) is constructed as in (10.4).}

\noindent PROOF:
The part of the proof of (a) related to
the satisfaction of (2.1) can be found
in Theorem~5.4.1 on p. 117 of [2] and in the Remark on p. 121 of [2],
by recalling that the integrability of the constructed potential
$V(x)$ in our case is assured by the integrability of
$F'(y)$ stated in $(\bold 2)$, as indicated in Proposition~16.10.
The fact that the constructed $f(k,x)$ satisfies
(9.1) follows from (10.6) by using
the fact that the solution $K(x,y)$
to the Marchenko equation (13.1) is integrable in $y\in[x,+\infty),$
as indicated in Proposition~16.1(b).
Hence, the proof of (e) is complete. Note that (9.4) indicates
that each column of the physical solution $\Psi(k,x)$ is a linear
combination of the columns of $f(-k,x)$ and of $f(k,x).$
By (a) we know that each column of $f(k,x)$ satisfies
(2.1), but then we conclude that each column of $f(-k,x)$ also
satisfies (2.1) because $k$ appears as $k^2$ in (2.1). Thus,
 from (9.4) it follows that (b) holds. From (9.8) we see that each column of $\Psi_j(x)$
is a linear combination of the columns of
$f(i\kappa_j,x),$ and by (a) we know that $f(k,x)$ satisfies
(2.1). Thus, the result stated in (c) holds. \qed

\newpage
\noindent {\bf 17. ADDITIONAL RESULTS RELATED TO THE INVERSE PROBLEM}
\vskip 3 pt

In this chapter we present certain auxiliary results needed in later chapters.

The next proposition presents certain properties of
the Jost solution $f(k,x),$ the
physical solution $\Psi(k,x),$ and the bound-state matrix
solutions $\Psi_j(x).$

\noindent {\bf Proposition 17.1} {\it Consider a scattering data set $\bold S$
as in (4.2), which consists of
an $n\times n$ scattering matrix $S(k)$ for $k\in\bR,$ a set of $N$ distinct
 positive constants $\kappa_j,$ and a set of
$N$ constant $n\times n$ hermitian and nonnegative matrices
$M_j$ with respective positive ranks $m_j,$ where $N$ is a nonnegative integer.
Assume that $\bold S$ satisfies $(\bold I)$
of Definition~4.3 and $(\bold 4_a)$ of Definition~4.2.
We then have the following:}

\item{(a)} {\it The Jost solution $f(k,x)$ constructed from
the scattering data as in (10.6) is uniformly bounded, i.e.}
$$|f(k,x)|\le C,\qquad k\in\bR,\quad x\ge 0,\tag 17.1$$
{\it for some constant $C$ independent of $k$ and $x.$}

\item{(b)} {\it The physical solution $\Psi(k,x)$ constructed from
the scattering data as in (9.4) is uniformly bounded, i.e.}
$$|\Psi(k,x)|\le C,\qquad k\in\bR,\quad x\ge 0,\tag 17.2$$
{\it for some constant $C$ independent of $k$ and $x.$}

\item{(c)} {\it The matrix $f(i\kappa_j,x)$ constructed from
the scattering data as in (10.6) is uniformly bounded, integrable,
and square integrable for $x\in[0,+\infty),$ and it satisfies}
$$|f(i\kappa_j,x)|\le C\, e^{-\kappa_j x},\quad x\ge 0,
\quad j=1,\dots,N.\tag 17.3$$

\item{(d)} {\it Each bound-state matrix solution $\Psi_j(x)$ constructed from
the scattering data as in (9.8) is uniformly bounded, integrable,
and square integrable for $x\in[0,+\infty),$ and it satisfies}
$$|\Psi_j(x)|\le C\, e^{-\kappa_j x},\quad x\ge 0,
\quad j=1,\dots,N.\tag 17.4$$

\noindent PROOF: Let us use $C$ for a generic
constant that may take different values in different appearances.
 From (10.6), we obtain
$$|f(k,x)|\le 1+\int_x^\infty dy\, |K(x,y)|,\tag 17.5$$
where $K(x,y)$ is the solution to the Marchenko equation.
 From Proposition~16.1(b) we know that $K(x,y)$
is integrable in $y\in(x,+\infty)$ for each fixed $x\ge 0.$
Thus, the integral in (17.5) is finite, and hence it (17.5)
yields (17.1).
%
%
%
%
Thus, (a) holds.
Let us now prove (b). From
(9.4) we get
$$|\Psi(k,x)|\le |f(-k,x)|+|f(k,x)|\,|S(k)|.\tag 17.6$$
Because $S(k)$ is unitary, as assumed in
$(\bold I),$ we have
$|S(k)|=1$ for $k\in\bR.$
Thus, using (17.1) in (17.6), we conclude (17.2), and hence
the proof of (b) is complete. Let us now turn to the proof of (c).
 From (10.6) we get
$$f(i\kappa_j,x)=e^{-\kappa_j x}I+\int_x^\infty dy\,K(x,y)\,e^{-\kappa_j y}.
\tag 17.7$$
 From (17.7) we obtain
$$|f(i\kappa_j,x)|\le  e^{-\kappa_j x}+e^{-\kappa_j x}\int_x^\infty dy\,
|K(x,y)|.\tag 17.8$$
Again, as a result of the integrability of $K(x,y)$
in $y\in(x,+\infty)$ for each fixed $x\ge 0,$
the integral in (17.8) is finite and hence
(17.8) yields (17.3).

Because each $\kappa_j$ is positive,
the right-hand side in (17.3)
is bounded, integrable, and square integrable
in $x\in[0,+\infty).$ Thus, we conclude that
$|f(i\kappa_j,x)|$ has the properties stated in (c).
Let us now prove (d). From (9.8), since each $M_j$
is a constant $n\times n$ matrix, we see that (17.3) implies (17.4)
and hence (c) implies (d). \qed

The next proposition is the analog of
Proposition~17.1 and presents certain properties of
the Jost solution $f'(k,x),$ the
physical solution $\Psi'(k,x),$ and the bound-state matrix
solutions $\Psi'_j(x).$

\noindent {\bf Proposition 17.2} {\it
Consider a scattering data set $\bold S$
as in (4.2), which consists of
an $n\times n$ scattering matrix $S(k)$ for $k\in\bR,$ a set of $N$ distinct
 positive constants $\kappa_j,$ and a set of
$N$ constant $n\times n$ hermitian and nonnegative matrices
$M_j$ with respective positive ranks $m_j,$ where $N$ is a nonnegative integer.
Assume that $\bold S$ satisfies $(\bold I)$
 of Definition~4.3 and $(\bold 4_a)$ of Definition~4.2. Assume also
that $F_s'(y)$ is integrable
in $y\in\bR^+,$ where $F_s(y)$ is the quantity
defined in (4.7).
We then have the following:}

\item{(a)} {\it The $x$-derivative of the Jost solution
$f(k,x)$ constructed as in
(10.6), i.e. the matrix $f'(k,x)$
satisfies}
$$|f'(k,x)|\le C\,(1+|k|),\qquad k\in\bR,\quad x\ge 0,\tag 17.9$$
{\it for some constant $C$ independent of $k$ and $x.$}

\item{(b)} {\it The matrix $\Psi'(k,x),$ i.e.
the $x$-derivative of the physical solution $\Psi(k,x)$ constructed from
the scattering data as in (9.4), satisfies}
$$|\Psi'(k,x)|\le C\,(1+|k|),\qquad k\in\bR,\quad x\ge 0,\tag 17.10$$
{\it for some constant $C$ independent of $k$ and $x.$}

\item{(c)} {\it Each matrix
$f'(i\kappa_j,x),$ i.e. the $x$-derivative of
the Jost solution $f(i\kappa_j,x)$ constructed from
the scattering data as in (10.6), is uniformly bounded, integrable,
and square integrable for $x\in[0,+\infty),$ and it satisfies}
$$|f'(i\kappa_j,x)|\le C\, e^{-\kappa_j x},\qquad x\ge 0,
\quad j=1,\dots,N.\tag 17.11$$

\item{(d)} {\it Each matrix
$\Psi_j'(x),$ i.e. the $x$-derivative of
 the bound-state matrix solution $\Psi_j(x)$ constructed from
the scattering data as in (9.8), is uniformly bounded, integrable,
and square integrable for $x\in[0,+\infty),$ and it satisfies}
$$|\Psi'_j(x)|\le C\, e^{-\kappa_j x},\qquad x\ge 0,
\quad j=1,\dots,N.\tag 17.12$$

\noindent PROOF: We again use $C$ to denote a generic
constant. Recall that $f'(k,x)$ is constructed from
the solution $K(x,y)$ to the Marchenko equation (13.1) and
its $x$-derivative given by $K_x(x,y).$ This is accomplished
by using $K(x,y)$ and $K_x(x,y)$ in
(15.37).
We obtain
$$|f'(k,x)|\le |k|+|K(x,x)|+\int_x^\infty dy\,|K_x(x,y)|,
\qquad x\ge 0.\tag 17.13$$
By Proposition~16.1(b) we know that
$K(x,y)$ is bounded when $0\le x\le y$ and hence
$|K(x,x)|\le C$ for $x\ge 0.$
By Proposition~16.5(c) and 16.5(d) we know
that $K_x(k,y)$ is integrable in $y\in(x,+\infty)$ for each $x\ge 0,$
and hence the integral in (17.13) is finite.
Thus, (17.13) yields (17.9), proving (a).
Let us turn to the proof of (b).
 From (9.4), we have
$$\Psi'(k,x)=f'(-k,x)+f'(k,x)\,S(k).\tag 17.14$$
and hence we obtain
$$|\Psi'(k,x)|\le |f'(-k,x)|+|f'(k,x)|\,|S(k)|,
\tag 17.15$$
Because $S(k)$ is unitary, we have $|S(k)|=1$ for $k\in\bR.$ Hence,
using (17.9) in (17.15) we obtain (17.10), proving (b).
Let us now turn to the proof of (c). From (15.37) we have
$$f'(i\kappa_j,x)=-\kappa_j\, e^{-\kappa_j x}I-K(x,x)\,e^{-\kappa_j x}+
\int_x^\infty K_x(x,y)\,e^{-\kappa_j y},
\qquad x\ge 0.
\tag 17.16$$
 From (17.16) we get
$$|f' (i\kappa_j,x)|\le \kappa_j\, e^{-\kappa_j x}+|K(x,x)|\,e^{-\kappa_j x}+
\int_x^\infty dy\,|K_x(x,y)|\,e^{-\kappa_j y},
\qquad x\ge 0,
\tag 17.17$$
which implies
$$|f' (i\kappa_j,x)|\le  e^{-\kappa_j x}
\left[ \kappa_j+|K(x,x)|+
\int_x^\infty dy\,|K_x(x,y)|\right],
\qquad x\ge 0,
\tag 17.18$$
As argued earlier, $|K(x,x)|\le C$ for every $x\ge 0$ and
$K_x(x,y)$ is integrable in $y\in(x,+\infty)$ for every $x\ge 0.$
Thus, (17.18) yields (17.11).
Since $\kappa_j$ is positive, from (17.11) we conclude
the properties stated in (c). Thus, the proof of (c) is completed.
%
%
%
We note that (d) directly follows from (c) because
as seen from (9.8), the matrices $\Psi'_j(x)$ and
$f'(i\kappa_j,x)$ are related to each other by the
constant matrix $M_j$ via
$$\Psi'_j(x)=f'(i\kappa_j,x)\,M_j.\tag 17.19$$
Thus, the proof is complete. \qed

The properties established in Proposition~16.1(b) for
the solution
$K(x,y)$ to the
Marchenko integral equation
enable us to define certain useful integral operators
related to $K(x,y).$ Toward that goal,
we first need the following result.

\noindent {\bf Proposition 17.3} {\it Consider a scattering data set $\bold S$
as in (4.2), which consists of
an $n\times n$ scattering matrix $S(k)$ for $k\in\bR,$ a set of $N$ distinct
 positive constants $\kappa_j,$ and a set of
$N$ constant $n\times n$ hermitian and nonnegative matrices
$M_j$ with respective positive ranks $m_j,$ where $N$ is a nonnegative integer.
Assume that $\bold S$ satisfies $(\bold I)$ of Definition~4.3 as well as
$(\bold 2)$ and $(\bold 4_a)$ of Definition~4.2.
Let $K(x,y)$ be the solution to the Marchenko equation (13.1), and
consider the integral equation}
$$M(x,y)+K(x,y)+\int_x^y dz\,M(x,z)\,K(z,y)=0,\qquad 0\le x< y,\tag 17.20$$
{where $M(x,y)$ is an $n\times n$ matrix. Then:}

\item{(a)} {\it The integral equation (17.20) is uniquely
solvable for $M(x,y).$}

\item{(b)} {\it The solution $M(x,y)$ to
(17.20) satisfies}
$$|M(x,y)|\le C\,\tau(x+y)\,
,\qquad 0\le x<y,\tag 17.21$$
{\it where $C$ is a generic constant and $\tau(x)$ is the scalar quantity
defined in (16.7).}

\item{(c)} {\it For each fixed $x\ge 0,$ the entries of the
matrix $M(x,y)$
belong to
$L^1(x<y<+\infty)$ and $L^\infty(x<y<+\infty),$
and hence they in particular belong to
$L^2(x<y<+\infty).$}

\item{(d)} {\it The integral equation associated with (17.21), which is
given by}
$$\tilde M(x,y)+K(x,y)+\int_x^y dz\,K(x,z)\,\tilde
M(z,y)=0,\qquad 0\le x< y,\tag 17.22$$
{\it is also uniquely solvable for $\tilde M(x,y).$
In fact, the unique
solution $\tilde M(x,y)$ to (17.22) satisfies}
$$\tilde M(x,y)\equiv M(x,y),\tag 17.23$$
{\it where $M(x,y)$ is the unique solution to (17.20).}

\noindent PROOF:
The Volterra equation (17.20) can be solved by using the method of
successive approximation as
$$M(x,y)=\sum_{j=0}^\infty M^{(j)}(x,y),\tag 17.24$$
where we have defined
$$M^{(0)}(x,y):=-K(x,y),\tag 17.25$$
$$M^{(j)}(x,y):=-\int_x^y dz\,M^{(j-1)}(x,z)\,K(z,y),\qquad j=1,2,\dots.
\tag 17.26$$
 From the
recurrence relations (17.25) and (17.26) we get
 $$\aligned
 M^{(j)}&(x,y)\\
 = &(-1)^{(j+1)}\int_x^y\,  dz_j\, \int_x^{z_j}\, d z_{j-1}\,  \cdots \int_x^{z_2} dz_1\, K(x,z_1)\, K(z_1, z_2)\,\cdots \, K(z_{j-1},z_j)\, K(z_j,y).
 \endaligned
\tag 17.27$$
Since $K(x,y)=0$ for $ x >y,$ we can write (17.27) as
 $$\aligned
M^{(j)}&(x,y)\\
=& (-1)^{(j+1)}\int_x^y\,  dz_j\, \int_x^{y}\, d z_{j-1}\,  \cdots \int_x^{y} dz_1\, K(x,z_1)\, K(z_1, z_2)\,\cdots \, K(z_{j-1},z_j)\, K(z_j,y).
\endaligned\tag 17.28
$$
With the help of (16.10), from (17.25) and (17.26)
we get
$$|M^{(0)}(x,y)|\le C \,\tau(x+y),\quad
|M^{(1)}(x,y)|\le  C\,\tau(x+y)\int_x^y
dz\,C\,\tau(x+z).\tag 17.29
$$
Using induction, from (17.29) we obtain
$$|M^{(j)}(x,y)|\le \ds\frac{1}{j!}\, C\,\tau(x+y)\left[\int_x^y
dz\,C\,\tau(x+z)\right]^j,\qquad j=1,2,\dots.\tag 17.30$$
Hence, the series in (17.24) converges uniformly and
yields $M(x,y)$ as the unique solution to (17.20).
Thus, the proof of (a) is complete.
Using (17.30) in (17.29), for $0\le x\le y$
we obtain
$$\aligned
|M(x,y)|\le &C\,\tau(x+y)\,
\exp\left(C\int_x^y dz\,\tau(x+z)\right)\\
\le
& C\,\tau(x+y)\,
\exp\left(C\int_x^\infty dz\,\tau(x+z)\right)\\
\le &
C\,\tau(x+y)\,
e^{C\,\tau_1(0)}
.\endaligned\tag 17.31$$
The property $(\bold 2)$ implies that $\tau_1(0)$ is finite.
Hence, (17.31) implies (17.21) with
some generic constant $C$ not necessarily
equal to $C$ appearing in (17.31). Thus, (b) is proved.
Let us now turn to the proof of (c).
Since we assume that
$(\bold 2)$ of Definition~4.2 holds, (16.8) implies that
$\tau(0)$ is also finite.
Then, using the second inequality in (16.8) and the fact that
$\tau(x)$ is a nonincreasing function of $x,$ we
see from (17.21) that $|M(x,y)|$ belongs to
$L^1(x<y<+\infty)$ and $L^\infty(x<y<+\infty)$ for each fixed $x\ge 0.$
Any function that belongs to $L^1$ and $L^\infty$ must also belong to
$L^p(x<y<+\infty)$ for $1<p<+\infty,$ and hence
$|M(x,y)|$ belongs to
$L^2(x<y<+\infty)$ as well. These properties satisfied by the
matrix norm $|M(x,y)|$ imply that
each entry of the matrix
also satisfy such properties related to
$L^1,$ $L^\infty,$ and $L^2.$ Thus, the proof of (c) is complete.
Let us now turn to the proof of
(d). We can solve (17.22) the same way we solve (17.20),
As in (17.24.)-(17.28),
we solve (17.22) by the method of successive approximations as
$$
 \tilde M(x,y)= \sum_{j=0}^\infty \, \tilde M^{(j)}(x,y),\tag 17.32
 $$
 with
 $$\tilde M^{(0)}(x,y):= -K(x,y),\tag 17.33$$
 $$
 \tilde{M}^{(j)}(x,y)= - \int_x^y\, dz \, K(x,z) \, \tilde{M}^{(j-1)}(z,y),
 \qquad j=1,2,\dots .\tag 17.34$$
The recurrence relations (17.32) and (17.33) yield
 $$\aligned
  \tilde{M}^{(j)}&(x,y)\\
  = &(-1)^{j+1} \int_x^y dz_1 \int_{z_1}^y  dz_2\cdots
  \int_{z_{j-1}}^y dz_j\, K(x,z_1)\, K(z_1,z_2)\,\cdots\,
  K(z_{j-1}, z_{j})\, K(z_j,y).\endaligned\tag 17.35
 $$
Since $K(x,y)=0$ for $ x >y,$ as in (17.27) and (17.28), from (17.35) we obtain
 $$\aligned
  \tilde{M}^{(j)}&(x,y)\\
  =& (-1)^{j+1} \int_x^y dz_1 \int_{x}^y  dz_2\,\cdots\,
  \int_{x}^y dz_j\, K(x,z_1)\, K(z_1,z_2)\,\cdots\,
  K(z_{j-1}, z_{j})\, K(z_j,y).
  \endaligned\tag 17.36
 $$
 Comparing the pair of equations (17.25) and (17.28) with
 the pair of equations (17.33) and (17.36), we see that
$$M^{(j)}(x,y)=  \tilde{M}^{(j)}(x,y).\qquad  j=0,1,\dots,\tag 17.37$$
Since the iterates in (17.20) and (17.22) coincide,
 from the unique solvability of (17.20) proved in (a)
 we conclude the unique solvability of
 (17.22). Furthermore, by comparing
 (17.24) and (17.32), from (17.32)
we also conclude (17.23).
 \qed

Associated with the unique solution $K(x,y)$ to the Marchenko equation
(13.1), let us define the operators
$\bold K$ and $\bold P$ as
$$\bold K: Y(x)\mapsto \int_x^\infty dy\,K(x,y)\,Y(y),\tag 17.38$$
$$\bold P: Y(x)\mapsto \int^x_0 dy\,K(y,x)^\dagger\,Y(y).\tag 17.39$$
Associated with the unique solution $M(x,y)$ to
(17.20) and also to (17.23), let us define
the operators $\bold M$ and $\bold Q$ as
$$\bold M: Y(x)\mapsto \int_x^\infty  dy\,M(x,y)\,Y(y),\tag 17.40$$
$$\bold Q: Y(x)\mapsto \int^x_0 dy\,M(y,x)^\dagger\,Y(y).\tag 17.41$$
The next proposition presents certain properties of the four operators
in (17.39)-(17.41) and some relationships
among them.

\noindent {\bf Proposition 17.4} {\it The four operators $\bold K,$ $\bold M,$ $\bold P,$ $\bold Q$ defined in (17.38)-(17.41), respectively, are related to each other as}
$$(I+\bold M)(I+\bold K)=I,\tag 17.42$$
$$(I+\bold K)(I+\bold M)=I.\tag 17.43$$
$$(I+\bold Q)(I+\bold P)=I,\tag 17.44$$
$$(I+\bold P)(I+\bold Q)=I.\tag 17.45$$
{\it The operator equations (17.42)-(17.45) hold on
$L^1(\bR^+).$ They also hold on $L^2(\bR^+).$}

\noindent PROOF: By postmultiplying (17.20) by $Y(y)$ and integrating the resulting
equation in $y\in(x,+\infty),$ we obtain
$$\int_x^\infty dy\,M(x,y)\,Y(y)+\int_x^\infty dy\,K(x,y)\,Y(y)
+\int_x^\infty dy\int_x^y dz\,M(x,z)\,K(z,y)\,Y(y)=0.\tag 17.46$$
By interchanging the order of integration in the double integral in
(17.46) we obtain
$$\int_x^\infty dy\,M(x,y)\,Y(y)+\int_x^\infty dy\,K(x,y)\,Y(y)
+\int_x^\infty dz\int_z^\infty dy\,M(x,z)\,K(z,y)\,Y(y)=0, \tag 17.47$$
which can be written as
$$(\bold M Y)(x)+(\bold K Y)(x)+(\bold M\,\bold K\, Y)(x)=0,\tag 17.48$$
yielding
$$\bold M+\bold K+\bold M\bold K=0,\tag 17.49$$
which is equivalent to (17.42).
In a similar way, since $M(x,y)$ solves (17.22),
we postmultiply (17.22), with $\tilde M(x,y)$ replaced with
$M(x,y)$ there, by $Y(y)$ and integrate the resulting
equation in $y\in(x,+\infty).$ By changing the order of integration
in the term involving the double integral, we get
$$(\bold M Y)(x)+(\bold K Y)(x)+(\bold K\bold M Y)(x)=0,\tag 17.50$$
which yields (17.43).
Note that, by taking the matrix adjoint of both sides of (17.20), we obtain
$$M(x,y)^\dagger+K(x,y)^\dagger+\int_x^y dz\,K(z,y)^\dagger\,M(x,z)^\dagger=0, \qquad 0\le x<y.\tag 17.51$$
Let us postmultiply (17.51) by $Y(x)$ and integrate
the resulting equation in $x\in(0,y).$ We then get
$$\int_0^y dx\, M(x,y)^\dagger\, Y(x)+\int_0^y dx\,K(x,y)^\dagger\, Y(x)+
\int_0^y dx\int_x^y dz\,K(z,y)^\dagger\,M(x,z)^\dagger\,Y(x)=0.\tag 17.52$$
By interchanging the order of integration
in the double integral in (17.52) we get
$$\int_0^y dx\, M(x,y)^\dagger\, Y(x)+\int_0^y dx\,K(x,y)^\dagger\, Y(x)+
\int_0^y dz\int_0^z dx\,K(z,y)^\dagger\,M(x,z)^\dagger\,Y(x)=0,\tag 17.53$$
yielding
$$(\bold Q Y)(y)+(\bold P Y)(y)+ ( \bold P\,\bold Q \,   Y)(y)=0,\tag 17.54$$
which is equivalent to (17.45).
In a similar way, since $M(x,y)$ solves (17.22),
we take the matrix adjoint of (17.22), with $\tilde M(x,y)$ replaced with
$M(x,y)$ there, and obtain
$$M(x,y)^\dagger+K(x,y)^\dagger+\int_x^y dz\,M(z,y)^\dagger\,K(x,z)^\dagger=0, \qquad 0\le x<y.\tag 17.55$$
By postmultiplying (17.55) with $Y(x)$ and integrating the resulting equation
in $x\in(0,y),$ after changing the order of integration in the term containing the
double integral, we obtain
$$\int_0^y dx\, M(x,y)^\dagger\, Y(x)+\int_0^y dx\,K(x,y)^\dagger\, Y(x)+
\int_0^y dz\int_0^z dx\,M(z,y)^\dagger\,K(x,z)^\dagger\,Y(x)=0.\tag 17.56$$
We recognize that (17.56) is equivalent to
$$(\bold Q Y)(y)+(\bold P\, Y)(y)+ ( \bold Q\,   \bold P\, Y)(y)=0,\tag 17.57$$
which yields (17.44). We remark that the change of the order of integration in the double
integrals is justified with the help of
(16.11), (17.21), and the analogous inequalities for
$K(x,y)^\dagger$ and $M(x,y)^\dagger.$ \qed

Some further properties of the four operators
$\bold K,$ $\bold P,$ $\bold M,$ and $\bold Q$
defined in (17.38)-(17.41), respectively, are
presented in the next proposition.

\noindent {\bf Proposition 17.5} {\it Consider a scattering data set $\bold S$
as in (4.2), which consists of
an $n\times n$ scattering matrix $S(k)$ for $k\in\bR,$ a set of $N$ distinct
 positive constants $\kappa_j,$ and a set of
$N$ constant $n\times n$ hermitian and nonnegative matrices
$M_j$ with respective positive ranks $m_j,$ where $N$ is a nonnegative integer.
Assume that $\bold S$ satisfies $(\bold I)$ of Definition~4.3 as well as
$(\bold 2)$ and $(\bold 4_a)$ of Definition~4.2.
Then, we have the following:}

\item{(a)} {\it Each of the four operators
$\bold K,$ $\bold P,$ $\bold M,$ and $\bold Q$
defined in (17.38)-(17.41), respectively,
is a bounded operator from $L^1(\bR^+)$ into $L^1(\bR^+).$}

\item{(b)} {\it Each of the four operators
$\bold K,$ $\bold P,$ $\bold M,$ and $\bold Q$
is a bounded operator from $L^2(\bR^+)$ into $L^2(\bR^+).$}

\item{(c)} {\it Each of the four operators
$(I+\bold K),$ $(I+\bold P),$ $(I+\bold M),$ and $(I+\bold Q)$
is invertible on $L^1(\bR^+).$ The corresponding inverses
$(I+\bold K)^{-1},$ $(I+\bold P)^{-1},$ $(I+\bold M)^{-1},$ and $(I+\bold Q)^{-1}$
are bounded on $L^1(\bR^+),$ and we have}
$$(I+\bold K)^{-1}=I+\bold M,\quad (I+\bold M)^{-1}=I+\bold K,\quad
(I+\bold P)^{-1}=I+\bold Q, \quad (I+\bold Q)^{-1}=I+\bold P.\tag 17.58$$

\item{(d)} {\it Each of the four operators
$(I+\bold K),$ $(I+\bold P),$ $(I+\bold M),$ and $(I+\bold Q)$
is invertible on $L^2(\bR^+).$ The corresponding inverses
$(I+\bold K)^{-1},$ $(I+\bold P)^{-1},$ $(I+\bold M)^{-1},$ and $(I+\bold Q)^{-1}$
are bounded on $L^2(\bR^+),$ and we have}
$$(I+\bold K)^{-1}=I+\bold M,\quad (I+\bold M)^{-1}=I+\bold K,\quad
(I+\bold P)^{-1}=I+\bold Q, \quad (I+\bold Q)^{-1}=I+\bold P.\tag 17.59$$

\item{(e)} {\it The operator $\bold P$ on $L^2(\bR^+)$ corresponds to
the adjoint operator for $\bold K$ on $L^2(\bR^+).$ Similarly,
the operator $\bold Q$ on $L^2(\bR^+)$ corresponds to
the adjoint operator for $\bold M$ on $L^2(\bR^+).$}

\noindent PROOF: Let us first remark that
(16.11) and (17.21) hold and
they also imply
$$|K(y,x)^\dagger|\le C\,\tau(x+y),\qquad 0\le x\le y,\tag 17.60$$
$$|M(y,x)^\dagger|\le C\,\tau(x+y),\qquad 0\le x\le y,\tag 17.61$$
for some generic constants $C,$ not necessarily having the same value
in different appearances, and where
$\tau(x)$ is the scalar function
defined in (16.7). In proving (a) and (b), it is enough to give the proof
for only one of the four operators because those four operators
satisfy the same upper bound in the four inequalities
(16.11), (17.21), (17.60), (17.61). Thus, let us start the proof
of (a) for the operator $\bold K$ appearing in (17.38).
It is enough to prove that $||\bold K Y||_1\le C\,||Y||_1,$
for some constant $C$ and $Y(x)\in L^1(\bR^+),$ where we recall
that $||\cdot||_1$ denotes the standard norm in $L^1(\bR^+).$
 From (17.38) we get
$$ ||\bold K Y||_1:=
\int_0^\infty dx\,|(\bold K Y)(x)|\le
\int_0^\infty dx\int_x^\infty dy\,|K(x,y)|\,|Y(y)|.\tag 17.62$$
Using (16.11) in (17.62) we get
$$\aligned
||\bold K Y||_1\le &C \int_0^\infty dx\int_x^\infty dy\,\tau(x+y)\,|Y(y)|\\
\le & C \int_0^\infty dx\,\tau(2x)\int_0^\infty dy\,|Y(y)|\\
=& \ds\frac{1}{2}\,C\, \tau_1(0)\,||Y||_1.\endaligned\tag 17.63$$
Since $\tau_1(0)$ is finite when $(\bold 2)$ holds, from (17.63) we conclude
that the operator norm of $\bold K$ on $L^1(\bR)$ is bounded and hence
(a) is proved for $\bold K.$ Since the proof of (a) can be repeated for
the other three operators in exactly the same way,
the proof of (a) is complete. Let us now return to the proof of (b)
for the operator $\bold K.$ The operator norm $|\bold K|_2$
on $L^2(\bR^+)$ satisfies
$$|\bold K|_2:=\sup_{||Y||_2=1} ||\bold K Y||_2=\sup_{||Y||_2=1}\left(\sup_{||\tilde Y||_2} |(\bold K Y,\tilde Y)_2|
\right),\tag 17.64$$
and hence we will consider
estimating $(\bold K Y,\tilde Y)_2$ when
$Y(x)$ and $\tilde Y(x)$ belong to
$L^2(\bR^+).$
 From (17.38) we get
$$\aligned |(\bold K Y,\tilde Y)_2|&\le
\int_0^\infty dx\int_x^\infty dy\,| Y(y)^\dagger\, K(x,y)^\dagger\, \tilde Y(x)|
\\ &\le
\int_0^\infty dx\int_x^\infty dy\,| Y(y)|\, | K(x,y)^\dagger|\, |\tilde Y(x)|.
\endaligned
\tag 17.65$$
Using Young's inequality
$$|Y(y)|\,|\tilde Y(x)|\le \ds\frac{1}{2}\,
\left[|Y(y)|^2+|\tilde Y(x)|^2
\right],\tag 17.66$$
 from (17.65) we obtain
 $$|(\bold K Y,\tilde Y)_2| \le\ds\frac{1}{2}
\int_0^\infty dx\int_x^\infty dy\,| K(x,y)^\dagger|\, |Y(y)|^2
+\ds\frac{1}{2}
\int_0^\infty dx\int_x^\infty dy\,| K(x,y)^\dagger|\, |\tilde Y(x)|^2.
\tag 17.67$$
Using (17.60) in (17.67) we get
$$|(\bold K Y,\tilde Y)_2| \le\ds\frac{C}{2}
\int_0^\infty dx\int_x^\infty dy\,\tau(x+y)\, |Y(y)|^2
+\ds\frac{C}{2}
\int_0^\infty dx\int_x^\infty dy\,\tau(x+y)\, |\tilde Y(x)|^2,
\tag 17.68$$
which yields
$$|(\bold K Y,\tilde Y)_2| \le\ds\frac{C}{2}
\int_0^\infty dx\,\tau(2x)\int_x^\infty dy\, |Y(y)|^2
+\ds\frac{C}{2}
\int_0^\infty dx\, |\tilde Y(x)|^2\int_x^\infty dy\,\tau(x+y).
\tag 17.69$$
 From (17.69) we obtain
 $$|(\bold K Y,\tilde Y)_2| \le\ds\frac{C}{4}\,\tau_1(0)\,||Y||_2^2+
 \ds\frac{C}{2}\,\tau_1(0)\,||\tilde Y||_2^2.\tag 17.70$$
Since $\tau_1(0)$ is finite when $(\bold 2)$ holds, from
(17.64) and (17.70) we get
$|\bold K|_2\le C,$ and hence
the operator $\bold K$ is bounded on $L^2(\bR^+).$
Thus, the proof of (b) is complete for $\bold K.$ The same proof can be repeated
to prove that the remaining three operators on $L^2(\bR^+)$ also have finite
operator norms, and hence the proof of (b) is complete.
Let us now turn to the proof of (c). From (a) it follows that
the four operators $(I+\bold K),$ $(I+\bold M),$ $(I+\bold P),$ $(I+\bold Q)$
are bounded on $L^1(\bR^+).$ Then, from (11.11)-(11.13) we conclude that
each of these four operators are invertible on $L^1(\bR^+)$ and their inverses are
also bounded on $L^1(\bR^+)$ and
(17.58) holds. The proof of (d) on $L^2(\bR^+)$ is similar to the proof of
(c) on $L^1(\bR^+).$ As for the proof of (e),
we observe from the kernels in (17.38) and (17.39) that
the operators $\bold K$ and $\bold P$ on $L^2(\bR^+)$ are adjoints of each.
Similarly, from the kernels in (17.40) and (17.41) we observe that
the operators $\bold M$ and $\bold Q$ on $L^2(\bR^+)$ are adjoints of each. \qed

\newpage
\noindent {\bf 18. FURTHER RESULTS RELATED TO THE INVERSE PROBLEM}
\vskip 3 pt

In this chapter
we provide some results related to
the characterization conditions
 presented in
Chapters~4-7.

Recall that there are two essential parts in solving the inverse scattering problem.
Given the scattering data set $\bold S$ as in (4.2), the first part involves
the construction of the corresponding input data set $\bold D$ as in (4.1), i.e.
to construct the corresponding potential $V(x)$ and the boundary
matrices $A$ and $B$ appearing in (4.1). The second part involves
proving that the physical solution $\Psi(k,x)$ constructed as in
(9.4) as well as the matrix bound-state solutions $\Psi_j(x)$ constructed
as in (9.8) satisfy the boundary conditions. Paraphrasing, the second part
involves showing that we have $\Delta(k)\equiv 0,$ where we have defined
$$\Delta(k):=-B^\dagger \Psi(k,0)+A^\dagger \Psi'(k,0),\tag 18.1$$
and showing that $\Delta_j=0$ for $j=1,\dots,N,$ where we have defined
$$\Delta_j:=-B^\dagger \Psi_j(0)+A^\dagger \Psi'_j(0),\qquad j=1,\dots,N.\tag 18.2$$
We know that the property $(\bold 3_a)$ of Definition~4.2 is equivalent to
saying $\Delta(k)\equiv 0$ and the condition $(\bold V_a)$ of Definition~4.3
is equivalent to having $\Delta_j=0$ for $j=1,\dots,N.$

We are interested in analyzing $\Delta(k)$ defined in (18.1) carefully
to understand it better. The following proposition
provides an integral representation of $\Delta(k)$ resembling a Fourier
transform. It expresses $\Delta(k)$ in terms of the
solution $K(x,y)$ to the Marchenko equation (13.1) and the solution
$K_x(x,y)$ to the derivative Marchenko equation
(13.7) as well as
$F_s(y)$ in (4.7) and its derivative $F_s'(y).$

\noindent {\bf Proposition 18.1} {\it Let $\bold S$ in (4.2) be the
scattering data set consisting of a scattering matrix $S(k),$
the constants $\kappa_j$ as distinct positive numbers, and
the matrices $M_j$ as $n\times n$ nonnegative hermitian
matrices. Assume that $\bold S$ satisfies the conditions
$(\bold 1)$, $(\bold 2)$, and $(\bold 4_a)$ stated in Definition~4.5. Let $F_s(y)$ be the quantity
given in (4.7) in such a way that
$F_s'(y)$ for $y\in\bR^-$ is the sum of an integrable
function and a square-integrable function. Then, we have the following:}

\item{(a)} {\it We can represent the quantity $\Delta(k)$ defined in (18.1) as}
$$\Delta(k)=\int_{-\infty}^\infty dy\,\hat \Delta(y)\,e^{-iky},\tag 18.3$$
{\it where $\hat \Delta(y)$ is the quantity given by}
$$\hat \Delta(y):=-B^\dagger \Gamma_{10}(y)
+A^\dagger \Gamma_{11}(y),\tag 18.4$$
{\it with $\Gamma_{10}(y)$ and $\Gamma_{11}(y)$ given by}
$$\Gamma_{10}(y):=K(0,y)+F_s(y)+K(0,-y)\,S_\infty+\int_{-\infty}^\infty dz\,K(0,z)\,F_s(z+y),
\tag 18.5$$
$$\Gamma_{11}(y):=K_x(0,y)+\overset{\circ}\to F'_s(y)+K_x(0,-y)\,S_\infty
-K(0,0)\,F_s(y)+\int_{-\infty}^\infty dz\,K_x(0,z)\,F_s(z+y).
\tag 18.6$$
{\it Here $A$ and $B$ are the boundary matrices
constructed as in Proposition~16.9, $S_\infty$ is the constant matrix
defined as in (4.6), $K(x,y)$ is the unique solution to the Marchenko equation
(13.1),
and $\overset{\circ}\to F'_s(y)$ is the quantity related to
$F_s'(y)$ as in (16.52). Because of (10.2) and (10.8), the lower integration limits
$x=-\infty$ in
(18.5) and (18.6) can be replaced with $x=0.$}

\item{(b)} {\it The matrix $\hat \Delta(y)$ is integrable in $y\in\bR^+,$
and it is the sum of an integrable function and a square-integrable function
in $\bR^-.$}

\item{(c)} {\it For $y\in\bR^+,$ the matrix $\hat \Delta(y)$ can be expressed in terms of
$K(0,y),$ $K_x(0,y),$ $F_s(y),$ and $F_s'(y)$ as}
$$\hat \Delta(y)=
-\Gamma_1(y)^\dagger\, B+\Gamma_2(y)^\dagger\, A,\qquad y\in\bR^+
,\tag 18.7$$
{\it where $\Gamma_1(y)$ and $\Gamma_2(y)$ are obtained
 from (18.5) and (18.6), respectively, by
using $K(0,-y)=0$ and $K_x(0,-y)=0$ and given by}
$$\Gamma_1(y):=K(0,y)+F_s(y)+\int_0^\infty dz\,K(0,z)\,F_s(z+y),\tag 18.8$$
$$\Gamma_2(y):=K_x(0,y)+F_s'(y)-K(0,0)\,F_s(y)+\int_0^\infty
dz\,K_x(0,z)\,F_s(z+y).\tag 18.9$$

\item{(d)} {\it For $y\in\bR^+,$ the matrix $\hat \Delta(y)$ can be expressed in terms of
the normalized bound-state matrices $\Psi_j(x)$ constructed as in (9.8) as}
$$\hat \Delta(y)=\ds\sum_{j=1}^N \left[B^\dagger \Psi_j(0)-A^\dagger \Psi'_j(0)
\right]M_j\,e^{-\kappa_j y},\qquad y\in\bR^+,\tag 18.10$$
{\it or equivalently expressed in terms of
the Jost matrix $J(k)$ constructed as in (9.2)
as}
$$\hat \Delta(y)=\ds\sum_{j=1}^n J(i\kappa_j)^\dagger\,M_j^2\,e^{-\kappa_j y},\qquad y\in\bR^+,\tag 18.11$$
{\it and hence $\hat \Delta(y)$ is in fact continuous in $y\in\bR^+$
and vanishes as $y\to+\infty.$}

\item{(e)} {\it The matrix $\hat \Delta(y)$ satisfies
the integral equation}
$$\int_{-\infty}^\infty dz\,\hat \Delta(z)\,F_s(z+y)=\hat \Delta(y)-\hat \Delta(-y)\,S_\infty,
\qquad y\in\bR.\tag 18.12$$

\noindent PROOF: When $(\bold 1)$ and $(\bold 4_a)$
of Definition~4.5 are satisfied, by Proposition~16.1
we are assured the existence and uniqueness of $K(x,y)$ as the solution to
the Marchenko equation (13.1). We also know that $K(x,y)$ is integrable
in $y\in(x,+\infty)$ for each fixed $x\ge 0.$
Similarly, if $(\bold 1)$, $(\bold 2)$, and $(\bold 4_a)$ are satisfied,
Proposition~16.5 implies that $K_x(x,y)$ is the unique solution
to (13.7). For each $x\ge 0,$ the quantity $K_x(x,y)$ is integrable
in $y\in(x,+\infty).$ By solving (13.1) and (13.7), we obtain
$K(0,y)$ and $K_x(0,y).$ Then, with the help of (10.6),
we construct $f(k,0)$ and $f'(k,0)$ in terms of
$K(0,y)$ and $K_x(0,y)$ as in (15.38) and (15.39),
respectively.
 From the physical solution $\Psi(k,x)$ constructed as in (9.4) we get
$$\Psi(k,0)=f(-k,0)+f(k,0)\,S(k),\quad \Psi'(k,0)=f'(-k,0)+f'(k,0)\,S(k).
\tag 18.13$$
 From (4.7) and (12.3), we respectively have
$$S(k)=S_\infty+\int_{-\infty}^\infty dy\,F_s(y)\,e^{-iky},\tag 18.14$$
$$ik\,S(k)=ik\,S_\infty+\int_{-\infty}^\infty dy\,F'_s(y)\,e^{-iky}.\tag 18.15$$
Using (15.38) and (15.39) in (18.13), with the help of
(11.37), (18.14), (18.15), and Proposition~16.7 we can express
$\Delta(k)$ given in
(18.1) in terms of $K(0,y)$ and $K_x(0,y)$ as
$$\Delta(k)=-B^\dagger\, \Gamma_{12}(k)+A^\dagger\, \Gamma_{13}(k),\tag 18.16$$
where we have defined
$$\Gamma_{12}(k):=I+S_\infty+\int_{-\infty}^\infty dy\,\Gamma_{10}(y)
\,e^{-iky},\tag 18.17$$
$$\Gamma_{13}:=-ik(I-S_\infty)-K(0,0)\,(I+S_\infty)
+\int_{-\infty}^\infty dy\,\left[G_1\,\delta(y)+\Gamma_{10}(y)\right]\,e^{-iky},
\tag 18.18$$
with $\Gamma_{10}(y)$ and $\Gamma_{11}(y)$ are the quantities defined in
(18.5) and (18.6), respectively.
Using (18.17) and (18.18) in (18.16), we obtain
$$\Delta(k)=\Gamma_{14}+\int_{-\infty}^\infty dy\,\left[-B^\dagger \, \Gamma_{10}+A^\dagger\, \Gamma_{11}(k)\right]\,e^{-iky},\tag 18.19$$
where we have defined the constant matrix $\Gamma_{14}$ as
$$\Gamma_{14}:=-ik\,A^\dagger (I-S_\infty)-B^\dagger (I+S_\infty)
+A^\dagger \left[G_1-K(0,0)-K(0,0)\,S_\infty
\right].\tag 18.20$$
Since the boundary matrices $A$ and $B$ satisfy (16.70),
the left-hand side of (18.20) vanishes and we get
$\Gamma_{14}=0.$
Using $\Gamma_{14}=0$ in (18.19), we see that with the help of (18.4) we can write
(18.19) as (18.3), completing the proof of (a).
Let us now turn to the proof of (b). For $y\in\bR^+,$ each term on the right-hand sides of (18.5) and (16.6) is
integrable, where this property of $F_s(y)$ follows from
$(\bold 1)$, that of
$K(0,y)$ follows from Proposition~16.1(a), $K(0,-y)=0$
as a result of (10.2), $K_x(0,y)$ is integrable as indicated in Proposition~16.5,
the integrability of
$\overset{\circ}\to F'_s(y)$ follows from $(\bold 2)$ and (16.52),
and each of the
integrals in (18.5) and (18.6) is integrable
in $y\in\bR^+$ because they are essentially the
convolution of two integrable functions.
Thus, from (18.4) we conclude that $\hat \Delta(y)$ is integrable
in $y\in\bR^+.$ Let us now show that for $y\in\bR^-$ the
quantity $\hat \Delta(y)$ is the sum of an integrable function and a square-integrable
function. From (18.4) we see that it is enough to show that
each term on the right-hand sides of (18.5) and (18.6)
is either integrable or square integrable in $y\in\bR^-.$
For $y\in\bR^-,$ by Proposition~4.4 we know that
$F_s(y)$ is square integrable, by Proposition~16.1 we know
that $K(0,-y)$ is integrable, by (10.2) we have
$K(0,y)=0,$ by Proposition~16.5 we know that
$K_x(0,-y)$ is integrable, by (10.8) we have $K_x(0,y)=0,$ as seen from
(16.3) the quantity
$\overset{\circ}\to F'_s(y)$ coincides with
$F'_s(y)$ for $y\in\bR^-,$ and as indicated in the statement of our
proposition
$F'_s(y)$
is assumed to be the sum of an integrable function and a square-integrable
function for $y\in\bR^-.$ As for the integral
in (18.5), $K(0,y)$ is square integrable
in $y\in\bR$ as result of (10.2) and Proposition~16.1(b), and
moreover
$F_s(y)$ is square integrable in $y\in\bR$ as implied by
Proposition~4.4. The integrals in (18.5) and (18.6) are each
essentially a convolution of an integrable and a square
integrable functions and hence from Young's inequality
for convolutions it follows that those two integrals
are each square integrable in $y\in\bR^-.$
Hence, the proof of (b) is complete.
Let us now turn to the proof of (c).
For $y\in\bR^+,$ from (10.2) and (10.8) it follows that
we have
$$\Gamma_1(y)=\Gamma_{10}(y),\quad  \Gamma_2(y)=\Gamma_{11}(y),\qquad
y\in\bR^+.\tag 18.21$$
Thus, using (18.21) in (18.4) we obtain (18.7),
completing the proof of (c).
Let us now turn to the proof of (d).
Using (4.7) and (4.12), we can write the Marchenko equation at $x=0$ given in
 (4.11) as
$$\aligned
K(0,y)+F_s(y)+\int_0^\infty dz\,&K(0,z)\,F_s(z+y)\\
& =
-\sum_{j=1}^N e^{-\kappa_j y}\left(I+\int_0^\infty dz\,K(0,z)\,e^{-\kappa_j z}\right)
M_j^2,
\qquad y>0.\endaligned\tag 18.22$$
Using (15.38) we recognize the quantity in the parentheses
on the right-hand side of (18.22) as
$f(i\kappa_j,0),$ and hence with the help of (9.8) we obtain
$$\left(I+\int_0^\infty dz\,K(0,z)\,e^{-\kappa_j z}\right)
M_j^2=\Psi_j(0)\,M_j,\qquad j=1,\dots,N,\tag 18.23$$
where $\Psi_j(x)$ is the bound-state matrix  solution constructed as in (9.8).
Using (18.23) in (18.22) we obtain
$$K(0,y)+F_s(y)+\int_0^\infty dz\,K(0,z)\,F_s(z+y)=
-\sum_{j=1}^N \Psi_j(0)\,M_j^2\,e^{-\kappa_j y},
\qquad y>0.\tag 18.24$$
We remark that (18.24) could also be obtained from
(13.4) at $x=0$ and by using (9.8).
Similarly, using (4.7) and (4.12) in (13.6) at $x=0,$ i.e.
in the derivative Marchenko equation at $x=0,$
we obtain
$$\aligned
K_x(0,y)+&F'_s(y)-K(0,0)\,F_s(y)+\int_0^\infty dz\,K(0,z)\,F_s(z+y)\\
& =
-\sum_{j=1}^N e^{-\kappa_j y}\left(-\kappa_j\,I
-K(0,0)+\int_0^\infty dz\,K_z(0,z)\,e^{-\kappa_j z}\right)
M_j^2,
\qquad y>0.\endaligned\tag 18.25$$
Using (15.39) we recognize the quantity in the parentheses
on the right-hand side of (18.25) as
$f'(i\kappa_j,0),$ and hence with the help of (9.8) we get
$$\left(-\kappa_j\,I
-K(0,0)+\int_0^\infty dz\,K_z(0,z)\,e^{-\kappa_j z}\right)
M_j^2=\Psi'_j(0)\,M_j,\qquad j=1,\dots,N,\tag 18.26$$
and hence, using (18.26) in (18.25), for $y\in\bR^-$ we obtain
$$K_x(0,y)+F'_s(y)-K(0,0)\,F_s(y)+\int_0^\infty dz\,K(0,z)\,F_s(z+y)
=
-\sum_{j=1}^N \Psi'_j(0)\,M_j\,e^{-\kappa_j y}.\tag 18.27$$
With the help of (18.8) and (18.9),
we see that
we can write (18.24) and (18.27), respectively, as
$$\Gamma_1(y)=
-\sum_{j=1}^N \Psi_j(0)\,M_j\,e^{-\kappa_j y},
\qquad y\in\bR^+,\tag 18.28$$
$$\Gamma_2(y)=
-\sum_{j=1}^N \Psi'_j(0)\,M_j\,e^{-\kappa_j y},
\qquad y\in\bR^+,\tag 18.29$$
where $\Psi_j(x)$ is the bound-state matrix  solution constructed as in (9.8).
Premultiplying (18.28) by $-B^\dagger$ and premultiplying (18.29) by
$A^\dagger$ and adding the resulting equations, we obtain
$$-B^\dagger\,\Gamma_1(y)+A^\dagger\,\Gamma_2(y)=
\ds\sum_{j=1}^n \left[B^\dagger\,\Psi_j(0)
-A^\dagger\,\Psi'_j(0)
\right]M_j \,e^{-\kappa_j y},\qquad y\in\bR^+.\tag 18.30$$
By taking the adjoint of both sides of (18.30)
we obtain
$$-\Gamma_1(y)^\dagger\,B+\Gamma_2(y)^\dagger\,A=
\ds\sum_{j=1}^n M_j \left[\Psi_j(0)^\dagger\,B
-\Psi'_j(0)^\dagger\,A
\right]\,e^{-\kappa_j y},\qquad y\in\bR^+,\tag 18.31$$
where we have used the hermitian property of
$M_j.$ Comparing (18.7) with (18.31) we obtain
(18.10).
Let us now show that (18.10) and (18.11) are
equivalent.
Using (9.2) and (9.8) we see that
$$\Psi_j(0)^\dagger \,B-\Psi_j'(0)^\dagger \,A=M_j\,J(i\kappa_j),
\qquad j=1,\dots,N,\tag 18.32$$
or equivalently, by taking the adjoint of (18.32) we have
$$B^\dagger\,\Psi_j(0)-A^\dagger\,\Psi_j'(0)^\dagger=J(i\kappa_j)^\dagger\,M_j,
\qquad j=1,\dots,N.\tag 18.33$$
Hence, using (18.33) in (18.10) we get (18.11).
$$-\Gamma_1(y)^\dagger\,B+\Gamma_2(y)^\dagger\,A=
\ds\sum_{j=1}^n M_j^2 \,J(i\kappa_j)\,e^{-\kappa_j y},\qquad y\in\bR^+.\tag 18.34$$
 From (18.10) or (18.11) we observe that the only $y$ dependence of
 $\hat\Delta(y)$ is through the exponential factors
$e^{-\kappa_j}$ and hence $\hat\Delta(y)$
is continuous in $y\in\bR^+$ and vanishes as
$y\to+\infty.$ Thus, the proof of (d) is complete.
Let us now turn to the proof of (e).
Using (18.5) and (18.6) in (18.4) and isolating the integral
terms, we obtain
$$\int_{-\infty}^\infty dz\,\left[-B^\dagger\,K(0,z)+A^\dagger\,
K_x(0,z)\right] F_s(z+y)=
\hat\Delta(y)+B^\dagger \,
\Gamma_{15}(y)-A^\dagger\,\Gamma_{16}(y),
\qquad y\in\bR,
\tag 18.35$$
where we have defined
$$\Gamma_{15}(y):=K(0,y)+F_s(y)+K(0,-y)\,S_\infty,\tag 18.36$$
$$\Gamma_{16}(y):=K_x(0,y)+\overset{\circ}\to F'_s(y)+K_x(0,-y)\,S_\infty
-K(0,0)\,F_s(y).\tag 18.37$$
 From (18.35), for $x\in\bR$ we obtain
$$\int_{-\infty}^\infty dz\,\left[-B^\dagger\,K(0,y)+A^\dagger\,
K_x(0,y)\right] F_s(y+x)
=
\hat\Delta(x)+B^\dagger \,
\Gamma_{15}(x)-A^\dagger\,\Gamma_{16}( x)
,\tag 18.38$$
$$\int_{-\infty}^\infty dz\,\left[-B^\dagger\,K(0,y)+A^\dagger\,
K_x(0,y)\right] F_s(y-x)
=
\hat\Delta(- x)+B^\dagger \,
\Gamma_{15}(- x)-A^\dagger\,\Gamma_{16}(- x)
.\tag 18.39$$
In order to prove (e), let us start with the left-hand side of (18.12).
Postmultiplying (18.4) by $F_s(y+x)$ and
integrating the resulting equality over $y\in\bR,$ we obtain
$$\int_{-\infty}^\infty dy\,\hat \Delta(y)\,F_s(y+x)=\int_{-\infty}^\infty dy\,
\left[-B^\dagger \Gamma_{10}(y)
+A^\dagger \Gamma_{11}(y)\right]\,F_s(y+x)
\qquad y\in\bR.\tag 18.40$$
We can evaluate the right-hand side of (18.40)
with the help of (16.63), (16.64), and (16.66), and we get
$$\int_{-\infty}^\infty dy\,\hat \Delta(y)\,F_s(y+x)=
\Gamma_{31}(x)+\Gamma_{32}(x)+\Gamma_{33}(x)+\Gamma_{34}(x)+\Gamma_{35}(x),
\qquad x\in\bR,\tag 18.41$$
where we have defined
$$\Gamma_{31}(x):=\int_{-\infty}^\infty dy\,\left[
-B^\dagger\,K(0,y)\,F_s(y+x)+
A^\dagger\,K_x(0,y)\,F_s(y+x) \right],\tag 18.42$$
$$\Gamma_{32}(x):= \int_{-\infty}^\infty dy\,\left[
-B^\dagger\,K(0,-y)\,S_\infty\,F_s(y+x)+
A^\dagger\,K_x(0,-y)\,S_\infty\,F_s(y+x) \right],\tag 18.43$$
$$\aligned
\Gamma_{33}(x):= &B^\dagger\left[F_s(-x)\,S_\infty+S_\infty\,F_s(x)\right]
+A^\dagger \left[-
\overset{\circ}\to F'_s(-x)\, S_\infty+S_\infty\,
\overset{\circ}\to F'_s(x)-G_1\,F_s(x)
\right]\\
&+A^\dagger\,K(0,0)\,
\left[F_s(-x)\,S_\infty+S_\infty\,F_s(x)
\right],\endaligned\tag 18.44$$
$$\Gamma_{34}(x):=\int_{-\infty}^\infty dz\,\left[
B^\dagger\,K(0,z)\,F_s(z-x)\,S_\infty-
A^\dagger\,K_x(0,z)\,F_s(z-x)\,S_\infty \right],\tag 18.45$$
$$\Gamma_{35}(x):= \int_{-\infty}^\infty dz\,\left[
B^\dagger\,K(0,z)\,S_\infty\,F_s(x-z)-
A^\dagger\,K_x(0,z)\,S_\infty\,F_s(x-z) \right].\tag 18.46$$
We observe that
$$\Gamma_{32}(x)=-\Gamma_{35}(x).\tag 18.47$$
Using (18.38) in (18.42), using (18.39) in (18.45), with the help of
(18.47), we rewrite (18.41) as
$$\aligned\int_{-\infty}^\infty dy\,\hat \Delta(y)\,F_s(y+x)=&\hat\Delta(x)+
B^\dagger\,\Gamma_{15}(x)-A^\dagger\,\Gamma_{16}(x)+\Gamma_{33}(x)
-\hat\Delta(-x)\,S_\infty\\
&-B^\dagger\,\Gamma_{15}(-x)+
A^\dagger\,\Gamma_{16}(-x)\,S_\infty,
\qquad x\in\bR,\endaligned\tag 18.48$$
Since $A$ and $B$ satisfy (16.70),
the adjoint of (16.70) yields
(16.71) and (16.72). With the help of (16.71) and
(16.72), we obtain
$$B^\dagger\,\Gamma_{15}(x)-A^\dagger\,\Gamma_{16}(x)+\Gamma_{33}(x)
-B^\dagger\,\Gamma_{15}(-x)+
A^\dagger\,\Gamma_{16}(-x)\,S_\infty=0,
\qquad x\in\bR,\tag 18.49$$
and hence using (18.49) in (18.48) we obtain
(18.12). Thus, the proof of (e) is complete. \qed

 From Proposition~18.1, we obtain several important results.
The following proposition indicates that
$(\bold 3_a)$ of Definition~4.2 implies
$(\bold V_a)$ of Definition~4.3. Furthermore,
it indicates that
$(\bold 3_a)$ is equivalent to
the combination of two properties, namely
$(\bold{III}_a)$ and
$(\bold V_a)$ of Definition~4.3.

\noindent {\bf Proposition 18.2} {\it Let $\bold S$ in (4.2) be the
scattering data set consisting of a scattering matrix $S(k),$
the constants $\kappa_j$ as distinct positive numbers, and
the matrices $M_j$ as $n\times n$ nonnegative hermitian
matrices.
Let $F_s(y)$ be the quantity defined in (4.7),
$A$ and $B$ be the boundary matrices constructed
as in Proposition~16.9 and $\Psi(k,x)$ be the
physical solution constructed as in (9.4). Then:}

\item{(a)} {\it If $(\bold 1)$, $(\bold 2)$, $(\bold 3_a)$, and $(\bold 4_a)$ of Definition~4.2 hold,
then $(\bold V_a)$ of Definition~4.3 holds.}


\item{(b)} {\it If $(\bold 1)$, $(\bold 2)$, $(\bold 4_a)$ of Definition~4.2 hold then
$(\bold 3_a)$ of Definition~4.2 is equivalent to the combination of
$(\bold{III}_a)$ and $(\bold V_a)$ of Definition~4.3.}

\noindent PROOF: Let us first argue that
Proposition~18.1 is applicable in both cases of (a) and (b)
because in both cases we have
$F_s'(y)$ for $y\in\bR^-$ is the sum of an integrable function and a square-integrable
function.  In (a), since $\bold S$ satisfies
$(\bold 1)$, $(\bold 2)$, $(\bold 3_a)$, and $(\bold 4_a)$ it belongs to the Marchenko
class, and hence by Theorem~5.1(b) there exists a corresponding unique
input data set $\bold D$ in the Faddeev class. Then, by Theorem~12.1(h)
we see that $F_s'(y)$ for $y\in\bR^-$ is the sum of an integrable function and a square-integrable
function. On the other hand, in (b) it is assumed that $(\bold{III}_a)$ holds
and hence, by the definition of $(\bold{III}_a)$,
it follows that
$F_s'(y)$ for $y\in\bR^-$ is the sum of an integrable function and a square-integrable
function. Thus, we are able to apply Proposition~18.1.
As seen from (18.3) of Proposition~18.1, $\Delta(k)$ and
$\hat\Delta(y)$ are Fourier transforms of each other. Hence, if $(\bold 3_a)$ is satisfied,
then we must have $\Delta(k)\equiv 0,$ yielding
$\hat\Delta(y)\equiv 0.$ Then, we must have the right-hand-side
of (18.10) vanishing for all $y\in\bR^+.$ Since the $\kappa_j$ are distinct,
 from (18.10) we conclude that
$$\left[-B^\dagger\,\Psi_j(0)+
A^\dagger\,\Psi'_j(0)\right] M_j
=0,\qquad j=1.\dots,N.\tag 18.50$$
Since $M_j$ is hermitian, (18.50) can also be written
as
$$\left[-B^\dagger\,\Psi_j(0)+
A^\dagger\,\Psi'_j(0)\right] M_j^\dagger
=0,\qquad j=1.\dots,N.\tag 18.51$$
 From (18.51) we get
$$a_j\,\left[-B^\dagger\,\Psi_j(0)+
A^\dagger\,\Psi'_j(0)\right] M_j^\dagger\, b_j^\dagger
=0,\qquad j=1,\dots,N,\tag 18.52$$
where $a_j$ and $b_j$ are arbitrary row vectors
with $n$ components each.
Let $f(k,x)$ be the Jost solution constructed from
the solution $K(x,y)$ to the Marchenko equation
as in (10.6).
Choosing $b_j$ in (18.52) as
$$b_j=a_j\,\left[-B^\dagger\,f(i\kappa_j,0)+
A^\dagger\,f'(i\kappa_j,0)\right],\tag 18.53$$
we see that (18.52) yields
$$a_j\,\left[-B^\dagger\,\Psi_j(0)+
A^\dagger\,\Psi'_j(0)\right] M_j^\dagger\, \left[-B^\dagger\,f(i\kappa_j,0)+
A^\dagger\,f'(i\kappa_j,0)\right]^\dagger a_j^\dagger
=0,\qquad j=1,\dots,N,\tag 18.54$$
or equivalently, after using (9.8), we have
$$a_j\,\left[-B^\dagger\,\Psi_j(0)+
A^\dagger\,\Psi'_j(0)\right] \left[-B^\dagger\,\Psi_j(0)+
A^\dagger\,\Psi'_j(0)\right]^\dagger a_j^\dagger=0,\qquad j=1,\dots,N.\tag 18.55$$
 From (18.55) we see that, for each $j=1,\dots,N,$ the row vector
$a_j\,\left[-B^\dagger\,\Psi_j(0)+
A^\dagger\,\Psi'_j(0)\right]$ has zero length
and hence it must be equal to the zero vector. On the other hand since
$a_j$ is arbitrary, we conclude that the matrix
$\left[-B^\dagger\,\Psi_j(0)+
A^\dagger\,\Psi'_j(0)\right]$ must be zero, yielding (4.20).
Hence, the proof of (a) is complete.
Let us now turn to the proof of (b).
Since $(\bold V_a)$ is assumed to be satisfied, we have the
right-hand side of (18.10) is zero and hence
(18.10) implies that
$\hat \Delta(y)=0$ for $y\in\bR^+.$
Then, from (18.12) we get
$$\hat\Delta(y)+\int_{-\infty}^0 dz\,\hat \Delta(z)\,F_s(z+y)=0,\qquad
y\in\bR^-.
\tag 18.56$$
Let us now prove that $\hat \Delta(y)$ belongs to
$L^2(\bR^-).$
%
%
%
%
This is justified as follows. From
Proposition~18.1(b) we know that
$\hat \Delta(y)$ is the sum of an integrable function and a square-integrable
function in $y\in\bR^-.$ With the help of
Propositions~3.3(c) and 3.4(c) we can conclude that
if a solution to (18.56) is the sum of an integrable function and
a square-integrable function then that solution must be bounded
and hence also be square integrable.
Since we assume that
$(\bold {III}_a)$ holds, we know that
the only solution in $L^2(\bR^-)$ to (18.56)
is the trivial solution $\hat \Delta(y)=0$ for $y\in\bR^-.$
However, then we have
$\hat \Delta(y)\equiv 0$ because we have already had $\hat\Delta(y)=0$
for $y\in\bR^+$ as a result of $(\bold V_a).$
Then, by (18.3) we see that $\Delta(k)\equiv 0,$ yielding
$(\bold 3_a)$ of Definition~4.2.
Conversely, if $(\bold 3_a)$ holds, then
$\hat \Delta(y)\equiv 0,$ and as a result
$(\bold V_a)$ holds because
$\hat \Delta(y)=0$ for $y\in\bR^+$ as indicated
by (18.10), and
$(\bold {III}_a)$ holds because
$\hat \Delta(y)=0$ for $y\in\bR^-$ as the trivial solution to (18.56).
\qed

With the help of Proposition~18.1 and 18.2, we get the following result,
which also indicates the equivalence of
$(\bold V_a)$ and $(\bold V_b)$ in Proposition~6.5.

\noindent {\bf Proposition 18.3} {\it Let $\bold S$ in (4.2) be the
scattering data set consisting of a scattering matrix $S(k),$
the constants $\kappa_j$ as distinct positive numbers, and
the matrices $M_j$ as $n\times n$ nonnegative hermitian
matrices. Assume that $\bold S$ satisfies the conditions
$(\bold 1)$, $(\bold 2)$, and $(\bold 4_a)$ stated in Definition~4.2.
Let $A$ and $B$ be the boundary matrices constructed
as in Proposition~16.9, $J(k)$ be the Jost matrix constructed as in
(9.2), $\Psi_j(x)$ be the bound-state matrix solutions
constructed as in (9.8),
$\hat \Delta(y)$ be the quantity
defined in (18.4) and represented as in (18.7)
for $y\in\bR^+,$ and
$\Gamma_1(y)$ and $\Gamma_2(y)$ be the matrices appearing
in (18.8) and (18.9), respectively.
Then, the following statements are equivalent:}

\item{(a)} {\it We have the matrix equality}
$$\hat\Delta(y)=0,\qquad y\in\bR^+.
\tag 18.57$$

\item{(b)} {\it We have the matrix equality}
$$-\Gamma_1(y)^\dagger\, B+\Gamma_2(y)^\dagger\, A=0,\qquad y\in\bR^+.
\tag 18.58$$

\item{(c)} {\it The bound-state matrix solutions $\Psi_j(x)$ satisfy (18.50).}

\item{(d)} {\it The Jost matrix $J(k)$ satisfies}
$$J(i\kappa_j)^\dagger \,M_j^2\,=0,\qquad j=1,\dots,N.\tag 18.59$$

\item{(e)} {\it The bound-state matrix solutions $\Psi_j(x)$ satisfy
the boundary condition (2.4). This is the same as saying that $(\bold V_a)$ of Proposition~4.6 holds, i.e. (4.20) holds.}

\item{(f)} {\it The Jost matrix $J(k)$ satisfies (4.21), i.e.
the scattering data set $\bold S$ satisfies
$(\bold V_b)$ of Definition~4.3.}

\noindent PROOF: From (18.7) we
know that (a) and (b) are equivalent.
 From (18.30) we see that
(b) holds if and only if (c) holds, as a result of the functions
$e^{-\kappa_j y}$ with $j=1,\dots,N$ being linearly independent in
$y\in\bR^+.$ Then, from (18.33) we see that (e)
and (f) are equivalent.
 From (18.33), after postmultiplying with $M_j,$ we see that (c) and (d) are equivalent.
Hence, it is enough to prove
the equivalence of (c) and (e).
Note that (e) implies (c), as seen by postmultiplying (4.20) with $M_j.$
Let us now prove that (c) implies (e).
Arguing as in (18.50)-(18.55) in the proof of
of Proposition~18.2, we prove that
$a_j\left[-B^\dagger\,\Psi_j(0)+A^\dagger\,\Psi'_j(0)\right]=0$
for any $a_j\in\bC^n$ when $j=1,\dots,N,$
which proves (4.20) and hence confirms (e). \qed

%

In the next proposition we explore a key feature of the solution
$X(y)$ to (4.22), which will be used later in Proposition~18.6.

\noindent {\bf Proposition 18.4} {\it Let $\bold S$ in (4.2) be the
scattering data set consisting of a scattering matrix $S(k),$
the constants $\kappa_j$ as distinct positive numbers, and
the matrices $M_j$ as $n\times n$ nonnegative hermitian
matrices. Assume that $\bold S$ satisfies the conditions
$(\bold 1)$, $(\bold 2)$, and $(\bold 4_a)$ of Definition~4.2. Let $F_s(y)$ be the quantity
given in (4.7) in such a way that
$F_s'(y)$ for $y\in\bR^-$ is the sum of an integrable
function and a square-integrable function. We have the following:}

\item{(a)} {\it Any solution $X(y)$ in $L^1(\bR^+)$ to
(4.22) satisfies}
$$\ds\int_0^\infty dy\, X(y)\,\hat\Delta(y)=0
,\tag 18.60$$
{\it where $\hat\Delta(y)$ is the quantity defined in (18.4)
and whose value for $y\in\bR^+$ is given by
(18.7).}

\item{(b)} {\it Any solution $\hat X(k)$ in $\hat L^1(\bCp)$ to
(4.23) satisfies}
$$\ds\sum_{j=1}^N \hat X(i\kappa_j)\,M_j^2\, J(i\kappa_j)=0
,\tag 18.61$$
{\it where $J(k)$ is the Jost matrix constructed as in (9.2).}

\noindent PROOF:
Note that $F_s(y)$ and $F(y)$ are hermitian as stated in
Proposition~16.4(b), and $K(0,0)$ is hermitian as stated in Proposition~16.4(d).
Thus, from (18.8) and (18.9), we respectively get
$$\Gamma_1(y)^\dagger=K(0,y)^\dagger+F_s(y)+\int_0^\infty dz\,F_s(y+z)\,K(0,z)^\dagger,\tag 18.62$$
$$\Gamma_2(y)^\dagger=K_x(0,y)^\dagger+F_s'(y)-F_s(y)\,K(0,0)+\int_0^\infty
dz\,F_s(y+z)\,K_x(0,z)^\dagger.\tag 18.63$$
Let $X(y)$ be the general solution in $L^1(\bR^+)$ to (4.22).
Using (18.62) and (18.63) we obtain
$$\aligned \ds\int_0^\infty dy\, X(y)\left[
-\Gamma_1(y)^\dagger\, B+\Gamma_2(y)^\dagger\, A
\right]=&-\ds\int_0^\infty dy\,
X(y)\,F_s(y)\,B+ \ds\int_0^\infty dy\,
X(y)\,F'_s(y)\,A\\
&-\ds\int_0^\infty dy\,
X(y)\,F_s(y)\,K(0,0)\,A-\Lambda_1+\Lambda_2,\endaligned\tag 18.64$$
where we have defined
$$\Lambda_1:=\ds\int_0^\infty dy\,\left[X(y)+\int_0^\infty dz\,
X(z)\,F_s(z+y)
\right] K(0,y)^\dagger \,B,\tag 18.65$$
$$\Lambda_2:=\ds\int_0^\infty dy\,\left[X(y)+\int_0^\infty dz\,
X(z)\,F_s(z+y)
\right] K_x(0,y)^\dagger \,A.\tag 18.66$$
Because $X(y)$ satisfies (4.22), from (18.65) and (18.66) we
see that
$$\Lambda_1=0,\quad \Lambda_2=0,\tag 18.67$$
and hence the right-hand side of (18.64) only consists of the first
three integrals there.
Recall that $X(y)=0$ for $y\in\bR^-$ and
$F'_s(y)$ and $\overset{\circ}\to F'_s(y)$
coincide when $y\in\bR^+,$ where
$\overset{\circ}\to F'_s(y)$ is the regular part of
$F'_s(y)$ defined in (16.52).
Thus, with the help of (18.67), we can write (18.64) as
$$\ds\int_0^\infty dy\, X(y)\left[
-\Gamma_1(y)^\dagger\, B+\Gamma_2(y)^\dagger\, A
\right]=\ds\int_0^\infty dy\,X(y)\,\Lambda_3(y)
,\tag 18.68$$
where we have defined
$$\Lambda_3(y):=-F_s(y)\,B+\overset{\circ}\to F'_s(y)\,A-F_s(y)\,K(0,0)\,A,
\qquad y\in\bR.\tag 18.69$$
Note that we can replace the lower integration
limit $y=0$ by $y=-\infty$ in the integral on the right-hand side
of (16.75) to have
$$\ds\int_0^\infty dy\,X(y)\,\Lambda_3(y)
=\ds\int_{-\infty}^\infty dy\,X(y)\,\Lambda_3(y)
.\tag 18.70$$
We justify (18.70) as follows. As asserted in Proposition~15.1(c),
the solution $X(y)$ in $L^1(\bR^+)$ to (4.22) is bounded
in $y\in\bR^+.$ Since $X(y)=0$ for $y\in\bR^-,$ it follows that
$X(y)$ is bounded and integrable
in $y\in\bR.$ By $(\bold 1)$ we know that $F_s(y)$ is bounded
in $y\in\bR.$ Thus the product
$X(y)\,F_s(y)$ is integrable in $y\in\bR$ and vanishes
for $y\in\bR^-.$
By $(\bold 2)$ we know that $F'_s(y)$ is integrable
in $y\in\bR^+,$ and for $y\in\bR^-$
it is the sum of an integrable function and
a square-integrable function. From (16.52) it follows that
$\overset{\circ}\to F'_s(y)$ is also
integrable
in $y\in\bR^+,$ and for $y\in\bR^-$
it is the sum of an integrable function and
a square-integrable function. Thus, the product
$X(y)\,\overset{\circ}\to F'_s(y)$ is
integrable in $y\in\bR^+,$ vanishes for $y\in\bR^-,$
and is integrable in any finite interval containing $y=0.$
Thus, (18.70) is justified.
With the help of (3.67), (4.7), (11.37) we get
$$\int_{-\infty}^\infty dy\,X(y)\,F_s(y)=\ds\frac{1}{2\pi}\int_{-\infty}^\infty dk\,
\hat X(k)\left[S(k)-S_\infty\right],\tag 18.71$$
and with the help of (3.67), (12.3), (16.51) we obtain
$$\int_{-\infty}^\infty dy\,X(y)\,\overset{\circ}\to F'_s(y)=\ds\frac{1}{2\pi}\int_{-\infty}^\infty dk\,
\hat X(k)\left[ik \left(S(k)-S_\infty\right)-G_1\right].\tag 18.72$$
Using (18.71) and (18.72) in (16.76), the integral on the right-hand
side of (18.70) yields
$$\ds\int_{-\infty}^\infty dy\,X(y)\,\Lambda_3(y)=
\ds\frac{1}{2\pi}
\ds\int_{-\infty}^\infty dk\,\hat X(k)\,\,\Lambda_4(k)
,\tag 18.73$$
where we have defined
$$\Lambda_4(k):=\left[-S(k)+S_\infty\right]B
+\left(ik\left[S(k)-S_\infty\right]-G_1\right)\,A-\left[S(k)-S_\infty\right] K(0,0)\, A.
\tag 18.74$$
Since $X(y)$ satisfies (4.22), from Proposition~15.7
it follows that $\hat X(k)$ satisfies (4.23); hence,
we can replace $\hat X(k)\,S(k)$ on the right-hand side
of (18.73) by $-\hat X(-k).$ We then obtain
$$\ds\int_0^\infty dy\, X(y)\left[
-\Gamma_1(y)^\dagger\, B+\Gamma_2(y)^\dagger\, A
\right]=\ds\frac{1}{2\pi}
\int_{-\infty}^\infty dk\,\left[\hat X(-k)\,\Lambda_5(k)+\hat X(k)\,\Lambda_6(k)\right]
,\tag 18.75$$
where we have defined
$$\Lambda_5(k):=B-ik\,A+K(0,0)\,A,\tag 18.76$$
$$\Lambda_6(k):=S_\infty\,B-ik\,S_\infty\,A-G_1\,A+S_\infty\,K(0,0)\,A.
\tag 18.77$$
We can replace $\hat X(-k)\,\Lambda_5(k)$ by
$\hat X(k)\,\Lambda_5(-k)$ in the integrand on the
right-hand side of (18.75). Then, using (18.76) and (18.77)
in (18.75) we obtain
$$\ds\int_0^\infty dy\, X(y)\left[
-\Gamma_1(y)^\dagger\, B+\Gamma_2(y)^\dagger\, A
\right]=\ds\frac{1}{2\pi}
\int_{-\infty}^\infty dk\,\hat X(k)\,\Lambda_7(k)
,\tag 18.78$$
where we have defined
$$\Lambda_7(k):=(I+S_\infty)\,B+ik\,(I-S_\infty)\,A-G_1\,A+K(0,0)\,A+S_\infty\,K(0,0)\,A.
\tag 18.79$$
When
$(\bold 1)$ and $(\bold 4_a)$ in
Definition~4.2 hold, we have (16.70) satisfied, which
makes
the right-hand side of (18.79) vanish.
Thus, using $\Lambda_7(k)\equiv 0$ on the right hand side of (18.78), we see that
 (18.78) yields (18.60). Thus, the proof of (a) is complete.
 Let us now turn to the proof of (b).
Using (18.34) in
(18.60), we obtain
$$\ds\sum_{j=1}^n \int_0^\infty dy\, X(y)\,e^{-\kappa_j y}\, M_j^2 \,J(i\kappa_j)=0.
\tag 18.80$$
 From (3.68) we see that
$$\int_0^\infty dy\,X(y)\, e^{-\kappa_j y}=\hat X(i\kappa_j),
\qquad j=1,\dots,N.\tag 18.81$$
and hence using (18.81) on the left-hand side of (18.80) we obtain
(18.61). Hence, the proof of (b) is complete. \qed

In the next proposition we show that the results stated in
Proposition~18.4 for the solutions $X(y)\in L^1(\bR^+)$ to
(4.22) and $\hat X(k)\in \hat L^1(\bCp)$ to (4.23)
actually hold for the solutions $X(y)\in L^2(\bR^+)$ to
(4.22) and $\hat X(k)\in \bold H^2(\bCp)$ to (4.23).

\noindent {\bf Proposition 18.5} {\it Let $\bold S$ in (4.2) be the
scattering data set consisting of a scattering matrix $S(k),$
the constants $\kappa_j$ as distinct positive numbers, and
the matrices $M_j$ as $n\times n$ nonnegative hermitian
matrices. Assume that $\bold S$ satisfies the conditions
$(\bold 1)$, $(\bold 2)$, and $(\bold 4_a)$ of Definition~4.2. Let $F_s(y)$ be the quantity
given in (4.7) in such a way that
$F_s'(y)$ for $y\in\bR^-$ is the sum of an integrable
function and a square-integrable function. We have the following:}

\item{(a)} {\it Any solution $X(y)$ in $L^2(\bR^+)$ to
(4.22) satisfies}
$$\ds\int_0^\infty dy\, X(y)\,\hat\Delta(y)=0
,\tag 18.82$$
{\it where $\hat\Delta(y)$ is the quantity defined in (18.4)
and whose value for $y\in\bR^+$ is given by
(18.7).}

\item{(b)} {\it Any solution $\hat X(k)$ in $\bold H^2(\bCp)$ to
(4.23) satisfies}
$$\ds\sum_{j=1}^N \hat X(i\kappa_j)\,M_j^2\, J(i\kappa_j)=0
,\tag 18.83$$
{\it where $J(k)$ is the Jost matrix constructed as in (9.2).}

\noindent PROOF: In the proof of Proposition~18.4, let us replace the solution
$X(y)\in L^1(\bR^+)$ to (4.22) with $X(y)\in L^2(\bR^+)$ to
(4.22). Then, the proof remains valid provided we can prove that
the integrability of
$X(y)\,F_s(y)$ in $y\in\bR^+$ as well as the integrability
of
$X(y)\,\overset{\circ}\to F'_s(y)$ in $y\in\bR^+$ are unaffected.
The following argument shows that those integrabilities are indeed
unaffected. We know from Proposition~14.6(a) that any solution
$X(y)\in L^2(\bR^+)$ to (4.22) must actually belong to
$L^2(\bR^+)\cap L^\infty(\bR^+).$ Thus,
$X(y)\,F_s(y)$ still remains integrable in $y\in\bR^+$ because
$F_s(y)\in L^1(\bR^+)$ and $X(y)\,F_s(y)$ is a product of a bounded
quantity with an integrable quantity. Similarly,
$X(y)\,\overset{\circ}\to F'_s(y)$ still remains
integrable in $y\in\bR^+$ because
$\overset{\circ}\to F'_s(y)\in L^1(\bR^+)$ and $X(y)\,F_s(y)$ is a product of a bounded
quantity with an integrable quantity. Then, with the minor
replacement that we use Proposition~15.6 instead of Proposition~15.7
in the proof of Proposition~18.4,
we know that the proof constitutes a proof for
Proposition~18.5. \qed

The next proposition shows that the property
$(\bold V_c)$ implies $(\bold V_b)$ in Proposition~6.6.

\noindent {\bf Proposition 18.6} {\it Let $\bold S$ in (4.2) be the
scattering data set consisting of a scattering matrix $S(k),$
the constants $\kappa_j$ as distinct positive numbers, and
the matrices $M_j$ as $n\times n$ nonnegative hermitian
matrices. Assume that $\bold S$ satisfies the conditions
$(\bold 1)$, $(\bold 2)$, and $(\bold 4_a)$ of Definition~4.2. Then, the condition
$(\bold V_c)$ of Proposition~4.1 implies $(\bold V_b)$
of Proposition~6.6.}

\noindent PROOF: Let $\Cal N$ be the nonnegative integer
defined in (4.3). We would like to show that,
if (4.22) has $\Cal N$ linearly independent solutions in $L^1(\bR^+),$ then
the Jost matrix constructed as in (9.2) from the input scattering data satisfies
(4.21). For the proof we proceed as follows. Let
$X^{(l)}(y)$ for $l=1,\dots,\Cal N$ be linearly independent solutions
in $L^1(\bR^+)$
to (4.22). By Proposition~15.7(a) we know that (4.23)
has also $\Cal N$ linearly independent solutions
given by
$\hat X^{(l)}(k)$ with
$l=1,\dots,\Cal N,$ where each
$\hat X^{(l)}(k)$ is related to $X^{(l)}(y)$
as in (3.67) and (3.68).
Thus, the general solution to (4.23)
can be written as a linear combination as
$$\hat X(k)=\sum_{l=1}^{\Cal N} \gamma_l\,  \hat X^{(l)}(k),\tag 18.84$$
where the $\gamma_l$ are arbitrary scalar coefficients.
From Proposition~6.1 we know that $(\bold 4_a)$ and $(\bold 4_d)$ of
Definition~4.2 are equivalent, and hence $(\bold 4_d)$ holds.
Note that (4.23) and the second line of (4.15) coincide.
Thus, the quantity $\hat X(k)$ given in (18.84) satisfies
the second line of (4.15).
Since $(\bold 4_d)$ holds, this means that
if $\hat X(k)$ given in (18.84) also satisfies
$$\hat X(i\kappa_j)\,M_j=0,\qquad j=1,\dots, N,
\tag 18.85$$
then we must have $\hat X(k)\equiv 0.$
Then, from (18.85) we can conclude that
the linear homogeneous system given by
$$\ds\sum_{l=1}^{\Cal N} \gamma_l\, \hat X^{(l)}(i\kappa_j)\,M_j=0,\qquad
j=1,\dots, N.\tag 18.86$$
can only have the trivial solution $\gamma_l=0$ for
$l=1,\dots.\Cal N.$
We can write (18.86) in the matrix notation as
$$\bm \gamma_1&\cdots&\gamma_{\Cal N}\endbm
\bm \hat X^{(1)}(i\kappa_1)\,M_1 & \dots &\hat X^{(1)}(i\kappa_N)\,M_N\\
\vdots &\ddots &\vdots\\
\hat X^{(\Cal N)}(i\kappa_1)\,M_1&\dots&\hat X^{(\Cal N)}(i\kappa_N)\,M_N\endbm=\bm 0&\cdots&0\endbm.\tag 18.87$$
We remark that in (18.87) the unknown is a row vector with $\Cal N$ entries,
the coefficient matrix has the matrix size
$\Cal N\times (nN),$ with $\hat X^{(l)}(i\kappa_j)\,M_j$ being an
$1\times n$ matrix, and in the zero vector on the right-hand side
of (18.87) each zero represents the zero row vector with $n$ entries.
Since the only solution to (18.87) must be the zero solution,
the rank of
the coefficient matrix in (18.87) must be the same as the
number of unknowns, i.e. must be equal to $\Cal N.$ This implies that
any nonhomogeneous system associated with (18.87) must have a
unique solution. In particular, let us
consider the nonhomogeneous system given by
$$\ds\sum_{l=1}^{\Cal N} \gamma_l\, \hat X^{(l)}(i\kappa_j)\,M_j=a^{(j)}\,M_j,\qquad
j=1,\dots, N,\tag 18.88$$
which we can write in the matrix notation as
$$\bm \gamma_1&\cdots&\gamma_{\Cal N}\endbm
\bm \hat X^{(1)}(i\kappa_1)\,M_1 & \dots &\hat X^{(1)}(i\kappa_N)\,M_N\\
\vdots &\ddots &\vdots\\
\hat X^{(\Cal N)}(i\kappa_1)\,M_1&\dots&\hat X^{(\Cal N)}(i\kappa_N)\,M_N\endbm=\bm a^{(1)}M_1&\cdots&a^{(N)}M_N\endbm,\tag 18.89$$
where each $a^{(j)}$ is a row vector with $n$ components.
Because of the full rank of the coefficient matrix in (18.89), the
corresponding nonhomogeneous system has a unique solution $\gamma_l$
for $l=1,\dots,\Cal N$ for any choice
of $a^{(j)}$ with $j=1,\dots,N.$
With those values of $\gamma_l,$ the row vector
$\hat X(k)$ given in (15.59)
is a solution to (4.23) and we have
$$\hat X(i\kappa_j)\,M_j=a^{(j)}\,M_j,\qquad
j=1,\dots, N.\tag 18.90$$
By Proposition~18.4(b) we know that any solution to (4.23) must
satisfy (18.61).
Using (18.90) in (18.61) we get
$$\ds\sum_{l=1}^N a^{(j)}\,M_j\,M_j^\dagger\,J(i\kappa_j)=0,
\tag 18.91$$
where we have used the fact that the matrix $M_j$ is hermitian.
Since the $a^{(j)}$ can be chosen arbitrarily,
 we conclude that each term in the summation in (18.91)
must vanish, i.e. we have
$$a^{(j)}\,M_j\,M_j^\dagger\,J(i\kappa_j)=0,
\qquad j=1,\dots, N.\tag 18.92$$
Let us multiply (18.92) by an arbitrary column vector
$b_j^\dagger$ having $n$ components. We get
$$a^{(j)}\,M_j\,M_j^\dagger\,J(i\kappa_j)\,b_j^\dagger=0,
\qquad j=1,\dots, N.\tag 18.93$$
Choosing $a^{(j)}$ as
$a^{(j)}=b_j\,J(i\kappa_j)^\dagger,$ from (18.93) we obtain
$$\left[b_j\,J(i\kappa_j)^\dagger\,M_j\right]\left[
M_j^\dagger \,J(i\kappa_j)\,b_j^\dagger\right]=0,
\qquad j=1,\dots, N,\tag 18.94$$
or equivalently
$$\left[
M_j^\dagger \,J(i\kappa_j)\,b_j^\dagger\right]^\dagger
\left[
M_j^\dagger \,J(i\kappa_j)\,b_j^\dagger\right]=0,
\qquad j=1,\dots, N.\tag 18.95$$
The left-hand side of (18.95) is the length of the
vector $M_j^\dagger \,J(i\kappa_j)\,b_j^\dagger,$ and hence
that vector must be the zero vector, yielding
$$M_j^\dagger \,J(i\kappa_j)\,b_j^\dagger=0,
\qquad j=1,\dots, N.\tag 18.96$$
Since $b_j^\dagger$ can be chosen arbitrarily,
(18.96) implies that
$$M_j^\dagger \,J(i\kappa_j)=0,
\qquad j=1,\dots, N.\tag 18.97$$
Comparing (18.97) with (4.21)
we conclude that $(\bold V_b)$ of Definition~4.3 is satisfied. \qed

The next proposition shows that the condition
$(\bold V_f)$ of Proposition~6.6 implies $(\bold V_b)$ there.

\noindent {\bf Proposition 18.7} {\it Let $\bold S$ in (4.2) be the
scattering data set consisting of a scattering matrix $S(k),$
the constants $\kappa_j$ as distinct positive numbers, and
the matrices $M_j$ as $n\times n$ nonnegative hermitian
matrices. Assume that $\bold S$ satisfies the conditions
$(\bold 1)$, $(\bold 2)$, and $(\bold 4_a)$ of Definition~4.2. Then, the condition
$(\bold V_f)$ implies $(\bold V_b)$
in Proposition~6.6.}

\noindent PROOF: The proof of Proposition~18.6 also constitutes
a proof for Proposition~18.7 with the following minor modifications.
In the proof of Proposition~18.6 we replace
$X^{(l)}(y)\in L^1(\bR^+)$ with $X^{(l)}(y)\in L^2(\bR^+),$
replace the reference to Proposition~15.7 by
the reference to Proposition~15.6,
replace the reference to Proposition~18.4(b) by
the reference to Proposition~18.5(b), and replace
the mention of $\hat X(k)\in \hat L^1(\bCp)$ by the
mention of $\hat X(k)\in \bold H^2(\bCp).$
Hence,
$(\bold V_f)$ implies $(\bold V_b)$. \qed

The following results shows the equivalence of $(\bold V_c)$ and
$(\bold V_f)$ when
the scattering data set $\bold S$ satisfies all the four conditions
$(\bold 1)$, $(\bold 2)$, $(\bold 3_a)$, and $(\bold 4_a)$ of Definition~4.5.

\noindent {\bf Proposition 18.8} {\it Let $\bold S$ in (4.2) be the
scattering data set consisting of a scattering matrix $S(k),$
the constants $\kappa_j$ as distinct positive numbers, and
the matrices $M_j$ as $n\times n$ nonnegative hermitian
matrices. Assume that $\bold S$ satisfies the conditions
$(\bold 1)$, $(\bold 2)$, $(\bold 3_a)$, and $(\bold 4_a)$ of Definition~4.5. Then, the properties $(\bold V_c)$ and
$(\bold V_f)$ of Definition~4.3 are equivalent.}

\noindent PROOF: The general solution $\hat X(k)\in \bold H^2(\bCp)$
to (4.23)
given in (15.59) and the general solution
$\hat X(k)\in L^1(\bCp)$ to (4.23) given in (15.60) are identical.
As argued in Propositions~15.8 and 15.9 they each contain
$\Cal N$ linearly independent row vector solutions. Thus, $(\bold V_c)$ and
$(\bold V_f)$ are equivalent. \qed

\newpage
\noindent {\bf 19. INVERSE PROBLEM BY USING ONLY THE SCATTERING MATRIX}
\vskip 3 pt

In this chapter we assume that our scattering data set
$\bold S$ given in (4.2) does not contain any information
on the bound states and it consists of the scattering
matrix $S(k)$ alone. We then investigate whether we can supplement
$S(k)$ with some appropriate bound-state data so that
the resulting $\bold S$ becomes a scattering data set for
some input data set $\bold D$ as in (4.1) belonging to
the Faddeev class.

The following result shows that, if the scattering matrix
$S(k)$ satisfies $(\bold I)$ of Definition~4.3, then we can always find some
bound-state data set so that the resulting $\bold S$ satisfies
also $(\bold 4_c)$ and $(\bold V_f)$ of Theorem~7.2. The analogous result
in the Dirichlet case is given in Corollary to Lemma~5.6.1 of [2].

\noindent {\bf Proposition 19.1} {\it If the scattering matrix
$S(k)$ satisfies $(\bold I)$ of Definition~4.3,
then there exists at least one bound-state data set $\{\kappa_l,M_l\}_{l=1}^p$
in such a way that the resulting scattering data set
$\{S,\{\kappa_l,M_l\}_{l=1}^p\}$ satisfies $(\bold 4_c)$ and $(\bold V_f)$
of Theorem~7.2. Here $p$ is either zero, in which case
the resulting scattering data set consists of $S(k)$ alone, or
$p$ is a positive integer
in such a way that the $\kappa_l$ are distinct
positive numbers and the
$n\times n$ matrices $M_l$ are each
nonnegative, hermitian, and of rank one.}

\noindent PROOF: Consider the vector space of
row vectors with $n$ components as functions
in $x$ belonging to $L^2(\bR^+).$ For any arbitrary positive
integer $p,$ consider a $p$-dimensional subspace
of the aforementioned vector space, which is spanned
by $p$ linearly independent vector-valued functions
in $x\in\bR^+.$ As indicated in Lemma~5.6.1 of [2], one
can explicitly construct at least one set of $p$
nonnegative, hermitian matrices $M_l,$ each of which
is an $n\times n$ constant matrix of rank one, and
$p$ distinct positive numbers $\kappa_l$ in such a way that
if $X(y)$ is any vector in the aforementioned
$p$-dimensional subspace and if we have
$$\int_0^\infty dy\,X(y)\,M_l\, e^{-\kappa_l y}=0,
\qquad l=1,\dots,p,\tag 19.1$$
then $X(y)\equiv 0.$ Note that the sum of
the ranks of $M_l$ is $p$ because each
$M_l$ has rank one.
Given $S(k)$ satisfying $(\bold I)$ of Definition~4.3, consider the integral
equation (4.22), where $F_s(y)$ is constructed as in
(4.7). Since $(\bold I)$ holds, by Proposition~4.4 we know that
$F_s(y)$ belongs to
$L^2(\bR^+).$ By Proposition~3.4(a),
the operator associated with
(4.22) is compact on $L^2(\bR^+).$ Any solution
to (4.22) in $L^2(\bR^+)$ must be an eigenfunction
of that operator with eigenvalue $-1.$ From Theorem~6.26 on p. 185 of [27] it follows that
any nonzero eigenvalue of a compact operator on a Hilbert space
must have finite multiplicity. Thus, the number of
linearly independent solutions in $L^2(\bR^+)$ to (4.22)
must be some finite nonzero integer $p.$
Thus, the solution space for (4.22) is a $p$-dimensional
subspace of row vectors with $n$ components that belong to
$L^2(\bR^+).$ If $p=0,$ then we assume that
there are no bound states and hence (4.22) becomes the same as
(4.14), as seen from (4.12). By Proposition~15.7(c) we know that
any solution in $L^1(\bR^+)$ to (4.14) must be bounded
and hence must belong to $L^2(\bR^+).$
Thus, if the only solution in $L^2(\bR^+)$ to (4.14)
is the trivial solution then the only solution
in $L^1(\bR^+)$ to (4.22) must also be the trivial
solution.
Then, $(\bold V_f)$ of Theorem~7.2 and $(\bold 4_c)$ are both satisfied
because the only solutions to (4.14) and (4.22) in $L^2(\bR^+)$
are $X(y)\equiv 0.$ If $p\ge 1,$ then (4.22) has $p$
linearly independent solutions in $L^2(\bR^+).$ Analogous to
(4.12), let us set
$$F(y)=F_s(y)+\sum_{l=1}^p M_l^2\,e^{-\kappa_l y},\qquad y\in\bR^+,\tag 19.2$$
where we recall that the sum of the ranks of
$M_l$ is equal to $p.$ Then, using (19.2) we see that
the left-hand side of (4.14) satisfies
$$\aligned
X(y)+\int_0^\infty dz\,X(z)&\,F(z+y)\\
=&X(y)+\int_0^\infty dz\,X(z)\,F_s(z+y)\\
&+
\sum_{l=1}^p \left[\int_0^\infty dz\, X(z)\,M_l\, e^{-\kappa_l z}
\right] M_l\, e^{-\kappa_l y},\qquad y\in\bR^+.\endaligned\tag 19.3$$
 From (19.3) and the conclusion associated with (19.1)
 we observe the following:
Any solution in $L^2(\bR^+)$ to (4.22) satisfying (19.1) must be a solution in $L^2(\bR^+)$ to
(4.14), and conversely any solution in $L^2(\bR^+)$ to (4.14) can be expressed
as a solution in $L^2(\bR^+)$ to (4.22) satisfying (19.1).
Note that, by Proposition~15.3(d), $X(y)$ satisfies (19.3)
if any only if its Fourier transform $\hat X(k)$ given in
(3.68) satisfies both lines of (4.15). The first line of (4.15)
implies that (19.1) is satisfied by $X(y).$
By Proposition~15.7(a), the second line of (4.15) implies, after using (3.67), that $X(y)$ satisfies (4.22).
Then, $X(y)$ satisfies (19.1) and (4.22), and furthermore it
belongs to $L^2(\bR^+)$ because it is in $L^1(\bR^+)\cap L^2(\bR^+)$ by
Proposition~15.1(c).
 By the construction of $\{\kappa_l,M_l\}_{l=1}^p,$ we know that
any solution $X(y)$ in $L^2(\bR^+)$ to (4.22) satisfying
(19.1) must be the trivial solution $X(y)\equiv 0.$
Thus, the scattering data set $\{S,\{\kappa_l,M_l\}_{l=1}^p\}$ satisfies
$(\bold 4_c)$ of Theorem~7.2. By construction, the number of
linearly independent solutions in $L^2(\bR^+)$ to (4.22) is
equal to $p,$ and hence $(\bold V_f)$ of Theorem~7.1 also holds. \qed

With the help of Proposition~19.1 we obtain the following result.

\noindent {\bf Proposition 19.2} {\it The properties $(\bold 1)$, $(\bold 2)$, and $(\bold{III}_a)$ are necessary and sufficient for an $n\times n$ scattering matrix $S(k)$
to be the scattering matrix for some input data set
$\bold D$ as in (4.1) belonging to the Faddeev class.}

\noindent PROOF: By Theorem~7.1 we know that there exists a unique input data set
$\bold D$ in the Faddeev class if the scattering data set $\bold S$ satisfies
$(\bold 1)$, $(\bold 2)$, $(\bold{III}_a)$, $(\bold 4_c)$, and $(\bold V_c)$. If the scattering matrix $S(k)$ in
$\bold S$ satisfies $(\bold 1)$, then
$(\bold I)$ of Definition~4.3
is automatically satisfied. Then, as Proposition~19.1 implies we can always complete
$S(k)$ to a scattering data set $\bold S$ satisfying $(\bold 4_c)$ and $(\bold V_c)$
by adding some bound-state data set $\{\kappa_l,M_l\}_{l=1}^p$
for some nonnegative integer $p,$ and such a procedure certainly does not
affect $(\bold 1)$, $(\bold 2)$, $(\bold{III}_a)$. Thus,
the resulting scattering data set
$\{S,\{\kappa_l,M_l\}_{l=1}^p\}$ satisfies all of
$(\bold 1)$, $(\bold 2)$, $(\bold{III}_a)$, $(\bold 4_c)$, $(\bold V_c)$ of Theorem~7.1,
and hence it corresponds to an input data set $\bold D$ in the Faddeev class.
\qed

The following result is related to the
characterization stated in Theorem~7.9, and it states
that unless a scattering matrix $S(k)$ satisfies Levinson's theorem,
it is impossible to
supplement it with any bound-state data set
so that the resulting scattering data set corresponds to an input
data set $\bold D$ in the Faddeev class.

\noindent {\bf Corollary 19.3} {\it Consider a scattering data set $\bold S$
as in (4.2), which consists of
an $n\times n$ scattering matrix $S(k)$ for $k\in\bR,$ a set of $N$ distinct
 positive constants $\kappa_j,$ and a set of
$N$ constant $n\times n$ hermitian and nonnegative matrices
$M_j$ with respective positive ranks $m_j,$ where $N$ is a nonnegative integer.
If $\Cal N$ defined in (4.3) does not satisfy (21.5), i.e.
if the scattering matrix does not satisfy Levinson's theorem,
then as
stated in Theorem~7.9, the associated
scattering data set $\bold S$ cannot belong to the Marchenko
class and hence
at least one of the conditions in any of
Theorems~5.1, 7.1-7.6 must be
violated.}

Based on an intrinsic property of
the scattering matrix $S(k),$ in some cases we might be able
to conclude that it is impossible that such a scattering matrix
may correspond to an input data set
$\Cal D$ in the Faddeev class. The following result directly follows from
Corollary~19.3.

\noindent {\bf Corollary 19.4} {\it Consider an $n\times n$ scattering matrix
$S(k)$ satisfying $(\bold I)$ of Definition~4.3
and $(\bold 2)$ of Theorem~7.1. Consider (21.5) related to
$S(k),$ where each term except $\Cal N$ is uniquely determined
by $S(k).$ If the value of
$\Cal N$ obtained by solving (21.5) algebraically is negative,
then the scattering matrix $S(k)$ cannot be a scattering matrix for some input data set $\bold D$ as in (4.1) belonging to the Faddeev class.}

\newpage
\noindent {\bf 20. PARSEVAL'S EQUALITY}
\vskip 3 pt

Parseval's equality
is the completeness relation for the physical solutions $\Psi(k,x)$
appearing in (9.4)
and the normalized bound-state solutions $\Psi_j(x)$ appearing
in (9.8), and it is formulated as
$$\ds\frac{1}{2\pi}\int_0^\infty dk\,\Psi(k,x)\,\Psi(k,y)^\dagger+\ds\sum_{j=1}^N
\Psi_j(x)\,\Psi_j(y)^\dagger=\delta(x-y)\,I,\qquad x,y\in\bR^+.\tag 20.1$$
In this chapter we prove that if the scattering data set
$\bold S$ given in (4.2)
satisfies $(\bold I)$ of Definition~4.3 and
$(\bold 2)$ of Definition~4.2
 then Parseval's equality holds.
In this chapter, we also show that
if Parseval's equality holds then
the Marchenko integral equation (13.1) is uniquely solvable.

The following proposition is needed in the proof of Parseval's equality.
Even though the physical solution $\Psi(k,x)$ appearing in (9.4) and
(9.6) is not an even function in $k\in\bR,$ the following result shows that
$\Psi(k,x)\,\Psi(k,y)^\dagger$ is even in $k$ for $k\in\bR.$

\noindent {\bf Proposition 20.1} {\it Consider a scattering data set $\bold S$
as in (4.2), which consists of
an $n\times n$ scattering matrix $S(k)$ for $k\in\bR,$ a set of $N$ distinct
 positive constants $\kappa_j,$ and a set of
$N$ constant $n\times n$ hermitian and nonnegative matrices
$M_j$ with respective positive ranks $m_j,$ where $N$ is a nonnegative integer.
Assume that $\bold S$ satisfies $(\bold I)$ of Definition~4.3.
Then, for any $x,y\in\bR^+,$ the
physical solution $\Psi(k,x)$ constructed from $\bold S$ as in
step (9.4) satisfies}
$$\Psi(-k,x)\,\Psi(-k,y)^\dagger=\Psi(k,x)\,\Psi(k,y)^\dagger,
\qquad k\in\bR.\tag 20.2$$

\noindent PROOF: The physical solution is constructed from
$f(k,x)$ and $S(k)$ via (9.4). In turn, $f(k,x)$ is constructed
 from the solution $K(x,y)$ to the Marchenko equation
(13.1) as in (10.6). Proposition~16.1(a) assures the existence
and the uniqueness of $K(x,y)$ for each $x\in\bR^+$
if $(\bold I)$ holds. Using
(9.4) on the right-hand side of (20.2) and simplifying
the result with the help of (4.4), we observe that
the right-hand side of (20.2) is an even function of
$k\in\bR.$ \qed

In the following theorem we show that
Parseval's equality holds if
the Marchenko equation (13.1) is uniquely solvable.

\noindent {\bf Proposition 20.2} {\it Consider a scattering data set $\bold S$
as in (4.2), which consists of
an $n\times n$ scattering matrix $S(k)$ for $k\in\bR,$ a set of $N$ distinct
 positive constants $\kappa_j,$ and a set of
$N$ constant $n\times n$ hermitian and nonnegative matrices
$M_j$ with respective positive ranks $m_j,$ where $N$ is a nonnegative integer.
Assume that $\bold S$ satisfies $(\bold I)$ of Definition~4.3.
Let $\Psi(k,x)$ be the corresponding physical solutions
constructed from $\bold S$ as in (9.4)
and $\Psi_j(x)$ be the normalized bound-state solutions constructed
as in (9.8). Then,
Parseval's relation (20.1) holds for $x,y\in\bR^+.$}

\noindent PROOF:
Because of (20.2) the integral in (20.1) can be
written in the equivalent form as
$$\ds\frac{1}{2\pi}\int_0^\infty dk\,\Psi(k,x)\,\Psi(k,y)^\dagger
=\ds\frac{1}{4\pi}\int_{-\infty}^\infty dk\,\Psi(k,x)\,\Psi(k,y)^\dagger.
\tag 20.3$$
 From (9.4) we obtain
$$\aligned
\Psi(k,x)\,\Psi(k,y)^\dagger=&
f(-k,x)\,f(-k,y)^\dagger+f(-k,x)\,S(k)^\dagger \, f(k,y)^\dagger\\
\stretch
&+f(k,x)\,S(k)\,f(-k,y)^\dagger+f(k,x)\,f(k,y)^\dagger,\endaligned
\tag 20.4$$
where we have used the unitarity of $S(k)$ expressed in the
second equality in (4.4).
With the help of (20.4) and the first equality in (4.4),
we obtain
$$\ds\frac{1}{4\pi}\int_{-\infty}^\infty dk\,\Psi(k,x)\,\Psi(k,y)^\dagger=
\ds\frac{1}{2\pi}\int_{-\infty}^\infty dk\,\left[f(k,x)\,f(k,y)^\dagger
+f(k,x)\,S(k)\,f(-k,y)^\dagger
\right].\tag 20.5$$
Let us replace $S(k)$ in (20.5) by
the sum of $S_\infty$ and $[S(k)-S_\infty],$ where
$S_\infty$ is the constant $n\times n$ matrix given in (4.6).
Furthermore, let us use (4.7) and
(10.6) so that we can write the right-hand side of (20.5)
in terms of $F_s(y)$ and $K(x,y).$
As indicated in Proposition~16.1(a) the existence and uniqueness
of $K(x,y)$ for $x,y\in\bR^+$ is assured by $(\bold I)$ of
Definition~4.3. With the help of
$$\aligned
\ds\frac{1}{4\pi}\int_{-\infty}^\infty dk\,\Psi(k,x)&\,\Psi(k,y)^\dagger\\
\stretch
=&
\delta(x-y)\,I+K(y,x)^\dagger+K(x,y)+
S_\infty \,\delta(x+y)\\
\stretch
&
+S_\infty\,K(y,-x)^\dagger
+K(x,-y)\,S_\infty+F_s(x+y)\\
\stretch
&
+\int_{-\infty}^\infty d\xi\, K(x,\xi)\,K(y,\xi)^\dagger
+\int_{-\infty}^\infty dz\, K(x,z)\,F_s(z+y)
\\
\stretch
&
+\int_{-\infty}^\infty d\xi\, F_s(x+\xi)\,K(y,\xi)^\dagger
+\int_{-\infty}^\infty d\xi\, K(x,\xi)\,S_\infty\,K(y,-\xi)^\dagger\\
\stretch
&
+\int_{-\infty}^\infty d\xi \int_{-\infty}^\infty dz\, K(x,z)\,F_s(z+\xi)\,K(y,\xi)^\dagger,
\endaligned\tag 20.6$$
where we have written the integration limits over
$\bR$ because, in the Marchenko equation (13.1), we know that $K(x,y)=0$ for $y<x.$
We will consider the case $0<x\le y$ and
and the case $0<y<x$ separately.
When we have $0<x\le y,$ we see that the second,
fourth, fifth, sixth, and eleventh
terms on the right-hand side in (20.6) vanish. We can group
the third, seventh, and ninth terms into one group
and the eighth, tenth, and twelfth terms into another group.
Then, from (20.6) we obtain
$$\aligned
\ds\frac{1}{4\pi}\int_{-\infty}^\infty dk\,\Psi(k,x)&\,\Psi(k,y)^\dagger
-\delta(x-y)\, I
\\
\stretch
=&
\left[
K(x,y)
+F_s(x+y)+\int_x^\infty dz\, K(x,z)\,F_s(z+y)\right]\\
\stretch
&
+\int_y^\infty d\xi\,\left[ K(x,\xi)
+F_s(x+\xi)+\int_x^\infty dz\, K(x,z)\,F_s(z+\xi)\right]\,K(y,\xi)^\dagger
,\endaligned\tag 20.7$$
where we have used
$K(x,y)=0$ for $y<x.$
Let us now consider the summation term in (20.1).
With the help of (9.8) we get
$$\Psi_j(x)\,\Psi_j(y)^\dagger=f(i\kappa_j,x)\,M_j^2\, f(i\kappa_j,y)^\dagger,
\tag 20.8$$
where we have used $M_j^\dagger=M_j.$
Using (10.6) on the right-hand side
of (20.8), we obtain
$$\aligned\Psi_j(x)&\,\Psi_j(y)^\dagger=
M_j^2 e^{-\kappa_j(x+y)}+\int_x^\infty dz\, K(x,z)\,M_j^2 e^{-\kappa_j(z+y)}
\\
\stretch
&
+\int_x^\infty d\xi\, M_j^2\,K(y,\xi)^\dagger\, e^{-\kappa_j(x+\xi)}
+\int_y^\infty d\xi \int_x^\infty dz\,
K(x,z)\, M_j^2\,K(y,\xi)^\dagger\, e^{-\kappa_j(z+\xi)}\endaligned\tag 20.9$$
Applying the summation $\sum_{j=1}^N$ on both sides of (20.9), we can add the resulting equation to (20.7).
With the help of (4.12), we can combine the resulting four
summation terms with the four terms on the right-hand side of (20.7).
This yields
$$\aligned
\ds\frac{1}{2\pi}\int_0^\infty dk\,\Psi(k,x)\,\Psi(k,y)^\dagger
+&\sum_{j=1}^N \Psi_j(x)\,\Psi_j(y)^\dagger
-\delta(x-y)\, I\\
\stretch
&=
P(x,y)
+\int_y^\infty d\xi\, P(x,\xi)\,K(y,\xi)^\dagger
,\endaligned\tag 20.10$$
where we have used (20.3) in the first term on the left-hand side
and have defined
$$P(x,y):=K(x,y)+F(x+y)+\int_x^\infty dz\,
K(x,z)\, F(z+y).\tag 20.11$$
Since $K(x,y)$ satisfies the Marchenko equation (13.1) for $0<x\le y,$
comparing (13.1) and (20.11) we see that
$P(x,y)=0$ for $0<x\le y,$ and hence (20.11) yields (20.1).
Let us now consider the case $0<y<x.$
In this case, we see that $K(x,y)=0$ for $x>y$ implies that
the third, fourth, fifth, sixth, and eleventh terms on the
right-hand side of (20.6) vanish. We can group the second, seventh,
and tenth terms into one group and the eighth, ninth, and twelfth
terms into another group. Then, from (20.6), instead of (20.7), we get
$$\aligned
\ds\frac{1}{4\pi}\int_{-\infty}^\infty dk\,\Psi(k,x)&\,\Psi(k,y)^\dagger
-\delta(x-y)\,I
\\
\stretch
=&
\left[
K(y,x)
+F_s(y+x)+\int_y^\infty d\xi\, K(y,\xi)\,F_s(\xi+x)\right]^\dagger\\
\stretch
&
+\int_y^\infty dz\,K(x,z)\left[ K(y,z)
+F_s(y+z)+\int_y^\infty d\xi\, K(y,\xi)\,F_s(\xi+z)\right]^\dagger
,\endaligned\tag 20.12$$
where we have used $K(x,y)=0$ for $x>y$
and also used
$F_s(y)^\dagger=F_s(y),$ where the hermitian property
of $F_s(y)$ follows
 from $(\bold I)$ as indicated in Proposition~16.4(b).
Proceeding as in the previous case $0<x\le y,$
after (20.8) and (20.9), instead of (20.10) we get
$$\aligned
\ds\frac{1}{2\pi}\int_0^\infty dk\,\Psi(k,x)\,\Psi(k,y)^\dagger
+&\sum_{j=1}^N \Psi_j(x)\,\Psi_j(y)^\dagger
-\delta(x-y)\, I\\
\stretch
&=
P(y,x)^\dagger
+\int_y^\infty dz\, K(x,z)\,P(y,z)^\dagger
,\endaligned\tag 20.13$$
where $P(x,y)$ is the quantity defined in (20.11).
In this case we have $0<y<x,$ and hence the Marchenko equation
(13.1) yields $P(y,x)=0$ for $0<y<x.$
Thus, the right-hand side of (20.10) vanishes
for $0<y<x.$ Hence, the proof is completed. \qed

In the next proposition we show that the Marchenko equation (13.1) is satisfied
if Parseval's equality (20.1) holds.

\noindent {\bf Proposition 20.3} {\it Consider a scattering data set $\bold S$
as in (4.2), which consists of
an $n\times n$ scattering matrix $S(k)$ for $k\in\bR,$ a set of $N$ distinct
 positive constants $\kappa_j,$ and a set of
$N$ constant $n\times n$ hermitian and nonnegative matrices
$M_j$ with respective positive ranks $m_j,$ where $N$ is a nonnegative integer.
Assume that $\bold S$ satisfies $(\bold I)$ of Definition~4.3
Let $\Psi(k,x)$ be the corresponding physical solutions
constructed from $\bold S$ as in (9.4)
and $\Psi_j(x)$ be the normalized bound-state solutions constructed
as in (9.8). Assume further that
Parseval's equality (20.1) holds for $x,y\in\bR^+.$
Then, for each $x\in\bR^+,$ the Marchenko equation (13.1) is
uniquely solvable for $y>x>0$ where the solution
$K(x,y)$ in $y$ belongs to $L^1(x,+\infty).$}

\noindent PROOF: Let us look for a solution
$K(x,y)$ in $y$
in $L^1(x,+\infty)$ to
(13.1) for $0<x<y.$ Thus, we can let $K(x,y)=0$ for $y<x.$

If Parseval's equality (20.1) holds for $x,y\in\bR^+,$ then the left-hand side
of (20.10) vanishes for $0<x<y,$ yielding
$$P(x,y)+\int_y^\infty d\xi\, P(x,\xi)\,K(y,\xi)^\dagger=0,\qquad 0<x<y.
\tag 20.14$$
Let us view (20.14) as a homogeneous integral equation
where $P(x,y)$ is the unknown in
$L^1(x<y<+\infty)$ for each $x>0$ and $K(y,\xi)^\dagger$ is the kernel
of the integral operator.
Since $K(x,y)$ is considered in $y$ to belong to $L^1(x,+\infty),$
 from Proposition~3.3(a), we conclude that
the integral operator associated with
(20.14) is compact on $L^1(x,+\infty).$
We notice that
(20.14) is a homogeneous Volterra integral equation.
Thus, it is uniquely solvable and its solution is given by
$P(x,y)=0$ for $0<x<y.$ Then, from (20.11)
we see that $K(x,y)$ must satisfy the Marchenko equation (13.1).
Since $(\bold I)$ of Definition~4.3 holds, from Proposition~16.1(a) it follows that
(13.1) must have a unique solution, and hence we know that
$K(x,y)$ must be the unique solution to (13.1). Thus, the proof is completed.
\qed

\newpage
\noindent {\bf 21. ALTERNATE CHARACTERIZATION VIA LEVINSON'S THEOREM}
\vskip 3 pt

In this chapter we give a new characterization where we use Levinson's theorem.  Let us first recall this theorem.

 For $ \theta_j \in (0,\pi]$ with $ j=1,2,\dots, n,$ let us denote
$$
 \tilde{A}:= -\text{\rm diag}\{ \sin\theta_1,\cdots, \sin\theta_n\}, \quad \tilde{B}:= \{\cos\theta_1, \cdots, \cos \theta_n \}.
 \tag 21.1$$
The matrix pair $(\tilde{A},\tilde{B})$ satisfy (2.5) and (2.6), and the boundary conditions (2.4) for these matrices is given by
$$
 \left(\cos\theta_j\right) \, \psi_j(0) +\left(\sin\theta_j\right) \, \psi'_j(0) =0, \qquad j=1,\dots n.
\tag 21.2$$
 The particular case $\theta_j=\pi$ corresponds to the Dirichlet boundary condition, and the case $\theta_j= \pi/2$ corresponds to the Neumann boundary condition. In  general there will be $n_D$ values with $\theta_j= \pi$, $n_N$ values with $\theta_j= \pi/2$, and then, $n_M=n- n_D-n_N$ values with $ \theta_j \in (0, \pi/2)\cup (\pi/2, \pi)$, that correspond to mixed boundary conditions.

 Let $(A,B)$ be any matrix pair satisfying (2.5) and (2.6). It is proven in Proposition~4.3 of [9] that there exists  a matrix pair  $(\tilde{A}, \tilde{B})$ as in (21.1),
a unitary matrix $U,$ and invertible matrices $T_1$ and $T_2$   such that
$$
A= U\,  \tilde{A}\, T_1\, U^\dagger T_2, \quad B=  U\,  \tilde{A}\, T_1\, U^\dagger T_2.
\tag 21.3$$
Furthermore, in Proposition~4.1 of [9] it is proven that under the transformation $(A,B) \mapsto (\tilde{A}, \tilde{B})$ with $(\tilde{A},\tilde{B})$ as in (21.1) and $U,$ $T_1,$ and $T_2$ as in (21.3), we have
the Jost matrix and the scattering matrix transforming, respectively, as
$$
J(k)=     U \,   \tilde{J}(k)\, T_1\, U^\dagger T_2 , \quad  S(k)=  U\, \tilde{S}(k)\, U^\dagger,
\tag 21.4$$
where $\tilde{J}(k)$ and $\tilde{S}(k)$
 denote the Jost matrix and the scattering matrix, respectively, for the potential $\tilde{V}(x):= U^\dagger \,V(x)\,  U$ and the boundary
conditions (2.4) with the matrix pair $(\tilde{A},\tilde{B})$ instead of
the pair $(A,B)$. Note that $\tilde{V}(x)$ satisfies (2.2) and (2.3) because $U$ is unitary and $V(x)$ satisfies (2.2) and (2.3).  This means that our problem with the general boundary condition specified by (2.4)-(2.6) with  the matrix
 pair $(A,B)$  is actually equivalent to a problem with the boundary conditions (21.2) given by the diagonal matrix pair $(\tilde{A},\tilde{B})$   and the potential $\tilde{V}(x):= U^\dagger\, V(x)\,  U.$
Note that the transformation $(A,B) \mapsto (A\, T,B\, T)$ with an invertible matrix $T$ is just a reparametrization of the boundary condition and the transformation $ V(x) \mapsto U\, V(x)\, U^\dagger$ with a unitary matrix $U$ is a change of representation in the quantum mechanical sense.

We present Levinson's theorem next.

\noindent {\bf Theorem 21.1} {\it
  Suppose that the input data set $\bold D := \{ V,A,B  \}$ belongs to the Faddeev class. Then, the number $\Cal N$ of bound states (including multiplicities) is related to the argument of the determinant of the scattering matrix as}
$$\arg[\det S(0^+)]-\arg[\det S(+\infty)]=\pi\left[ 2\Cal N+\mu-(n-n_{\text{\rm D}})
\right],\tag 21.5$$
{\it  where $\mu$ is the algebraic (and geometric) multiplicity of the eigenvalue $+1$  of the zero-energy scattering matrix $S(0),$ and $n_D$ is the number defined after
(21.2), namely  it is the number of Dirichlet boundary conditions in the representation where the boundary conditions are  given  as in (21.2) by matrices $\tilde{A},
  \tilde{B}$ that satisfy (21.1) and (21.3).   The quantity $n_D$ is also equal to the algebraic (and geometric) multiplicity of the eigenvalue $-1$  of the hermitian matrix $S_\infty$ defined in (4.6). Furthermore, the number $\Cal N$ of bound states (including multiplicities) is equal to the sum of the ranks $m_j$ of the matrices $M_j$ for $j=1,\dots,N$ that
   appear in the definition (4.2) of the unique data set $\bold S$ that corresponds to the input data set $\bold  D$, according to Theorem 5.1,   $ \Cal N= \sum_{j=1}^N m_j.$}

  \noindent PROOF: This result is proven in Theorem~9.3 of [9]. \qed

  Levinson's theorem is a remarkable result, linking scattering information encoded in the scattering matrix to bound state  information.

Next we show that the three properties $(\bold 4_{c,2}),$
 $(\bold 4_{d,2}),$ and $(\bold 4_{e,2})$ of Definition~7.8 are equivalent.

\noindent {\bf Proposition 21.2} {\it Consider a scattering data set $\bold S $
as in (4.2), which consists of
an $n\times n$ scattering matrix $S(k)$ for $k\in \bR,$ a set of $N$ distinct
 positive constants $\kappa_j,$ and a set of
$N$ constant $n\times n$ nonnegative, hermitian matrices
$M_j$ with respective positive ranks $m_j,$ where $N$ is a nonnegative integer.
Assume that $\bold S$ satisfies $(\bold  I)$ of Definition~4.3. Then, the three conditions $(\bold 4_{c,2}),$
 $(\bold 4_{d,2}),$ and $(\bold 4_{e,2})$ of Definition~7.8 are equivalent.
}

\noindent  PROOF: The proposition is proved the same way as in (c) and (d) of Proposition~15.3. Note that the proof of Proposition~15.3(d) amounts to reduce the problem to the case of $L^2(\bR^+)$ and of $\bold H^2(\bCp),$ which we consider here. \qed

Our main goal in this chapter is to prove the characterization
results stated in Theorems~7.9 and 7.10.
The strategy of the proof of Theorem~7.9 is to prove that all the conditions of Theorem~7.6 are satisfied. Then, the proof of
   Theorem~7.10 would follow as a result of the equivalence
   indicated in Proposition~21.2.

   Before we prove Theorem~7.9 we first obtain
certain preliminary results that we need.
As in Lemma~1 on p. 265 of [2], the Corollary and Lemma~2 on page 268 of [2], we prove that if conditions $(\bold 1)$ and  $(\bold 2)$ hold then there is a
nonnegative integer $q$ such that  the matrix $S_q(k)$ defined by
$$
   S_q(k):= S(k)\, \left(  \frac{k+i}{k-i} \right)^{2q} ,
\tag 21.6$$
  satisfies the conditions $(\bold 1),$ $(\bold 2),$ and $(\bold {III}_a)$ of Proposition~19.2. Then, this proposition implies that there is  a set of $N_q$ distinct positive constants $\kappa_{q,j}$ and a set of $N_q$ nonnegative, hermitian $n \times n$ matrices  $M_{q,j}$ with respective positive ranks $m_{q,j}$,  where $N_q$ is a nonnegative integer, such that  $ \bold S_q=\{S_q(k), \{\kappa_{q,j}, M_{q,j}\}_{j=1}^{N_{q}} \}$ is the scattering data set of a unique input data set $ \Cal D: =\{V_q, A_q,B_q\}$ in the Faddeev class.

\noindent {\bf Remark 21.3} {\it
 Note  that by (21.6) $S_q(0)= S(0)$. Then, the algebraic (and geometric) multiplicity of the eigenvalue
$+1$ of $S_q(0)$ is the number $\mu$ that appears in condition $(\bold L)$ in Definition~7.7.
Furthermore, by (21.6),}
$$
S_{q,\infty}:= \lim_{k \rightarrow \infty} S_q(k)= \lim_{k \rightarrow \infty} S(k) = S_\infty.
\tag 21.7$$
{\it Then,  the algebraic (and geometric) multiplicity of the eigenvalue $-1$ of  the hermitian matrix
   $S_{q,_\infty}$ is the number  $n_D$  that appears   in condition $(\bold L)$ in Definition~7.7.}

   Let us denote by $J_q(k)$ the Jost matrix for $\bold S_q$.    We have
$$
   S_q(k)= - J_q(-k)\, [J_q(k)]^{-1}, \qquad k \in \bold R.
\tag 21.8$$
   By (21.6) and (21.8), we have
$$
   S(k)=  - \left(  \frac{k-i}{k+i} \right)^{2q}\, J_q(-k)\, [J_q(k)]^{-1}, \qquad k \in \bold R.
\tag 21.9$$

\noindent {\bf Definition 21.4} {\it
Let  $h(k)$ be a column vector with $n$ components, that is defined  for $ k \in \bCpb$ in such a way that
it is analytic   in $ k \in \bold C^+$. We say that $h(k)$  is of finite order if for some real number  $ \nu$ and
some constant $C_\nu$ we have}
$$
   |h(k)| \leq C_\nu\, (1+   |k|^\nu),\qquad k\in\bCpb.
\tag 21.10$$
 {\it
The order of $h(k)$ is the infimum of the $\nu$-values such that (21.10) holds for some  constant $C_\nu$. The definition for order also applies
for functions defined for $k\in\bCmb,$ analytic in $\bCm,$ and (21.10) holds in
$k\in\bCmb.$}

  We now consider the solutions to the following equation that appears in condition $(\bold V_h)$ in Theorem~7.6:
$$
  h(-k)+S(k)\, h(k)=0, \qquad k \in \bold R,
\tag 21.11$$
  where  $h(k)$ is a column vector  that is analytic in $ \bold C^+$  and of finite order.

It follows from Proposition~10.2(a) that   $k\, [J_q(k)]^{-1}$ is continuous in $ k \in \bold R $ and
$$
   J_q(k)=B_q  -ik A_q + O(1), \qquad k \to \infty \text{ in } \bCpb.
\tag 21.12$$
  Then,  using  (21.9) and (21.12) we prove as in Lemma~3 in page 270 of [2] that every solution to (21.11)  with finite order is of the form
$$
  h(k)=  (k+i)^{2q}\,  J_q(k) \, \left[ \sum_{j=1}^{N_q} \ds\frac{2\, i \, \kappa_{q,j}}{k^2+\kappa_{q,j}^2}\,  N_{q,j}\,  d^{(j)} +p(k^2) \right],
\tag 21.13$$
  where $ i \kappa_{q,j}$ with $ j= 1,\dots, N_q$ are the poles of $[J_q(k)]^{-1}$ and $N_{q,j}$ is the residue of   $[J_q(k)]^{-1}$ at $i \kappa_{q,j}$. Moreover, $p(k)$ is a column
 vector with $n$ components that are polynomials in $k$, and $d^{(j)}$ is the column vector with $n$ components given by
$$
  d^{(j)}:= \bm (k+i)^{-2q}\, h_1(i \kappa_{q,j}) &\cdots & (k+i)^{-2q}\, h_n(i \kappa_{q,j})\endbm^T
\tag 21.14$$
where we recall that the superscript $T$ denotes
  the matrix transpose. By (21.12) and (21.13), each  component $h_l(k)$ with $ l=1,\dots,n,$  of a solution of finite order  behaves as $ C_l\, k^{\beta_{l}}$ as  $ |k| \to \infty$, where $C_l$ is a complex number different from zero, and $\beta_{l}$ is an integer. The order, $\beta$,  of the solution $h(k)$ is equal to $ \beta:= \text{max} \{\beta_l: \ l=1,\dots, n\}$. Note that from (21.13)
    it follows that $\beta$ can be an even or an odd integer.

  We have the following result.

\noindent {\bf Proposition 21.5} {\it
Suppose that conditions  $(\bold 1)$ and $ (\bold 2)$ of Definition 4.2 are satisfied. Then, there are $n$ solutions $h^{(j)}(k)$ of finite order for
  $j=1,\dots n$ to (21.11) with the following properties:}

\item {(a)} {\it Each $h^{(j)}(k)$ is of order  $-\beta_j$,  with $ - \beta_1 \geq - \beta_2 \geq \cdots\geq -\beta_n$. Moreover, $ -\beta_1$ is the smallest possible order  of all  solutions to (21.11) with finite order. Note that from
     (21.13) it follows that  $\beta_1$ is finite.}

\item {(b)} {\it For every $j=1,\dots, n,$ the solutions $h^{(j)}(k)$ cannot be represented in the form}
$$
h^{(j)}(k)= \sum_{l=1}^{j-1} \, p_l(k)\,  h^{(l)}(k),
\tag 21.15$$
{\it where the $p_l(k)$ are polynomials.}

\item {(c)} {\it All the  solutions $h(k)$  to (21.11)  with finite order  can be represented as}
$$
h(k)= \sum_{j=1}^n \, p_j(k^2)\, h^{(j)}(k),
\tag 21.16$$
{\it for some polynomials $p_j$ with $ j=1,\dots,n.$
Furthermore, any solution $h(k)$ to (21.11) of order smaller than $ -\beta_l$ can be represented as}
 $$
 h(k)= \sum_{j=1}^{l-1}\, p_j(k^2)\, h^{(j)}(k),\tag 21.17
 $$
{\it for some polynomials $p_j(k)$ with $j=1,\dots, l-1.$}

\noindent PROOF: This proposition is proven as in Appendix I of [2] (see also pages 393--404 of [39]). \qed

By (21.13), each of the $h^{(j)}(k)$ is of the form,
$$
 h^{(j)}(k)=  J_q(k)\, r^{(j)}(k),\qquad  j=1,\dots,n,
\tag 21.18$$
where $r^{(j)}(k)$ is a column  vector with $n$ components that are rational functions of $k$. Let
$$
\omega^{(j)}(k):= k^{\beta_j}\, h^{(j)}(k),\qquad j=1,\dots, n.
\tag 21.19$$
Moreover, we define following  $ n\times n$ matrices:
$$
\bold Z(k) := \left[  z_{i,j}(k)   \right],  \quad z_{i,j}(k):= h^{(j)}_i(k),
 \qquad 1 \leq i,j \leq n,
\tag 21.20$$
$$
\bold \Omega(k):= \left[  \omega_{i,j}(k) \right], \quad \omega_{i,j}(k):= \omega^{(j)}_i(k), \qquad 1 \leq i,j \leq n,
\tag 21.21$$
$$
\bold R(k):= \left[ r_{i,j}(k)\right], \quad r_{i,j}(k):= r^{(j)}_i(k),
\qquad 1 \leq i,j \leq n.
\tag 21.22$$
Then,
$$
\bold Z(-k)+ S(k) \,\bold Z(k)=0, \qquad k \in \bold R,
\tag 21.23$$
and
$$
\bold Z(k)= J_q(k)\, \bold R(k),\qquad k \in \overline{\bold C^+}.
\tag 21.24$$

\noindent {\bf Proposition 21.6} {\it
Suppose that conditions  $(\bold 1)$ and $ (\bold 2)$ of Definition 4.2 are satisfied. Then, the  following statements hold:}

\item {(a)} {\it The determinant of $\bold Z(k)$ is different from zero for all $ k \in \overline{\bold C^+}\setminus\{0\}$.}

\item {(b)} {\it
The following two limits exist:}
$$
 \omega^{(j)}(\infty):= \lim_{k \to \infty}  \omega^{(j)}(k),
\tag 21.25$$
$$\left[\Omega(\infty)\right]= \lim_{k \to \infty}\left[  \Omega(k) \right],
\tag 21.26$$
{\it and the determinant of $\Omega(\infty)$ is different from zero.}

\item {(c)} {\it The order  of}
$$
\sum_{j=1}^n\, p_j(k)\, h^{(j)}(k),
\tag 21.27$$
{\it where the $p_j(k)$ for
$j=1,\dots,n$  are polynomials, is equal to the highest of the orders of the individual terms in the summation in (21.27).}

\item {(d)} {\it The determinant of $\bold R(k)$ is different from zero  for all $ k \in \bCpb.$}

\noindent PROOF: This proposition is proven as in Appendix I of [2] (see also pages
393--404 of [39]).
We give some details for the reader's convenience. Let us prove that (d) holds. Suppose that for some $k_0 \in \overline{\bCp}$ the determinant of $\bold R(k_0)$ is zero. Then, the columns of  $\bold R(k_0)$ are linearly dependent and there are some $\lambda_j$ with $j=1,\dots,n$ that are not all equal to zero,  such that
 $$\sum_{j=1}^n \, \lambda_j \, r^{(j)}(k_0)=0.\tag 21.28$$
 Let
 $$h(k):= \frac{1}{k^2-k_0^2} J_q(k)\,  \sum_{j=1}^n \, \lambda_j\, r^{(j)}(k)= \frac{1}{k^2-k_0^2}\,  \sum_{j=1}^n \, \lambda_j\, h^{(j)}(k).\tag 21.29$$
 By (21.13) and (21.18) we have
$$ h(k)= \frac{1}{k^2-k_0^2}   (k+i)^{2q}\, J_q(k)\, \left( (k^2-k_0^2) Q+ O((k^2-k_0^2)^2\right), \qquad k \to k_0,\tag 21.30$$
 for some constant column vector $Q$. Then, $h(k$) is a solution to (21.11). Let $ \lambda_m$ be the last of the coefficients in (21.29) that is different from zero. Hence, $h(k)$ is a solution to (21.11) of order $( -\beta_m-2)$, and by
 Proposition~21.5(c) it can be represented as
$$ h(k)= \sum_{j=1}^{m-1}\, p_j(k^2)\, h^{(j)}(k).\tag 21.31$$
 By (21.29) and (21.31) we have
$$ h^{(m)}(k)=
 \frac{1}{\lambda_m}\,    \left(\sum_{j=1}^{m-1}\, (k^2-k_0^2) p_j(k^2)\, h^{(j)}(k)-   \sum_{j=1}^{m-1} \, \lambda_j\, h^{(j)}(k)\right).\tag 21.32$$
 However, by Proposition~21.5(b) this is not possible, and then the determinant of $\bold R(k)$ never vanishes. We prove in the same way that the determinant of $\bold Z(k)$ does not vanish for $ k \in \bCp$ using that $\bold Z(k)$ is analytic on $\bCp$. Moreover, by (21.24) the determinant of  $\bold Z(k)$ is different from zero on  $ \bR \setminus\{ 0  \}$, because the determinant of $\bold R(k)$ and of $J_q(k)$ are nonzero on $ \bR \setminus\{ 0  \}$.
 \qed

\noindent {\bf Remark 21.7} {\it
Let  $\tilde{A_q}$ and $\tilde{B_q}$ be the matrices as in (21.1) related to $A_q$
and $ B_q$ as in (21.3), i.e.}
$$\aligned
 \tilde{A_q}:= -\,\text{\rm diag}\,\{ \sin\theta_{q,1},\cdots, \sin\theta_{q,n}\}, \quad \tilde{B_q}:= \,\text{\rm diag}\, &\{\cos\theta_{q,1}, \cdots, \cos \theta_{q,n} \},\\
 & 0 < \theta_{q,j} \leq \pi,\quad  j=1,\dots,n,\endaligned
\tag 21.33$$
{\it where $n_{D}$ of the $\theta_{q,j}$ are equal to $\pi$ and the $(n-n_D)$ remaining  $ \theta_{q,j}$ are different from $\pi$.  We can always reorder the $\theta_{q,j}$ in such a way that the first $n_D$ of the $\theta_{q,j}$ are equal to $ \pi$.
By Theorem 7.6 of [9] the number of $ \theta_{q,j}= \pi$ that appear in (21.33) is equal to the algebraic (and geometric) multiplicity of the eigenvalue equal to $-1$ of the matrix $S_{q,\infty}$, where $S_q(k)$
is the matrix defined in (21.8). By Remark~21.7 this is precisely the number, $n_D$, of eigenvalues $-1$ of the matrix $S_\infty$ that appears in condition $(\bold L)$ of Definition~7.7.
}

\noindent {\bf Proposition 21.8} {\it
Suppose that conditions  $(\bold 1)$ and $ (\bold 2)$ of Definition 4.2 are satisfied. Then, among the
solutions $h^{(j)}(k)$ with $j=1,\dots,n,$
 we have $n_D$ of them having even order   and $n-n_D$ of them having odd order.}

\noindent PROOF:
Let $\tilde{A}_q$ and $\tilde{B}_q$ be the matrices that appear in Remark~21.7. Let $P_1$ and $P_2$ be the diagonal matrices defined as
$$
P_1:=\, \text{\rm diag}\{1,\cdots,1,0,\cdots,0\},\tag 21.34
$$
with the first $n_D$ diagonal entries equal to one and the remaining $(n-n_D)$ diagonal entries equal to zero, and
$$
P_2:= \,\text{\rm diag}\{0,\cdots,0,1,\cdots,1\},\tag 21.35
$$
with the first $n_D$ diagonal entries equal to zero  and the remaining $n-n_D$ diagonal entries equal to one.
 Then,
$$
 \tilde{B}_q- ik \tilde{A}_q= \,\text{\rm diag}\,\left\{-1,-1,\cdots,-1, \cos \theta_{q, n_D+1}+ i k\sin\theta_{q,n_D+2}, \cdots, \cos\theta_{q, n}+ i k\sin\theta_{q,n}       \right\},
\tag 21.36$$
with $-1$ in the first $n_D$ diagonal entries and the remaining ones different from zero.
Hence,
$$
 P_1 ( \tilde{B}_q- ik \tilde{A}_q)   = \text{\rm diag}\,\left\{ -1,\cdots,  -1, 0, \cdots,0 \right\},
\tag 21.37$$
has  $-1$ in the first $n_D$ diagonal entries and $0$ in the remaining $n-n_D$ diagonal entries. Furthermore,
$$
  P_2 (  \tilde{B}_q- ik \tilde{A}_q )  = \text{\rm diag}\,\left\{ 0, \cdots, 0, \cos\theta_{q, n_D+1}+ ik \sin\theta_{q,n_D+1}, \cdots, \cos\theta_{q, n}+ i k\,\sin\theta_{q,n}   \right\},
\tag 21.38$$
has zeros in the first $n_D$ diagonal entries and the remaining $n-n_D$ diagonal entries are nonzero.
By (21.13) we have
$$
  k^\beta_j\, h^{(j)}(k)= k^{\beta_j + 2q} \left(1+ \frac{i}{k}\right)^{2q}\,  J_q(k) \, c^{(j)}(k^2),
\tag 21.39$$
  for some  column vector $c^{j}(k)$ with components that are rational functions.
   By  (21.3) for $A_q, B_q$ and $\tilde{A_q}, \tilde{B}_q,$  and (21.12), we have
$$
 U^\dagger\, J_q(k)  = \left(\tilde{B}_q-ik \tilde{A}_q + O(1)\right)\, T_1 \, U^\dagger\, T_2   , \qquad  k \to \infty \text{ in } \bCpb.
\tag 21.40$$
Hence, by (21.19), (21.25), and (21.37)-(21.40) we get
$$P_1\, U^\dagger\,  \omega^{(j)}(\infty) = \lim_{k \to \infty}\, \left[
 \bold \Theta_1 + O(1) \right] T_1\, U^\dagger\,  T_2\,  k^{\beta_j + 2q} \left(1+ \frac{i}{k}\right)^{2q}\, c^{(j)}(k^2),
\tag 21.41$$
 and,
$$
  P_2\, U^\dagger\,  \omega^{(j)}(\infty) = \lim_{k \to \infty}\left[ \bold \Theta(k)
  + O(1) \right] T_1\, U^\dagger\,  T_2\,  k^{\beta_j + 2q} \left(1+ \frac{i}{k}\right)^{2q}\, c^{(j)}(k^2)
 \tag 21.42$$
where
$$\bold \Theta_1:=\text{\rm diag}\,\left\{ -1,\cdots,  -1, 0, \cdots,0 \right\},$$
$$\bold \Theta(k):=
   \text{\rm diag}\,\left\{ 0, \cdots, 0, \cos\theta_{q, n_D+1}+ ik \sin\theta_{q,n_D+1}, \cdots, \cos\theta_{q, n}+ i k\sin\theta_{q,n}   \right\}.
$$
 Suppose that the order of $h^{(j)}(k)$ is even. Then, the only way in which   the limit on the right-hand side of (21.42) can be finite is if that limit is zero. It follows that
$$
  P_2\, U^\dagger \,  \omega^{(j)}(\infty)   =0,\qquad \text{ if }\beta_j \text{ is even}.
\tag 21.43$$
 Similarly if  the order of $h^{(j)}(k)$ is odd, then the only way in which   the limit on the right-hand side of (21.41) can be finite is if it is zero. As a
 result, we have
$$
 P_1 \,U^\dagger \,  \omega^{(j)}(\infty) =0,\, \qquad \text{ if }\beta_j \text{ is odd}.
\tag 21.44$$
Let us introduce the following subspaces of $\bold C^n:$
$$
 \bold V_{\text{even}}:=\left\{ v\in   \bold C^n :\ v= \sum_{(j:\ \beta_j
   \text{ is even})}  \lambda_j  \, U^\dagger \,\omega^{(j)}(\infty), \quad \lambda_j \in \bold C \right\},
\tag 21.45$$
$$
 \bold V_{\text{\rm  odd}}:=\left\{ v\in   \bold C^n : \ v= \sum_{(j:\ \beta_j  \text{ is odd})}  \lambda_j   \,U^\dagger\,  \omega^{(j)}(\infty),\quad \lambda_j \in \bold C \right\}.
\tag 21.46$$
The subspace  $\bold V_{\text{\rm even}}$ is the subspace of all linear combinations of the $ U^\dagger\, \omega^{(j)}(\infty)$  with $ h^{(j)}(k)$ of even order, and   $\bold V_{\text{\rm odd}}$ is the subspace of all linear combinations of the $U^\dagger\,  \omega^{(j)}(\infty)$  with $ h^{(j)}(k)$ of  odd order.  Let us denote by $n_{\text{\rm even}}$,  the number of $ h^{(j)}(k)$ of even order  and by $n_{\text{\rm odd}}$,  the number of $ h^{(j)}(k)$ of odd order.  As by Proposition~21.6 the determinant of $\Omega(\infty)$ is different from zero, the $ \omega^{(j)}(\infty)$ for $j= 1,\dots,n$ are linearly independent. Then,
 from the unitarity of $U,$ (21.43), and (21.44) we get
$$
\bold V_{\text{even}} \subset  P_1\, \bold C^n, \quad \text{dim} \,[\bold V_{\text{even}}]= n_{\text{even}} \leq   \text{dim}  [P_1\, \bold C^n] =   n_D,
\tag 21.47$$
$$
\bold V_{\text{odd}} \subset  P_2 \,\bold C^n, \quad \text{dim} \,[\bold V_{\text{odd}}]= n_{\text{odd}} \leq   \text{dim}  [P_2\, \bold C^n] = n-  n_D.
\tag 21.48$$
Then,   $\bold V_{\text{even}} \cap \bold V_{\text{odd}}= \{0\}$ , and consequently we have
$$
\text{dim}\left[\bold V_{\text{even}} +  \bold V_{\text{odd}}  \right]= \text{dim}[ \bold V_{\text{even}}]+ \text{dim} [\bold V_{\text{odd}}]
=  n_{\text{even}}+ n_{\text{odd}}.
\tag 21.49$$
Suppose that $n_{\text{even}} < n_D.$ Then, by (21.47) and (21.48) we have
$$
\text{dim}\left[\bold V_{\text{even}} + \bold V_{\text{odd}}  \right] < n_D + n-n_D < n.
\tag 21.50$$
This implies that the vectors $ U^\dagger\, \omega^{(j)}(\infty)$
with $j=1,\dots,n$ generate a subspace of $\bC^n$ of dimension smaller than $n$. This is impossible because as the  $\omega^{(j)}(\infty)$
with $j=1,\dots,n$ are linearly independent,  and  $U$ is unitary, the vectors $ U^\dagger \,\omega^{(j)}(\infty)$ with $j=1,\dots,n$ are linearly independent.  Consequently,
we have $n_{\text{\rm even}}=n_D$ and $n_{\text{\rm odd}}=n-n_D$. \qed

We can now   compute the number of linearly independent solutions  to (21.11)  that have negative order. Let us assume that there are
$d\geq 0$ orders $\beta_j$ that are positive, that is to say
$$
\beta_1 \geq \beta_2 \geq \cdots \geq \beta_d >0  \geq \beta_{d+1} \geq \cdots \geq \beta_n.
\tag 21.51$$

\noindent {\bf Proposition 21.9} {\it
Suppose that conditions  $(\bold 1)$ and $ (\bold 2)$ of Definition 4.2 are satisfied. Denote by $n_{\{\text{\rm odd, +}\}}$ the number of the  $\beta_{j}$-values for
$j=1,\dots,d,$ that are odd. Then, the number, $N_+$, of linearly independent solutions to (21.11)
that are of negative order satisfies}
$$
2 N_+= \sum_{j=1}^{d}  \beta_j +  n_{\{\text{\rm odd, +}\}}.
\tag 21.52$$

\noindent PROOF: It follows from Proposition~21.5(c) and Proposition~21.6(c)
 that all  solutions to (21.11)  of negative order, i.e. of order not exceeding $-1$, are of the form
$$
h(k)= \sum_{j=1}^d p_j(k^2)\, h^{(j)}(k),
\tag 21.53$$
where $p_j(k)$ is a polynomial of degree $\gamma_j$ so that $ 2 \gamma_j < \beta_j$. If $\beta_j$ is even this implies that $ \gamma_j  \leq \frac{\beta_j}{2}-1$. Since a polynomial of order $\frac{\beta_j}{2}-1$ has $\frac{\beta_j}{2}$ independent coefficients, for each  $h^{(j)}(k)$ of even order, $\beta_j,$ there are  $\frac{\beta_j}{2}$ linearly independent solutions to (21.11) of negative order. If $\beta_j$ is odd, the inequality $ 2 \gamma_j < \beta_j$ implies that $ \gamma_j \leq \frac{\beta_j}{2}- \frac{1}{2}$. Then,   for each  $h^{(j)}(k)$ of odd order $\beta_j,$ there are  $\frac{\beta_j}{2}+ \frac{1}{2}$ linearly independent solutions to (21.11) of negative order.
Then, the number, $N_+$, of linearly independent solutions of negative order is given by
$$
\sum_{(j:\ 1\leq j\leq d,\  \beta_j \text{ is even}) } \frac{\beta_j}{2}+ \sum_{(j:\ 1\leq j\leq d,\  \beta_j \text{ is odd}) } \left(\frac{\beta_j}{2}+\frac{1}{2}\right)= \sum_{j=d+1}^n \frac{\beta_j}{2}+ \frac{n_{\{\text{\rm odd}, +\}}}{2}.
\tag 21.54$$
Then, from (21.54) we obtain (21.52). \qed

We now consider the equation that appears in condition $(\bold {III_c})$  in Theorem 7.6, i.e. consider
$$
- h(-k)+ S(k)\, h(k)=0, \qquad  k \in \bold R,
\tag 21.55$$
where $ h(k)$ is a column vector with $n$ components in $ \bold H^2(\bCm)$. Taking the adjoint of (21.23), then taking the inverse, using $ S(k)^\dagger=S(k)^{-1}$
which appears in (4.4), and multiplying by $k$ we obtain
$$
  k  \,\left[\bold Z(-k)^\dagger \right]^{-1} + k \, S(k) \,  \left[\bold Z( k)^\dagger \right]^{-1} \,=0, \qquad k \in \bold R \setminus\{0\}.
\tag 21.56$$
Let us define
$$
\bold Y(k):=  k     \left[\bold Z(  k^\ast)^\dagger \right]^{-1}, \qquad k \in
\bCm \setminus \{0\}.
\tag 21.57$$
Then, by (21.56) we have
$$
-\bold Y(-k)+S(k) \,\bold Y(k) \,S(k)=0, \qquad k \in \bold R \setminus\{0\}.
\tag 21.58$$
Letting
$$
\bold Y(k) := \left[  y_{i,j}  \right], \quad y_{i,j}(k):= y_i^{(j)}(k),    \qquad  1 \leq i,j \leq n,
\tag 21.59$$
we see that the columns $ y_i^{(j)}(k)$ of $\bold Y(k)$ are  solutions of (21.55) analytic for $ k \in \bold C^-$. Recall that Proposition~10.2(a)
  indicates that $k\, J_q(k)^{-1}$ is continuous for $ k \in \bold R. $
  Moreover, from Proposition~21.6(d) and (21.24) we
conclude that $\bold Y(k)$ is continuous at $k=0.$
  We prove as in Appendix I of [2] that $y_i^{(j)}$ are  solutions of  order  equal to  $\beta_j+1$ with $j=1,\dots, n$  that  $\bold Y(k)$ satisfies the analogs of (a) and (b) of Proposition~21.5, and  that any solution to (21.55) of finite order   can be represented as
$$
h(k)= \sum_{j=1}^n \, p_j(k^2)\, y^{(j)}(k),
\tag 21.60$$
for some polynomials $p_j$ with $j=1,\dots,n$, and  the order of
$$
\sum_{j=1}^n\, p_j(k)\, y^{(j)}(k),
\tag 21.61$$
where the $p_j(k)$ for $ j=1,\dots,n$  are polynomials, is equal to the highest of the orders of the individual terms in the sum in (21.61).

\noindent {\bf Proposition 21.10} {\it
Suppose that conditions  $(\bold 1)$ and $ (\bold 2)$ of Definition 4.2 are satisfied.
 Denote by $n_{\{\text{\rm odd, -}\}}$ the number of the $\beta_{j}$ with $ j=d+1,\dots,n$ that are odd. Then, the number, $N_-$, of linearly independent  solutions to (21.56)  that   are of negative order satisfies}
$$
2 N_-= \sum_{j=d+1}^d |\beta_j| - n_{\{\text{\rm odd, -}\}}.
\tag 21.62$$

  \noindent PROOF: Since any solution to (21.55) of finite order   can be represented as  (21.60),  and since the order of  (21.61) is equal to the highest of the orders of the individual terms in the sum, it follows that  any  solution to (21.55)  of negative   order   is of the form
$$
h(k)= \sum_{j=d+1}^n p_j(k^2)\, y^{(j)}(k),
\tag 21.63$$
where $p_j(k)$ is a polynomial of degree, $\gamma_j,$ so that, $ 2 \gamma_j + \beta_j+1 <0$. If $\beta_j$ is even this implies that $ \gamma_j  \leq \frac{|\beta_j|}{2}-1$.
Since a polynomial of order $\frac{|\beta_j|}{2}-1$ has $\frac{|\beta_j|}{2}$ independent coefficients, for each  $y^{(j)}(k)$ of even order, $\beta_j,$ there are  $\frac{|\beta_j|}{2}$ linearly independent solutions to (21.55) of negative order. If $\beta_j$ is  odd,   $ 2 \gamma_j + \beta_j+1 < 0$  implies that $ \gamma_j < \frac{|\beta_j|}{2}- \frac{1}{2}-1$. Then,   for each  $h^{(j)}(k)$ of odd order, $\beta_j,$ there are  $\frac{|\beta_j|}{2}- \frac{1}{2}$ linearly independent solutions to (21.55) of negative  order. Hence, the number, $N_-$, of linearly independent solutions of negative   order is given by,
$$
\sum_{(j: \ d+1  \leq j \leq  n,\  \beta_j \text{ is even}) } \frac{|\beta_j|}{2}+ \sum_{(j: \ d+1  \leq j \leq  n, \ \beta_j\text{ is odd}) } \left(\frac{|\beta_j|}{2}-\frac{1}{2}\right)= \sum_{j=1}^d \frac{|\beta_j|}{2} - \frac{n_{\text{odd, -}\}}}{2}.
\tag 21.64$$
 Equation (21.62) follows from (21.64). \qed

 Let us define the total index, $\beta$,  of $S(k)$ as
$$
 \beta:= \sum_{j=1}^n \beta_j.
\tag 21.65$$
 By  Propositions~21.8, 21.9, and 21.10 and since $ n_{\{\text{\rm odd, +}\}}+ n_{\{\text{\rm odd, -}\}}= n-n_D,$ we have
$$
 \beta = 2N_+- 2  N_-  - (n-n_D) .
\tag 21.66$$
With the help of (21.19)-(21.21), (21.65), and Proposition~21.6(b), we obtain
$$\det[\bold Z(k)]=C\,k^{-\beta}\left[1+o(1)\right],\qquad k\to\infty
\text{ in } \bCpb.\tag 21.67$$

\noindent {\bf Proposition 21.11} {\it
Suppose that conditions  $(\bold 1)$ and $(\bold 2)$ of Definition 4.2 are satisfied. Then:}

 \item {(a)} {\it Every column vector $ h(k)$ that is a solution to (21.11) in $\bold H^2(\bold C^+)$  is of  negative order.}

   \item {(b)} {\it  Every  column vector $ h(k)$ that is a solution to (21.55) in $\bold H^2(\bold C^-)$  is of  negative order.}

 \noindent PROOF: Suppose that $h(k)$ is a solution to (21.11)   in $\bold H^2(\bold C^+)$.
Let $\hat{X}(k):=h(-k^\ast)^\dagger$. Then  $\hat{X}(k)$ is a solution to (4.23)  in $ \bold H^2(\bold C^+)$. Then, by (21.9) we have
$$
\hat{X}(-k)= \left(  \frac{k-i}{k+i} \right)^{2q}\,  \hat{X}(k)\,
J_q(-k)\, J_q^{-1}(k), \qquad k \in \bold R,
\tag 21.68$$
and as in the proof of (15.25)  we obtain
$$
(k-i)^{2q}\,\hat X(k)\, [J(-k^\ast)^\dagger]^{-1}=  (k+i)^{2q} \,\hat X(-k)\, [J(k^\ast)^\dagger]^{-1},
\qquad k\in\bold R\setminus\{0 \}.
\tag 21.69$$
As in (15.27) we let
$$
\Xi(k):=\cases
    (k-i)^{2q}  \left(k\,\hat X(k)\, [J(-k^\ast)^\dagger]^{-1}+ k\,\displaystyle \sum_{j=1}^{N_q} \ds\frac{
2i\kappa_{q,j}\,\hat X(i\kappa_{q,j})\,N_{q,j}^\dagger}{k^2+\kappa_{q,j}^2}\right),\qquad k\in\bCp,\\
(k+i)^{2q} \left(k\,\hat X(-k)\, [J(k^\ast)^\dagger]^{-1}+k\, \displaystyle\sum_{j=1}^{N_q} \displaystyle\frac{
2i\kappa_{q,j}\,\hat X(i\kappa_{q,j})\,N_{q,j}^\dagger}{k^2+\kappa_{q,j}^2}\right)
,\qquad k\in \bCm.\endcases
\tag 21.70$$
As in the proof of Proposition~15.8 we prove that $\Xi(k)$ is an entire odd function and that $\Xi^{(p)}(0)=0$ for $p$ odd, $p\ge 2q-1$. Then, $\Xi(k)$ is a polynomial  order  smaller or equal to $ 2q-1$. Hence,  $\hat{X}(k)$ is of finite order, but since it is in    $\bold H^2(\bold C^+)$  the order has to be negative.   In consequence $\hat{X}(k)$ is of negative order, which proves that $h(k)$ is of  negative order.
Let us now prove (b). Suppose that $h(k)$ is a solution to (21.55) in $ \bold H^2(\bold C^-)$.  Then, by (21.6) and since
  $S_q(k)= S_q(-k)^\dagger$,  which   follows from (4.4),  we have
$$
(-k-i)^{2q}\, J_q(-k)^\dagger \, h(-k)= - (k-i)^{2q} \,J_q(k)^\dagger\, h(k), \qquad k \in \bold R.
\tag 21.71$$
Let us define the sectionally analytic function
$$
g(k):= \cases (-k-i)^{2q}\, J_q(-k^\ast)^\dagger \, h(-k), \qquad  k \in \bCp, \\ \\ - (k-i)^{2q} \,J_q(k^\ast)^\dagger\, h(k), \qquad k \in \bCm. \endcases
\tag 21.72$$
  As in the proof of Proposition~15.8  we prove that $g(k)$ is an odd entire function and that $g^{(p)}(0)=0$ for $p$ odd, $p \ge2q-1$. Then, $g(k)$ is a polynomial  order  smaller or equal to $ 2q-1$. Hence,  $h(k)$ is of finite order, but since it is in    $\bold H^2(\bCm)$  that order has to be negative. \qed

Next, we provide a proof of the characterization stated in Theorem~7.9.

\noindent{\bf Proof of Theorem~7.9:} If  the input data set $\bold D := \{ V,A,B  \}$ belongs to the Faddeev class it is proven in Theorem~7.6 that  conditions  of $(\bold 1)$ and $(\bold 2)$  hold. The continuity of  $S(k)$   in $ k \in \bold R$ is proven in Proposition~10.3(a).  The proof of  Levinson's theorem (21.5) is given in Theorem~9.3 of [9]. It remains to prove that $(\bold 4_{e,2})$ holds. By Proposition~21.2 we can, equivalently, prove that $( \bold 4_{d,2})$ is satisfied. By Theorem~5.1 we know that the conditions of Proposition~15.4 hold. As in the proof of Proposition~15.4(a) we prove that (15.25) holds and we define $\Xi(k)$ as in (15.27). Then, as in Proposition~15.8 we prove that $\Xi(k)\equiv 0$. We complete the proof that $\hat{X}(k)\equiv 0$ as in the proof of Proposition 15.4(a).
We now prove that if conditions  of $(\bold 1),$
 $(\bold 2),$ $(\bold 4_{e,2}),$  and $(\bold L)$
   are satisfied, then $\bold S$ is the scattering data  of a unique input data $ \bold D:=\{V, A, ,B\}$  in the Faddeev class. For this purpose we verify that all the conditions of Theorem~7.6 are satisfied.
Conditions  $(\bold 1)$ and  $(\bold 2)$ hold by assumption.
By Corollary~6.2 of [9], we have
$$
  \det[J_q(k)]= C\, k^\mu \left(1+o(1)\right), \qquad k \rightarrow 0.
\tag 21.73$$
  By Proposition~21.6(d), (21.23), (21.24), (21.67), (21.73), and using contour integration  we prove, as in Appendix I of [2], that
$$\text{arg} [\det[S(0^+)]]-\text{arg}[\det[S(\infty)]]= \pi \, (\beta +\mu ).
\tag 21.74$$
Furthermore, since by assumption $(\bold L)$  Levinson's theorem (21.5) holds, and using (21.65) we obtain
$$
\Cal N= N_+-N_-.
\tag 21.75$$
Recall that   $ N_+$ is the number of linearly independent  solutions  to (21.11)  of   negative order.   Then,  arguing as in   Appendix I of [2] we prove that $N_+$
cannot be larger than $\Cal N$ because otherwise $(\bold 4_{e,2})$ would be violated. Then, $N_+= \Cal N$ and $N_-=0$.
 The order of any solution to (21.11) of negative order is an integer
 not exceeding $-1.$  Then, it follows by a contour integration  that the Fourier transform (3.5)  of any solution to (21.11) of negative order is zero for
 $y\in\bR^-$. Hence, any solution to (21.11)  that is of   negative order  is  in $\bold H^2(\bold C^+)$. Since by Proposition~21.8 any solution to (21.11) that is in  $\bold H^2(\bold C^+)$ is of negative order, it follows that $N_+$ is  the number of linearly independent solutions to (21.11) that are in  $\bold H^2(\bold C^+)$. Hence,  $(\bold V_h)$  of Theorem~7.6 holds. Finally $N_-$ is the number of linearly independent solutions to (21.54) that are of  negative order.
 Again, by a contour integration
 we can show that
 these solutions are in $\bold H^2(\bold C^-)$ and since by Proposition~21.8 any solution to  (21.55) that is in $\bold H^2(\bold C^-)$ is of negative order, $N_-$ is the number of linearly independent solutions to (21.55)  that are in $\bold H^2(\bold C^-)$. Since we have proven that $N_-=0$
   it follows that $(\bold {III}_c)$  holds.
It only remains to prove that $(\bold 4_e)$ is satisfied. By Proposition~6.1 it is enough to prove that condition $(\bold 4_c)$ hods. On the contrary, suppose that
(4.14) has a nontrivial integrable solution, $X(y)$. Since by condition $(\bold I)$ of Definition~4.3 the matrix  $F_s(y)$ is bounded in $ y \in \bold R$ and it is integrable for $ y \in \bold R^+$, $X(y)$, is bounded, and then $ X(y) \in L^2(\bold R^+)$. Then, condition $(\bold 4_{c,2})$ will be violated, and by Proposition~21.2 also condition   $(\bold 4_{e,2})$
 will be violated. In consequence, condition $(\bold 4_c)$ holds. \qed

\newpage
\noindent {\bf 22. THE GENERALIZED FOURIER MAP}
\vskip 3 pt

The Fourier transform between square-integrable functions of $x$ and
square-integrable functions
of $k$ is an essential tool in the analysis of inverse scattering and spectral
problems. In the case of the Schr\"odinger equation
with a matrix-valued potential the
corresponding Fourier transform acts between
vectors with $n$ components that
are square-integrable functions of $x$ and vectors
with $n$ components that are square-integrable functions
of $k.$
It is possible to generalize such a Fourier transform and
introduce a generalized Fourier map $\bold F$
acting from a Hilbert space involving functions of
$x$ into another Hilbert space involving functions of $k.$

We use $C_0(\bR^+)$ to denote the subspace  of $L^2(\bR^+)$
consisting of column vectors with $n$ components
that are continuous functions of $x$ with compact support in $\bR^+.$
Thus, any function $Y(x)$ in $C_0(\bR^+)$
is continuous, bounded, integrable, and square integrable
in $x\in\bR^+.$ It is known that $C_0(\bR^+)$ is a dense subspace
of $L^2(\bR^+).$

We define the Hilbert space $\Cal R$ as the direct sum given by
$$\Cal R:=\text{\rm{Ran}} [M_1]\oplus \cdots \oplus \text{\rm{Ran}} [M_N]\oplus L^2(\bR^+),\tag 22.1$$
where $N$ is the number of bound states appearing in (4.2) and
$\text{\rm{Ran}}[M_j]$ denotes the range of the matrix $M_j$ appearing
in (4.3) and (10.22). Let us use
$(Z_1,\dots,Z_N,Z)$ to denote an element in $\Cal R,$
where $Z_j$ is a constant column vector with $n$ components and
has the form $M_j v_j$ for some vector $v_j$ in $\bC^n,$ and
$Z$ is a column vector with $n$ components that are square-integrable
functions of $k\in\bR^+.$
The scalar product in $\Cal R$ is defined as
$$\left((Z_1,\dots,Z_N,Z),(\tilde Z_1,\dots,\tilde Z_N,\tilde Z)\right)_{\Cal R}:=
Z_1^\dagger \tilde Z_1+\dots +Z_N^\dagger \tilde Z_N+
\int_0^\infty dk\,Z(k)^\dagger \tilde Z(k),\tag 22.2$$
and the norm $||\cdot ||_{\Cal R}$ is defined to be the norm induced by the scalar
product in (22.2).

Inspired by [42], we introduce the generalized Fourier map
$\bold F$ acting on the subspace $C_0(\bR^+)$
in terms of the components $\bold F_1,\dots,\bold F_N,\bold F_c$
as
$$\bold F Y=(\bold F_1 Y,\dots,\bold F_N Y,\bold F_c Y),\tag 22.3$$
where each
component $\bold F_j$ is associated with the bound state at $k=i\kappa_j,$
and $\bold F_c$ is associated with the continuous spectrum of the matrix
Schr\"odinger operator. Here, the component $\bold F_j$ is defined with the help
of the bound-state matrix solution $\Psi_j(x)$ appearing in (9.8) as
$$Z_j:=F_j Y=:\int_0^\infty dx\,\Psi_j(x)^\dagger \,Y(x),\qquad
j=1,\dots,N,\tag 22.4$$
and the component $\bold F_c$ is defined with the
help of the physical solution $\Psi(k,x)$ appearing in (9.4)
as
$$Z(k):=(\bold F_c Y)(k):=\ds\frac{1}{\sqrt{2\pi}}\int_0^\infty dx\,\Psi(k,x)^\dagger \,Y(x),\qquad
k\in\bR^+.\tag 22.5$$

\noindent {\bf Proposition 22.1} {\it Consider a scattering data set $\bold S$
as in (4.2), which consists of
an $n\times n$ scattering matrix $S(k)$ for $k\in\bR,$ a set of $N$ distinct
 positive constants $\kappa_j,$ and a set of
$N$ constant $n\times n$ hermitian and nonnegative matrices
$M_j$ with respective positive ranks $m_j,$ where $N$ is a nonnegative integer.
Assume that $\bold S$ satisfies $(\bold I)$ of Definition~4.3
and $(\bold 4_a)$ of Definition~4.2.
Then, we have the following:}

\item{(a)} {\it The map $\bold F_j$ defined in
(22.4) maps $C_0(\bR^+)$ into
$\text{\rm{Ran}}[M_j]$ appearing in (22.1).
It can be extended to a bounded map from
$L^2(\bR^+)$ into $\text{\rm{Ran}}[M_j].$}

\item{(b)} {\it The Fourier map $\bold F$ defined in (22.3)
maps the subspace $C_0(\bR^+)$ into the Hilbert space
$\Cal R$ defined in (22.1).
It is an isometry, i.e. $(FY,FY)_{\Cal R}=(Y,Y)_2$ for any
$Y(x)\in C_0(\bR^+).$ It can be extended to an
isometric map from $L^2(\bR^+)$ into $\Cal R.$}

\item{(c)} {\it The map $\bold F_c$ defined in
(22.5) maps $C_0(\bR^+)$ into
$L^2(\bR^+).$
It can be extended to a bounded map from
$L^2(\bR^+)$ into $L^2(\bR^+).$}

\noindent PROOF: The integrand in (22.4) is
integrable when $Y(x)$ belongs to $C_0(\bR^+)$
because $\Psi_j(x)$ is bounded in $x\in\bR^+,$ as
stated in Proposition~17.1(d). That integrand remains integrable
when $Y(x)$ belongs $L^2(\bR^+)$ because
the integrand now becomes a product of
two square-integrable quantities, as a result of
$\Psi_j(x)$ being
square integrable in $x\in\bR^+,$ as
stated in Proposition~17.1(d). Thus, the extension of $\bold F_j$ from
the domain of $C_0(\bR^+)$ to the domain of $L^2(\bR^+)$
is immediate. For any $Y(x)\in L^2(\bR^+),$
 from (9.8) and the fact that
$M_j$ is hermitian, it follows that $Z_j$ given in (22.4) has the
form $Z_j=M_j v_j,$ where $v_j$ is the constant vector in $\bC^n$
given by
$$v_j=\int_0^\infty dx\,f(i\kappa_j,x)^\dagger \,Y(x).\tag 22.6$$
Thus, we have shown that $\bold F_j$ maps $C_0(\bR^+)$ into
$\text{\rm{Ran}}[M_j],$ and its extension maps
$L^2(\bR^+)$ into
$\text{\rm{Ran}}[M_j].$ Let us now show that the extended map is bounded from
$L^2(\bR^+),$ i.e. let us show that $|\bold F_j Y|\le C\,||Y||_2$
for some constant $C.$ Using the inequality
$$|\Psi_j(x)^\dagger\,Y(x)|\le |\Psi_j(x)|\, |Y(x)|,\qquad x\in\bR^+,\tag 22.7$$
in (22.4) we obtain
$$|\bold F_j Y|\le \int_0^\infty dx\, |\Psi_j(x)^\dagger\, Y(x)|\le
\int_0^\infty dx\,|\Psi_j(x)|\,|Y(x)|.\tag 22.8$$
Applying the Schwarz inequality on the last integral term in (22.8)
we get
$$|\bold F_j Y|^2\le \int_0^\infty dy\,|\Psi_j(y)|^2 \int_0^\infty dz\,|Y(z)|^2.
\tag 22.9$$
As stated in Proposition~17.1(d), $\Psi_j(x)$ is square integrable in $x\in\bR^+,$ and hence the first integral in (22.9) is bounded by a constant. Thus, we get
$|\bold F_j Y|\le C\,||Y||_2,$
completing the proof of (a).
Let us now turn to the proof of
(b). For $Y(x)\in C_0(\bR^+),$ using (22.4) and (22.5) in (22.3)
we obtain
$$\aligned
(\bold F Y,\bold F Y)_{\Cal R}=\sum_{j=1}^N &
\int_0^\infty dx\, Y(x)^\dagger \,\Psi_j(x)
\int_0^\infty dy\, \Psi_j(y)^\dagger Y(y)\\
+&
\ds\frac{1}{2\pi}\int_0^\infty dk\left[
\int_0^\infty dx\, Y(x)^\dagger \,\Psi(k,x)
\int_0^\infty dy\, \Psi(k,y)^\dagger\, Y(y)\right].\endaligned\tag 22.10
$$
Since $Y(x)$ belongs to $C^\infty_0(\bR^+),$ the order of the integrations
in (22.10) can be interchanged, and hence
the right-hand side in (22.10) can be rearranged and we get
$$(\bold F Y,\bold F Y)_{\Cal R}=\int_0^\infty dx\int_0^\infty dy
\, Y(x)^\dagger \,q(x,y)\,
 Y(y),\tag 22.11$$
 where we have defined
$$q(x,y):=\sum_{j=1}^N
\Psi_j(x)
\, \Psi_j(y)^\dagger+
\ds\frac{1}{2\pi}\int_0^\infty dk\,
\Psi(k,x)
\, \Psi(k,y)^\dagger.\tag 22.12$$
 From Parseval's identity given in
 (20.1), we see that $q(x,y)=\delta(x-y)\,I$ and hence
we get
$$(\bold F Y,\bold F Y)_{\Cal R}=(Y,Y)_2,\qquad Y\in C_0(\bR^+),\tag 22.13$$
which proves that the Fourier map $\bold F$ is an isometry from
$C_0(\bR^+)$ into $\Cal R.$
Since the subspace $C_0(\bR^+)$ is dense in
$L^2(\bR^+),$ we can extend $\bold F$ to an isometry
 from $L^2(\bR^+)$ into $\Cal R$ so that
(22.13) remains valid when $Y(x)\in L^2(\bR^+),$ i.e. we
$$(\bold F Y,\bold F Y)_{\Cal R}=(Y,Y)_2,\qquad Y\in L^2(\bR^+).\tag 22.14$$
Thus, we have completed the proof of
 (b). Let us now turn to the proof of (c).
Using (22.2) in (22.14) we obtain
$$Z_1^\dagger \tilde Z_1+\dots +Z_N^\dagger \tilde Z_N+
(Z,Z)_2=(Y,Y)_2,\qquad Y\in C_0(\bR^+).\tag 22.15$$
Thus, (22.15) implies that
$$(Z,Z)_2\le (Y,Y)_2,\qquad Y\in C_0(\bR^+),\tag 22.16$$
and hence $\bold F_c$ defined in (22.5) is a bounded
operator from $C_0(\bR^+)$ into $L^2(\bR^+).$
Since $C_0(\bR^+)$ is a dense subspace of $L^2(\bR^+),$
 from (22.16) we conclude that $\bold F_c$
 uniquely extends to a bounded map from $L^2(\bR^+)$ into $L^2(\bR^+).$
Let us elaborate on such an extension. For any
$Y(x)$ in $L^2(\bR^+)$ there exists a sequence
$\{Y^{(j)}(x)\}_{j=1}^\infty$ converging to
$Y(x)$ in the $L^2$-sense with $Y^{(j)}(x)$
belonging to $C_0(\bR^+)$ for all $j\ge 1.$
Using (22.5) we then define
$$Z^{(j)}(k):=\ds\frac{1}{\sqrt{2\pi}} \int_0^\infty dx\, \Psi(k,x)^\dagger \,
Y^{(j)}(x),\qquad j=1,2,\dots.\tag 22.17$$
 From (22.16) we know that
$$(Z^{(j)},Z^{(j)})_2\le (Y^{(j)},Y^{(j)})_2,\qquad j=1,2,\dots,\tag 22.18$$
and hence $Z^{(j)}(k)$ is square integrable in $k\in\bR^+.$
Since the sequence $\{Y^{(j)}(x)\}_{j=1}^\infty$
is convergent in the $L^2$-sense, it follows from
(22.18) that also the sequence $\{Z^{(j)}(k)\}_{j=1}^\infty$
is convergent in the $L^2$-sense to some limit
$Z(k)\in L^2(\bR^+).$ Using that limit,
we then
let $(\bold F_c Y)(k):=Z(k).$
This approach allows us
interpret (22.5) as the extension of
$\bold F_c$ as a bounded map on $L^2(\bR^+)$ even though
the integral on the right-hand side of (22.5) may not literally exist
for $Y(x)\in L^2(\bR^+).$
Thus, the proof of (c) is completed.
\qed

 From Proposition~22.1 it follows that (22.3), (22.4), and
(22.5) can also be viewed as the extensions of $\bold F,$
$\bold F_j,$ and
$\bold F_c,$ respectively, where their domains are now
$L^2(\bR^+).$ In fact, unless otherwise stated,
we will use $\bold F,$
$\bold F_j,$ and
$\bold F_c,$ to denote such extensions.

Next, we analyze the component $\bold F_c$ of the Fourier map $\bold F$
further.
Toward this goal we first introduce the point subspace $\Cal P$
associated with our Schr\"odinger operator
related to (2.1) and (2.4). By definition, the point subspace
$\Cal P$ is the span of all eigenfunctions
of the Schr\"odinger operator associated with
(2.1) and (2.4). By Proposition~11.4(e)
we know that the eigenfunctions of
the Schr\"odinger operator correspond to the eigenvalues
$-\kappa_j^2$ for $j=1,\dots,N,$
and hence
we can write $\Cal P$ in the form of a direct sum as
$$\Cal P=\Cal P_1\oplus \cdots \oplus \Cal P_N.\tag 22.19$$
It follows from
Proposition~11.4 that
each $\Cal P_j$ is the column space of the
 bound-state matrix solution $\Psi_j(x)$
 appearing in (9.8), the dimension of
 $\Cal P_j$ is equal to the rank $m_j$ of $M_j,$
and $\Cal P_j$ is equal to the subspace of
$L^2(\bR^+)$ consisting of all column vectors of
the form $\Psi_j(x)\, v$ for all
$v\in\bC^n.$

\noindent {\bf Proposition 22.2} {\it Assume that the input data
set $\bold D$ in (4.1)
belongs to the Faddeev class specified in
Definition~4.1. Let
$\bold F_c$ be the map from $L^2(\bR^+)$ into $L^2(\bR^+)$
appearing in (22.5).}

\item{(a)} {\it The kernel of $\bold F_c$ is equal to
the point space $\Cal P$ defined in (22.19), i.e.}
$$\text{\rm{Ker}}[\bold F_c]=\Cal P.\tag 22.20$$
{\it Thus, $\text{\rm{Ker}}[\bold F_c]$ is
consists of linear combinations of columns of all
bound-state matrix solutions $\Psi_j(x)$ for
$j=1,\dots,N$ defined in
(10.7) and hence the dimension of
$\text{\rm{Ker}}[\bold F_c]$
is equal to the nonnegative integer $\Cal N$ appearing in (4.3).}

\item{(b)} {\it The subspace
$\text{\rm{Ker}}[\bold F_c]$
of $L^2(\bR^+)$ is a subspace of $L^1(\bR^+)\cap L^\infty(\bR^+).$}

\item{(c)} {\it The map $\bold F_c$ is unitary from
$\left(\text{\rm{Ker}}[\bold F_c]\right)^\perp$
onto $L^2(\bR^+).$}

\item{(d)} {\it The generalized Fourier map $\bold F$ defined in
(22.5) is a unitary map from $L^2(\bR^+)$ into $\Cal R,$ i.e.
it is an isometry and onto $\Cal R.$}

\noindent PROOF: For the proof of (22.20) and the unitarity
property stated in (c), we refer the reader to [42],
where the proof is given in Theorem~6.7, (6.37), and (6.38) there.
Actually, the notation $F_{A,B}^-$ is used
in [42] to denote our map $\bold F_c.$
Hence, $\text{\rm{Ker}}[\bold F_c]$ consists of
the columns of $\Psi_j(x)$ for all $j=1,\dots,N$ and
since the number of such columns contain exactly
$\Cal N$ linearly independent columns,
the dimension of
the kernel of $\bold F_c$ is also $\Cal N.$ Thus, the proof of
(a) is complete. From Proposition~17.1(d) we know that
each column of $\Psi_j(x)$ is bounded and integrable
in $x\in\bR^+$ besides being square integrable there.
Then, from (a) we conclude that any column
vector in $L^2(\bR^+)$ belonging to
$\text{\rm{Ker}}[\bold F_c]$
must be bounded and integrable in $x\in\bR^+.$
Thus, the proof of (b) is complete. Let us now prove (d). To prove the unitarity of
$\bold F$ from
$L^2(\bR^+)$ into $\Cal R$ we need to prove that
$\bold F$ is an isometry and is also onto $\Cal R.$
The former property is assured by Proposition~22.1(b)
and hence we only need to prove that
$\bold F$ is onto. Since
the kernel of the map $\bold F$
is a subspace
of $L^2(\bR^+),$ we have the orthogonal decomposition
$$L^2(\bR^+)=\text{\rm{Ker}}[\bold F_c]\oplus \left(\text{\rm{Ker}}[\bold F_c] \right)^\perp.\tag 22.21$$
 From (22.1), (22.3), and (22.21) we see that $\bold F$ is onto
if we can show that the map given by
$$(\bold F_1,\dots,\bold F_N): \
\text{\rm{Ker}}[\bold F_c]\to
\text{\rm{Ran}} [M_1]\oplus \cdots \oplus \text{\rm{Ran}} [M_N],\tag 22.22$$
is onto and also the map $\bold F_c$ maps $L^2(\bR^+)$ onto $L^2(\bR^+).$
The latter map, i.e. $\bold F_c$ is already onto because it is
unitary, as stated in (c). On the other hand, the map given in
(22.22) is onto if and only if the map
$\bold F_j$ given in (22.4) maps $\text{\rm{Ker}}[\bold F_c]$ onto
$\text{\rm{Ran}}[M_j]$ for $j=1,\dots,N.$ On the other hand,
from (22.19) and (22.20) we see that the map given in
(22.22) is onto if and only if $\Cal P_j$ is isomorphic to
$\text{\rm{Ran}} [M_j]$ for $j=1,\dots,N.$ The latter property
follows from Proposition~11.4(d) and Proposition~11.4(e). Thus, the proof of (d) is complete.
\qed

As shown in the next proposition, the adjoint of the Fourier map, denoted by
$\bold F^\dagger,$ can be expressed explicitly in terms of the physical solution $\Psi(k,x)$ and the bound-state matrix solutions $\Psi_j(x).$

\noindent {\bf Proposition 22.3} {\it Assume that the scattering data set $\bold S$ appearing in (4.2) satisfies $(\bold I)$
of Definition~4.3 and $(\bold 4_a)$
of Definition~4.2. Further, assume that
the generalized Fourier map $\bold F$ defined in (22.3) is unitary from
$L^2(\bR^+)$ onto $\Cal R,$ the Hilbert space
specified in (22.1). Then, we have the following:}

\item{(a)} {\it The adjoint of the map $\bold F_j$ given in (22.4), denoted by
$\bold F_j^\dagger,$ maps
$\text{\rm{Ran}}[M_j]$ into $L^2(\bR^+)$ and is given by}
$$\bold F_j^\dagger Z_j:=\Psi_j(x)\, Z_j,\qquad j=1,\dots,N,\tag 22.23$$
{\it where $\Psi_j(x)$ is the bound-state matrix solution
appearing in (9.8).}

\item{(b)} {\it The adjoint of the map $\bold F_c$ given in (22.5),
denoted by
$\bold F_c^\dagger,$ maps $L^2(\bR^+)$ onto $\left(\text{\rm{Ker}}[\bold F_c]\right)^\perp$ and is given by}
$$(\bold F_c^\dagger Z)(x):=\ds\frac{1}{\sqrt{2\pi}}
\int_0^\infty dk\, \Psi(k,x)\, Z(k),\qquad x\in\bR^+,\tag 22.24$$
{\it  where
$\Psi(k,x)$ is the physical solution in (9.4).}

\item{(c)} {\it The adjoint of
the Fourier map $\bold F$ given in (22.3)-(22.5), denoted by $\bold F,$
maps the Hilbert space $\Cal R$ given in (22.1)
onto $L^2(\bR^+)$ and described as}
$$\bold F^\dagger (Z_1,\dots,Z_N,Z)=\bold F_1^\dagger Z_1+
\dots +\bold F_N^\dagger Z_N+\bold F_c^\dagger Z,\tag 22.25$$
{\it where $\bold F_j^\dagger$ and $\bold F_c^\dagger$ are as in
(22.23) and (22.24), respectively.}

\noindent PROOF: Because $\bold F$ is unitary, it is onto. Hence,
its respective components $\bold F_j$ are each onto $\text{\rm{Ran}} [M_j]$
and $\bold F_c$ is onto $L^2(\bR^+).$ Thus, the domain of
$\bold F^\dagger$ is $\Cal R,$ the domain of
$\bold F_j^\dagger$ is $\text{\rm{Ran}} [M_j],$ and the domain of
$\bold F_c^\dagger$ is $L^2(\bR^+).$
Using the definition of the adjoint map
$\langle Z_j,\bold F_j \tilde Y\rangle=\left(\bold F_j^\dagger Z_j,\tilde Y\right)_2,$ where $Z_j\in \text{\rm{Ran}}[M_j],$ $\tilde Y(x)\in L^2(\bR^+),$ and
$\bold F_j$ as in (22.4), we obtain (22.23), and hence (a) holds.
Furthermore, we have $\left( Z,\bold F_c \tilde Y\right)_2=\left( \bold F_c ^\dagger Z,
\tilde Y\right)_2,$ where $Z(k)\in L^2(\bR^+),$ $\tilde Y(x)\in L^2(\bR^+),$ and
$\bold F_c$ as in (22.5). Thus, we obtain (22.24), and hence (b) holds.
%
Moreover, by using the definition of the operator adjoint given by
$$((Z_1,\dots,Z_N,Z),\bold F \tilde Y)_{\Cal R}=(\bold F^\dagger\, (Z_1,\dots,Z_N,Z),\tilde Y)_2,\tag 22.26$$
where $(Z_1,\dots,Z_N,Z)$ and $\tilde Y$ are some arbitrary vectors
in $\Cal R$ and $L^2(\bR^+),$ respectively.
With the help of (22.2)-(22.5), one can verify directly that (22.26)
yields (22.25).
We see from
(22.23) that, for any $Z_j\in \text{\rm{Ran}} [M_j]$
its image under $\bold F_j^\dagger$ given by
$\Psi_j Z_j$ belongs to $L^2(\bR^+),$ as seen from (9.8)
and Proposition~17.1(d).
On the other hand, the integral in (22.24) does not exist
in the usual sense
when $Z(k)\in L^2(\bR^+)$ and needs to be understood
in the $L^2$-sense, analogous to
a similar description given in the proof of
Proposition~22.1. We provide a brief elaboration.
For any $Z(k)\in L^2(\bR^+)$
we can find a sequence
$\{Z^{(l)}(k)\}_{l=1}^\infty$ converging to
$Z(k)$ in the $L^2$-sense with $Z^{(l)}(k)$
belonging to $L^1(\bR^+)\cap L^2(\bR^+)$ for all $l\ge 1.$
Using (22.24) we let
$$Y^{(l)}(x):=(\bold F_c^\dagger\, Z^{(l)})(x)=
\ds\frac{1}{\sqrt{2\pi}} \int_0^\infty dk\, \Psi(k,x) \,
Z^{(l)}(k),\qquad l=1,2,\dots.\tag 22.27$$
By Proposition~22.2(c) we know that $\bold F_c$ is a unitary
map from $\left(\text{\rm{Ker}}[\bold F_c]\right)^\perp$
onto $L^2(\bR^+),$ and hence
$\bold F_c^\dagger$ is unitary from
$L^2(\bR^+)$ onto $\left(\text{\rm{Ker}}[\bold F_c]\right)^\perp.$
Thus, from (22.27) we obtain
$$||Y^{(l)}||_2=||\bold F_c^\dagger \, Z^{(l)}||_2=||Z^{(l)}||_2.\tag 22.28$$
Since $Z^{(l)}(k)$ converges to $Z(k)$ in the $L^2$-sense, it follows from (22.28) that
$Y^{(l)}(x)$ must converge in the $L^2$-sense to some $Y(x)\in L^2(\bR^+),$
where we have let
$$Y(x):=\ds\lim_{l\to +\infty}(\bold F_c^\dagger\, Z^{(l)})(x)=
\ds\lim_{l\to +\infty}\ds\frac{1}{\sqrt{2\pi}} \int_0^\infty dk\, \Psi(k,x) \,
Z^{(l)}(k).\tag 22.29$$
Thus, the proof is complete. \qed

The following result indicates that the physical solution $\Psi(k,x)$
appearing in (9.4) and (9.6)
along with
the normalized bound-state matrix solutions $\Psi_j(x)$ defined in (9.8) satisfy
certain orthonormality properties.

\noindent {\bf Proposition 22.4} {\it Assume that the input data set $\bold D$ appearing in (4.1) belongs to
the Faddeev class specified
in Definition~4.1. Then, the physical solution $\Psi(k,x)$
appearing in (9.4) and (9.6)
and
the normalized bound-state
matrix solutions $\Psi_j(x)$ appearing in (9.8) satisfy}
$$\int_0^\infty dx\, \Psi_l(x)^\dagger\, \Psi_j(x)=\delta_{lj}\,P_j,\qquad
l,j=1,\dots,N,\tag 22.30$$
$$\int_0^\infty dx\, \Psi_l(x)^\dagger\, \Psi(k,x)=0,\qquad
l=1,\dots,N,\quad k\in\bR^+,\tag 22.31$$
$$\int_0^\infty dx\, \Psi(k,x)^\dagger\, \Psi_l(x)=0,\qquad
l=1,\dots,N,\quad k\in\bR^+,\tag 22.32$$
$$\ds\frac{1}{2\pi}\int_0^\infty dx\, \Psi(k,x)^\dagger \Psi(\ell,x)=\delta(k-\ell)\, I,\qquad
k,\ell\in\bR^+,\tag 22.33$$
{\it where $P_j$ is the projection matrix in (11.1),
$\delta_{jl}$ is the Kronocker delta, and
$\delta(k)$ is the Dirac delta distribution.}

\noindent PROOF: The normality in (22.30), i.e. (22.30) when $j=l,$
 directly follows from (11.23). Let us prove
 (22.30) when $j\ne l.$
 We know from Proposition~11.4(b) that $\Psi_j(x)$ satisfies
 (2.1) with $k=i\kappa_j,$ i.e.
$$-\Psi''_j(x)+V(x)\,\Psi_j(x)=-\kappa_j^2\,\Psi_j(x),\qquad x\in\bR^+,
\tag 22.34$$
Since the potential $V$ satisfies (2.2), (22.34) yields
$$-\Psi''_l(x)^\dagger+\Psi_l(x)^\dagger \,V(x)=-\kappa_l^2\,\Psi_l(x)^\dagger,\qquad x\in\bR^+,
\tag 22.35$$
Premultiplying (22.34) by $\Psi_l(x)^\dagger$ and postmultiplying
(22.35) by $\Psi_j(x)$ and subtracting the resulting matrix equations, we
obtain
$$(\kappa_l^2-\kappa_j^2)\,\Psi_l(x)^\dagger\,\Psi_j(x)=
\Psi''_l(x)^\dagger\,\Psi_j(x)-\Psi_l(x)^\dagger\,\Psi''_j(x),\qquad x\in\bR^+,\tag 22.36$$
or equivalently we have
$$(\kappa_l^2-\kappa_j^2)\,\Psi_l(x)^\dagger\,\Psi_j(x)=\ds\frac{d}{dx}\left[
\Psi'_l(x)^\dagger\,\Psi_j(x)-\Psi_l(x)^\dagger\,\Psi'_j(x)\right],\qquad x\in\bR^+.\tag 22.37$$
By integrating over $x\in\bR^+,$ from (22.37) we obtain
$$(\kappa_l^2-\kappa_j^2)\int_0^\infty dx\,\Psi_l(x)^\dagger\,\Psi_j(x)=
\Psi_l(0)^\dagger\,\Psi'_j(0)-\Psi'_l(0)^\dagger\,\Psi_j(0),\tag 22.38$$
where we have used the fact that
$\Psi_j(x)$ and $\Psi_j'(x)$ vanishes as $x\to+\infty$ for
$j=1,\dots, N,$ as indicated by Proposition~17.1(d) and Proposition~17.2(d), respectively. By Proposition~14.2, the right-hand side of (22.38) vanishes because
$\Psi_l(x)$ and $\Psi_j(x)$ each satisfy the boundary condition (2.4),
as indicated in Proposition~11.4(b).
Hence, we obtain
$$(\kappa_l^2-\kappa_j^2)\int_0^\infty dx\,\Psi_l(x)^\dagger\,\Psi_j(x)=0.
\tag 22.39$$
Since $\kappa_l\ne \kappa_j,$ from (22.39) we obtain (22.30).
Let us next prove (22.31)
By Proposition~9.6(c), the physical solution $\Psi(k,x)$ satisfies (2.1) , i.e. we have
$$-\Psi''(k,x)+V(x)\,\Psi(k,x)=k^2\,\Psi(k,x),\qquad x\in\bR^+.
\tag 22.40$$
Postmultiplying (22.35) with $\Psi(k,x)$ and premultiplying
(22.40) with $\Psi_l(x)^\dagger$ and subtracting the resulting equations
we obtain the analog of (22.37), i.e. we obtain
$$(\kappa_l^2+k^2)\,\Psi_l(x)^\dagger\,\Psi(k,x)=\ds\frac{d}{dx}\left[
\Psi'_l(x)^\dagger\,\Psi(k,x)-\Psi_l(x)^\dagger\,\Psi'(k,x)\right],\qquad x\in\bR^+.\tag 22.41$$
When $k\in\bR,$ from (17.2), (17.4), (17.10), and (17.12)
it follows that, for each $k\in\bR,$ we have
$$\Psi'_l(x)^\dagger\,\Psi(k,x)-\Psi_l(x)^\dagger\,\Psi'(k,x)=o(1), \qquad
x\to+\infty,\tag 22.42$$
and that $\Psi_l(x)^\dagger\,\Psi(k,x)$ is integrable in $x\in\bR^+.$
Thus, integrating over $x\in\bR^+,$ from (22.41) we obtain
$$(\kappa_l^2+k^2)\int_0^\infty dx\,\Psi_l(x)^\dagger\,\Psi(k,x)=
\Psi_l(0)^\dagger\,\Psi'(k,0)-\Psi'_l(0)^\dagger\,\Psi(k,0),
\qquad k\in\bR.\tag 22.43$$
By Proposition~9.6(b) we know that $\Psi(k,x)$ satisfies (2.4), and
by Proposition~11.4(b) we know that $\Psi_l(x)$ satisfies (2.4). Hence,
Proposition~14.2 indicates that the right-hand side of (22.43) must vanish.
 For $k\in\bR,$ we have $\kappa_l^2+k^2>0$ and hence
 (22.43), with the right-hand side being zero, yields (22.31).
 Note that (22.32) can be obtained by taking the matrix adjoint of
 both sides of (22.31), and hence
 (22.32) is valid. It only remains to
 prove (22.33).
As stated in Proposition~22.2(c) the map $\bold F_c$ from $\left(\text{\rm{Ker}}[\bold F_c]\right)^\perp$
onto $L^2(\bR^+)$
given in
(22.5) is unitary. This implies [27] that
we have $\bold F_c^\dagger \bold F_c =I$ on $\left(\text{\rm{Ker}}[\bold F_c]\right)^\perp$
and $\bold F_c \bold F_c^\dagger  =I$ on $L^2(\bR^+).$
The latter can be written as $\bold F_c \bold F_c^\dagger Z =Z$ for $Z\in L^2(\bR^+),$
or equivalently as
$$\ds\frac{1}{2\pi}\int_0^\infty dx\,\Psi(k,x)^\dagger\int_0^\infty dl\,
\Psi(l,x)\,Z(l)=Z(k),\qquad Z(k)\in L^2(\bR^+).
\tag 22.44$$
Since $C_0(\bR^+)$ is a subspace of $L^2(\bR^+),$
we can use (22.44) with $Z(k)\in C_0(\bR^+).$ Then,
the integral on the left-hand side of (22.44) exists
and we are allowed to interchange the order of integration and
obtain
$$\ds\frac{1}{2\pi}\int_0^\infty dl\int_0^\infty dx\,\Psi(k,x)^\dagger\,
\Psi(l,x)\,Z(l)=Z(k),\qquad Z(k)\in C_0(\bR^+).
\tag 22.45$$
Then, from (22.45)
we conclude the formal expression (22.33)
understood in the distribution sense.
\qed

In the following proposition we present some relations among
the maps $\bold F_j,$ $\bold F_c,$ $\bold F_j^\dagger,$ $\bold F_c^\dagger$
appearing in (22.4), (22.5), (22.23), (22.24), respectively.

\noindent {\bf Proposition 22.5} {\it Assume that the input data set
$\bold D$ appearing in (4.1) belongs to the Faddeev class. Then,
the components $\bold F_j$ and $\bold F_c$ of the Fourier map
$\bold F$ defined in (22.3)-(22.5) satisfy}
$$\cases\bold F_l \bold F_j^\dagger Z_j=\delta_{lj} Z_j,
\qquad j,l=1,\dots,N,\quad Z_j\in\text{\rm{Ran}} [M_j],\\
\bold F_l \bold F_c^\dagger Z=0,\qquad l=1,\dots,N,\quad Z\in L^2(\bR^+),\\
\bold F_c \bold F_l^\dagger Z_l=0,\qquad l=1,\dots,N,\quad Z_l\in\text{\rm{Ran}} [M_l],\\
\bold F_c\bold F_c^\dagger Z=Z,\quad Z\in L^2(\bR^+),\endcases\tag 22.46$$
{\it where we recall that $\delta_{jl}$ is the Kronocker delta,
$\bold F_j^\dagger$ is the adjoint map in (22.23),
$\bold F_c^\dagger$ is the adjoint map in (22.24),
and $Z$ is a column vector
with $n$ components that are square-integrable functions of
$k\in\bR^+.$}

\noindent PROOF:
We know from Proposition~22.2(d) that the generalized
 Fourier map $\bold F$ from $L^2(\bR^+)$ into
$\Cal R$ is unitary. Consequently, we have $\bold F^\dagger \bold F=I$ on
$L^2(\bR^+)$ and we also have
$\bold F \bold F^\dagger=I$ on $\Cal R.$
The latter property is equivalent to
$$\bold F\bold F^\dagger (Z_1,\dots,Z_N,Z)=(Z_1,\dots,Z_N,Z).\tag 22.47$$
where $(Z_1,\dots,Z_N,Z)$ denotes a typical element in $\Cal R.$
With the help of (22.25) we see that (22.47) is equivalent to
$$\bold F(
\bold F_1^\dagger Z_1+
\dots +\bold F_N^\dagger Z_N+\bold F_c^\dagger Z)=(Z_1,\dots,Z_N,Z),\tag 22.48$$
 From (22.3) we see that (22.48) is equivalent to
$$\cases\bold F_l\left(\sum_{j=1}^N \bold F_j^\dagger Z_j+\bold F_c^\dagger Z\right)=Z_l,
\qquad l=1,\dots,N,\\
\bold F_c\left(\sum_{j=1}^N \bold F_j^\dagger Z_j+\bold F_c^\dagger Z\right)=Z,
\endcases\tag 22.49$$
where $Z_l\in \text{\rm{Ran}} [M_l]$ and $Z\in L^2(\bR^+).$
Using (22.4) and (22.23), from (22.30) we obtain
$$\bold F_l\,\bold F_j^\dagger Z_j=\delta_{lj}P_j\,Z_j,\qquad Z_j\in\text{\rm{Ran}} [M_j].\tag 22.50$$
Since $Z_j=M_j v_j$ for some vector $v_j\in\bC^n,$
we have $P_j Z_j=P_j M_j v_j.$
On the other hand, using (11.1),
the definition of $M_j$ given in
(11.22),  and the fact that $B_j^{-1/2}$ commutes with $P_j,$
as asserted by Proposition~11.2(d), we get
$$P_j M_j=M_j,\qquad j=1,\dots,N,\tag 22.51$$
yielding $P_j Z_j=M_j v_j$ or equivalently,
$$P_j Z_j=Z_j,\qquad j=1,\dots,N.\tag 22.52$$
 From (22.50) and (22.52) we see that the first
 line of (22.46) is satisfied.
Using the first line of (22.46) in the first line of (22.49),
we get $\bold F_l \bold F_c^\dagger Z=0$ and hence the second line of
(22.30) is satisfied.
By postmultiplying (22.32) by $Z_l$ in the range
of $M_l,$ we obtain the third line of (22.46).
Then, using the third line of (22.46) in the second line of
(22.49) we get
the fourth line of (22.46). \qed

\newpage
\noindent {\bf 23. AN ALTERNATE METHOD TO SOLVE THE INVERSE PROBLEM}
\vskip 3 pt

In this chapter, we present an alternate solution to
the inverse problem and this is related to
the characterization given in Theorem~8.1.

The part of the solution to the inverse problem involving the construction of the
potential is practically the same as the solution outlined in Chapter~16.
However, the part of the solution related to the
boundary condition is different than the procedure outlined
in Chapter~16.
We summarize the construction of $\bold D$ from $\bold S$ in this
alternate method,
where the existence and uniqueness are implicit at each step:

\item{(a)} From the large-$k$ asymptotics of the
scattering matrix $S(k),$ with the help of
(4.6), we determine the $n\times n$ constant
matrix $S_\infty.$ Contrary to the method
of Chapter~16, we do not deal with the determination of
the constant
$n\times n$ matrix $G_1$ specified in
via (14.1).
It follows from (4.4) that the matrix
$S_\infty$ is hermitian when $\bold S$ satisfies
the condition $(\bold I).$

\item{(b)} In terms of the quantities in $\bold S,$ we uniquely construct
the $n\times n$ matrix $F_s(y)$ by using (4.7) and
the $n\times n$ matrix $F(y)$ by using (4.12). This step is the same as
(b) of the summary of the method outlined in Chapter~16.

\item{(c)} If the condition $(\bold 4_c)$
is also satisfied, then one uses the matrix $F(y)$ as input to the
Marchenko integral equation (13.1).
When $F(y)$ is integrable in $y\in(x,+\infty)$ for each $x\ge 0,$
as shown in Proposition~16.1, for each fixed $x\ge 0,$ there exists a solution $K(x,y)$ integrable
in $y\in(x,+\infty)$ to (13.1)
and such a solution is unique.
The solution $K(x,y)$ can be constructed by iterating (13.1).
Even though $K(x,y)$ is constructed for $y>x\ge 0,$ one can extend
$K(x,y)$ to $y\in\bR^+$ by letting $K(x,y)=0$ for $0\le y<x.$
This step is the same as (c) of the summary of the method outlined in Chapter~16.

\item{(d)} Having obtained $K(x,y)$ uniquely from $\bold S,$ one
constructs the potential $V(x)$ via (10.4)
and also constructs the Jost solution
$f(k,x)$ via (10.6). Then, it follows
 from Proposition~16.11 that,
by using $(\bold I),$ $(\bold 2)$, and $(\bold 4_c)$ of Theorem~8.1, one proves that the constructed $V(x)$ satisfies
(2.2) and (2.3) and that
the constructed $f(k,x)$ satisfies (2.1)
with the constructed potential $V(x).$
This step is the same as (d) of the summary of the method outlined in Chapter~16.

\item{(e)} Having constructed
the Jost solution $f(k,x),$ one then
constructs the physical solution $\Psi(k,x)$ via
(9.4) and the normalized bound-state matrices
$\Psi_j(x)$ via (9.8). One then proves that the
constructed matrix $\Psi(k,x)$ satisfies (2.1) and
and that the constructed $\Psi_j(x)$
satisfies (2.1) at $k=i\kappa_j.$

\item{(f)} Having constructed the potential
$V(x),$ one forms a matrix-valued
differential operator denoted by $\Cal L_{\text{min}},$
which acts as $(-D_x^2 I+V)$ with $D_x:=d/dx,$
with a domain that is a dense subset
of $L^2(\bR^+).$ More precisely, the domain of $\Cal L_{\text{min}}$
consists of the column vectors with $n$ components
each of which is a function of $x$ belonging to
a dense subset of $L^2(\bR^+).$
The constructed
operator $\Cal L_{\text{min}}$ is symmetric, i.e.
it satisfies $\Cal L_{\text{min}}\subset \Cal L_{\text{min}}^\dagger,$
but is not selfadjoint, i.e. it
does satisfy $\Cal L_{\text{min}}=\Cal L_{\text{min}}^\dagger.$
For the meaning of the operator inclusion, we refer the reader
to Chapter~3.

\item{(g)} One then constructs a selfadjoint realization of
$\Cal L_{\text{min}},$ namely an operator $\Cal L$ in such a way that
$\Cal L_{\text{min}}\subset \Cal L$ and $\Cal L=\Cal L^\dagger.$
The constructed operator
$\Cal L$ is a restriction of $\Cal L_{\text{min}}^\dagger,$ i.e.
we have $\Cal L\subset \Cal L_{\text{min}}^\dagger$
but not $\Cal L=\Cal L_{\text{min}}^\dagger.$

\item{(h)} The construction of the operator $\Cal L$ is achieved
by using the generalized Fourier map $\bold F$ and its adjoint
$\bold F^\dagger$ introduced in Chapter~22, inspired by [42].

\item{(i)} Once the selfadjoint operator $\Cal L$ is constructed,
it follows from the results in [45] that the domain of $\Cal L$
is a maximal isotropic subspace, which is sometimes also called a Lagrange plane.
Once we know that the domain of $\Cal L$ is a maximal isotropic subspace,
then it follows from Lemma~2.2 of [24] and Theorem~2.1
of [5] that the functions in the
domain of $\Cal L$ must satisfy the boundary condition (2.4)
for some boundary matrices $A$ and $B$ satisfying
(2.5) and (2.6), where $A$ and $B$ are uniquely determined up to
a postmultiplication by an invertible matrix $T.$

\item{(j)} Finally, one proves that the constructed physical solution
$\Psi(k,x)$ and the bound-state matrix solutions $\Psi_j(x)$ satisfy
the boundary condition (2.4) with the boundary matrices $A$ and
$B$ specified in the previous step; however, such a proof is
different in nature than the proofs for any of the previous characterizations. For the constructed matrices
$\Psi_j(x),$ it is immediate that they satisfy the
boundary condition because they belong to the domain of $\Cal L.$
Thus, it remains to prove that the constructed
$\Psi(k,x)$ satisfies the boundary condition.
We note that the matrix $\Psi(k,x)$ does not belong to
the domain of $\Cal L$ because its entries do not
belong to $L^2(\bR^+).$ On the other hand,
$\Psi(k,x)$ is locally square integrable in $x\in[0,+\infty),$ i.e.
it is square integrable in every compact subset of $[0,+\infty).$
Hence, it is possible to use a simple limiting argument to
prove that $\Psi(k,x)$ satisfies the boundary condition,
and the condition $(\bold {VI})$ is utilized
in that limiting argument.

\item{(k)} As in the previous characterization, we still need to prove that
the input data set
$\bold D$ of (4.1) constructed from the scattering data set
$\bold S$ of (4.2) yields $\bold S.$
The proof of this step is the same as the proof given
for the earlier characterizations and it is
given in the proof of Theorem~5.1.

\noindent {\bf Theorem 23.1} {\it For any input data set
$\bold D$ in the Faddeev class specified in Definition~4.1,
there exists and uniquely exists a corresponding scattering data set $\bold S$
as in (4.2) satisfying the properties
$(\bold I),$ $(\bold 2),$
$(\bold A),$
$(\bold 4_c)$, either one of $(\bold V_e)$ or
$(\bold V_h),$
and
$(\bold{VI}),$ of Theorem~8.1.}

\noindent PROOF: By Theorem~15.10 we know that
for any input data set
$\bold D$ in the Faddeev class,
there exists and uniquely exists a corresponding scattering data set $\bold S$
in the Marchenko class, i.e.
satisfying the properties $(\bold 1)$, $(\bold 2)$,
$(\bold 3_a)$, $(\bold 4_a)$ of Definition~4.5. Hence, in order to prove
our theorem, it is enough to prove that
those four conditions imply
$(\bold I),$ $(\bold 4_c)$,
$(\overset{\circ}\to 5)',$
$(\overset{\circ}\to 5)''$,
and $(\bold{VI}).$
The property $(\bold 1)$ implies $(\bold I),$
the confirmation of $(\bold {VI})$ follows from
Proposition~10.3(a), and Proposition~6.6 indicates that
$(\bold 4_c)$, $(\bold V_e),$ and $(\bold V_h)$
hold.
Thus, we only need to prove that $(\bold A)$
holds when $\bold D$ belongs to the Faddeev class.
In other words, we must prove that for any $g(k)$ belonging to a dense subset
$\overset{\circ}\to \Upsilon$ of the vector space
$\Upsilon$ of column vectors with $n$ components and satisfying
$g(-k)=S(k)\,g(k)$ for $k\in\bR,$ the corresponding equation (8.1)
has at least one solution $h(k)\in \bold H^2(\bCp).$
Since $\overset{\circ}\to \Upsilon$ can be any dense subspace of
$\Upsilon,$ we can certainly choose $\overset{\circ}\to \Upsilon$
as $\Upsilon.$ So, let us start with $g(k)\in L^2(\bR)$ satisfying
$g(-k)=S(k)\,g(k)$ for $k\in\bR$ and prove the existence of
some $h(k)$ satisfying (8.1).
Because $g(k)\in L^2(\bR),$ there exists $Y(x)\in
\left(\text{\rm{Ker}}[\bold F_c]\right)^\perp\subset L^2(\bR^+)$ such that
$$(\bold F_c Y)(k)=g(k),\qquad k\in\bR^+.\tag 23.1$$
Here $\bold F_c$ is the map
defined in (22.5), and
the existence of the corresponding $Y(x)$
is guaranteed because $\bold F_c$ onto $L^2(\bR^+)$ from
$\left(\text{\rm{Ker}}[\bold F_c]\right)^\perp$
as indicated
by Proposition~22.2(c).
We will construct a solution $h(k)$ to (8.1) with the
help of $Y(x).$ We proceed
as follows. Using (22.5) we evaluate the left-hand side of (23.1)
as
$$(\bold F_c Y)(k)=\ds\frac{1}{\sqrt{2\pi}}\int_0^\infty dx\,\Psi(k,x)^\dagger \,Y(x),
\tag 23.2$$
where $\Psi(k,x)$ is the physical solution constructed from
$\bold D$ via the procedure outlined in Chapter~9,
i.e. constructed as in (9.4). Thus, using (9.4) on the right-hand side of
(23.2) we obtain
$$(\bold F_c Y)(k)=\ds\frac{1}{\sqrt{2\pi}}\int_0^\infty dx\,
\left[f(-k,x)^\dagger+S(k)^\dagger\,f(k,x)^\dagger\right]
\,Y(x).
\tag 23.3$$
Recall that $f(k,x)$ is constructed from $\bold D$ via (10.6) and in turn
$K(x,y)$ is obtained as the unique solution to the
Marchenko equation. Thus, using (10.6) on the right-hand side of
(23.3) we obtain
$$\aligned
(\bold F_c Y)(k)=&\ds\frac{1}{\sqrt{2\pi}}\int_0^\infty dx\,
\left[e^{ikx}+S(-k)\,e^{-ikx}\right]Y(x)\\
&+
\ds\frac{1}{\sqrt{2\pi}}\int_0^\infty dx
\int_x^\infty dy\,\left[e^{iky}+S(-k)\,e^{-iky}\right]K(x,y)^\dagger\,Y(x)
,\endaligned
\tag 23.4$$
where we have also replaced $S(k)^\dagger$ by
$S(-k),$ which follows from (4.4).
Changing the order of integrations in the second integral in (23.4) we obtain
$$\aligned
\int_0^\infty dx
\int_x^\infty dy\,&\left[e^{iky}+S(-k)\,e^{-iky}\right]K(x,y)^\dagger\,Y(x)\\
&=
\int_0^\infty dy
\int_0^y dx\,\left[e^{iky}+S(-k)\,e^{-iky}\right]K(x,y)^\dagger\,Y(x).\endaligned
\tag 23.5$$
We remark that the change of the order of integrations
in (23.24) can be justified
by using an argument similar to the one given at the end of the proof of
Proposition~22.1, i.e. by approximating $Y(x)\in L^2(\bR^+)$ appearing in
(23.4) with a convergent sequence in $C_0(\bR^+).$
In terms of the operator $\bold K^\dagger$ related to (17.38), we recognize
the right-hand side of (23.4) and obtain
$$\aligned
\int_0^\infty dy
\int_0^y dx\,&\left[e^{iky}+S(-k)\,e^{-iky}\right]K(x,y)^\dagger\,Y(x)\\
=&
\int_0^\infty dy \,\left[e^{iky}+S(-k)\,e^{-iky}\right]
(\bold K^\dagger Y)(y).\endaligned\tag 23.6$$
Using (23.6) in (23.4), we get
$$
(\bold F_c Y)(k)=
\ds\frac{1}{\sqrt{2\pi}}
\int_0^\infty dy \,\left[e^{iky}+S(-k)\,e^{-iky}\right]
\left[\left(I+\bold K^\dagger) Y\right(y)\right],\tag 23.7$$
or equivalently we obtain
$$(\bold F_c Y)(k)=h(k)+S(-k)\,h(-k),\qquad k\in\bR^+,\tag 23.8$$
where we have defined
$$h(k):=\ds\frac{1}{\sqrt{2\pi}}
\int_0^\infty dy \,e^{iky}\,\left[(I+\bold K^\dagger) Y(y)\right].\tag 23.9$$
 From Proposition~17.5(e) we know that
$( I+\bold K^\dagger)$ is a bounded operator
on $L^2(\bR^+)$ and hence
$( I+\bold K^\dagger)Y$ belongs to $L^2(\bR^+)$ because
$Y\in L^2(\bR^+).$ Because the integrand in (23.9) has support in $y\in\bR^+,$ we conclude
that $h(k)$ defined in (23.9) belongs to the Hardy space
$\bold H^2(\bCp).$ Comparing (23.1) and (23.8) we see that for any
$g(k)\in \Upsilon\subset L^2(\bR),$ there exists $h(k)\in \bold H^2(\bCp)$ such that
$$h(k)+S(-k)\,h(-k)=g(k),\qquad k\in\bR^+.\tag 23.10$$
By postmultiplying (23.10) with $S(k)$ we get
$$S(k)\,h(k)+S(k)\,S(-k)\,h(-k)=S(k)\,g(k),\qquad k\in\bR^+.\tag 23.11$$
 From (4.4) we know that $S(k)\,S(-k)=I$ for $k\in\bR,$ and
 because $g(k)\in\Upsilon$ we have $S(k)\,g(k)=g(-k)$ for
 $k\in\bR.$ Thus, (23.11) is equivalent to
$$S(-k)\,h(-k)+h(k)=g(k),\qquad k\in\bR^-.\tag 23.12$$
 From (23.11) and (23.12) we conclude that
 for any $g(k)\in \Upsilon\subset L^2(\bR),$ there exists $h(k)\in \bold H^2(\bCp)$ such that
$$h(k)+S(-k)\,h(-k)=g(k),\qquad k\in\bR,\tag 23.13$$
which proves
$(\bold A).$
Thus, we have shown that if $\bold D$ belongs to the Faddeev class, then
there exists and uniquely exists a corresponding scattering data set $\bold S$
as in (4.2) satisfying the conditions
$(\bold I),$ $(\bold 2),$
$(\bold A),$
$(\bold 4_c),$ either one of
$(\bold V_e)$ or $(\bold V_h),$ and
$(\bold {VI})$ of Theorem~8.1. \qed

\noindent {\bf Proposition 23.2} {\it
 Let $ \overset{\circ}\to\Upsilon$ be a dense set in $\Upsilon$, where  $\Upsilon$ is defined in
 the statement of Theorem~8.1. Let us denote by $ \overset{\circ}
 \to \Upsilon_+$ the set of all the restrictions of vectors in  $ \overset{\circ}\to\Upsilon$ to $\bR^+$, i.e.,}
$$
 \overset{\circ}\to\Upsilon_+:= \left\{ g(k) \in L^2(\bR^+): g(k)= f(k), k \in \bR^+, \, \text{\rm for some}\, f(k) \in     \overset{\circ}\to \Upsilon  \right\}.
 \tag 23.14$$
{\it Then, $ \overset{\circ}\to \Upsilon_+$ is dense in $L^2(\bR^+).$
 }

\noindent PROOF:  Suppose that $ h(k)\in L^2(\bR^+)$ and that,
$$\left( h(k), g(k) \right)_2=0,\qquad \forall g(k) \in \overset{\circ}\to \Upsilon_+.
\tag 23.15$$
 We will prove that $h(k)\equiv 0$, which implies that  $ \overset{\circ}
 \to\Upsilon_+$ is dense in $L^2(\bR^+)$. We extend $h(k)$ to a function in $\Upsilon$ defining $h(k):= S(-k)\, h(-k)$ when $k \leq 0.$ Since by (4.4)
 we have $S(-k)\, S(k)= I$, it follows that $h(k)$ is indeed a vector in $\Upsilon$. Then,
$$\aligned
\left( h(k),g(k)  \right)_2&=  \left( h(k),g(k) \right)_{L^2(\bR^+ )}+  \left( h(k),g(k) \right)_ {L^2(\bR^- )}\\
&=
  \left( h(k),g(k) \right)_ {L^2(\bR^+ )}+  \left( S(k) h(k), S(k) g(k) \right)_ {L^2(\bR^+ )}\\
  &= 2 \left( h(k),g(k) \right)_ {L^2(\bR^+ )}=0,
\endaligned
\tag 23.16$$
 where we used that by (4.4) $S(k)^\dagger\, S(k)=I.$ Hence, $h(k)$ is orthogonal in $\Upsilon$ to $\overset{\circ}\to \Upsilon$, and since
  $\overset{\circ}\to \Upsilon$ is dense in $\Upsilon$ we have $ h(k)=0$
  for $k \in \bR$. \qed

\noindent {\bf Proposition 23.3} {\it
Suppose that the following conditions are satisfied: Conditions $(\bold 2)$ and $(\bold 4_c)$ of Definition 4.2,  $(\bold I)$, $(\bold{VI}),$ and either one of $(\bold V_e)$ or $(\bold V_h)$ of Definition 4.3 and  $(\bold A)$ of
Theorem~8.1. Then, the generalized Fourier map, $\bold F,$ defined in (22.3) is a unitary operator from $L^2(\bold R^+)$ onto $ \Cal R$, where $\Cal R$ is the Hilbert space defined in (22.1).}

\noindent PROOF:  Under our assumptions  Proposition~6.1 and   Proposition~16.10 apply. Hence, we construct the potential $V(x)$ satisfying (2.2) and (2.3),  the Jost solution $f(k,x)$, the physical solution, $\Psi(k,x),$  and the normalized
bound-state matrix solutions $ \Psi_j(x)$ for $j=1,\cdots, N$. In consequence, the generalized Fourier map $\bold F$  is well defined. By  Proposition~22.1~(b),
the Fourier map
$\bold F$ is isometric  from $L^2(\bR^+)$ into $ \Cal R$. Hence, to prove that it is unitary we only need to prove that $\bold F$ is onto
$\Cal R$. But as the range of any isometric operator is closed, it is enough to prove that  the  range of $\bold F$ is dense in $\Cal R$.
We first consider the case when condition $(\bold V_h)$ of Definition~4.3 holds.
Suppose that $ Y(x) \in \text{\rm Ker}\, [\bold F_c]$, i.e. we have
$$
(\bold F_c Y)(k)=0, \qquad k \in \bR^+.
\tag 23.17$$
Then, (23.1) holds with $ g(k)=0$. Furthermore, arguing exactly as in the proof of Theorem~23.1 we prove that equations (23.8) and (23.9) hold, i.e. we have
$$
 (\bold F_c Y)(k)= h(k)+ S(-k) h(k)=0,\qquad k \in \bR^+,
\tag 23.18$$
$$
 h(k ):= \frac{1}{\sqrt{2\pi}}\, \int_0^\infty\, dy \, e^{iky}\, \left[ (I+ \bold K^\dagger) Y(y)    \right].
\tag 23.19$$
Then, exactly in the same way as in the proof of Theorem 23.1 we prove that (23.13) holds with $g(k)=0$, i.e., that,
$$
h(k)+ S(-k)\, h(-k)=0, \qquad k \in \bR.
\tag 23.20$$
By Proposition~17.5(d) and the equivalence of $(\bold 4_a)$ and $(\bold 4_c)$  given in Proposition~6.1, $(I+ \bold K^\dagger)$ is a bijection on $L^2(\bR^+)$ and since the Fourier transform is a bijection from $L^2(\bR^+)$ onto $ \bold H^2(\bCp)$, we have that $ h(k) \in \bold H^2(\bCp)$. Hence, $Y(x) \in \text{\rm Ker}\,[ \bold F_c]$ if and only if $ h(k) \in \bold H^2(\bCp)$  is a solution to (23.20). Then,  the dimension of the kernel of $\bold F_c$ is equal to the number of linearly independent solutions to (23.20) that are in $\bold H^2(\bCp)$.

By condition  $(\bold V_h)$ there are $\Cal N$ linearly independent solutions to (23.20) in $\bold H^2(\bCp)$ and in consequence  the kernel of $\bold F_c$ has dimension $\Cal N$.

Let us define the following map from $L^2(\bR^+)$ into  $\text{\rm Ran}\,[M_1]\oplus \cdots \oplus\text{\rm Ran}\,[M_N],$
$$
\bold F_p Y:= \left( \bold F_1 Y, \cdots, \bold F_N Y \right),\qquad Y(x) \in L^2(\bR^+),
\tag 23.21$$
where the $\bold F_j,j=1,\cdots,N,$ are defined in (22.4). It follows from  (22.3) that,
$$
\bold F Y = \left(\bold F_p Y, \bold F_c Y \right),\qquad Y(x) \in L^2(\bR^+).
\tag 23.22$$
 By Proposition~22.1(b), the map  $ \bold F_p$ is isometric from $\text{\rm Ker}[\bold F_c]$ into $\text{\rm Ran}\,[M_1]\oplus \cdots \oplus\text{\rm Ran}\,[M_N]$ and since the dimension of  $\text{\rm Ker} \,[\bold F_c]$ and of   $\text{\rm Ran}\,[M_1]\oplus \cdots \oplus
 \text{\rm Ran}\,[M_N]$ is $\Cal N$, we have that  $\bold F_p$ is unitary from
 $\text{\rm Ker}\, [\bold F_c]$ onto  $\text{\rm Ran}\,[M_1]\oplus \cdots\oplus
 \text{\rm Ran}\,[M_N]$. Furthermore,
$$
L^2= \text{\rm Ker}\,[ \bold F_c] \oplus \left(\text{\rm Ker}\, [\bold F_c]\right)^{\perp}.
\tag 23.23$$
Since  $\bold F_p$ is unitary from $\text{\rm Ker}\, F_c$ onto
$\text{\rm Ran}\,[M_1]\oplus \cdots \text{\rm Ran}\,[M_N]$,
to prove that the range of $\bold F$ is dense  in $\Cal R$ it is enough to prove that  the range of   $\bold F_c$ is dense in  $L^2(\bR^+)$. By Proposition~23.2 it is sufficient to prove that  for every $ g(k) \in   \overset{\circ}\to\Upsilon_+ $ there is a function $ Y(x) \in  (\text{\rm  Ker}\, \bold F_{c})^{\perp}$ such that,
$$
g(k) = (\bold F_c Y)(k), \qquad k >0.
\tag 23.24$$
Since  $g(k) \in  \overset{\circ}\to\Upsilon_+$,  there is a $ f(k)\in  \overset{\circ}\to\Upsilon$ such that $g(k)=f(k), k \in \bR^+$. Furthermore,  by  condition  $(\bold A)$ there
is a $h \in \bold H^2_+(\bCp)$ such that,
$$
f(k)= h(k)+ S(-k)\, h(-k), \qquad k \in \bR.
\tag 23.25$$

Denote,
$$
m(x):= (I+\bold K^\dagger)^{-1} \frac{1}{\sqrt{2 \pi}}\, \int_{-\infty}^\infty\, e^{-ik x} h(k)\, dk.
\tag 23.26$$
The vector $m(x)$ belongs to  $L^2(\bR^+)$, because the inverse Fourier transform  is a bijection from  $\bold H^2(\bCp)$ onto  $L^2(\bR^+)$    and  $(I+\bold K^\dagger)$ is a bijection on  $L^2(\bR^+)$ .
By (23.25) and (23.36) we have
$$
f(k)=  \frac{1}{\sqrt{2\pi}}\int_0^\infty \, dx   \left(e^{i k x} + S(-k)\,  e^{-i k x} \right)   \left[(I+\,\bold K^\dagger)\right] m(x), \qquad k \in \bR.
\tag 23.27$$
As in the proof of (23.7) we prove that the right-hand side of (23.27) is equal to $(\bold F_c m )(k)$, for $ k \in \bR^+$,    and then,
$$
  f(k) = (\bold F_c m )(k), \qquad k \in \bR^+.
\tag 23.28$$
Let us decompose $m(x)$ as,
$$
m(x)= m_1(x)+ m_2(x),\quad  m_1(x) \in\, \text{\rm Ker}\, [\bold F_c],
 \quad m_2(x) \in \left(\text{ Ker}\,[ \bold F_c]\right)^{\perp}.
\tag 23.29$$
Hence, by  (23.28) and (23.29)
$$
f(k)=  (\bold F_c m_2)(k), k >0,
\tag 23.30$$
and since $ g(k)= f(k), k \in \bR^+,$ we obtain that,
$$
g(k)=  (\bold F_c m_2)(k), k >0.
\tag 23.31$$
Then, (23.24) holds with $Y(x)=m_2(x)$, what proves that the range of $\bold F_c$ is dense in $L^2(\bR^+)$.
Suppose now that condition $(\bold V_e)$ of Definition 4.3 holds. Then, there are $\Cal N$ linearly independent solutions to (23.20) that are in $\hat{L}^1(\bCp),$ but by Proposition~15.7(e) these solutions are in   $\hat{L}^1_\infty(\bCp)$ and as   $\hat{L}_\infty^1(\bCp) \subset \bold H^2(\bCp)$, there are $\Cal N$ linearly independent solutions to (23.20)  in $ \bold H^2(\bCp)$. We complete the proof as in the case when condition $(\bold V_e)$ of Definition 4.3 is satisfied. \qed

Let us define the operator $ \hat{\Cal L}$ in the Hilbert space $\Cal R$  defined in (22.1),
$$
\hat{\Cal L}\left(Z_1,\cdots,Z_N, Z\right):= \left( -\kappa_1^2 Z_1,\cdots, -\kappa_N^2 Z_N, k^2 Z(k)\right), \quad \left(Z_1,\cdots,Z_N, Z\right) \in D(\hat{\Cal L}),
\tag 23.32$$
where the domain of $\hat{\Cal L}$, that we denote by $D(\hat{\Cal L}),$ is defined as follows,
$$
D(\hat{\Cal L}):=\left\{  \left(Z_1,\cdots, Z_N, Z\right)\in \Cal R: k^2 Z(k) \in L^2(\bR^+)  \right\}.
\tag 23.33$$
Since $\hat{\Cal L}$ is a multiplication operator defined on its maximal domain, it is selfadjoint, i.e.  $\hat{\Cal L}^\dagger=\hat{\Cal L}.$

Using the generalized Fourier map $\bold F$ defined in (22.3), we define operator $\Cal L$ in $L^2(\bR^+),$
$$
\Cal L:= \bold F^\dagger\, \hat{\Cal L}\, \bold F.
\tag 23.34$$
Then, we have that,
$$
(\Cal L Y)(x)= (\bold F^\dagger \,\hat{\Cal L}\, \bold F Y)(x), Y(x) \in D(\Cal L):=\left\{ Y(x)\in L^2(\bR^+): (\bold  F Y)(k) \in D(\hat{\Cal L})  \right\},
\tag 23.35$$
where by $D(\Cal L)$ we denote the domain of $\Cal L$.

\noindent {\bf Proposition 23.4} {\it
Suppose that the following conditions are satisfied: Conditions $(\bold 2)$ and $(\bold 4_c)$ of Definition 4.2,  $(\bold I)$, $(\bold{VI} ),$ and either one of $(\bold V_e)$ or $(\bold V_h)$ of Definition 4.3 and  $(\bold A)$ of Theorem 8.1.
Then, the operator $\Cal L$ defined in (23.34) is selfadjoint and its domain, $D(\Cal L),$ is contained  in $\bold H^1(\bR^+)$,
$$
 D(\Cal L)  \subset\bold H^1(\bR^+),
\tag 23.36$$
where by   $\bold H^1(\bR^+)$ we denote the Sobolev space of all vectors in $L^2(\bR^+)$ with first derivative in  $L^2(\bR^+)$.}

\noindent PROOF:  By Proposition~23.3 the generalized Fourier map $\bold F$ is unitary from $ L^2(\bR^+)$ onto $\Cal R$. Then, $ \Cal L$ is selfadjoint because  by its definition in (23.34),  $\Cal L$ is unitarily equivalent to $\hat{\Cal L}$ that is selfadjoint because it is a multiplication operator defined on its maximal domain. Let us prove (23.36). By definition  $ D(\Cal L)  \subset L^2(\bR^+)$. Hence, we only need to prove that for any $Y(x)\in   D(\Cal L)$, we have that, $Y'(x)\in  L^2(\bR^+)$.
Recall that by Proposition 6.1 conditions $(\bold 4_a)$,  and $(\bold 4_e)$ are equivalent. Let $ Y(x)$ belongs to    $D(\Cal L)$. Then by (23.35), and since $\bold F^\dagger \bold F=I,$  there is a $ \left(Z_1,\cdots, Z_N, Z\right) \in D(\hat{\Cal L})$ such that
$$
Y'(x)=( \bold F^\dagger\, (Z_1,\cdots, Z_N, Z))' (x)= \sum_{j=1}^N (\bold F_j^\dagger \,Z_j)'(x)+ (\bold F^\dagger_c \,Z)'(x).
\tag 23.37$$
By (17.12) and (22.4),
$$
 (\bold F_j^\dagger \,Z_j)'(x )\in  L^2(\bR^+), j=1.\cdots,N.
 \tag 23.38$$
 By (9.4) and (22.24)
$$
  (\bold F^\dagger_c \,Z)'(x)= Y_1(x)+Y_2(x),
\tag 23.39$$
where
$$
Y_{1}(x):= - \frac{1}{\sqrt{2\pi}}\, \int_0^\infty\, f'(-k,x)\, Z(k)\, dk,
\tag 23.40$$
and
$$
Y_2(x):=\frac{1}{\sqrt{2\pi}}\, \int_0^\infty\, f'(k,x)\, S(k)\, Z(k)\, dk.
\tag 23.41$$
Note that the derivation under the integral sign  in (23.40), (23.41)  is justified by Lebesque dominated convergence theorem, since the integrands in (23.40), (23.41) are integrable column vectors. This is proven  using (17.12), that by (4.4), $\|S(k) \|=1$, and that  as $ k^2 Z(k) \in  L^2(\bR^+)$, we have that  $ (1+k)\, Z(k)\in L^1(\bR^+),$
$$
\int_0^\infty (1+k)\, |Z(k)|\, dk \leq \left( \int_0^\infty \frac{1}{(1+k)^2}\, dk\right)^{\frac{1}{2}}\,  \left( \int_0^\infty\,(1+k)^4\,|Z(k)|^2 \,dk \right)^{\frac{1}{2}}\, < \infty.
\tag 23.42$$
Let us prove that $ Y_1(x)\in L^2(\bR^+)$.
Recall that, as in (5.1), the Jost solution can be characterized as the unique solution of the integral equation,
$$
f(k,x)= e^{ikx} I_n + \frac{1}{k}\, \int_x^\infty\, \sin k (y-x)\, V(y)\, f(k,y)\, dy.
\tag 23.43$$
Here, $V(x)$ is the reconstructed potential obtained from (10.4).

We have that,
$$
f'(-k,x)=- i k \,e^{ikx} I_n- \int_x^\infty \cos k(y-x) \, V(y)\, f(k,y)\, dy.
\tag 23.44$$
The differentiation under the integral sign is justified because by (17.11) and  (2.3) the integrand in (23.44) is integrable.

  By (23.44),
$$
 Y_1(x)=- \frac{1}{\sqrt{2\pi}} \,\int_0^\infty\, f'(-k,x)\,  Z(k)\, dk = g_1(x)+ g_2(x),
\tag 23.45$$
   where,
$$
   g_1(x):= \frac{1}{\sqrt{2\pi}}\, \int_0^\infty \, e^{ik x} \, i k Z(k)\, dk \in L^2(\bR^+), \text{since}\,   k  \,Z(k) \in L^2(\bR^+),
\tag 23.46$$
   and
$$
   g_2(x):=  \frac{1}{\sqrt{2\pi}}\, \int_0^\infty\, dk \,   Z(k) \, \int_x^\infty \cos k(y-x)\,  V(y)\, f(k,y)\, dy.
\tag 23.47$$
   By Proposition~6.1, (10.4), (16.78),  (17.1) and (23.42),
$$\aligned
   |g_2(x)| &\leq C \left( \int_0^\infty |Z(k)| \, dk \right)\, \int_x^\infty |V(y)| \, dy\\
    &\leq C (1+x)^{-1}\, \left(\int_0^\infty\, (1+y)\, |V(y)|\, dy\right)  \in L^2(\bR^+).\endaligned
\tag 23.48$$
   Then, by (2.3), Proposition~7.1, (17.1), and
   (23.42), we have
$$
     Y_1(x)   \in L^2(\bR^+).
\tag 23.49$$
   Recall that by (4.4)  $\| S(k) \|=1$. This allows us to  prove, as above,
$$
   Y_2(x) \,\in L^2(\bR^+).
\tag 23.50$$
Hence, by (23.39) and (23.49) we get
$$
   ( F^\dagger_{ c}\, Z )'(x) \in L^2(\bR^+).
\tag 23.51$$
By (23.37), (23.38), and (23.51),
we conclude that $Y'(x)\in  L^2(\bR^+).$ \qed

We now introduce some concepts  from the spectral theory of differential operators [4].  We denote by $\Cal L_{\text{\rm max}}$ the maximal operator associated to $ - D^2_x I +V$, where $D_x$ is the derivative operator $d/dx$, namely
$$
  \Cal L_{\text{\rm max}} Y(x) := (-D^2_x +V(x))\, Y(x),\qquad  Y(x) \in D( \Cal L_{\text{\rm max}}),
\tag 23.52$$
  where the domain of $ \Cal L_{\text{\rm max}}$, that we denote by $ D( \Cal L_{\text{\rm max}}), $ is defined as follows,
$$\aligned
D(  \Cal L_{\text{\rm max}} )=\left\{  Y(x) \in L^2(\bR^+): Y(x),Y'(x) \, \text{ are absolutely continuous on  }\, (0, \infty)\right.  \\
\left. \hbox{and}   ( - D^2_x +V(x))\, Y(x)
  \in L^2 (\bR^+)  \right\}.\endaligned
\tag 23.53$$
We designate by $\Cal L_{\text{\rm min}}$ the minimal operator associated to
$ - D^2_x
\,I+V$,
$$
  \Cal L_{\text{\rm min}} Y(x) := (-D^2_x +V(x))\, Y(x),\qquad  Y(x) \in D( \Cal L_{\text{\rm min}}),
\tag 23.54$$
  where the domain of $ \Cal L_{\text{\rm min}}$, that we denote by $ D( \Cal L_{\text{\rm min}}), $ is defined as follows,
$$
D(  \Cal L_{\text{\rm min}} )=\left\{  Y(x) \in D(  \Cal L_{\text{\rm max}} ): Y(x) ,\,\text{\rm has compact support in }\, (0,\infty)  \right\}.
\tag 23.55$$
Clearly, $ \Cal L_{\text{\rm min}}  \subset   \Cal L_{\text{\rm max}}.$ By Theorems 3.7 in page 47 and Theorem 3.9 in page 49 of  [42]   $ D(  \Cal L_{\text{\rm min}} )$ is dense in $L^2(\bR^+)$,  $\Cal L_{\text{\rm min}}$ is symmetric  and $ \Cal L_{\text{\rm min}} \subset \Cal L_{\text{\rm min}}^\dagger = \Cal L_{\text{\rm max}.}$ We denote by $\overline{\Cal L_{\text{\rm min}}}$ the closure of  $\Cal L_{\text{\rm min}}.$
Note the difference in notation. In [42]   $\Cal L_{\text{\rm min}}$ is called  $T'_0$,  $\overline{\Cal L_{\text{\rm max}}}$ is called  $T_0$ and
 $\Cal L_{\text{\rm max}}$ is called $T$.

\noindent {\bf Proposition 23.5} {\it
Suppose that the following conditions are satisfied: Conditions $(\bold 2)$ and $(\bold 4_c)$ of Definition 4.2,  $(\bold I)$, $(\bold{VI} ),$ and either one of $(\bold V_e)$ or $(\bold V_h)$ of Definition 4.3 and  $(\bold A)$ of Theorem 8.1. Then,
$$
 \overline{\Cal L_{\text{\rm min}}} \subset \Cal L \subset \Cal L_{\text{\rm max}},
\tag 23.56$$
where $ \overline{\Cal L_{\text{\rm min}}}$ is the closure of the operator $ \Cal L_{\text{\rm min}}$ defined in (23.54), $ \Cal L $ is the selfadjoint operator defined in (23.34) and $\Cal L_{\text{\rm max}}$ is the operator defined in (23.52).
}

\noindent PROOF:  We first prove that $\Cal L \subset \Cal L_{\text{max}}.$ For this purpose we define $\Cal L_0,$ a restriction of $\Cal L,$ as
$$(\Cal L_0 Y)(x):= (\Cal L Y)(x), \qquad Y(x) \in \text{\rm Dom}[\Cal L_0],
\tag 23.57$$
where the domain of $\Cal L_0$ is defined as
$$\aligned
\text{\rm Dom}[\Cal L_0]:= \left\{ Y(x)\in L^2(\bR^+): \,
Y(x)= \bold F^\dagger(Z_1,\cdots, Z_N, Z)\  \text { with } \right.
\\
\left. (Z_1,\cdots,Z_N,Z)\in\Cal R, \text{ where }
Z(k) \in L^2(\bR^+) \text { and } Z(k)\text{ has compact support}   \right\}.
\endaligned
\tag 23.58$$
As $Z(k)$ has compact support, $ k^2 Z(k)\in L^2(\bR^+)$ and then, $\Cal L_0 \subset \Cal L$. Moreover, $\overline{\Cal L_0}= \Cal L$. Furthermore, as $\Cal L$ is closed because it is selfadjoint and since $ \Cal L_{\text{\rm max}}$ is closed because it is the adjoint of  $\Cal L_{\text{\rm min}},$ it is enough to prove that $\Cal L_0 \subset \Cal L_{\text{\rm max}}$.
Suppose that $Y(x)\in \text{\rm Dom}[\Cal L_0]$. Then,
$$
\Cal L_0Y(x)= \bold F^\dagger  \left( -\kappa_1^2 Z_1,\cdots, -\kappa_N^2 Z_N, k^2 Z(k)\right)= \sum_{j=1}^N \bold F_j^\dagger(- \kappa_j^2) Z_j+ \bold F_c^\dagger k^2 Z(k).
\tag 23.59$$
But,
$$
\bold F_j^\dagger(- \kappa_j^2) Z_j= \Psi_j(x) (- \kappa_j^2) Z_j=  (-D^2_x +V(x))\,  \bold F_j^\dagger Z_j,\qquad j=1,\dots,N,
\tag 23.60$$
because $\Psi_j(x)$ is a solution to (2.1) with $ k= i \kappa_j$
for $ j=1,\dots,N$.
Moreover,
$$\aligned
\left(\bold F_c^\dagger\, k^2\, Z(k)\right)(x)&= \frac{1}{\sqrt{2 \pi}}\,\int_0^\infty \Psi(k,x)\, k^2 \, Z(k)\, dk\\
&=   (-D^2_x +V(x)) \frac{1}{\sqrt{2 \pi}} \int_0^\infty \Psi(k,x)\,  Z(k)\, dk\\
&=  (-D^2_x +V(x))\left(\bold F_c^\dagger Z(k)\right)(x),\endaligned
\tag 23.61$$
where we used that $\Psi(k,x)$ is a solution of (2.1) and that $Z(k)$ has compact support. By (23.35), (23.59), (23.60), (23.61)
$$
\Cal L_0\, Y(x)=  \left(-D^2_x +V(x)\right) Y(x) \in L^2(\bR^+) , \qquad Y(x)\in \text{\rm Dom}[\Cal L_0].
\tag 23.62$$
Furthermore,
$$
D^2_x\, Y(x)= V(x)\, Y(x)-\Cal L_0\, Y(x).
\tag 23.63$$
By Proposition~23.4 $Y(x)$ and $Y'(x)$ belong to  $L^2(\bR^+)$. Then,
$$
|Y(x)|^2= - \int_x^\infty \left(|Y(x)|^2\right)' \, dx= -   \int_x^\infty( (Y^\dagger)'(x) Y(x)+  Y^\dagger (x) Y'(x)) \, dx,
\tag 23.64$$
and then, by Schwarz inequality,
$$
|Y(x)|^2 \leq 2 \,\| Y(x) \|_2\, \| Y'(x)\|_2.
\tag 23.65$$
Since $V(x)$ satisfies (2.3), $Y(x)$ is bounded by (23.65), and $\Cal L_0 Y(x) \in  L^2(\bR^+),$ we have that $D^2_x Y(x) \in L^1(0,R)$ for any $ R >0$. In consequence $ Y(x)$ and $Y'(x)$ are absolutely continuous and since by (23.62)  $(-D^2_x +V(x)) Y(x) \in L^2(\bR^+)$ it follows that $Y(x)$ is in the domain of
 $\Cal L_{\text{\rm max}}$, and we have proved that,
$$
\text{\rm Dom}[\Cal L_0] \subset \text{\rm Dom}[\Cal L_{\text{max}}].
\tag 23.66$$
Equation (23.62) and (23.66) imply that  $\Cal L_0 \subset \Cal L_{\text{\rm max}}$.
Let us now prove that $  \overline{\Cal L_{\text{\rm min}}} \subset \Cal L.$ Since $\Cal L$ is closed, it is enough to prove that  $\Cal L_{\text{\rm min}} \subset \Cal L$. Suppose that $Y(x) \in \text{\rm Dom}[\Cal L_{\text{min}}]$. Then, denote
$$
Z_j:= \bold F_j\, Y, \qquad j=1,\dots, N, \quad Z(k):= (\bold F_c Y)(k).
\tag 23.67$$
We have that,
$$
\bold F\,Y= \left( Z_1,\cdots, Z_N, Z(k)\right).
\tag 23.68$$
Moreover,
$$\aligned
\bold F_j \Cal L_{\text{\rm min}} Y(x)&= \bold F_j  (-D^2_x +V(x)) Y(x)\\
&= \int_0^\infty \Psi_j^\dagger(x)  (-D^2_x +V(x)) Y(x) dx\\
&= - \kappa_j^2 \,Z_j, \qquad j=1,\dots,N,\endaligned
\tag 23.69$$
where we used that $\Psi_j(x)$ satisfies (2.1) and that $Y(x)$ has compact support in $\bR^+$ in order to integrate by parts. In a similar way we prove that
$$\aligned
(\bold F_c ( \Cal L_{\text{\rm min}} Y)(k)&=(\bold F_c (-D^2_x +V(x)) Y(x))(k)\\
&= \frac{1}{\sqrt{2\pi}}\,\int_0^\infty\, \Psi(k,x)^\dagger\,(-D^2_x +V(x)) Y(x)\, dx\\
&= k^2 Z(k).\endaligned
\tag 23.70$$
By (23.69), (23.70)
$$\aligned
\bold F (  \Cal L_{\text{\rm min}}Y(x)(k)&=\left(\bold F (-D^2_x +V(x)) Y(x)\right)\\
&= \left(  - \kappa_1^2 Z_1,\cdots, - \kappa_N^2 Z_N, k^2 Z(k) \right)\\
&= \hat{\Cal L}\left(Z_1,\cdots,Z_N, Z(k)   \right) \in \Cal R,\endaligned
\tag 23.71$$
where we used that $ \Cal L_{\text{\rm min}}Y(x)= (-D^2_x +V(x)) Y(x) \in L^2(\bR^+)$ and that $\bold F$ is unitary from  $L^2(\bR^+)$ onto $\Cal R$. In particular, this implies that
$k^2 Z(k) \in  L^2(\bR^+)$. This means that $\bold F Y \in \text{\rm Dom}[\hat\Cal L]$, and then, $Y(x)\in \text{\rm Dom}[\Cal L]$. It follows that,
$$
\text{\rm Dom}[\Cal L_0] \subset \text{\rm Dom}[\Cal L].
\tag 23.72$$
Multiplying by $\bold F^\dagger$ in both sides of (23.71), using $\bold F^\dagger\bold F=I $ and (23.68) we obtain that,
$$
  \Cal L_{\text{\rm min}}Y(x)= \left(\bold F^\dagger \hat{\Cal L}\, \bold F Y\right)(x)= \Cal L Y(x),
\tag 23.73$$
   where we used (23.34). Equations (23.72) and (23.73) imply that $\Cal L_{\text{\rm min}} \subset \Cal L$. \qed

\noindent {\bf Proposition 23.6} {\it
Suppose that the following conditions are satisfied: Conditions $(\bold 2)$ and $(\bold 4_c)$ of Definition 4.2,  $(\bold I)$,$(\bold{VI} ),$ and either one of $(\bold V_e)$ or $(\bold V_h)$ of Definition 4.3 and  $(\bold A)$ of Theorem 8.1. Then,
there is a pair of matrices $(A,B)$, unique up to post multiplication by an invertible matrix $T$,  that satisfy (2.6), (2.7),  and such that, all vectors $Y(x)$ in the domain of $\Cal L$ satisfy}
$$
- B^\dagger\, Y(0)+ A^\dagger\, Y'(0)=0, \qquad Y(x)\in \text{\rm Dom}[\Cal L].
\tag 23.74$$

\noindent PROOF:  As in the proof of Proposition~23.5 we prove that  for any $ Y(x)\in \text{\rm Dom}[\Cal L]$, $D^2_x\, Y(x) \in L^1(0,R)$ for any $ R >0$. Then, $Y(x)$ and $Y'(x)$ are absolutely continuo on $[0,\infty)$ and in consequence, $Y(0)$ and $Y'(0)$ are well defined so that (23.74) makes sense.
  Let us define the  quadratic form,
$$
  [Y,G]_x:= \sum_{j=1}^n \left(Y_j(x)^\ast\,G'_j(x) - Y'(x)_j^\ast \, G_j(x)\right), \qquad Y,G \in \text{\rm Dom}[\Cal L_{\text{max}}].
\tag 23.75$$
  Furthermore (see  page 28 of [2] and  Section~II of [9]), the matrix Schr\"odinger equation (2.1) has the $ n\times n$ matrix solution  $g(k,x)$ that satisfies for each $ k \in \overline{ \bCp} \setminus \{0\}$ the asymptotics
$$\cases
  g(k,x)= e^{-ik x}\left(I+ o\left(\ds\frac{1}{x}\right)\right),   \qquad x \to  +\infty\\
   g'(k,x)= -i k e^{-ik x}\left(I+ o\left(\ds\frac{1}{x}\right)\right), \qquad x \to  + \infty.\endcases
\tag 23.76$$
  Moreover, the combined $ 2 n$ columns of $f(k,x)$ and $g(k,x)$ form a fundamental system of solutions to (2.1). Then, any column vector solution  $Y(k,x)$  to (2.1) can be written as follows,
$$
   Y(k,x)= f(k,x)\, Q_1+ g(k,x)\, Q_2,
\tag 23.77$$
    for some constant column vectors $Q_1,Q_2$. It follows from (23.76) that each of the $n$ columns of $g(k,x)$ grows exponentially as $ x \to +\infty$ for each fixed $ k \in \bCp$. Then, all solutions to (2.1) with  $ k \in \bCp$ (or equivalently with $ \text{\rm Im} \,k^2 \neq 0$ )  that belong to $ L^2(  R < x < \infty)$ for $ R \geq 0$ must  have $Q_2=0$. Note that it follows from (10.6) that  the $n$ columns of $ f(k,x)$ are linearly independent solution to (2.1) and that they decay exponentially as $ x \to +\infty$ for each fixed $ k \in \bCp$. Hence, for each fixed $ k^2$   with $\text{Im} [k^2]\neq 0$ there are exactly $n$ linearly independent solution to (2.1) that are in  $ L^2(  R < x < \infty)$ for $ R \geq 0$. Then,  it follows from  Proposition~23.5 and Theorem~4.8 in page 61 of [42] that,
$$[Y,G]_\infty:= \lim_{x \to +\infty}\,  [Y,G]_x=0, \qquad \forall  \,Y(x),\, G(x) \in \text{\rm Dom}[\Cal L_{\text{max}}].
\tag 23.78$$
  Hence, by  Theorem~4.1 on page 53 of [42], we have that
  for $Y(x), G(x) \in \text{\rm Dom}[\Cal L]$,
 $$
 [Y,G]_0= (\Cal L Y,G)-(Y, \Cal L G)=0,\tag 23.79
 $$
( what is obvious in our case because $\Cal L$ is selfadjoint), and   moreover,  if $Y \in  \text{\rm Dom}[\Cal L_{\text{max}}]$ and $[Y,G]_0=0$ for all $G \in \text{\rm Dom}[\Cal L]$, then, $f \in \text{\rm Dom}[\Cal L]$.
  This means that $ D(\Cal L)$ is  a maximal  isotropic subspace for $[\cdot,\cdot ]_0$.  In consequence, by   Lemma~2.2 of [29] and Theorem 2.1 of [5] there exist matrices $A,B$ that satisfy (2.6), (2.7)  such that all the vectors
    on the  domain of $\Cal L$ satisfy the boundary conditions (23.74). \qed


\noindent{\bf Proof of Theorem 8.1} By Theorem~23.1 if the input data set $\Cal D$ belongs to the Faddeev class the conditions   $(\bold 2)$ and $(\bold 4_c)$ of Definition 4.2,  $(\bold I)$,
$(\bold{VI} ),$ and either one of $(\bold V_e)$ or $(\bold V_h)$ of Definition 4.3 and  $(\bold A)$, are satisfied.
Let us assume that the scattering data set $ \bold S$ satisfies the conditions:  $(\bold 2)$ and $(\bold 4_c)$ of Definition~4.2, $(\bold I)$, $(\bold{VI} ),$ and either one of $(\bold V_e)$ or $(\bold V_h)$ of Definition 4.3 and  $(\bold A)$.
 Using Proposition~6.1 and   Proposition~16.10 we construct the potential $V(x)$, that satisfies (2.2), (2.3),  the Jost solution, $f(k,x)$, the physical solution, $\Psi(k,x),$  and the normalized bound state matrix solutions $ \Psi_j(x)$
 for $j=1,\dots, N$. We have to prove that   the $\Psi_j(x)$
 for $j=1,\dots,N,$ and the physical solution $\Psi(k,x)$ satisfy  the boundary condition  (2.4).
By Proposition~23.4 the operator, $\Cal L$ defined in (23.34), (23.35), is selfadjoint and by Proposition~23.6 the vectors in the domain of $\Cal L$ satisfy the boundary condition (23.74).
For any constant column vector$ Z_j \in \text{\rm Ran}\, M_j$
with $j=1,\dots,N,$ let us define
$$
\left( 0,\cdots,0, Z_j ,0\cdots,0 \right) \in \Cal R,
\tag 23.80$$
where the component  at the position $j$ is equal to $Z_j$ and all the others are equal to zero. Then,
$$
(0,\cdots,0,Z_j,0,\cdots,0) \in \text{\rm Dom}[\hat\Cal L],
\tag 23.81$$
where the operator $ \hat{\Cal L}$ is defined in (23.32), (23.33). Then,
$$
\Psi_j(x)Z_j= \bold F_j^\dagger Z_j = \bold F(0,\cdots, Z_j,0\cdots,0)
\in \text{\rm Dom}[\Cal L],
\tag 23.82$$
where we used (23.35). Then, we have
$$
- B^\dagger \Psi_j(0)Z_j+ A^\dagger \Psi_j'(0)Z_j=0,\qquad
 \forall Z_j   \in \text{\rm Ran} [M_j],   \quad j=1,\dots,N.
\tag 23.83$$
Given any constant column vector $ Q \in \bC^n$ let us decompose it as follows,
$$
Q= Z_j+\tilde{Q}, \quad Z_j \in  \text{\rm Ran}[M_j],
\quad \tilde{Q} \,\in \text{\rm Ker}[M_j],\qquad j=1,\dots,N.
\tag 23.84$$
Hence, by the definition of $\Psi_j$  in (9.8) and (23.82)
$$\aligned
- B^\dagger\, \Psi_j(0)\,Q+ A^\dagger\, \Psi_j'(0)\,Q&=   - B^\dagger\, \Psi_j(0)\,Z_j+ A^\dagger\, \Psi_j'(0)\,Z_j\\
&=0, \qquad  j=1,\dots,N, \quad \forall Q \in \bC^n.\endaligned
\tag 23.85$$
But then,
$$
- B^\dagger\, \Psi_j(0)+ A^\dagger\, \Psi_j'(0)=0,\qquad   j=1,\dots,N.
\tag 23.86$$
Let us now prove that the physical solution $\Psi(k,x)$ satisfies the boundary condition. For any   $ k_0 \in (0,\infty),$ $\varepsilon >0$, we denote by $\chi_{[k_0-\varepsilon, k_0+\varepsilon]} (k)$ the characteristic function of $[k_0-\varepsilon, k_0+\varepsilon]$, i.e.,  $\chi_{[k_0-\varepsilon, k_0+\varepsilon]} (k)=1$ for $ k \in  [k_0-\varepsilon, k_0+\varepsilon]$ and   $\chi_{[k_0-\varepsilon, k_0+\varepsilon]} (k)=0$  if $ k \in [0,\infty]\setminus [k_0-\varepsilon, k_0+\varepsilon]$. Let us define,
$$
(0,\cdots,0,\frac{1}{2\varepsilon}\chi_{[k_0-\varepsilon, k_0+\varepsilon]}(k)) \in \Cal R.
\tag 23.87$$
Since $ k^2\, \chi_{[k_0-\varepsilon, k_0+\varepsilon]}(k) \in L^2(\bR^+)$, we have
$$
(0,\cdots,0,\frac{1}{2\varepsilon}\,\chi_{[k_0-\varepsilon, k_0+\varepsilon]}(k)) \in \text{\rm Dom}[\hat \Cal L] .
\tag 23.88$$
Then,
$$\aligned
  \psi_{\varepsilon}(k_0,x):&= \frac{1}{\sqrt{2 \pi}} \, \int_{k_0-\varepsilon}^{k_0+\varepsilon}    \psi(k,x)\,   \frac{1}{2 \varepsilon} \, dk  \\
  &= \bold F^\dagger (0,\cdots,0,\frac{1}{2\varepsilon}\chi_{[k_0-\varepsilon, k_0+\varepsilon]}(k))   \in \text{\rm Dom}[\Cal L].\endaligned
 \tag 23.89$$
 In consequence,
$$
  \left(  -B^\dagger\psi_\varepsilon(k_0,0)+A^\dagger\psi_\varepsilon'(k_0,0) \right)=0.
\tag 23.90$$
 By  the definition (9.4) of the physical solution and since $f(k,x)$ is continuous in $k$ for each fixed $x$ and $S(k$ is continuous in $k$ , it follows from (23.89) and the mean value theorem that,
$$
 -B^\dagger\psi(k_0,0)+A^\dagger\psi'(k_0,0)= \lim_{\varepsilon \rightarrow 0}  \left(  -B^\dagger\psi_\varepsilon(k_0,0)+A^\dagger\psi_\varepsilon'(k_0,0) \right)=0,
\tag 23.91$$
 which proves that the physical solution $\Psi(k,x)$ satisfies   the boundary condition (2.4) for $k >0$, and by continuity also for $k=0$.
 We prove that, in the direct problem, the reconstructed input data set, $\bold D :=\{V,A,B\}$ yields the same scattering data set $\bold S$ used as input of the inverse scattering problem as in the proof of Theorem~5.1. The fact $V$ is unique and that $(A,B)$ are unique up to the transformation $(A,B) \rightarrow (A T, B T)$ for any invertible matrix $T$ follows from Proposition~16.2. \qed

\newpage
\noindent {\bf 24. A STAR GRAPH}
\vskip 3 pt

In this chapter we show that a matrix Schr\"odinger equation with a diagonal potential matrix is unitarily equivalent to a star graph.
 A star graph is a quantum graph with only one vertex and a finite number, $n,$ of semi infinite edges. The Hilbert space is given by
$$
 \Cal H:= \oplus_{j=1}^n L^2(\bR^+, \bC),
\tag 24.1$$
 where by $  L^2(\bR^+, \bC)$ we denote the Hilbert space of square-integrable functions defined in $\bR^+$ and with values in $\bC$. An element
 $\bold Y$ of $\Cal H$ is a finite sequence
 $$
  \bold Y:=\{Y_1(x), \cdots, Y_n(x)\},
\tag 24.2$$
 and the scalar product of $\bold Y$ and $\bold W$ in
 $ \Cal H$  is described as
$$
( \bold Y, \bold W) := \sum_{j=1}^n (Y_j,W_j)_{ L^2(\bR^+, \bC)}.
\tag 24.3$$
 The Schr\"odinger equation on the star graph is given by
$$
 \bold L\, \bold Y= k^2 \,\bold Y,
\tag 24.4$$
 where
$$
 \bold L\, \bold Y:= ( -Y_1''(x)+ V_1(x) Y_1(x),\cdots ,-Y_n''(x)+ V_n(x) Y_n(x)),
\tag 24.5$$
with the potentials $V_j(x)$ for $j=1,\dots,n$ being real-valued functions satisfying (2.3).
  The boundary conditions are given by
$$
  -B^\dagger\, Y(0)+ A^\dagger\, Y'(0)=0,
 \tag 24.6$$
  where the $n \times n$ matrices $A$ and $B$ satisfy (2.5) and (2.6).

  Let us prove that the star graph is unitarily equivalent to a matrix Schr\"odinger equation  with the diagonal matrix potential given by
$$
  V(x):= \hbox{\rm diag}\{V_1(x), \cdots, V_n(x)\},
\tag 24.7$$
  and with the boundary condition (2.4) given by the same matrices $A$
  and $B$ as in (24.6). We define the unitary operator $U$ from $\Cal H$ onto $L^2(\bR^+)$ as
$$
  \psi (x)= U\, \bold Y:= \bm
  Y_1(x)\\
  \vdots \\ Y_n(x)
\endbm.
\tag 24.8$$
  Clearly we have
$$
||U\, \bold Y||_2= ||Y||_{\Cal H},\qquad  Y(x) \in \Cal H,
\tag 24.9$$
   and $U$ is onto $\Cal H$. Moreover, $\bold Y(x)$ is a solution to the system
  consisting of (24.4)
   and (24.5) if and only if $\Psi(x):= U\, \bold Y(x)$ is a solution to (2.1) and $Y(x)$ satisfies the boundary condition (24.6) if and only if  $\Psi(x):= U \bold Y(x)$ satisfies the boundary condition (2.4) with the same matrices $A$
   and $B$. Then $U$ establishes a unitary equivalence between both problems.

\newpage
\noindent {\bf 25. THE SCHR\"ODINGER EQUATION ON THE FULL LINE}
\vskip 3 pt

   A $2\times2$ matrix Schr\"odinger equation  is unitarily equivalent to a Schr\"odinger equation on the full line with a point interaction at $x=0$. The Hilbert space for the Schr\"odinger equation on the line is $L^2(\bR, \bC),$ where by  $L^2(\bR, \bC)$ we denote the Hilbert space of square integrable functions defined on $\bR$    and with values in $\bC$. We define the unitary operator $U$ from    $L^2(\bR^+)$ onto  $L^2(\bR, \bC)$ as
$$
   Y(x)= U \psi:= \cases  \psi_1(x),\qquad  x \geq 0, \\
   \stretch
    \psi_2(-x),\qquad  x < 0.\endcases\tag 25.1$$
    Clearly we have
$$
   \| U \psi \|_{L^2(\bold R, \bold C)}= \| \psi\|_{L^2(\bold R^+)},
    \qquad \psi(x) \in L^2(\bold R^+),
\tag 25.2$$
   and $U$ is onto $ L^2(\bold R, \bold C).$
   Suppose that the potential matrix $V(x)$ is diagonal, i.e.
$$
  V(x):= \text{\rm diag}\{V_1(x), V_2(x)\}.
\tag 25.3$$

We conclude that $\psi(x)$ satisfies the Schr\"odinger equation (2.1) if and only if $Y(x):= U\, \psi(x)$ satisfies the Schr\"odinger equation on the line given by
$$
   - Y''(x)+ Q(x)\, Y(x)= k^2\, Y(x),
\tag 25.4$$
   with the potential,
$$
   Q(x):= \cases  V_1(x), \qquad x \geq 0, \\
   \stretch
    V_2(-x),\qquad  x < 0.
   \endcases
\tag 25.5$$
   Moreover, $\psi(x)$ satisfies the boundary condition (2.4) if and only if $Y(x)$ satisfies the point-interaction  condition
$$
\cases
   -(B^\dagger)_{11}\, Y(0^+)- (B^\dagger)_{12}\, Y(0^-)+ (A^\dagger)_{11}\, Y'(0^+)- (A^\dagger)_{12}\, Y'(0^-) =0,   \\
   \stretch
    -(B^\dagger)_{21}\, Y(0^+)- (B^\dagger)_{,2}\, Y(0^-) +(A^\dagger)_{21}\, Y'(0^+) - (A^\dagger)_{22}\, Y'(0^-) =0,
    \endcases
\tag 25.6$$
where $(A^\dagger)_{ij}$ and $(B^\dagger)_{ij}$ denote
the $(i,j)$-th entry of the matrices
$A^\dagger$ and $B^\dagger,$ respectively.
    For example, suppose that $\psi(x)$ satisfies the $\delta$-type boundary condition $\psi_1(0)= \psi_2(0)$ and $ \psi'_1(0)+ \psi_2'(0)=\lambda\,\psi_1(0),$ where $\lambda$ is a real number,
    for which the special case $\lambda=0$ corresponds to
    the Kirchhoff boundary condition. In this case, the matrices appearing in
    (25.6) are given by
$$
    A:= \bm 0 &1 \\ 0&1\endbm,
\tag 25.7$$
$$
    B:= \bm -1 &\lambda \\ 1&0\endbm.
\tag 25.8$$
The boundary conditions (25.6) corresponding to
the matrices $A$ and $B$ given in (25.7) and
(25.8) are $Y(0^-)= Y(0^+)$ and  $Y'(0^+)-Y'(0^-)=\lambda\, Y(0^+)$, which is
related to the Schr\"odinger equation on the line with a
$\delta$-point interaction.

\newpage
\noindent {\bf 26. SOME EXPLICIT EXAMPLES}
\vskip 3 pt

In this chapter, through some explicit examples,
we illustrate our theoretical results obtained in previous chapters.

The explicit examples for the direct and inverse scattering problems
mainly involve scattering matrices $S(k)$ that are rational functions
of $k$ in the complex plane. In this case, the quantity $F(y)$
defined in (4.12) yields
a separable kernel for the Marchenko integral equation
(13.1). Then, it is possible to solve (13.1) explicitly
by using methods of linear algebra.

In the first example, we show that, by using a special case of the method
of [3,4,10], one can easily produce explicit
examples for the input data set $\bold D$ and the corresponding scattering
data set $\bold S$ for any values of $n$ when the scattering matrix $S(k)$
is a rational function of $k.$ This amounts to [3,4,10]
choosing $F(y)$ as a matrix-valued function containing a matrix exponential.

\noindent {\bf Example 26.1} For any constant $n\times n$ hermitian matrix $\bold a$
with positive eigenvalues, let us choose $F(y)$ given in (4.12)
as
$$F(y)=\bold c\,e^{-\bold a y}\,\bold c,\qquad y\in\bR^+,\tag 26.1$$
where $\bold c$ is a constant $n\times n$ hermitian matrix.
We remark that we use a matrix exponential in (26.1) because
$\bold a$ is a matrix.
The corresponding Marchenko equation has a separable kernel and hence
can be solved explicitly by algebraic means. Let $\bold m$ be the solution to
the linear system
$$\bold a \bold m+\bold m \bold a =\bold c^2,\tag 26.2$$
where the existence and uniqueness of $\bold m$ is assured [17]
and in fact it can be evaluated explicitly by using
$$\bold m=\int_0^\infty dy\,e^{-\bold a y}\,\bold c^2 \,e^{-\bold a y}.\tag 26.3$$
Using $F(y)$ given in (26.1) as
input to (13.1), one obtains
the solution to (13.1) explicitly as
$$K(x,y)=-\bold c \left(\bold m+e^{2 \bold a x}\right)^{-1} e^{\bold a (x-y)}\bold c,\tag 26.4$$
yielding
$$K(x,x)=-\bold c \left(\bold m+\ds e^{2 \bold a x}\right)^{-1} \bold c,
\quad K(0,0)=-\bold c \left(\bold m+I\right)^{-1} \bold c.\tag 26.5$$
Using (26.4) in (10.6) and using the first equality of (26.5) in (10.4),
we obtain the corresponding Jost solution $f(k,x)$ and the potential
$V(x)$ as
$$f(k,x)=e^{ikx}\left[I-\bold c\,\left(\bold m+\ds e^{2 \bold a x}\right)^{-1}
\left(\bold a-ik I\right)^{-1} \bold c\right],\tag 26.6$$
$$V(x)=-4\,\bold c \left(\bold m+e^{2 \bold a x}\right)^{-1}
\bold a \,e^{2 \bold a x}
\, \left(\bold m+e^{2 \bold a x}\right)^{-1}\bold c.\tag 26.7$$
 From (26.6) we get
$$f(k,0)=I- \bold c\,\left(\bold m+I\right)^{-1}
\left(\bold a-ik I\right)^{-1} \bold c,\tag 26.8$$
$$f'(k,0)=ikI+\bold c\,\left(\bold m+I\right)^{-1}
\left[2\bold a-ik I+2\bold a \left(\bold m+I\right)^{-1}
\right]
\left(\bold a-ik I\right)^{-1} \bold c,\tag 26.9$$
Along with any pair of constant $n\times n$ matrices $A$ and $B$ satisfying
(2.5) and (2.6), using (26.8) and (26.9) in (9.2) we obtain the corresponding
Jost matrix $J(k)$ and then obtain the corresponding scattering
matrix $S(k)$ by using (9.3). The corresponding physical solution
$\Psi(k,x)$ can be obtained via (9.4).
One can certainly enhance this method
by choosing $F_s(y)$ for $y\in\bR^-$ in a form similar to (26.1)
by replacing $\bold a$
with some constant $n\times n$ hermitian matrix with negative
eigenvalues. The use of matrix exponentials allows us to write
the explicit solutions in the direct and inverse problems
in a compact form. For instance, by choosing
$$\bold a=\bm 3&-1&0\\
\stretch
-1&3&0\\
\stretch
0&0&2\endbm,\quad \bold c=\bm 1&0&0\\
\stretch
0&2&1\\
\stretch
0&1&1\endbm,\tag 26.10$$
we get an example where $\bold a$ has eigenvalues $4,$ $2,$ and $2$ and the matrix
$\bold m$ is given by
$$\bold m=\bm 11/48&3/16&1/8\\
\stretch
3/16&43/48&5/8\\
\stretch
1/8&5/8&1/2\endbm,\quad e^{-\bold a y}=\ds\frac{1}{2}\,
\bm e^{-2y}+e^{-4y}& e^{-2y}-e^{-4y}&0\\
\stretch
e^{-2y}-e^{-4y} &  e^{-2y}+e^{-4y}&0\\
\stretch
0&0&2 e^{-2y}\endbm,\tag 26.11$$
$$(\bold a-ik I)^{-1}=\ds\frac{1}{(k+2i)^2(k+4i)}\bm i(k+2i)(k+3i)&-(k+2i)&0
\\
\stretch
-(k+2i)&i(k+2i)(k+3i)&0\\
\stretch
0&0&i(k+2i)(k+4i)\endbm.\tag 26.12$$

The purpose of the next example is to emphasize the
fact that, in seeking a solution to an integral
equation such as (4.17), (4.22), and (7.1), it is important
for us to state whether
we look for a square-integrable solution
or an integrable solution.
For example, the class
$L^1(\bR^+)$ is different from $L^2(\bR^+),$ and
the class $L^1(\bR^+)\cap L^\infty(\bR^+)$
is a proper subspace of the class $L^2(\bR^+)\cap L^\infty(\bR^+).$


\noindent {\bf Example 26.2} Assume that $n=1$ and that we have
$$X(y)=\cases \ds\frac{1}{1+y},\qquad y\in\bR^+,\\
\stretch
0,\qquad y\in\bR^-.\endcases\tag 26.13$$
We notice that $X(y)$ is square integrable in
$y\in\bR^+$ but not integrable there. Using (3.68) we can evaluate
$\hat X(k)$ explicitly as
$$\hat X(k)=-i\,e^{-ik}\left[ \ds\frac{\pi}{2}-\text{Si}(k)
-i\,\text{Ci}(k)\right],\tag 26.14$$
in terms of the sine integral function $\text{Si}(k)$ and the cosine integral function $\text{Ci}(k)$ defined as
$$\text{Si}(k):=\int_0^k dt\,\ds\frac{\sin t}{t} ,\quad
\text{Ci}(k):=\int_k^\infty dt\,\ds\frac{\cos t}{t}.\tag 26.15$$
Even though Si$(k)$ is well behaved at $k=0,$ the function Ci$(k)$ has a logarithmic
singularity at $k=0,$ which is seen from its representation
$$\text{Ci}(k)=\gamma+\log k+\ds\sum_{j=1}^\infty \ds\frac{(-k^2)^j}{2j\,(2j)!},
\tag 26.16$$
where $\gamma$ is the Euler-Mascheroni constant.
Thus, $\hat X(k)$ given in
(2.16b) is not continuous at $0.$
This example indicates that the
proof given in [2] for
Theorem~3.5.1 of [2] needs to be improved. For example, one can use
a procedure such as that given in our own Propositions~15.12 to prove
Theorem~3.5.1 of [2].

The purpose of the next example is to emphasize the
fact that, in seeking a solution to a Riemann-Hilbert problem
such as (4.19), (4.24), and (7.2), it is important
for us to state whether
we look for a solution in the Hardy space
$\bold H^2(\bCp)$ or in $\hat L^1(\bCp).$
In the next example we present a function that belongs to the Hardy space
$\bold H^2(\bCp)$ but not continuous on $\bR$ and hence
not in $\hat L^1(\bCp).$

\noindent {\bf Example 26.3} Let
$$\hat X(k)=\ds\frac{\log k}{k+i},\qquad k\in\bCp,\tag 26.17$$
where $\log$ denotes the principal branch of the logarithm function
with the argument of $k$ limited to the interval $(-\pi,\pi).$
From (26.17), we see that $\hat X(k)$ is analytic in $\bCp$
but it is not continuous at $k=0$ and hence it is not continuous
in $k\in\bCpb.$
Letting $k=|k|\, e^{i\theta}$ with $\theta\in(-\pi,\pi),$
 we can write $\hat X(k)$ in $\bCp$ as
 $$\hat X(k)=\ds\frac{\ln |k|+i\theta}{|k|\,\cos \theta+i(1+|k|\,\sin \theta)},
 \qquad |k|>0,\quad \theta\in(-\pi,\pi).\tag 26.18$$
 Let us use $k_R$ and $k_I$ for the real and imaginary parts
 of $k.$
Because $\hat X(k)$ is continuous on the line $k=k_R+i k_I$ with
$k_R\in\bR$ for any fixed positive
$k_I,$ in order to prove
the integrability of $|\hat X(k_R+i k_I)|^2$ in
$k_R\in\bR$ for each fixed $k_I>0,$
it is enough to check the integrability
of $|\hat X(k_R+i k_I)|^2$ as $k_R\to\pm\infty.$ From (26.18), we
$$|\hat X(k_R+i k_I)|^2=\ds\frac{\left(\ln|k|\right)^2+\theta^2}{|k|^2+2\,|k|\,\sin \theta+1},\tag 26.19$$
and hence
we can find some constant $c>1$ such that
$$|\hat X(k_R+i k_I)|^2\le \ds\frac {C\,\left(\ln|k|\right)^2}{|k|^2},
\qquad |k_R|\ge c,\tag 26.20$$
for some generic constant $C.$
Then, with the help of (26.20) we get
$$\int_{-\infty}^\infty dk_R\, |\hat X(k_R+i k_I)|^2
\le C+C\int_1^\infty d|k|\,\ds\frac {\left(\ln|k|\right)^2}{|k|^2}
\le C+C\int_0^\infty d\alpha\, \ds\frac{\alpha^2}{e^\alpha}<+\infty,
\tag 26.21$$
where we have used the substitution $\alpha=\ln |k|.$
Thus, $\hat X(k)$ given in (26.17) belongs to
the Hardy space $\bold H^2(\bCp).$
On the other hand, $\hat X(k)$ cannot be in
$\hat L^1(\bCp)$ because the Riemann-Lebesgue lemma requires
that any function in $\hat L^1(\bCp)$ must be continuous
in $k\in\bR,$ whereas from (26.17) we know that
$\hat X(k)$ is not continuous at $k=0$ on the real
axis.

In the next example, we present a scattering data set $\bold S$ for which all the
characterization conditions given in Theorem~5.1 are satisfied.

\noindent {\bf Example 26.4} Assume that $n=1$ and that the scattering matrix $S(k)$ is given by
$$S(k)=-\ds\frac{k+i}{k-i},\qquad k\in\bR,\tag 26.22$$
and that there are no bound states.
Thus, from (4.3) we get $\Cal N=0.$ This is compatible with (21.5)
because $S(0)=1$ and hence $\mu=1,$ $S_\infty=-1$ and hence
$n_{\text{\rm D}}=1,$ and the left-hand side of (21.5) is equal to
$\pi.$
By using the construction process outlined in the
beginning of Chapter~16, we uniquely obtain
$$S_\infty=-1,\quad G_1=2,\tag 26.23$$
$$F_s(y)=\cases 2e^{-y},\qquad y\in\bR^+,\\
\stretch
0,\qquad y\in\bR^-,\endcases\tag 26.25$$
$$F_s'(y)-G_1\,\delta(y)=
\cases -2e^{-y},\qquad y\in\bR^+,\\
\stretch
0,\qquad y\in\bR^-,\endcases\tag 26.25$$
$$K(x,y)=\cases -e^{-y}\,{\text{sech}\, x},\qquad y>x\ge 0,\\
\stretch
0,\qquad y<x,\endcases\tag 26.26$$
$$K_x(x,y)=\cases e^{-y}\, (\tanh x)({\text{sech}\, x}),\qquad y>x\ge 0,\\
\stretch
0,\qquad y<x,\endcases\tag 26.27$$
$$K(0,0)=-1,\quad f(k,x)=e^{ikx}\left[1-\ds\frac{i}{k+i}\,\ds\frac{e^{-x}}{\cosh x}\right],\tag 26.28$$
$$V(x)=-2\,{\text{sech}^2\, x} ,\quad
\Psi(k,x)=-\ds\frac{2ik}{k-i}\,\sin kx-\ds\frac{1}{k-i}\,(\cos kx)(\tanh x),\tag 26.29$$
$$A=0,\quad J(k)=\ds\frac{k}{k+i}\,B,\quad \Psi(k,0)=0,\quad \Psi'(k,0)=-2i(k+i),\tag 26.30$$
where $B$ is an arbitrary nonzero constant.
One can directly verify that
each of the four conditions $(\bold 1)$,
$(\bold 2)$,
$(\bold 3_a)$, $(\bold 4_a)$ of Theorem~5.1 is satisfied.

In the following example, we present a scattering
data set $\bold S,$ for which, except for the second equality
in (4.4), all
the remaining characterization conditions in Theorem~7.1 are satisfied.

\noindent {\bf Example 26.5} Assume that $n=1$ and that the scattering matrix $S(k)$ is given by
$$S(k)=\ds\frac{k}{k+i},\qquad k\in\bR,\tag 26.31$$
and that there are no bound states.
 From (26.31) we get
$S(0)=0,$ which indicates that $S(k)$ cannot be unitary for
$k\in\bR.$ Nevertheless,
using the construction process outlined in Chapter~16 we obtain
$$S_\infty=1,\quad G_1=1,\tag 26.32$$
$$F_s(y)=\cases 0,\qquad y\in\bR^+,\\
\stretch
-e^{y},\qquad y\in\bR^-,\endcases\tag 26.33$$
$$F_s'(y)-G_1\,\delta(y)=
\cases 0,\qquad y\in\bR^+,\\
\stretch
-e^{y},\qquad y\in\bR^-,\endcases\tag 26.34$$
$$F(y)=0,\qquad y\in\bR^+.
\tag 26.35$$
Using the second line of (26.33) we determine that
the only solution in $L^2(\bR^-)$ to (4.17)
is the trivial solution, and by using
(26.35) we see that the only solution in $L^1(\bR^+)$
to (4.14) is also the trivial solution.
Since it is assumed that there are no bound states,
(4.22) and (4.14) coincide and hence
the only solution in $L^1(\bR^+)$
to (4.22) is also the trivial solution.
Then, we conclude that,
in Theorem~7.1, all the conditions are satisfied, except for the unitarity of $S(k)$ in $(\bold 1)$ stated in the second equality in
(4.4). Equivalently stated, we have
$(\bold 2),$ $(\bold{III}_c),$ $(\bold 4_c),$ and
$(\bold V_c)$ are all satisfied and only $(\bold 1)$
is not satisfied. Nevertheless, if we continue
using the method of Chapter~16 to construct the corresponding input data set $\bold D,$ we find that
$$K(x,y)\equiv 0,\quad V(x)\equiv 0,\quad f(k,x)=e^{ikx},\quad B=\ds\frac{1}{2}\,A,\tag 26.36$$
where $A$ is an arbitrary nonzero constant. For the
input data set $\bold D$ given in (26.36),
by using the method of Chapter~9 we find that the
corresponding Jost matrix $J(k)$ and
the scattering matrix $S(k)$ are given by
$$J(k)=\left(\ds\frac{1}{2}-ik\right)A,
\quad S(k)=\ds\frac{k-i/2}{k+i/2}.\tag 26.37$$
However, $S(k)$ given in
(26.37) is not compatible with Levinson's theorem.
This is because for
the scattering matrix $S(k)$ in
(26.37), the left-hand side of (21.5) is
$-\pi,$ $\mu=1$ because $S(0)=1,$ $n_D=0$ because
$S_\infty=1,$ and hence (21.5) yields
$\Cal N=-1/2,$ which is not a nonnegative
integer.

In the following example, we present a scattering
data set $\bold S,$ for which,
except for the first equality in (4.4), all
the remaining characterization conditions in Theorem~7.1 are satisfied.

\noindent {\bf Example 26.6} Assume that $n=1$ and that the scattering matrix $S(k)$ is given by
$$S(k)=i\,\ds\frac{k-i}{k+i},\qquad k\in\bR,\tag 26.38$$
and that there are no bound states. We observe that
the property $S(-k)=S(k)^\dagger$ in (4.4) is not satisfied
by $S(k)$ given in (26.38). On the other hand, $S(k)$ given in
(26.38) satisfies the second equality in (4.4) and hence it
is unitary.
Nevertheless, let us show that,
in Theorem~7.1, all the remaining characterization conditions are satisfied.
Using the construction process outlined in Chapter~16 we obtain
$$S_\infty=i,\quad G_1=2i,\tag 26.39$$
$$F_s(y)=\cases 0,\qquad y\in\bR^+,\\
\stretch
-2\,i\,e^{y},\qquad y\in\bR^-,\endcases\tag 26.40$$
$$F_s'(y)-G_1\,\delta(y)=
\cases 0,\qquad y\in\bR^+,\\
\stretch
-2\,i\,e^{y},\qquad y\in\bR^-,\endcases\tag 26.41$$
$$F(y)=0,\qquad y\in\bR^+.
\tag 26.42$$
Using the second line of (26.40) we determine that
the only solution in $L^2(\bR^-)$ to (4.17)
is the trivial solution, and by using
(26.42) we see that the only solution in $L^1(\bR^+)$
to (4.14) is also the trivial solution.
Since it is assumed that there are no bound states,
(4.22) and (4.14) coincide and hence
the only solution in $L^1(\bR^+)$
to (4.22) is also the trivial solution.
Then, we conclude that,
in Theorem~7.1, all the conditions are satisfied, except for the symmetry of $S(k)$ in $(\bold 1)$ stated in the first equality in
(4.4). Equivalently stated, we have
$(\bold 2),$ $(\bold{III}_c),$ $(\bold 4_c),$ and
$(\bold V_c)$ are all satisfied and only $(\bold 1)$
is not satisfied.
We can continue to use the method of Chapter~16 to construct the corresponding input data set $\bold D,$ and we find that
$$K(x,y)\equiv 0,\quad V(x)\equiv 0,\quad f(k,x)=e^{ikx},\quad A=B=0.\tag 26.43$$
Because the constructed $A$ and $B$ does not satisfy (2.6),
we do not have a selfadjoint boundary condition as in (2.4).
Nevertheless, let us
also explicitly construct the corresponding physical solution
$\Psi(k,x)$ via (9.4)
as
$$\Psi(k,x)=e^{-ikx}+i\,e^{ikx}\,\ds\frac{k-i}{k+i}.\tag 26.44$$
 From (26.7) we obtain
$$\Psi(k,0)=\ds\frac{(1+i)(k+1)}{k+i},\quad \Psi'(k,0)=
-\ds\frac{(1+i)k(k-1)}{k+i},\tag 26.45$$
and hence from (2.4) we see that
(2.4) cannot be satisfied
unless $A=B=0.$
One can directly verify that
$(\bold 2)$ and $(\bold 4_a)$ in Definition~4.2 are satisfied but neither of
$(\bold 1)$ and $(\bold 3_a)$ there are satisfied.
This example indicates that the first equality of (4.4) given in $(\bold 1)$
is necessary in the characterization stated in Theorem~7.1.

In the following example, we present a scattering
data set $\bold S$ for which some of the characterization conditions
are not satisfied.

\noindent {\bf Example 26.7} Assume that $n=1$ and that the scattering matrix $S(k)$ is given by
$$S(k)=\ds\frac{k+i}{k-i},\qquad k\in\bR,\tag 26.46$$
and that there are no bound states.
We remark that the scattering matrix in (26.46) differs from that given in (26.22)
only in the sign.
By using the construction process outlined in the
beginning of Chapter~16, we uniquely obtain
$$S_\infty=1,\quad G_1=-2,\tag 26.47$$
$$F_s(y)=\cases -2e^{-y},\qquad y\in\bR^+,\\
\stretch
0,\qquad y\in\bR^-,\endcases\tag 26.48$$
$$F_s'(y)-G_1\,\delta(y)=
\cases 2e^{-y},\qquad y\in\bR^+,\\
\stretch
0,\qquad y\in\bR^-,\endcases\tag 26.49$$
$$F(y)=-2\,e^{-y},\qquad y\in\bR^+.\tag 26.50$$
Using the second line of (26.48) in (4.17) we
see that the only solution in $L^2(\bR^-)$ to
(4.17) is the trivial solution.
On the other hand, using (26.50) in (4.14)
we observe that
(4.14) has a one-parameter family of solutions
given by $X(y)=\alpha\,e^{-y}$ for $y\in\bR^+,$
where $\alpha$ is an arbitrary parameter.
Since there are no bound states, we also
obtain that (4.22) has a one-parameter family of solutions
given by $X(y)=\alpha\,e^{-y}$ for $y\in\bR^+,$
where $\alpha$ is an arbitrary parameter.
We conclude that our scattering data set
satisfies $(\bold 1),$ $(\bold 2),$ and $(\bold{III}_c)$
in Theorem~7.1, but $(\bold 4_c)$ and
$(\bold V_c)$ there both fail.
Using the method of Chapter~16, we construct
$$K(x,y)=\cases \ds\frac{e^{-y}}{\sinh x},\qquad y>x> 0,\\
\stretch
0,\qquad y<x,\endcases\tag 26.51$$
$$K_x(x,y)=\cases -\ds\frac{\cosh x}{\sinh^2 x}\,e^{-y},\qquad y>x> 0,\\
\stretch
0,\qquad y<x,\endcases\tag 26.52$$
$$K(0,0)=-\infty,\quad f(k,x)=e^{ikx}\left[1+\ds\frac{i}{k+i}\,\ds\frac{e^{-x}}{\sinh x}\right],\tag 26.53$$
$$V(x)=\ds\frac{8 e^{2x}}{(e^{2x}-1)^2},\quad
\Psi(k,x)=\ds\frac{2k}{k-i}\,\cos kx-\ds\frac{1}{k-i}\,(\sin kx)(\coth x),\tag 26.54$$
$$V(x)=\ds\frac{2}{x^2}-\ds\frac{2}{3}+\ds\frac{2x^2}{15}+O(x^4),\qquad x\to 0.\tag 26.55$$
In this example, the boundary parameters $A$ and $B$ do not exist because
$K(0,0)$ is not finite. Both $\Psi(k,0)$ and $\Psi'(k,0)$ blow up at $x=0$ and hence
$\Psi(k,x)$ does not satisfy a boundary condition like (2.4).
In the characterization conditions specified in Theorem~5.1,
we find that $(\bold 1)$ and $(\bold 2)$ are satisfied but $(\bold 3_a)$ and $(\bold 4_a)$ are not satisfied.
Let us now check the equivalences indicated
in Chapter~6.
We cannot have $(\bold 3_b)$ of Definition~4.2
or $(\bold V_b)$ of Definition~4.3 satisfied because
we cannot construct a Jost matrix $J(k)$ in this example as we cannot construct
the boundary matrices $A$ and $B.$ In the absence of
bound states, none of $(\bold 4_d),$ $(\bold 4_e),$ $(\bold V_d)$, $(\bold V_e),$
$(\bold V_g)$, $(\bold V_h)$
are satisfied
because each of (4.15), (4.16), (4.23), (4.24)
has the one-parameter
family of solutions of the form $c/(k+i)$ for $k\in\bCpb$
with $c$ being the
arbitrary parameter.
Note that neither $(\bold 4_c)$ nor $(\bold V_c)$ is satisfied because each of
(4.14) and
(4.22)
has the
one-parameter family of solutions given by $c\,e^{-y}$ for
$y\in\bR^+,$ with $c$ being the
arbitrary parameter.
Ironically, even though $(\bold 3_a)$
of Definition~4.2 fails, each of
$(\bold{III}_a)$, $(\bold{III}_b)$, $(\bold{III}_c)$ in Definition~4.3 is satisfied
because (4.17), (4.18), (4.19) each have
only the trivial solution.
%
%
%
%
In this example, the potential $V(x)$
constructed, although hermitian, does not satisfy (2.3).
In summary, in
Theorem~7.1, $(\bold 1)$, $(\bold 2)$, and $(\bold{III}_a)$ are satisfied, but
$(\bold 4_c)$ and $(\bold V_c)$ are not satisfied. In Theorem~5.1,
$(\bold 1)$ and $(\bold 2)$ are satisfied, but $(\bold 3_a)$ and
$(\bold 4_a)$ each fail.

In Example~26.7 presented above, we have observed that each of
(4.22), (4.23), (4.24) has one linearly independent
solution. Thus, in order for $S(k)$ given in (26.46) to satisfy
the characterization conditions, there has to be exactly one bound state
associated with the corresponding scattering data.
In the following example, we supplement the scattering matrix
of (26.46) with one bound state
and obtain an example where all the characterization conditions are met.

\noindent {\bf Example 26.8} Assume that $n=1$ and that the scattering matrix $S(k)$ is given by (26.46) and that there is exactly one bound state at $k=i\kappa_1$ with
$\kappa_1=1$ and
the Marchenko normalization constant $M_1=\sqrt{2}.$ The number of
bound states is consistent with Levinson's theorem.
This is because from (26.8) we see that the left-hand side
of (21.5) is $\pi,$ $\mu=0$ because $S(0)=-1,$ $n_D=0$ because
$S_\infty=1$ and hence
the value of $\Cal N$ predicted by Levinson's theorem
is equal to one. Since
(26.46)-(26.49) still hold, and
we also have
$$F(y)=0,\qquad y\in\bR^+,\tag 26.56$$
we see that (4.14) now has only the trivial solution
and hence $(\bold 4_c)$ is satisfied.
Using the method of Chapter~16, we construct
$$K(x,y)\equiv 0,\qquad K(0,0)=0,\quad V(x)\equiv 0,\quad f(k,x)=e^{ikx},\tag 26.57$$
$$\Psi(k,x)=\ds\frac{2(k\,\cos kx-\sin kx)}{k-i},\quad \Psi_1(x)=\sqrt{2}\,e^{-x},
\quad
J(k)=-iA (k+i),
\tag 26.58$$
$$\Psi(k,0)=\ds\frac{2k}{k-i},\quad \Psi'(k,0)=-\ds\frac{2k}{k-i},\quad
\Psi_1(0)=\sqrt{2},\quad \Psi'_1(0)=-\sqrt{2},\tag 26.59$$
with the boundary parameters $B=A,$ where $A$ is an arbitrary nonzero constant.
One can directly verify that
each of the four conditions stated in Definition~4.5 is satisfied
and that each of the five conditions in Theorem~7.1 is satisfied.

Next, we present an example of the scattering data set $\bold S$
failing only $(\bold{III}_a)$ in Theorem~7.1, failing only
$(\bold 3_a)$ of Theorem~5.1, and failing only $(\bold L)$ of Theorem~7.9,
 whereas all the remaining conditions in those three theorems
 are satisfied.

\noindent {\bf Example 26.9} Assume that
$n=1$ and $\bold S$ consists of the scattering matrix
$$S(k)=\left(\ds\frac{k-i}{k+i}\right)^2,\qquad k\in\bR,\tag 26.60$$
and that there are no bound states. Using (26.60) in (10.14) we see
that the constants $S_\infty$ and $G_1$ appearing in
(4.5) are given by
$$S_\infty=1,\quad G_1=4 .\tag 26.61$$
Using (26.60) and (26.61) in (4.7), through some residue computations,
we obtain
$$F_s(y)=\cases 0,\qquad y>0,\\
\stretch
-4(1+y)\,e^y,\qquad y<0.\endcases\tag 26.62$$
Since there are no bound states, using (26.62) in (4.12) we
obtain
$$F(y)=0,\qquad y>0.\tag 26.63$$
Using (26.60) and (26.62)
we find that $(\bold 1)$ and $(\bold 2)$ in
Theorem~7.1 are satisfied. Using (26.63) in (4.14) we see that
the only solution in $L^1(\bR^+)$ to (4.14) is the trivial solution and
hence $(\bold 4_c)$ in Theorem~7.1 is satisfied. Similarly, using the first line of
(26.62) in (4.22) we see that the only solution in $L^1(\bR^+)$ to (4.22) is the trivial solution and
hence $(\bold V_c)$ in Theorem~7.1 is satisfied. Let us use the second line of (26.62) in
(4.17), which yields
$$-X(y)+\int_{-\infty}^0 dz\,X(z)\left[-4(1+z+y)e^{z+y}\right]=0,\qquad y\in\bR^-.
\tag 26.64$$
 From (26.64) we see that its solution must have the form
$$X(y)=\alpha e^y+\beta y e^y,\tag 26.65$$
for some constants $\alpha$ and $\beta.$ Using (26.65) in (26.64)
we find that $\alpha=0$ and $\beta$ is arbitrary. Thus,
(26.64) has the nontrivial solution in $L^2(\bR^-)$ given by
$X(y)=\beta\, y \,e^y,$ where $\beta$ is arbitrary. Thus, $(\bold{III}_a)$
in Theorem~7.1 is
violated, although $(\bold 1),$ $(\bold 2),$ $(\bold 4_c),$ and
$(\bold V_c)$ are satisfied. In this example, the construction outlined
in Chapter~16 by starting with (26.63) yields
$$K(x,y)=0,\quad K(0,0)=0,\quad V(x)\equiv 0,\quad f(k,x)=e^{ikx},\tag 26.66$$
$$\Psi(k,0)=\ds\frac{2k^2-2}{(k+i)^2},\quad \Psi'(k,0)=\ds\frac{4k^2}{(k+i)^2}.
\tag 26.67$$
Using $S_\infty=1,$ $G_1=4,$ $K(0,0)=0$ in (14.2) we obtain
$B=2A$ with $A$ being an arbitrary nonzero constant.
Thus, the boundary condition (2.4) is given by
$$\psi'(0)-2\psi(0)=0.\tag 26.68$$
However, using (26.67) in (26.68) we see that the physical solution
$\Psi(k,x)$ does not satisfy the boundary condition, and hence
$(\bold 3_a)$ in Theorem~5.1 does not hold. On the other hand, the remaining
characterization conditions $(\bold 1),$ $(\bold 2)$, and $(\bold 4_a)$ in Theorem~5.1
all hold.
Let us now check the compatibility of $S(k)$ given in
(26.60) with Levinson's theorem. In other words, let us
check if $(\bold L)$ of Theorem~7.9 is satisfied.
Using (26.60), we find that the left-hand side of (21.5) is
$-2\pi,$ $\mu=1$ because $S(0)=1,$ $n_D=0$ because $S_\infty=1,$ and
hence (21.5) predicts $\Cal N=-1,$ which is not a nonnegative integer.
Thus, $(\bold L)$ in Theorem~7.9 does not hold. One can directly
verify that the only solution in $\bold H^2(\bCp)$ to (7.3) is the trivial
solution and hence all the four conditions in Theorem~7.9
are satisfied with the exception of $(\bold L).$
One can use the procedure
described in Chapter~9 to solve the direct problem and show
that the input data set $\bold D$ consisting of
the zero potential and the boundary condition (26.68)
does not correspond to the scattering matrix
$S(k)$ given in (26.60) with no bound states. In fact, with the help
of (9.1)-(9.3) we get
$$f(k,x)= e^{ikx},\quad J(k)=-i(k+2i)A,\quad
S(k)=\ds\frac{k-2i}
{k+2i},\tag 26.69$$
and there are no bound states because $J(k)$ does not vanish
on the positive imaginary axis and it vanishes only at $k=-2i.$

As we have seen in Example~26.9, for $S(k)$ given in (26.60) Levinson's
theorem predicts $\Cal N=-1.$ Hence, as indicated in Corollary~19.4,
it is impossible to supplement $S(k)$ with any bound-state
data set so that the resulting scattering data set $\bold S$ can belong
to the Marchenko class.

Next, we present
an example satisfying all the characterization conditions
in Theorem~7.1 except $(\bold 4_c)$ and $(\bold V_c)$.

\noindent {\bf Example 26.10} Assume that
$n=1$ and $\bold S$ consists of the scattering matrix
$$S(k)=\left(\ds\frac{k+i}{k-i}\right)^2,\qquad k\in\bR,\tag 26.70$$
and that there are no bound states. Let us use the construction
outlined in Chapter~16.
Using (26.60) in (10.14) we see
that the constants $S_\infty$ and $G_1$ appearing in
(4.5) are given by
$$S_\infty=1,\quad G_1=-4,\tag 26.71$$
$$F_s(y)=\cases 4(-1+y)\,e^{-y},\qquad y\in\bR^+,\\
\stretch
0,\qquad y\in\bR^-,\endcases\tag 26.72$$
$$F(y)=4(-1+y)\,e^{-y},\qquad y\in\bR^+.\tag 26.73$$
One can directly verify that $(\bold 1)$ and $(\bold 2)$ in
Theorem~7.1 are satisfied. Using the second line of
(26.72) in (4.17) we see that the only
solution in $L^2(\bR^-)$ to (4.17) is the trivial solution and hence $(\bold{III}_a)$ of Theorem~7.1 is satisfied.
Using (26.69) in (4.14) we see that
the general solution in $L^1(\bR^+)$ to (4.14) is
$X(y)=\beta (-1+y)\,e^{-y},$ where $\beta$ is arbitrary, and hence
$(\bold 4_c)$ of Theorem~7.1 is not satisfied.
In the absence of bound states, (4.22) and (4.14)
coincide and their general solutions must also coincide.
Thus, $(\bold V_c)$ of Theorem~7.1 is not satisfied because
(4.22) has one linearly independent solution although there
are no bound states.
Therefore,
unless the scattering matrix of (26.70) is accompanied with
exactly one bound state, the corresponding scattering data set
cannot correspond to an input data set $\bold D$ in the Faddeev class.
Continuing with the method of Chapter~16, we obtain
$$K(x,y)=\ds\frac{4\, e^{x-y}\left[1+x+e^{2 x}-x\,e^{2 x}+y\,(1+e^{2x}) \right]}{-1+4\, x\, e^{2 x}+e^{4 x}},\tag 26.74$$
$$K(x,x)=\ds\frac{4+(4-8x)\,e^{2x}}{-1+4x\,e^{2x}+e^{4x}},\tag 26.75$$
$$V(x)=\ds\frac{-32 \,e^{2x}(1+e^{2x})\left(-1-x+(-1+x)\,e^{2x}\right)}
{\left(-1+4\, x\, e^{2 x}+e^{4 x}\right)^2},\tag 26.76$$
$$f(k,x)=e^{ikx}\ds\frac{2\left( k^2 \,x+ik+x\right)+2ik\,\cosh(2x)+(k^2-1)\,\sinh(2x)}{(k+i)^2 \left(2x+\sinh(2x)\right) }.
\tag 26.77$$
We have
$$K(x,x)=\ds\frac{1}{x}-2+O(x),\quad V(x)=\ds\frac{2}{x^2}-\ds\frac{4}{3}+O(x^2),
\qquad x\to 0,\tag 26.78$$
$$\Psi(k,x)=O\left(\ds\frac{1}{x}\right),\quad \Psi'(k,x)=O\left(\ds\frac{1}{x}\right),
\qquad x\to 0.\tag 26.79$$
Because $K(0,0)$ is not finite, the boundary matrices $A$ and $B$ do not exist.
Hence, $(\bold 3_a)$ of Theorem~5.1 does not hold. By Proposition~6.1
we have the equivalence of $(\bold 4_a)$ and $(\bold 4_c),$ and
hence $(\bold 4_a)$ fails because $(\bold 4_c).$
In summary, in Theorem~5.1 we have $(\bold 1)$ and $(\bold 2)$ satisfied
and we have $(\bold{III}_c)$, $(\bold 4_c$, and $(\bold V_c)$ not satisfied.
In Theorem~5.1, we have
$(\bold 1)$ and $(\bold 2)$ satisfied
and we have $(\bold 3_a)$ and $(\bold 4_a)$ not satisfied.
In this example, the number of bound states predicted by Levinson's
theorem is not consistent with the absence of bound states.
This is because from (26.70) we see that the left-hand side
of (21.5) is $2\pi,$ $\mu=1$ because $S(0)=1,$ $n_D=0$ because
$S_\infty=1$ and hence
the value of $\Cal N$ predicted by Levinson's theorem
is equal to one, inconsistent with the absence of bound states
in this example.

In the next example we use the scattering matrix of Example~26.10
and we choose the number of bound states compatible withh Levinson's
 theorem.
We illustrate that
we can supplement the scattering data set of Example~26.10 with one bound state at $k=i\kappa_1$
for any $\kappa_1>0$ and with the normalization matrix
$M_1$ for any positive constant $M_1.$ The resulting scattering data
set belongs to the Marchenko class.

\noindent {\bf Example 26.11} Assume that
$n=1$ and $\bold S$ consists of the scattering matrix $S(k)$
given in (26.70) and that there is one bound state at
$k=i\kappa_1$ for some $\kappa_1>0$ and $M_1>0.$ Then,
(26.71) and (26.72) still hold, but instead of
(26.73) we get
$$F(y)=4(-1+y)\,e^{-y}+M_1^2\,e^{-\kappa_1 y},\qquad y\in\bR^+.\tag 26.80$$
Using (26.79) as input to (4.14) we obtain the fact that
the only integrable solution to (4.14) is the trivial solution and hence
$(\bold 4_c)$ is satisfied. We already know from Example~26.10 that
(4.22) has the one-parameter family of solutions
$X(y)=\beta\,(-1+y)\,e^{-y}$ for $y\in\bR^+$ and hence
in this example $(\bold V_c)$ is satisfied.
Thus,
all the five characterization conditions given in Theorem~7.1 hold
Then, based on the equivalence results of Chapter~6, all the four
characterization conditions in
Theorem~5.1 are satisfied, and hence the corresponding
scattering data set $\bold S$ belongs to the Marchenko class
for any choice of positive $\kappa_1$ and positive $M_1.$
Since the explicit expressions for the constructed quantities
are fairly lengthy, we display below the constructed quantities
in the simplest choice of $\kappa_1=1$ and $M_1=2.$ In this case we have
$$F(y)=4y\,e^{-y},\qquad y\in\bR^+,\tag 26.81$$
$$K(x,y)=\ds\frac{2\, e^{-x-y}\left[1+x-y-(x+y)\, e^{2x}\right]}
{1+2x+\sinh(2x)},\tag 26.82$$
$$K(0,0)=2,\quad V(x)=\ds\frac{4 \cosh x\left[2 \,\cosh x-(1+2x)\,\sinh x\right]}
{\left[1+2x+\sinh(2x)\right]^2},\tag 26.83$$
$$f(k,x)=e^{ikx}\left[1+\ds\frac{2}{(k+i)^2} \,\ds\frac{ik(e^{-2x}-2x)+1+2x}
{1+2x+\sinh(2x)}
\right],\tag 26.84$$
$$\Psi(k,0)=2\,\ds\frac{k+i}{k-i},\quad\Psi'(k,0)=-8\,\ds\frac{k+i}{k-i},\quad
\Psi_1(0)=2,\quad \Psi_1'(0)=-8,\quad B=-4A,\tag 26.85$$
where $A$ is an arbitrary nonzero constant.

In the next example, we elaborate on the compatibility with Levinson's theorem.

\noindent {\bf Example 26.12} Let the scattering matrix be given by
$$S(k)=\left(\ds\frac{k+i}{k-i}\right)^p,\qquad k\in\bR,\tag 26.86$$
where $p$ is an integer. The left-hand side of (21.5) is then equal to $p\pi.$
We have $S_\infty=1$ and hence $n_D=0.$ We have $S(0)=(-1)^p$ and hence
$\mu=1$ if $p$ is even and $\mu=0$ if $p$ is odd. Thus, Levinson's theorem
predicts that
$$\Cal N=\cases \ds\frac{p}{2},\qquad p\text{ even integer, }\\
\stretch
\ds\frac{p-1}{2},\qquad p\text{ odd integer. }\endcases\tag 26.87$$
Thus, $\Cal N$ predicted by Levinson's theorem
is always an integer. However, if $p$ is negative then
$\Cal N$ in (26.87) is negative, and hence the scattering matrix
cannot be a part of a scattering data set $\bold S$ in the Marchenko class.
In Example~26.5 the predicted $\Cal N$ is $-1/2,$  but in that
example the scattering matrix given in (26.31) is not unitary.

In the next example we elaborate on the choices of supplementing
a scattering matrix with an appropriate bound-state data set
so that the corresponding scattering data set can belong to
the Marchenko class. The elaborations can be found in Chapter~19.
This also provides
an example where all the characterization conditions
except $(\bold 4_c)$ may be satisfied in Theorem~7.1.
It also illustrates a case where in Theorem~7.9
we have all the properties are satisfied
except
$(\bold 4_{e,2})$
there.

\noindent {\bf Example 26.13} Consider the scattering matrix given by
$$S(k)=\bm \left(\ds\frac{k+i}{k-i}\right)^4 &0\\
\stretch 0&1\endbm,\qquad k\in\bR.\tag 26.88$$
Note that
associated with $S(k)$ in (26.88), using the procedure
outlined in Chapter~16, we obtain
$$F_s(y)=\cases \bm \left(-8+24\,y-16\,y^2+\ds\frac{8}{3}\,y^3\right) e^{-y}& 0\\
\stretch 0&0\endbm,\qquad y\in\bR^+,\\
\bm 0&0\\
\stretch 0&0\endbm,\qquad y\in\bR^-.
\endcases\tag 26.89
$$
Using (26.89) in (4.17), as a solution to
(4.17) we obtain $X(y)=\bm 0&0\endbm$ for $y\in\bR^+.$
Thus, $(\bold{III}_c)$ of Theorem~7.1 is satisfied.
Using (26.89) in (4.22) we get the two parameter family
of solutions given by
$$X(y)=\bm \alpha \left(1-3y+\ds\frac{1}{3}\,y^3\right)\,e^{-y}+\beta \,\left(y^2-\ds\frac{1}{3}\,y^3\right) \,e^{-y}&\ \  0\endbm,
\qquad y\in\bR^+,
\tag 26.90$$
where $\alpha$ and $\beta$ are arbitrary parameters.
The fact that $X(y)$ in (26.90) contains two arbitrary
parameters suggests that we must have $\Cal N=2.$
Thus, the property $(\bold V_c)$ of
Theorem~7.1 is satisfied if and only if
we have $\Cal N=2.$
We can also get the same conclusion from Levinson's
theorem stated in (21.5). We see this as follows.
We have $S_\infty=I$ and $S(0)=I,$ and hence
$\mu=2$ and $n_D=0$ in (21.5). Note that
$I$ here denotes the $2\times 2$ unit matrix.
The left-hand side of (21.5)
is equal to $4\pi.$ Thus, Levinson's theorem predicts that
we must have $\Cal N=2.$ In other words, if $S(k)$ given in
(26.88) is a part of a scattering data $\bold S$
as in (4.2), then (4.2) must be compatible with
(4.3). Note that we can achieve having $\Cal N=2$
in essentially two ways. The first way is to have one bound state
at some $k=i\kappa_1$ of multiplicity two, in which case
the normalization matrix
$M_1$ must have rank two. The second way is to have two simple
bound states, say at $k=i\kappa_1$ and $k=i\kappa_2$
with $\kappa_1\ne \kappa_2,$ in which case
the corresponding normalization matrices $M_1$ and $M_2$ each must
have rank one. In this specific example, we will show that
the first way does not necessarily always lead to a scattering data set in the
Marchenko class. We see this as follows.
Let us add
one bound state at $k=i$ with
$M_1=\sqrt{8}\,I.$ Since $M_1$ has rank two, this ensures that
the property $(\bold L)$ of Theorem~7.9 is satisfies. Using (26.89) in (4.7)
we then have
$$F(y)= \bm \left(24\,y-16\,y^2+\ds\frac{8}{3}\,y^3\right) e^{-y}& 0\\
\stretch 0&8\,e^{-y}\endbm,\qquad y\in\bR^+.\tag 26.91$$
Using (26.91) as input to (4.14) we obtain the general solution
$X(y)$ to (4.14) as
$$X(y)=\bm \gamma \left(1-3y+y^2\right)\,e^{-y}&\ \  0\endbm,
\qquad y\in\bR^+,
\tag 26.92$$
where $\gamma$ is an arbitrary parameter.
Thus, using the
scattering data set
$\bold S$ consisting of
$S(k)$ in (26.88), $N=1,$ $\kappa_1=1,$ $M_1=\sqrt{8}\,I,$
we see that the properties $(\bold 1),$ $(\bold 2),$
$(\bold {III}_c),$ $(\bold V_c)$ of Theorem~7.1 are satisfied
but $(\bold 4_c)$ is not satisfied. We also conclude that
in Theorem~7.9 the conditions $(\bold 1),$ $(\bold 2),$ $(\bold L)$
are satisfied but not $(\bold 4_{e,2}).$ Similarly,
in Theorem~7.10 the conditions $(\bold 1),$ $(\bold 2),$ $(\bold L)$
are satisfied but not $(\bold 4_{c,2})$ or
$(\bold 4_{d,2}).$

Next, we present an example that does not satisfy all the characterization
conditions because the number of bound states predicted
by the solution to (4.22) is inconsistent with the number of bound states
in the scattering data.

\noindent {\bf Example 26.14} Let $S(k)$ be the scattering matrix
$$S(k)=\ds\frac{1}{(k-i)\left(k-\ds\frac{i}{3}\right)}\bm k(k+i)&\ds\frac{i}{3}(k+i)\\
\stretch
\ds\frac{i}{3}(k+i)&k(k+i)\endbm,\tag 26.93$$
and assume that there are no bound states.
 From (26.93) we get
$$S_\infty=\bm 1&0\\
\stretch 0&1\endbm,\quad S(0)=
\bm 0&1\\
\stretch
1&0\endbm,\quad G_1=\bm -\ds\frac{7}{3} & -\ds\frac{1}{3}\\
\stretch  -\ds\frac{1}{3}& -\ds\frac{7}{3}\endbm,\tag 26.94$$
The determinant of $S(k)$ given in (26.93) is given by
$$\det[S(k)]=\ds\frac{(k+i)^2(k+i/3)}{(k-i)^2(k-i/3)},\qquad k\in\bR.\tag 26.95$$
Using the first two equalities of (26.94) and
also using (26.95) in (21.5),
we see that the left-hand side of (21.5) is equal to
$3\pi,$ $\mu=1$ because $S(0)$ has eigenvalues $1$ with
multiplicity one, $n_D=0$ because
$S_\infty=I$ with $I$ denoting the
$2\times 2$ identity matrix. Thus,
Levinson's theorem predicts $\Cal N=2,$
which is not compatible with the assumption of
no bound states.
Using the construction process
described in Chapter~16, from (26.93) we obtain
$$F_s(y)=\cases \bm  -3e^{-y}+\ds\frac{2}{3}\, e^{-y/3}  & -e^{-y}+\ds\frac{2}{3}\, e^{-y/3}   \\
- e^{-y}+\ds\frac{2}{3}\, e^{-y/3} & -3e^{-y}+\ds\frac{2}{3}\, e^{-y/3} \endbm,\qquad y\in\bR^+,\\
  \stretch
\bm  0  & 0   \\
0  &0  \endbm,\qquad y\in\bR^-.\endcases\tag 26.96$$
$$F(y)=\bm  -3e^{-y}+\ds\frac{2}{3}\, e^{-y/3}  & -e^{-y}+\ds\frac{2}{3}\, e^{-y/3}   \\
- e^{-y}+\ds\frac{2}{3}\, e^{-y/3} & -3e^{-y}+\ds\frac{2}{3}\, e^{-y/3} \endbm,\qquad y\in\bR^+,\tag 26.97$$
$$F'_s(y)-G_1\,\delta(y)=\cases \bm  3e^{-y}-\ds\frac{2}{9}\, e^{-y/3}  & e^{-y}-\ds\frac{2}{9}\, e^{-y/3}   \\
 e^{-y}-\ds\frac{2}{9}\, e^{-y/3} & 3e^{-y}-\ds\frac{2}{9}\, e^{-y/3}  \endbm,\qquad y\in\bR^+,\\
  \stretch
\bm  0  & 0   \\
0  &0  \endbm,\qquad y\in\bR^-.\endcases\tag 26.98$$
Using (26.96) in (4.22) we obtain a two-parameter
family of solutions to (4.22) given by
$$X(y)=\bm -(\alpha_1+6\,\alpha_2)\,e^{-y}+\alpha_2\, e^{-y/3}& \alpha_1\, e^{-y}+\alpha_2\, e^{-y/3}\endbm,
\tag 26.99$$
where $\alpha_1$ and $\alpha_2$ are arbitrary constants.
Hence the number of bound states
including the multiplicities must be two. In the absence
of bound states, (4.22) and (4.14) coincide and hence
(4.22) has the general solution given in (26.99), which is not the trivial
solution.
Thus, if we assume that
there are no bound states, even though $(\bold 1)$, $(\bold 2)$, and $(\bold{III}_a)$ in Theorem~7.1
are satisfied, neither $(\bold 4_c)$ nor $(\bold V_c)$ are satisfied.
Using the method of Chapter~16, we construct
$$K(x,y)=\bm \beta_1(x)&\beta_2(x)\\
\stretch
\beta_2(x)&\beta_1(x)\endbm e^{-y}+\beta_7(x)
\bm 1&1\\
\stretch
1&1\endbm e^{-y/3},\qquad y>x,\tag 26.100$$
$$K(x,x)=\bm \beta_3(x)&\beta_4(x)\\
\stretch
\beta_4(x)&\beta_3(x)\endbm,\qquad V(x)=\bm \beta_5(x)&\beta_6(x)\\
\stretch \beta_6(x)&\beta_5(x)\endbm,\tag 26.101
$$
$$f(k,x)=e^{ikx}\left(\bm 1&0\\
\stretch 0&1\endbm+\ds\frac{i\,e^{-x}}{k+i}\,
\bm \beta_1(x)&\beta_2(x)\\
\stretch
\beta_2(x)&\beta_1(x)\endbm+\ds\frac{i\,e^{-x/3}\,\beta_7(x)}{k+i/3}\,\bm 1&1\\
\stretch
1&1\endbm
\right),\tag 26.102$$
where we have defined
$$\beta_1(x):=\ds\frac{3\,e^x+4 \,e^{5x/3}+3\,e^{7x/3}}{(e^{2x}-1)(e^{2x/3}+1)^2},
\quad \beta_2(x):=\ds\frac{7+12\,e^{2x/3}+7\, e^{4x/3}-2\,e^{8x/3}}{(e^{2x}-1)(e^{2x/3}+1)^2},\tag 26.103$$
$$\beta_3(x):=\ds\frac{1+e^{4x/3}-2\,e^{8x/3}}{3(e^{2x}-1)(e^{2x/3}+1)^2},
\quad \beta_4(x):=\ds\frac{e^x+e^{7x/3}}{3(e^{2x}-1)(e^{2x/3}+1)^2},\tag 26.104$$
$$\beta_5(x):=\ds\frac{-8\,e^{2x/3}+16\,e^{4 x/3}+68\,e^{2x}+136\,e^{8x/3}+68\,e^{10 x/3}+16\,e^{4x}-8\,e^{14 x/3}}{9(e^{2x}-1)^2(e^{2x/3}+1)^2},\tag 26.105$$
$$\beta_6(x):=\ds\frac{-8\,e^{2x/3}+4\,e^{2x}-8\,e^{10 x/3}}{9(1+2\,e^{2x/3}+2\,e^{4x/3}+e^{2x})^2},\qquad\beta_7(x):=\ds\frac{-2\,e^{x/3}+2\,e^x-2 \,e^{5x/3}}{3(e^{2x}-1)(e^{2x/3}+1)^2}.\tag 26.106
$$
We have
$$K(x,x)=\ds\frac{1}{x}\,\bm 1&0\\
\stretch 0&1\endbm-\ds\frac{1}{6}\bm 7&1\\
\stretch 1&7\endbm +O(x),\qquad x\to 0,\tag 26.107$$
$$V(x)=\ds\frac{2}{x^2}\,\bm 1&0\\
\stretch 0&1\endbm-\ds\frac{1}{27}\bm 19&1\\
\stretch 1&7\endbm +O(x^2),\qquad x\to 0,\tag 26.108$$
$$\Psi(k,x)=-\ds\frac{x^2}{9} \bm 6k^2+7ik-1& i(k+i)\\
\stretch i(k+i)&6k^2+7ik-1\endbm+O(x^3),\qquad x\to 0,\tag 26.109$$
$$\Psi'(k,x)=-\ds\frac{2x}{9} \bm 6k^2+7ik-1& i(k+i)\\
\stretch i(k+i)&6k^2+7ik-1\endbm+O(x^3),\qquad x\to 0.\tag 26.110$$
Even though $\Psi(k,0)=0$ and $\Psi'(k,0)=0,$ there is no selfadjoint
boundary condition of the form (2.4). This is because $K(0,0)$ is not finite and
hence the boundary matrices $A$ and $B$ satisfying (16.70) with
the further property (2.6) do not exist.
As a result, $(\bold 1)$ and $(\bold 2)$ in Theorem~5.1 are satisfied, but
$(\bold 3_a)$ and $(\bold 4_a)$ there are violated.
In Theorem~7.9, the condition $(\bold L)$ is not satisfied
but the remaining conditions $(\bold 1),$ $(\bold 2),$ $(\bold 4_{e,2})$
are satisfied.

Next, we use the scattering matrix in Example~26.14 with one bound state of
multiplicity two in order to have an example where all the five characterization conditions in Theorem~7.1
are satisfied.

\noindent {\bf Example 26.15} Let $S(k)$ be the scattering matrix given in
(26.93) and assume that we have one bound state of multiplicity two
at $k=i\kappa_1$ with $\kappa_1=1$ and the rank-two
normalization matrix $M_1$ given by
$$M_1=\bm 1+\ds\frac{1}{\sqrt{2}}&1-\ds\frac{1}{\sqrt{2}}\\
\stretch
1-\ds\frac{1}{\sqrt{2}} &1+\ds\frac{1}{\sqrt{2}}\endbm,\quad
M_1^2=\bm 3
& 1\\
\stretch
1& 3\endbm .\tag 26.111$$
Thus, using (26.70) and (26.94), from (4.12) we obtain
$$F(y)=\ds\frac{2}{3}\, e^{-y/3} \bm 1&1\\
\stretch 1&1\endbm,\qquad y\in\bR^+.\tag 26.112$$
Then, all the characterization conditions are met and we obtain
$$K(x,y)=-\ds\frac{2}{3}\,\ds\frac{e^{-(x+y)/3}}{1+2e^{-2x/3}}\,\bm 1&1\\
\stretch 1&1\endbm,\quad V(x)=-\ds\frac{8e^{2x/3}}{9(2+e^{2x/3})^2}\,\bm 1&1\\
\stretch 1&1\endbm,\tag 26.113$$
$$f(k,x)=e^{i kx}\left(\bm 1&0\\
\stretch 0&1\endbm-\ds\frac{2i}{(3k+i)(2+e^{2x/3})}\,\bm 1&1\\
\stretch 1&1\endbm\right),\tag 26.114$$
$$K(0,0)=-\ds\frac{2}{9}\,\bm 1&1\\
\stretch 1&1\endbm,\quad
B=\ds\frac{1}{18} \,\bm -17&1\\
\stretch1&-17\endbm \,A,\tag 26.115$$
$$\Psi(k,0)=\ds\frac{1}{3(k-i)}\bm 6k+i&
i\\
\stretch i&6k+i\endbm,\quad
\Psi'(k,0)=\ds\frac{1}{27(k-i)}\bm -51k-8i&
3k-8i\\
\stretch 3k-8i&-51k-8i\endbm,\tag 26.116$$
$$\Psi_1(0)=\ds\frac{2}{3}\,\bm 1&1\\
\stretch 1&1\endbm+\ds\frac{1}{\sqrt{2}}\,\bm 1&-1\\
\stretch -1&1\endbm,\quad
\Psi'_1(0)=-\ds\frac{16}{27}\,\bm 1&1\\
\stretch 1&1\endbm+\ds\frac{1}{\sqrt{2}}\,\bm 1&-1\\
\stretch -1&1\endbm,\tag 26.117$$
with the boundary matrix $A$ being any invertible $2\times 2$ matrix.
One can directly verify that all the five conditions in
Theorem~7.1 and all the four conditions in Theorem~5.1 are all
satisfied.

Next, we use the scattering matrix in Example~26.14 with two bound states, each with
multiplicity one in order to have an example where all five the characterization conditions
in Theorem~7.1 are satisfied.

\noindent {\bf Example 26.16} Let $S(k)$ be the scattering matrix given in
(26.93) and assume that we have two bound states
at $k=i\kappa_1$ and $k=i\kappa_2$ with $\kappa_1=1$ and $\kappa_2=1/3$
and with respective rank-one normalization matrices
$M_1$ and $M_2$ given by
$$M_1=\ds\frac{1}{\sqrt{2}}\,\bm 1&1\\
\stretch 1&1\endbm,\quad M_2=\ds\frac{1}{\sqrt{3}}\,\bm 1&-1\\
\stretch -1&1\endbm.\tag 26.118$$
With the help of (4.12), (26.96), and (26.118) we get
$$F(y)=\left(-2 e^{-y}+\ds\frac{4}{3}\,e^{-y/3}\right) I,\qquad y\in\bR^+,\tag 26.119$$
where $I$ is the $2\times 2$ identity matrix.
The construction procedure described in Chapter~16 yields
$$K(x,y)=\left(\alpha(x)\,e^{-(x+y)}+\beta(x)\,e^{-(x+y)/3}
\right) I,\quad K(0,0)=\ds\frac{4}{3}\,I,\tag 26.120$$
$$f(k,x)=e^{ikx}\left[1+\ds\frac{i\,\alpha(x)\,e^{-x}}{k+i}+\ds\frac{i\,\beta(x)\,e^{-x/3}}{k+i/3}
\right] I,\quad
B=-\ds\frac{1}{6}\,\bm 15&1\\
\stretch1&15\endbm A,\tag 26.121$$
$$V(x)=\left(\ds\frac{16 e^{2x/3}\left[18 e^{4x/3}+32 e^{2x}+18 e^{8x/3}-4 e^{-4x}-1\right]}{9\left[ 4e^{2x}+2 e^{8x/3}-2 e^{2x/3}-1\right]^2}
\right) I,\tag 26.122$$
where $A$ is any invertible $2\times 2$ matrix, and the quantities
$\alpha(x)$ and $\beta(x)$ are given by
$$\alpha(x):=\ds\frac{2e^{-x}+2 e^{-5x/3}}{1+2 e^{-2x/3}-e^{-2x}-\ds\frac{1}{2}
\,e^{-8x/3}},\quad \beta(x):=\ds\frac{-\ds\frac{4}{3}\,e^{-x/3}-\ds\frac{2}{3}\, e^{-7x/3}}{1+2 e^{-2x/3}-e^{-2x}-\ds\frac{1}{2}
\,e^{-8x/3}}.\tag 26.123$$
One can directly verify that all the five conditions in
Theorem~7.1, all the four conditions in Theorem~5.1,
and all the four conditions in Theorem~7.9 are all
satisfied.

In the next example we present a scattering matrix, which does not satisfy
$(\bold 1)$ in Theorem~7.1 because it is not unitary.

\noindent {\bf Example 26.17} Let $S(k)$ be the scattering matrix given
$$S(k)=\ds\frac{1}{(k+i)\left(k+\ds\frac{i}{3}\right)}\bm k(k-i)&\ds\frac{i}{3}(k-i)\\
\stretch
\ds\frac{i}{3}(k-i)&-k(k-i)\endbm,\tag 26.124$$
and assume that there are no bound states.
The matrix in (26.124) is not unitary
although it satisfies $S(-k)=S(k)^\dagger$ for $k\in\bR.$
Going through the construction procedure of Chapter~16, we obtain
$$F_s(y)=\cases \bm  0  & 0   \\
0  &0  \endbm
,\qquad y\in\bR^+,\\
  \stretch
  \bm  -3e^y+\ds\frac{2}{3}\, e^{y/3}  & e^y-\ds\frac{2}{3}\, e^{y/3}   \\
e^y-\ds\frac{2}{3}\, e^{y/3} & 3e^y-\ds\frac{2}{3}\, e^{y/3} \endbm
,\qquad y\in\bR^-.\endcases\tag 26.125$$
$$F(y)=\bm  0  & 0   \\
0  &0  \endbm
,\qquad y\in\bR^+,\tag 26.126$$
$$S_\infty=\bm 1&0\\
\stretch 0&-1\endbm,\quad G_1=\bm \ds\frac{7}{3} & -\ds\frac{1}{3}\\
\stretch  -\ds\frac{1}{3}& -\ds\frac{7}{3}\endbm,\quad K(x,y)\equiv 0,\quad K(0,0)=0,\tag 26.127$$
$$V(x)\equiv 0,\quad A=0,\quad B=0.\tag 26.128$$
In this example, it is impossible to construct the two boundary matrices
$A$ and $B$
with the rank of $\bm A\\
B\endbm$ being two.
Using the second line of (26.125) in
(4.17), one determines that (4.17) has
only the trivial solution and hence
$(\bold {III}_a)$ is satisfied.
In summary, in Theorem~5.1, the property $(\bold 1)$ is
not satisfied
because $S(k)^\dagger\ne S(k)^{-1}$ for $k\in\bR,$ $(\bold 2)$ is satisfied, $(\bold 3_a)$ is violated because there does not exist a corresponding selfadjoint boundary condition.
From the first line of (29.96) we conclude that
(4.22) has only the trivial solution and hence $(\bold V_c)$ is satisfied.
 From (26.126) we conclude that
 (4.14) has only the trivial solution and
hence $(\bold 4_c)$ is satisfied.
In Theorem~7.1, only $(\bold 1)$ is violated because $S(k)$ is not unitary,
while the remaining conditions $(\bold 2)$, $(\bold{III}_a)$, $(\bold 4_c)$, and $(\bold V_c)$ are all satisfied,

In the next example we present a unitary scattering matrix, which does not satisfy
$(\bold 1)$ in Theorem~7.1 because the symmetry
property $S(-k)=S(k)^\dagger$ for $k\in\bR$ does not hold.

\noindent {\bf Example 26.18} Let $S(k)$ be the scattering matrix given
$$S(k)=\ds\frac{1}{(k+i)\left(k+\ds\frac{i}{3}\right)}\bm k(k-i)&\ds\frac{i}{3}(k-i)\\
\stretch
-\ds\frac{i}{3}(k-i)&-k(k-i)\endbm,\tag 26.129$$
and assume that there are no bound states.
Note that $S(k)$ of (26.129) differs from $S(k)$ of (26.124)
only in the sign of the $(2,1)$-entry.
The matrix in (26.129) is unitary
but it does not satisfy $S(-k)=S(k)^\dagger$ for $k\in\bR.$

 From (26.129) we get
$$S_\infty=\bm 1&0\\
\stretch 0&-1\endbm,\quad S(0)=
\bm 0&-1\\
\stretch
1&0\endbm.\tag 26.130$$
The determinant of $S(k)$ given in (26.129) is given by
$$\det[S(k)]=-\ds\frac{(k-i)^2(k-i/3)}{(k+i)^2(k+i/3)},\qquad k\in\bR.\tag 26.131$$
Using (26.130) and
(26.131) in (21.5),
we see that the left-hand side of (21.5) is equal to
$-3\pi,$ $\mu=0$ because $S(0)$ has eigenvalues $i$ and $-i,$ $n_D=1.$ Thus,
Levinson's theorem predicts $\Cal N=-1,$
which contradicts the expectation that $\Cal N$ is a nonnegative integer.
Going through the construction procedure of Chapter~16, we obtain
$$F_s(y)=\cases \bm  0  & 0   \\
0  &0  \endbm
,\qquad y\in\bR^+,\\
  \stretch
  \bm  -3e^y+\ds\frac{2}{3}\, e^{y/3}  & e^y-\ds\frac{2}{3}\, e^{y/3}   \\
-e^y+\ds\frac{2}{3}\, e^{y/3} & 3e^y-\ds\frac{2}{3}\, e^{y/3} \endbm
,\qquad y\in\bR^-.\endcases\tag 26.132$$
$$F(y)=\bm  0  & 0   \\
0  &0  \endbm
,\qquad y\in\bR^+,\tag 26.133$$
$$G_1=\bm \ds\frac{7}{3} & -\ds\frac{1}{3}\\
\stretch  \ds\frac{1}{3}& -\ds\frac{7}{3}\endbm,\quad K(x,y)\equiv 0,\quad K(0,0)=0,\tag 26.134$$
$$V(x)\equiv 0,\quad A=0,\quad B=0.\tag 26.135$$
In this example, $F_s(y)$ is not hermitian for $y\in\bR^-$
and the matrix $G_1$ is not hermitian.
One cannot construct the two boundary matrices
$A$ and $B$
with the rank of $\bm A\\
B\endbm$ being two. Using the second line of (26.132) in
(4.17), one determines that (4.17) has
only the trivial solution and hence
$(\bold {III}_a)$ is satisfied.
In summary, in Theorem~5.1, $(\bold 1)$ is violated
because $S(-k)\ne S(k)^\dagger$ for $k\in\bR,$ $(\bold 2)$ is satisfied, $(\bold 3_a)$ is violated because there does not exist a corresponding selfadjoint boundary condition. From the first line of (29.100) we conclude that
(4.22) has only the trivial solution and hence $(\bold V_c)$ is satisfied.
 From (26.133) we conclude that
 (4.14) has only the trivial solution and
hence $(\bold 4_c)$ is satisfied.
In Theorem~7.1, only $(\bold 1)$ is violated because $S(-k)\ne S(k)^\dagger$ for $k\in\bR,$
while the remaining conditions $(\bold 2)$, $(\bold{III}_a)$, $(\bold 4_c)$, and $(\bold V_c)$ are all satisfied. In Theorem~7.9, $(\bold 1)$ and
$(\bold L)$ are not satisfied
while the remaining two properties
$(\bold 2)$ and $(\bold 4_{e,2})$ are satisfied.

In the next example, we present a scattering data set that satisfies all the five
conditions in Theorem~7.1, except for the unitarity of the scattering matrix.

\noindent {\bf Example 26.19} Let $S(k)$ be the scattering matrix given
$$S(k)=\bm \ds\frac{k-i}{k+i}& \ds\frac{1}{k^2+1}\\
\stretch
\ds\frac{1}{k^2+1}&\ds\frac{k-2i}{k+2i}\endbm,\tag 26.136$$
and assume that there are no bound states.
The scattering matrix given in (26.136) satisfies the first equality
in (4.4), but it is not unitary and hence it does not satisfy the second equality
in (4.4).
Going through the construction procedure of Chapter~16, we obtain
$$F_s(y)=\cases \bm  0 & \ds\frac{1}{2}\,e^{-y}   \\
\ds\frac{1}{2}\,e^{-y}  & 0  \endbm
,\qquad y\in\bR^+,\\
  \stretch
  \bm  -2 e^y  & \ds\frac{1}{2}\, e^y   \\
\ds\frac{1}{2}\, e^y & -4 e^{2y} \endbm
,\qquad y\in\bR^-.\endcases\tag 26.137$$
$$F(y)=\bm  0 & \ds\frac{1}{2}\,e^{-y}   \\
\ds\frac{1}{2}\,e^{-y}  & 0  \endbm
,\qquad y\in\bR^+,\tag 26.138$$
$$S_\infty=\bm 1&0\\
\stretch 0&1\endbm,\quad G_1=\bm 2 & 0\\
\stretch  0& 4\endbm,\quad K(0,0)=\bm \ds\frac{2}{15}&
-\ds\frac{8}{15}\\
\stretch
-\ds\frac{8}{15} & \ds\frac{2}{15}\endbm
,\tag 26.139$$
$$K(x,y)=\ds\frac{e^{x-y}}
{16 e^{4x}-1}\,\bm
2& -8e^{2x}\\
\stretch
-8e^{2x}&2\endbm,\quad B=\bm \ds\frac{13}{15}&
\ds\frac{8}{15}\\
\stretch
\ds\frac{8}{15} & \ds\frac{28}{15}\endbm A,\tag 26.140$$
$$V(x)=\ds\frac{1}{\left(16 e^{4x}-1\right)^2}\bm 256\, e^{4x}&-32\, e^{2x}-512\, e^{6x}\\
\stretch
-32\, e^{2x}-512 e^{6x}&256\, e^{4x}\endbm,\tag 26.141$$
$$f(k,x)=e^{ikx} \left(\bm 1&0\\
\stretch 0&1\endbm+\ds\frac{i}{(k+i)(16 e^{4x}-1)}\,
\bm
2& -8e^{2x}\\
\stretch
-8e^{2x}&2\endbm
\right),\tag 26.142$$
where $A$ is any invertible $2\times 2$ constant matrix.
One can directly verify that all the five conditions in
Theorem~7.1 are satisfied except for the unitarity
of $S(k)$ in $(\bold 1)$. This is because from the first line of (26.137) it is seen that
$(\bold 2)$ is satisfied. Using (26.138) in (4.14) one determines
that (4.14) has only the trivial solution and hence
$(\bold 4_c)$ is satisfied. Using the second line of
(26.137) in (4.17) one determines that
(4.17) has only the trivial solution and hence
$(\bold {III}_c)$ is satisfied.
Using the first line of (26.137) in
(4.22) we determine that
(4.22) has only the trivial solution and hence
$(\bold V_c)$ is satisfied.
Similarly, one can determine that
$(\bold 2)$ and $(\bold 4_a)$ in Definition~4.5 are satisfied but
neither $(\bold 1)$ nor $(\bold 3_a)$ are satisfied. As a result, the
scattering data in this example does not belong to
the Marchenko class. One can evaluate the
scattering data set corresponding to the potential $V(x)$ is
(26.141) and the boundary matrices $A$ and $B$ specified in the
second equality in (26.140). Using (26.142) in (9.2), with the help of
(9.3) one can construct the corresponding Jost matrix and the scattering matrix and finds
that
$$J(k)=\ds\frac{1}{225(k+i)}\,\bm
-i(225 k^2-510 ik+931)&16(15k+34 i)\\
\stretch
16(15k+34 i)&
-i(225 k^2-510 ik+931)\endbm A,\tag 26.143$$
$$S(k)=\ds\frac{\bm (k+i)(k-i)^2(225 k^2+2537)& 480 ik(k-i)(k^2-1)\\
\stretch
480 ik(k-i)(k^2-1)&(k+i)(k-i)^2(225 k^2+2537)\endbm}{(k+i)(9k^2-30ik+59)(25k^2-30ik+43)}
.\tag 26.144$$
One can verify that $S(k)$ given in (26.144) unitary.
The determinant of $J(k)$ given in (26.143)
is given by
$$\det[J(k)]=-\ds\frac{(9k^2-30ik+59)(25k^2-30ik+43)}{225(k+i)^2},\tag 26.145$$
and hence it vanishes in $\bCp$
at the two points $k=i\kappa_1$ and $k=i\kappa_2,$ where
$$\kappa_1=\ds\frac{3+2\sqrt{13}}{5},\quad
\kappa_2=\ds\frac{5+2\sqrt{21}}{3}.\tag 26.146$$
Using (11.1)-(11.3) we obtain
$$A_1=\bm \ds\frac{5(583-145\sqrt{13})}{1161}& -\ds\frac{1}{3(4+\sqrt{13})^2}\\
\stretch -\ds\frac{1}{3(4+\sqrt{13})^2}&\ds\frac{5(583-145\sqrt{13})}{1161}\endbm,\tag 26.147$$
$$
B_1=
\bm \ds\frac{905\sqrt{13}-2798}{774}& -\ds\frac{3572-905\sqrt{13}}{774}\\
\stretch -\ds\frac{3572-905\sqrt{13}}{774}&\ds\frac{905\sqrt{13}-2798}{774}\endbm,\tag 26.148$$
$$A_2=\bm \ds\frac{3(779-111\sqrt{21})}{7375}& -\ds\frac{6}{5(4+\sqrt{21})^2}\\
\stretch -\ds\frac{6}{5(4+\sqrt{21})^2}&\ds\frac{3(779-111\sqrt{21})}{7375}\endbm,\tag 26.149$$
$$B_2=\bm \ds\frac{4562-633\sqrt{21}}{2950}& \ds\frac{633\sqrt{21}-1612}{2950}\\
\stretch \ds\frac{633\sqrt{21}-1612}{2950}&\ds\frac{4562-633\sqrt{21}}{2950}\endbm,\tag 26.150$$
and via (11.22) we explicitly evaluate
the normalization matrices $M_1$ and $M_2$ as
$$M_1=\ds\frac{1}{4\sqrt{5}}\,\sqrt{ 49+\ds\frac{181}{\sqrt{13}}}\,\bm 1&1\\
\stretch 1&1
\endbm,\quad M_2=\ds\frac{1}{12}\,\sqrt{147+211\sqrt{\ds\frac{3}{7}}}\,\bm 1&-1\\
\stretch -1&1\endbm.\tag 26.151$$
The scattering data set $\bold S$
consisting of $S(k)$ in (26.144) with the
corresponding two bound states specified in (26.146) and (26.151) now belongs to the Marchenko class, and it corresponds to the input data set $\bold D$ consisting of
$\{V,A,B\}$ appearing in (26.141) and the second equality in (26.140).

In the following example, we check the characterization conditions
of Theorem~8.1

\noindent {\bf Example 26.20}
 Assume that $n=1$ and that the scattering matrix $S(k)$ is given by
$$
S(k)=-\ds\frac{(k+i)(k+2i)}{(k-i)(k-2i)},\qquad k\in\bR,
\tag 26.152$$
and that we have one bound state at $k=i\kappa_1$ with $\kappa_1=1$ and
the normalization constant $M_1=\sqrt{6}.$
We will prove that the conditions $(\bold I)$, $(\bold 2)$, $(\bold{III}_e)$,   $(\bold 4_c)$ ,  $(\bold V_h)$, and  $(\bold{VI})$ of Theorem 8.1 are satisfied.
 From (26.152) we get
$$
S_\infty:= \lim_{k \rightarrow \pm \infty} S(k)= -1.
\tag 26.153$$
 Using (26.152) in (4.7), with the help of a contour integration, we get
$$
F_s(y)=\cases  -6 e^{-y}+12 e^{-2y},\qquad y>0,\\
\stretch
0,\qquad y<0.
\endcases
\tag 26.154$$
Then,  from (4.12) we obtain
$$
F(y)=12 e^{-2 y},\qquad y>0.
\tag 26.155$$
  From (26.152), (26.153), and
  (26.154) we conclude that
$(\bold I)$ and $(\bold 2)$ are satisfied.
Note that (4.14) is given by
$$
X(y)+\int_0^\infty dz\,X(z)\, 12 e^{-2z-2y}=0, \qquad y\in\bR^+,
\tag 26.156$$
and hence any solution to (26.156) must be of the form
$$
X(y)=\alpha\,e^{-2y},\qquad y\in\bR^+,
\tag 26.157$$
where $\alpha$ is a constant to be determined.
We see that the solution given in  (26.157)  belongs to $L^1(\bR^+)$. Using  (26.157)
 in (26.156), we obtain
$$
 4 \alpha\,e^{-2y}=0,\qquad y\in\bR^+,
\tag 26.158$$
and hence we must have $\alpha=0,$ yielding $X(y)=0$ for $y\in\bR^+.$
Thus, $(\bold 4_c)$  is also satisfied.
We now prove that conditions $(\bold{III}_e)$ and $(\bold V_h)$ hold.
 We take as $\overset{\circ}\to\Upsilon$ the set given by
$$
 \overset{\circ}\to \Upsilon:= \left\{ X(k): X(k)=  \frac{k-2i}{k+i}\, (U(k)-U(-k)), U(k) \in \bold H^2(\bold C^+) \right \}.
\tag 26.159$$
Note that
$$
X(-k)=S(k)\, X(k),\qquad k \in \bold R,\quad X(k) \in \overset{\circ}\to \Upsilon,
\tag 26.160$$
and then we have $ \overset{\circ}\to \Upsilon \subset \Upsilon$.
Let us prove that $ \overset{\circ} \to\Upsilon$ is a dense set in $\Upsilon$. Suppose that some $ W(k) \in \Upsilon$ is orthogonal to  $ \overset{\circ}\to \Upsilon$, i.e.
for any $X(k)\in \overset{\circ} \to \Upsilon$ we have
$$
\left(  W(k), X(k) \right)_2=0,
\qquad X(k) \in  \overset{\circ} \to \Upsilon.
\tag 26.161$$
However, by (26.159), for any $U(k)$ in the
Hardy space $\bold H^2(\bCp),$ we have the scalar product
in $L^2(\bR)$ given by
$$
\left(  \frac{k+2 i}{k-i} W(k)-  \frac{k-2 i}{k+i} W(-k),   U(k) \right)_2=0, \qquad U(k) \in \bold H^2(\bold C^+).
\tag 26.162$$
Since $\bold H^2(\bold C^+)$ is dense in $L^2(\bold R),$ from (26.162) we obtain
$$
\frac{k+2 i}{k-i} W(k)-  \frac{k-2 i}{k+i} W(-k)=0.
\tag 26.163$$
On the other hand, since $W(-k)= S(k) W(k)$
it follows that (26.163) implies that $W(k)= - W(k)$ and then, $W(k)=0$ Since the only vector in $\Upsilon$ that is orthogonal to $\overset{\circ}\to \Upsilon$ is the zero vector,  $\overset{\circ}\to \Upsilon$ is dense in $\Upsilon$.
For each $ X(k) = \ds\frac{k-2i}{k+i}\, (U(k)-U(-k)) \in \overset{\circ}\to\Upsilon$  we need to find a column
vector $h(k)\in \bold H^2(\bold C^+)$  that solves the equation,
$$
h(k)+ S(-k)\, h(-k)= \frac{k-2i}{k+i}\, (U(k)-U(-k)), \qquad  k \in \bold R.
\tag 26.164$$
The solution is given by
$$
h(k)= \frac{k-2i}{k+i}\, U(k) \in \bold H^2(\bold C^+).
\tag 26.165$$
 This prove that the property $(\bold{III}_e)$ is satisfied.
Let us prove that the property $(\bold V_h)$ is satisfied. Note that
$$
S(k)= - J(-k)\, J^{-1}(k), \qquad\,   \text{ where }\ J(k)= \frac{k-i}{k+2i}.
\tag 26.166$$
Then, by  (21.13) with $q=0$   and Proposition~21.11,  every solution  to (4.24) that is in  $\bold H^2(\bold C^+)$
is of the form given by
$$
  h(k)=  \alpha\, J(k) \,  \frac{1 }{k^2+1}, \qquad \text{ for some } \alpha \in \bold C.
\tag 26.167$$
Hence, there is a one-parameter family of solutions and, as there is only one bound state with multiplicity one,   the property $(\bold V_h)$ holds. The property $(\bold{VI})$ is also satisfied by the definition of $S(k)$ in (26.152).

In the following example, we elaborate on the necessity
of the integrability of the potential when non-Dirichlet
boundary conditions are used.

\noindent {\bf Example 26.21} From Theorem~1.2.1 of [2], we know that
if the potential $V(x)$ satisfies
$\int_0^\infty dx\,x\,|V(x)|<+\infty,$ then
(2.1) has two linearly independent $n\times n$ matrix-valued solutions
$G(k,x)$ and $H(k,x)$ satisfying
the initial conditions
$$G(k,0)=x [I+o(1)],\quad G'(k,0)=I+o(1),\qquad x\to 0^+,
\tag 26.168$$
$$H(k,0)=I+o(1),\quad H'(k,0)=o\left(\ds\frac{1}{x}\right),\qquad x\to 0^+.
\tag 26.169$$
Thus, any solution to (2.1) can be expressed as a linear combination of
$G(k,x)$ and $H(k,x).$ Let us express the regular solution $\varphi(k,x)$
appearing in (9.5) as a linear combination of
$G(k,x)$ and $H(k,x)$ as
$$\varphi(k,x)=G(k,x)\,\alpha+H(k,x)\,\beta,\tag 26.170$$
where $\alpha$ and $\beta$ are two constant
$n\times n$ matrices to be determined by the boundary condition
at $x=0.$ From the $x$-derivative of (26.170) we obtain
$$\varphi'(k,x)=G'(k,x)\,\alpha+H'(k,x)\,\beta,\tag 26.171$$
If the boundary condition (2.4) is the Dirichlet condition,
which is the case considered in [2],
then we have $A=0$ and $B=I.$ From (26.168)-(26.171) we see that
by choosing $\alpha=I$ and $\beta=0$ in (26.170) and
(26.171), the regular solution $\varphi(k,x)$ satisfies the
Dirichlet boundary condition. Then, as a result of (9.6),
the physical solution $\Psi(k,x)$ also satisfies
the Dirichlet boundary condition. On the other hand, if
the boundary condition (2.4) is non-Dirichlet, then the
regular solution $\varphi(k,x)$ and hence also the
physical solution $\Psi(k,x)$ cannot be obtained from only $G(k,x)$ appearing
in (168.171). In the non-Dirichlet case, the involvement of
$H(k,x)$ with the behavior given in (26.169) makes it impossible
to define a selfadjoint boundary condition at $x=0$
if the potential $V(x)$ only satisfies
$\int_0^\infty dx\,x\,|V(x)|<+\infty$
but not (2.3). Such potentials are nevertheless important
in physical applications. For example, in the scalar case
the truncated Coulomb potential given by
$$V(x)=\cases \ds\frac{1}{x},\qquad 0<x<1,\\
\stretch
0,\qquad x>1,\endcases\tag 26.172$$
although satisfying $\int_0^\infty dx\,x\,|V(x)|<+\infty,$
is a potential not integrable at $x=0$ and hence
we cannot use a selfadjoint non-Dirichlet
boundary condition of the form (2.4)-(2.6) if (26.172) is used
in the Schr\"odinger equation (2.1).
In fact, the Schr\"odinger equation (2.1) when $k=0$
with $V(x)$ given in (26.172) has two linearly
independent solutions, one of which is regular at $x=0$ and
can be expressed in terms of the Bessel function of the first kind $I_1(\sqrt{2}\,x)$
and the other has a singular derivative at $x=0$ and 
can be expressed in terms of the modified
Bessel function of the second kind $K_1(\sqrt{2}\,x).$ So, the general solution
to (2.1) at $k=0$ has the behavior as $x\to 0^+$ given by
$$\aligned
\psi(0,x)=\alpha&\left[-x-\ds\frac{x^2}{2}+O(x^{5/2})\right]\\
&+
\beta\left[1+\left(-1+2\gamma+\log x\right)x+
\ds\frac{x^2}{2}\,\log x+O(x^2)\right],
\qquad x\to 0^+,\endaligned\tag 26.173$$
where $\alpha$ and $\beta$ are arbitrary constants
and $\gamma$ is the Euler-Mascheroni constant appearing in (26.16).
The derivative $\psi'(0,x)$ has the behavior as 
$x\to 0^+$ given by
$$\aligned
\psi'(0,x)=&\alpha\left[-1-x-\ds\frac{x^2}{4}+O(x^{5/2})\right]\\
&+
\beta\left[2\gamma+\log x+\left(-2+2\gamma+\log x\right)x+
+O(x^2 \log x)\right],\qquad x\to 0^+.
\endaligned\tag 26.174$$
When $k\ne 0,$ the Schr\"odinger equation (2.1)
with $V(x)$ given in (26.172) has two linearly
independent solutions, one of which is regular at $x=0$ and
can be expressed in terms of the Kummer confluent hypergeometric function 
${}_1F_1(a,b,z),$ with some appropriate $a,$ $b,$ $z$ 
expressed in terms of $k$ and $x,$
and the other has a singular derivative at $x=0$ and
can be expressed in terms of the Tricomi confluent hypergeometric function
$U(a,b,z)$ So, the general solution
to (2.1) with $k\ne 0$ has the behavior as $x\to 0^+$ given by
$$\aligned
\psi(k,x)=\alpha&\left[x+\ds\frac{x^2}{2}+
\left(\ds\frac{1}{12}-\ds\frac{k^2}{6}\right)x^3+O(x^{4})\right]\\
&+
\beta\left[\ds\frac{1}{\Gamma(1/(2ik))}+O(x)\right],
\qquad x\to 0^+,\endaligned\tag 26.175$$
where $\Gamma(1/(2ik))$ is the gamma function evaluated at $1/(2ik).$
The derivative $\psi'(k,x)$ has the behavior as
$x\to 0^+$ given by
$$\aligned
\psi'(k,x)=&\alpha\left[1+x+\left(\ds\frac{1}{4}-\ds\frac{k^2}{2}\right)
x^2+O(x^3)\right]\\
&+
\beta\left[\ds\frac{\log x}{\Gamma(1/(2ik))}+O(1)\right],\qquad x\to 0^+.
\endaligned\tag 26.176$$
As argued earlier, in the example of the truncated Coulomb potential 
(26.172), only in the Dirichlet case with $\beta=0$
we can have the regular and physical solutions to the Schr\"odinger
equation satisfy the self adjoint boundary 
condition at $x=0.$ In the non-Dirichlet case we must have 
$\beta\ne 0,$ and hence there are neither regular nor
 physical nontrivial solutions to (2.1)
 with $V(x)$ as in (26.172) satisfying a non-Dirichlet
boundary condition of the form (2.4)-(2.6).

\newpage

\vskip 10 pt

\noindent {\bf Acknowledgments.}
The research leading to this
article was supported in part by CONACYT under project CB2015,
254062, and by Project PAPIIT-DGAPA-UNAM IN102215.

\newpage

\noindent {\bf{References}}

\vskip 3 pt

\item{[1]} R. A. Adams and J. F. Fournier, {\it Sobolev spaces,}
2nd, ed., Academic Press, Amsterdam, 2005.

\item{[2]} Z. S. Agranovich and V. A. Marchenko, {\it The inverse problem of
scattering theory,} Gordon and Breach, New York, 1963.

\item{[3]} T. Aktosun, F. Demontis, and C. van der Mee,
{\it Exact solutions to the focusing
nonlinear Schr\"odinger equation,}
Inverse Problems {\bf 23}, 2171--2195 (2007).

\item{[4]} T. Aktosun, M. Klaus, and C. van der Mee,
{\it Small-energy asymptotics of the scattering matrix for the matrix
Schr\"odinger equation on the line,}
J. Math. Phys. {\bf 42}, 4627--4652 (2001).

\item{[5]} T. Aktosun, M. Klaus, and R. Weder,
{\it Small-energy analysis for the self-adjoint matrix
Schr\"odinger operator on the half line,}
J. Math. Phys. {\bf 52}, 102101 (2011).

\item{[6]} T. Aktosun, M. Klaus, and R. Weder,
{\it Small-energy analysis for the self-adjoint matrix
Schr\"odinger operator on the half line. II,}
J. Math. Phys. {\bf 55}, 032103 (2014).

\item{[7]} T. Aktosun, P. Sacks, and M. Unlu,
{\it Inverse problems for selfadjoint Schr\"odinger operators on the half line with compactly supported potentials,}
J. Math. Phys. {\bf 56}, 022106 (2015).

\item{[8]} T. Aktosun and R. Weder,
{\it Inverse spectral-scattering problem with two sets
of discrete spectra for the radial Schr\"odinger equation,}
Inverse Problems {\bf 22}, 89--114 (2006).

\item{[9]} T. Aktosun and R. Weder,
{\it High-energy analysis and Levinson's theorem for the self-adjoint matrix Schr\"odinger operator on the half line,}
J. Math. Phys. {\bf 54}, 112108 (2013).

\item{[10]} T. Aktosun and C. van der Mee,
{\it Explicit solutions to the Korteweg-de Vries equation on the half line,}
Inverse Problems {\bf 22}, 2165--2174 (2006).

\item{[11]} G. Berkolaiko, R. Carlson, S. A. Fulling, and P. Kuchment (eds.),
{\it Quantum graphs and their applications,} Contemporary Mathematics, 415,
Amer. Math. Soc., Providence, RI, 2006.

\item{[12]} G. Berkolaiko and Wen Liu,
{\it Simplicity of eigenvalues and non-vanishing of eigenfunctions of a quantum graph,}
J. Math. Anal. Appl. {\bf 445}, 803--818 (2017).

\item{[13]} J. Boman and P. Kurasov,
{\it Symmetries of quantum graphs and the inverse scattering problem,}
Adv. Appl. Math. {\bf 35}, 58--70 (2005).

\item{[14]} K. Chadan and P. C. Sabatier, {\it Inverse problems in quantum
scattering theory,} 2nd ed., Springer, New York, 1989.

\item{[15]} Y. Colin de Verdi\`ere,
{\it Semi-classical measures on quantum graphs and the Gauss map of the determinant manifold,}
Ann. Henri. Poincar\'e {\bf 16}, 347--364 (2015).

\item{[16]} P. Deift and E. Trubowitz, {\it Inverse scattering
    on the line,} Commun. Pure Appl. Math. {\bf 32}, 121--251
    (1979).

\item{[17]} H. Dym, {\it Linear algebra in action,} Amer. Math. Soc., Providence, R.I., 2006.

\item{[18]} P. Exner, J. P. Keating, P. Kuchment, T. Sunada, and A. Teplyaev (eds.),
{\it Analysis on graphs and its applications,}
Proc. Symposia in Pure Mathematics, 77,
Amer. Math. Soc., Providence, RI, 2008.

\item{[19]} P. Exner and J. Lipovsk\'y,
{\it Pseudo-orbit approach to trajectories of resonances in quantum graphs with general vertex coupling: Fermi rule and high-energy asymptotics,}
J. Math. Phys. {\bf 58}, 042101 (2017).

\item{[20]} L. D. Faddeev, {\it Properties of the $S$-matrix of
    the one-dimensional Schr\"odinger equation,} Amer. Math.
    Soc. Transl. {\bf 65} (ser. 2), 139--166 (1967).


\item{[21]} N. I. Gerasimenko and B. S. Pavlov,
{\it A scattering problem on noncompact graphs,}
Theoret. Math. Phys. {\bf 74}, 230--240 (1988).

\item{[22]} B. Gutkin and U. Smilansky,
{\it Can one hear the shape of a graph?}
J. Phys. A {\bf 34}, 6061--6068 (2001).

\item{[23]} M. S. Harmer, {\it Inverse scattering for the matrix Schr\"odinger
operator and Schr\"odinger operator on
graphs with general self-adjoint boundary conditions,}
ANZIAM J. {\bf 44}, 161--168 (2002).

\item{[24]} M. S. Harmer, {\it The matrix Schr\"odinger operator
and Schr\"odinger operator on graphs,} Ph.D. thesis, University of
Auckland, New Zealand, 2004.

\item{[25]} M. Harmer, {\it Inverse scattering on matrices with boundary conditions,}
J. Phys. A {\bf 38}, 4875--4885 (2005).

\item{[26]} A. Hora and N. Obata, {\it Quantum probability
and spectral analysis of graphs,}
Springer, Berlin, 2007.

%

\item{[27]} T. Kato, {\it Perturbation theory for linear operators,}
2nd ed., Springer-Verlag, New York, 1976.

\item{[28]} J. B. Kennedy, P. Kurasov, G. Malenov\'a, and
D. Mugnolo,
{\it On the spectral gap of a quantum graph,}
Ann. Henri. Poincar\'e {\bf 17}, 2439--2473 (2016).

\item{[29]} V. Kostrykin and R. Schrader,
{\it Kirchhoff's rule for quantum wires,} J. Phys. A {\bf 32}, 595--630
(1999).

\item{[30]} V. Kostrykin and R. Schrader,
{\it Kirchhoff's rule for quantum wires. II: The inverse problem with possible applications to quantum computers,} Fortschr. Phys. {\bf 48}, 703--716
(2000).

\item{[31]} P. Kuchment,
{\it Quantum graphs. I. Some basic structures,}
Waves Random Media {\bf 14}, S107--S128 (2004).

\item{[32]} P. Kuchment,
{\it Quantum graphs. II. Some spectral properties of quantum and combinatorial graphs,}
J. Phys. A {\bf 38}, 4887--4900 (2005).

\item{[33]} P. Kurasov and B. Majidzadeh Garjani,
{\it Quantum graphs: $\Cal{PT}$-symmetry and reflection symmetry of the spectrum,}
J. Math. Phys. {\bf 58}, 023506 (2017).

\item{[34]} P. Kurasov and M. Nowaczyk,
{\it Inverse spectral problem for quantum graphs,}
J. Phys. A {\bf 38}, 4901--4915 (2005).

\item{[35]} P. Kurasov and F. Stenberg,
{\it On the inverse scattering problem on branching graphs,}
J. Phys. A {\bf 35}, 101--121 (2002).

\item{[36]} M. Lee and M. Zworski,
{\it A Fermi golden rule for quantum graphs,}
J. Math. Phys. {\bf 57}, 092101 (2016).

\item{[37]} B. M. Levitan, {\it Inverse Sturm-Liouville problems,}
VNU Science Press, Utrecht, 1987.

\item{[38]}  V.\; A.\; Marchenko, {\it Sturm-Liouville operators and
applications,} Birk\-h\"au\-ser, Basel, 1986.

\item{[39]} N. I. Muskhelishvili, {\it Singular Integral Equations,}
P. Noordhoff, Groningen, 1953.

\item{[40]} R. G. Newton and R. Jost,
{\it The construction of potentials from the $S$-matrix
for systems of differential equations,} Nuovo Cim. {\bf 1}, 590--622
(1955).

\item{[41]} E. M. Stein and G. L Weiss, {\it Introduction to Fourier analysis on Euclidean spaces,}
Princeton Univ. Press, Princeton, New Jersey, 1971.

\item{[42]} R. Weder,
{\it Scattering theory for the matrix Schr\"odinger operator on the half line with general boundary conditions,}
J. Math. Phys. {\bf 56}, 092103 (2015).

\item{[43]} R. Weder,
{\it Trace formulas for the matrix Schr\"odinger operator on the half-line with general boundary conditions,}
J. Math. Phys. {\bf 57}, 112101 (2016).

\item{[44]} R. Weder,
{\it The number of eigenvalues of the matrix Schr\"odinger operator on the half line with general boundary conditions,}
arXiv:1705.03157 [math-ph] (2017).

\item{[45]} J. Weidmann, {\it spectral theory of ordinary
differential operators,} Springer-Verlag, Berlin, 1987.

\end